\documentclass{aa}  

\usepackage{graphicx}
\usepackage{txfonts}
\usepackage[utf8]{inputenc}
\usepackage[english]{babel}
\usepackage[dvipsnames]{xcolor}
\usepackage{lscape}
\usepackage{caption}
\usepackage{longtable}
\usepackage{xcolor}
\usepackage{rotating}
\usepackage{float}
\usepackage{subfig}
\usepackage{multicol} 
\begin{document}

   \title{Gaia EDR3 comparative study of protoplanetary disk fractions in young stellar clusters}

\titlerunning{Gaia EDR3 comparative study of disk fractions}
   \author{I. Mendigutía\inst{1} \and E. Solano\inst{1,2} \and M. Vioque\inst{3,4} \and L. Balaguer-Nuñez\inst{5} \and A. Ribas\inst{6} \and N. Huélamo\inst{1} \and C. Rodrigo\inst{1,2}
          }

   \institute{$^{1}$Centro de Astrobiología (CAB), CSIC-INTA, Camino Bajo del Castillo s/n, 28692, Villanueva de la Cañada, Madrid, Spain.\\$^{2}$Spanish Virtual Observatory\\$^{3}$Joint ALMA Observatory, Alonso de Córdova 3107, Vitacura, Santiago 763-0355, Chile\\$^{4}$National Radio Astronomy Observatory, 520 Edgemont Road, Charlottesville, VA 22903, USA\\$^{5}$Institut de Ci\`encies del Cosmos, Universitat de Barcelona (IEEC-UB), Martí i Franqu\`es 1, E-08028 Barcelona, Spain\\$^{6}$European Southern Observatory (ESO), Alonso de Córdova 3107, Vitacura, Casilla 19001, Santiago de Chile, Chile\\}

   \date{Received January 18, 2022; accepted June 5, 2022}

 
  \abstract
   {The lifetime of protoplanetary disks around young stars limits the timescale when planets form. A disk dissipation timescale $\leq$ 10 Myr was inferred from surveys counting the relative number of stars with disks -the disk fraction- in young stellar clusters with different ages. However, most previous surveys focused on the compact region within $\sim$ 2 pc from the clusters' centers, for which the disk fraction information considering the outer part is practically absent.}
   {We aim to test if disk fraction estimates change when inferred from an extended region around the clusters' centers.}
   {Gaia EDR3 data and a best-suited, Virtual Observatory (VO)-based tool -\emph{Clusterix}-, are used to identify member stars for a representative sample of 19 young stellar clusters considering two concentric fields of view (FOV) with radii $\sim$ 20 pc and $\sim$ 2 pc. Inner-disk fractions associated to each FOV are identically derived from 2MASS color-color diagrams and compared to each other.}
   {Although the density of members is smaller in the periphery, the absolute number of member stars is typically $\sim$ 5 times larger outside $\sim$ 2 pc from the clusters' centers. In turn, our analysis reveals that the inner disk fractions inferred from the compact and the extended regions are equal within $\sim$ $\pm$ 10$\%$. A list of member and disk stars identified in each cluster is provided and stored in a VO-compliant archive, along with their membership probabilities, angular distances to the centre, Gaia and near-infrared data. Averaged values and plots characterizing the whole clusters are also provided, including HR diagrams based on Gaia colors and absolute magnitudes for the sources with known extinction.}
   {Our results cover the largest fields ever probed when dealing with disk fractions for all clusters analysed, and imply that their complete characterization requires the use of wide FOVs. However, the comparative study does not support a previous hypothesis proposing that disk fractions should be significantly larger considering extended regions. The resulting database is a benchmark for future detailed studies of young clusters, whose disk fractions must be accurately determined by using multi-wavelength analysis potentially combined with data from coming Gaia releases.}


   \maketitle
%

\section{Introduction}
\label{Sect:intro}

Planets form from protoplanetary disks surrounding pre-Main Sequence (PMS) stars. Thus, the lifetime of such disks constitutes a stringent upper limit to the timescale when planets assemble. The typical disk dissipation timescale is inferred by surveying young stellar clusters in order to account for the ratio between the number of stars with disks and the total number of members -the disk fraction-. Based on the observed temporal decay of the disk fraction, less than $\sim$ 10$\%$ of stars retain their disks after 2-6 Myr and almost all stars are diskless after $\sim$ 10 Myr \citep[e.g.][]{Haisch01,Hernandez08,Mamajek09,Muzerolle10,Fedele10,Ribas15}. 

However, the previous surveys are limited by the narrow field of view (FOV) of the telescopes used -mainly Spitzer. Being focused on the compact regions within $\sim$ 15$\arcmin$, they sample the central 2-3 pc at the typical distances to the clusters, with little information about the outer parts. Indeed, although the above mentioned numbers characterizing the disk dissipation timescale are widely accepted, they are not free of some controversial. For instance, they make our own planetary system somewhat an exception. Based on metheoritic analysis, Jupiter accreted a significant amount of gas along more than 4 Myr, similar to the timescale when the gas in the solar protoplanetary disk dissipated \citep[see e.g.][]{Desch18,Schiller18}. Moreover, the use of a narrow FOV may constitute a crucial observational bias playing a fundamental role in our general understanding of disk dissipation. \citet{Pfalzner14} suggested that the use of wide FOVs should lead to disk fractions significantly larger than inferred in previous surveys that focused on the central regions of the clusters, indicating that $\geq$ 50$\%$ of stars should have disks at ages of $\sim$ 10 Myr. The reason for this strong discrepancy with respect to the normally accepted disk dissipation timescale is mainly based on the assumption that young clusters significantly expand during the first Myr. In particular, the use of relatively small FOVs to survey clusters that have already expanded after $\sim$ 2-3 Myr allows us to probe only the stars initially located in the central, densest regions of the clusters. A large stellar density would in turn cause strong interactions leading to disk fractions smaller than if the stars located in periphery are also considered. 

The Gaia mission \citep{Prusti16} is providing a new view of stellar clusters thanks to accurate proper motions and parallaxes not limited to narrow fields or specific star forming regions. In particular, Gaia's data have served to confirm previous suggestions indicating that most clusters expand during the critical period of disk dissipation \citep[][]{Kuhn19}, showing that their sizes can be much larger than previously assumed \citep[e.g.][]{Grosschedl21,Meingast2021}. Under this new perspective, it is worth testing whether disk fraction estimates depend on the use of different FOVs or not, given its potential implications in our view of planet formation and our own planetary system. This work aims to use Gaia EDR3 data to compare the relative disk fractions of a representative sample of young stellar clusters when those are inferred considering two very different fields, compact and extended, around the center of each cluster. Section \ref{Sect:method} details the scope and the methodology. Section  \ref{Sect:results} describes the results of this work, which are critically discussed in Sect. \ref{Sect:analisis}. Finally, Sect. \ref{Sect:conclusions} includes some concluding remarks.

\section{Scope and Methodology}
\label{Sect:method}
Our aim is to test whether sampling fields with different sizes around the center of stellar clusters lead to significantly different disk fraction estimates or not. To this end, a sample of 19 young, nearby clusters listed in Table \ref{table:regions_properties} has been analysed. In particular, the vast majority of the stellar population within each cluster is in the PMS phase, and 15 out of the 19 clusters are closer than 1.5 kpc. In addition, all them were studied in previous reference works dealing with the typical disk dissipation timescale (based on relatively compact FOVs), and show a range of properties in terms of stellar masses, densities, and cluster expansion. 

Because this work is based on stellar members identified through kinematic and spatial segregation using large regions around the clusters' centers (Sect. \ref{Sect:members}), those clusters that cannot be unambiguously distinguished from each other were discarded. Thus, the clusters studied are separated by a wide enough angular distance, showing clearly different distances and proper motions otherwise (e.g. NGC 1333 and IC 348). This leaves well known regions such as the \object{Orion Nebula Cluster} and \object{$\sigma$ Ori} excluded from the analysis. Indeed, the separation between the previous clusters is relatively small, showing very similar distances and proper motions likely indicating a physical relation \citep{Kuhn19,Grosschedl21}. 

It is also out of the scope of this work to derive accurate numbers representing absolute disk fractions for each cluster (see e.g. the references in Col. 2 of Table \ref{table:regions_properties}). Instead, the comparison between relative disk fractions estimated from two concentric fields with very different sizes but using identical methodology is enough for our purposes. Such a methodology is described next, and its associated caveats and limitations are discussed in Sect. \ref{Sect:analisis}. 

\begin{table*}
\centering
\caption{General properties of the young stellar clusters}
\label{table:regions_properties}
\centering
\begin{tabular}{l l l l l l l l l}
\hline\hline
Name & References & Age & Massive? & Expanding? & distance & $\Sigma$$_{2pc}$ & $\Sigma$$_{20pc}$ \\
... & ... & [Myr] & (y/n) & (y/n) & [pc] & [$\times$10$^{-2}$arcmin$^{-2}$] & [$\times$10$^{-2}$arcmin$^{-2}$] \\ 
\hline\hline
NGC 1333 & $^1$ $^2$ $^3$ $^4$ $^5$ $^6$ & 0-2 & n & n & 291$\pm$7 & 1.6 & 0.052 \\
IC 348 & $^1$ $^2$ $^3$ $^4$ $^5$ $^6$ $^7$ $^8$ & 0-5 & n & n & 314$\pm$8 & 5.4 & 0.12 \\
NGC 1893 & $^6$ & 1-2 & n & y & 2814$\pm$302 & 80 & 5.9 \\
$\lambda$ Ori & $^1$ $^2$ $^4$ $^5$ $^6$ & 1-11 & n &  & 393$\pm$8 & 6.2 & 0.37 \\
NGC 1960 & $^7$ & 15-39$^{\dagger1}$ & n &  & 1131$\pm$37 & 100 & 2.4 \\
NGC 2169 & $^8$ & 6-12 & n &  & 943$\pm$43 & 17 & 0.67 \\
NGC 2244 & $^1$ $^3$ $^4$ $^6$ & 0-4 & y & y & 1414$\pm$85 & 57 & 6.8 \\
NGC 2264 & $^1$ $^2$ $^4$ $^6$ $^7$ & 0-6 & n & n & 711$\pm$32 & 31 & 1.0 \\
NGC 2353 & $^8$ & 8-16 & n &  & 1187$\pm$122 & 80 & 4.8 \\
NGC 2362 & $^1$ $^2$ $^3$ $^4$ $^6$ $^7$ $^8$ & 0-6 & y & n & 1271$\pm$88 & 110 & 5.8 \\
$\gamma$ Vel & $^1$ $^2$ $^4$ $^6$ & 0-16 & n &  & 359$\pm$19 & 9.7 & 0.80 \\
ASCC 58 & $^8$ & 7-13$^*$ & n &  & 470$\pm$15 & 3.5 & 0.35 \\
Tr 15 & $^4$ & 4-12 & y?$^{\dagger2}$ & y? & 2358$\pm$261 & 440 & 40 \\
NGC 6231 & $^6$ $^8$ & 1-9 & y & n & 1588$\pm$115 & 230 & 15 \\
NGC 6531 & $^8$ & 5-10 & n &  & 1214$\pm$68 & 96 & 7.4 \\
NGC 6611 & $^4$ $^6$ & 1-4 & y & y & 1602$\pm$126 & 150 & 16\\
Tr 37 & $^1$ $^2$ $^3$ $^4$ $^6$ & 0-6 & y$^{\dagger3}$ &  & 914$\pm$32 & 31 & 2.3 \\
NGC 7129 & $^1$ $^2$ $^3$ $^4$ $^6$ & 0-6 & n &  & 915$\pm$56 & 12 & 0.38 \\
NGC 7160 & $^1$ $^2$ $^3$ $^4$ $^8$ & 8-15 & n & n & 920$\pm$48 & 17 & 1.2 \\
\hline
\hline
\end{tabular}
\begin{minipage}{18cm}
\textbf{Notes:} Column 2 lists disk dissipation studies where the clusters in Col. 1 were included. The numbers in Cols 3 and 4 are based on such studies and references therein. The age range is indicated in Col. 3. Following \citet{Mamajek09}, an uncertainty of 30$\%$ has been assigned to ages without errorbars. Column 4 indicates whether the cluster has a total stellar mass above or below 500-1000M$_{\odot}$ (yes or not), which is in turn indicative of the presence of a significant number of massive OB stars within the clusters \citep[but note that most are part of OB associations when larger scales are considered; see e.g.][]{Wright20}. The presence or absence of evidence indicating cluster expansion (yes or not) is indicated in Col. 5, whenever available in \citet{Kuhn19}. The last three columns are based on our work. The average Gaia EDR3 distance (and standard deviation) to each cluster is shown in Col. 6 (see Appendix \ref{app_comparison}). Columns 7 and 8 show the surface density of member stars for the inner FOV$_{2pc}$ and the whole FOV$_{20pc}$, as inferred from Table \ref{table:general_stats}. 
\textbf{References:} $^1$\citet{Hernandez08}, $^2$\citet{Mamajek09}, $^3$\citet{Muzerolle10} $^4$\citet{Pfalzner14}, $^5$\citet{Ribas15}, $^6$\citet{Guarcello21}, $^7$\citet{Haisch01}, $^8$\citet{Fedele10}. $^{\dagger1}$\citet{Panja21}, $^{\dagger2}$\citet{Wang11}  $^{\dagger3}$\citet{Saurin12}.

\end{minipage}
\end{table*}

\subsection{Searching for member candidates}
\label{Sect:members}
The Virtual Observatory (VO) compliant tool \emph{Clusterix 2.0}\footnote{http://clusterix.cab.inta-csic.es/clusterix/} was used to find member candidates for each cluster. This is an interactive tool that determines membership probabilities using a fully non-parametric method based on proper motions. Clusterix carries out an empirical determination of the frequency functions from the vector point diagram without previous assumptions about their profiles. The complete description and performance of Clusterix are included in \citet{Balaguer20}.

Clusterix was applied to the stars within a circular field centered on each cluster and radius equivalent to $\sim$ 20 pc at the corresponding distances. This radius covers the large area potentially having sources unaffected by stellar interactions after cluster expansion, eventually showing larger disk fractions than in the compact area with radius $\sim$ 2 pc covered by most previous surveys \citep[see][and references in Table \ref{table:regions_properties}]{Pfalzner14}. Both previous areas are different by a factor 100, will be called FOV$_{20pc}$ and FOV$_{2pc}$ hereafter, and are analysed independently.

Gaia EDR3 \citep{Brown21} stellar magnitudes \citep{Riello21}, parallaxes and proper motions \citep{Lindegren21} were used for all stars. Only the ones with magnitude G $\leq$ 17 were considered, given that the accuracy of their proper motions is the best currently possible (up to $\sim$ 0.01 mas yr$^-1$). Such a magnitude limit constraints the minimum stellar mass that can be probed, which increases as the clusters are further away. For instance, the smallest MS stellar mass that could be probed at the distance of the furthest cluster ($\sim$ 2814 pc; NGC 1893) is $\sim$ 1.0 M$_{\odot}$, whereas for the closest one ($\sim$ 291 pc; NGC 1333) goes down to $\sim$ 0.4 M$_{\odot}$. The previous limits are larger as extinction increases and smaller for younger sources, thus strictly depending on the specific cluster. Nevertheless, the lack of stellar mass completeness should not significantly affect the intra-cluster comparison carried out here (but see Sect. \ref{Sect:mass_and_age}). In addition, only the sources with parallax $>$ 0, relative parallax error $<$ 0.1, and renormalised unit weight error (RUWE) $<$ 1.4 were taken into account. This way the sources with potentially spurious parallaxes are excluded and the presence of unresolved binaries minimized \citep{Lindegren21,ElBadry21}.

The stars whose proper motion-based membership probability is null according to Clusterix were discarded. Then, the stars whose parallax is not consistent with that associated to the cluster $\pm$3$\sigma$ \citep[being $\sigma$ the Gaia-based standard deviations in][]{Cantat20} were filtered out. Finally, potential sources showing proper motions different from the median by more than 3 times the standard deviation were considered outliers and removed from the analysis too. In short, members were identified based on kinematic segregation from proper motions, which have associated a non-null membership probability given by Clusterix, and spatial segregation from parallaxes. These criteria were equally applied to all clusters, allowing a direct comparison between the results obtained from the extended region (FOV$_{20pc}$) with those from the compact region within it (FOV$_{2pc}$). Additional filtering was applied to remove potential contamination by non-members (Sect \ref{Sect:consistency}).

\subsection{Estimating inner disk fractions}
\label{Sect:disk fractions}
A near-infrared (IR) H-K vs J-H diagram based on 2MASS photometry \citep{Skrutskie06} was built for the members in each cluster to estimate the disk fractions associated to FOV$_{20pc}$ and FOV$_{2pc}$. JHK photometry within 2" from the Gaia coordinates of each member were filtered by discarding magnitudes without errorbars and colors with errorbars $>$ 0.5 magnitudes. Members were classified as disk sources when they show J-H $<$ -0.007 + 1.698 $\times$ H-K, J-H $>$ 0.20, and  H-K $>$ 0.35 \citep[e.g.][]{Lada92}. For a given cluster and FOV, the disk fraction is f$_{disk}^{JHK}$ = 100 $\times$ (N$_{disk}^{JHK}$/N$_{memb}^{JHK}$), where N$_{memb}^{JHK}$ and N$_{disk}^{JHK}$ are the total number of member stars within the FOV with JHK photometry and the corresponding number of stars classified as disks sources, respectively. Errorbars for the disk fractions were derived from the different N$_{disk}^{JHK}$ values associated to the colors' uncertainties.

Alternative diagrams based on IRAS, AKARI, or Spitzer colors and magnitudes are commonly used to identify disk sources. Nevertheless, they are not ideal for our comparison purposes because their coverage is partial or limited to compact regions around the center of most clusters analyzed here. The situation is similar using the more complete all-sky WISE survey, once potentially spurious magnitudes are filtered out \citep{Koenig14}. In contrast, the 2MASS survey typically covers more than 90$\%$ of the identified members within both FOV$_{20pc}$ and FOV$_{2pc}$ simultaneously. This allows us to carry out a statistically significant comparison between the disk fractions inferred from the corresponding color-color diagram for all clusters and FOVs, guarantying that the inevitable exclusion of the few members without JHK counterparts has no major effect. It is remarked that the near-IR photometry used in this work allows us to identify sources with hot inner disks at distances $\lesssim$ 1 au from the central star, but not colder disks at longer distances. The caveats and limitations associated to the JHK color-color diagram are discussed in more detail in Sect. \ref{Sect:limitations}.   

\section{Results}
\label{Sect:results}
Appendix \ref{appendix_clusters_general} includes the figures and tables summarizing the results. The distributions of stars in the proper motion plane and the projected directions of motion of all members identified in each cluster are shown in Figs. \ref{figure:ppmra_vs_ppmdec} and \ref{figure:ppm_sky}. Their on-sky positions and the sizes of the compact and the extended FOVs are overplotted to an image of the region in the IR in Fig. \ref{figure:FOVS}. The JHK color-color diagrams used to infer the inner disk fractions are in Figure \ref{figure:disk_fractions}. Additionally, Fig. \ref{figure:CMDs} shows the Gaia color-magnitude diagram of each cluster, although its representativeness depends on how well the extinction is characterized (see the legends in that figure and Sect. \ref{Sect:consistency}). Table \ref{table:general_stats} lists the central coordinates of each cluster, along with the angular radius, fraction of members identified with respect to the field stars, mean parallaxes and proper motions derived using FOV$_{20pc}$ and FOV$_{2pc}$. Table \ref{table:members} lists the Gaia EDR3 coordinates, proper motions, parallax, and magnitudes, as well as the angular distance to the center, membership probability and 2MASS colors -when available- of all member stars. Disk stars identified through the procedure described in Sect. \ref{Sect:disk fractions} are indicated. Finally, a VO-compliant archive where all previous data are available is described in Appendix \ref{app_VO_archive}, and a comparison with previous Gaia-based results is in Appendix \ref{app_comparison}. Next we focus on the main results that will be analysed in this work.

\begin{figure}
   \centering
   \includegraphics[width=9cm]{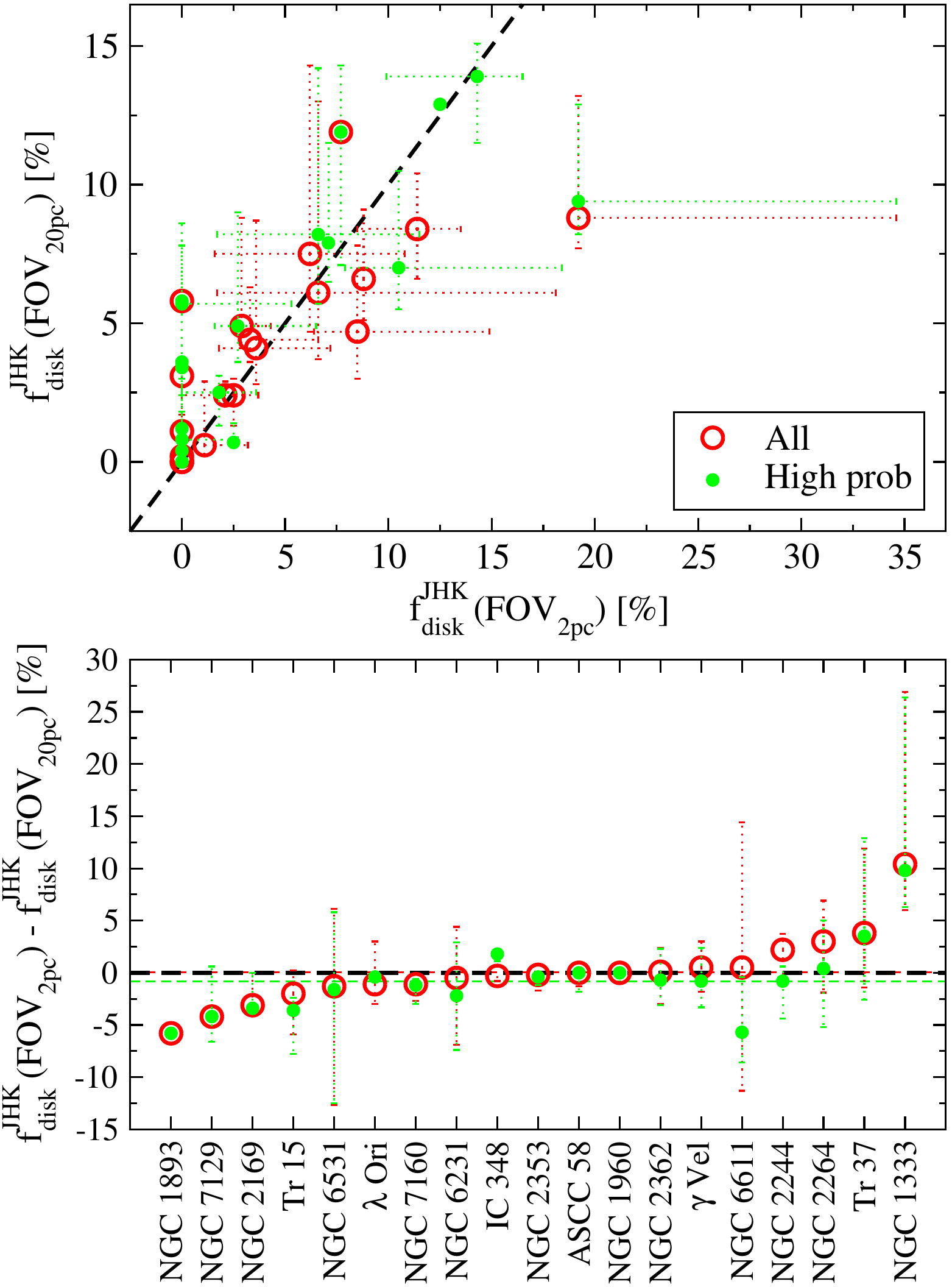}
      \caption{Inner disk fractions from the compact and extended regions are plotted against each other in the top panel, and their differences in the bottom panel. The red and green circles are results based on all members identified per cluster or on the high-probability members, respectively. The black dashed lines indicate equal values in both panels, and the mean differences are shown in the bottom panel with red and green dashed lines.} 
         \label{fig:disk_fractions_comparison}
   \end{figure}

Firstly, although the surface density of members closer than 2 pc is larger than that further out by a factor ranging from $\sim$ 10 to 86 (typically $\sim$ 28 on average), the absolute number of members in the periphery can be up to $\sim$ 11 (typically $\sim$ 5) times larger. Thus, the number of members associated to clusters significantly differs from previous surveys focused on narrow FOVs within the first $\sim$ 2 pc, which need to be updated to better account for the whole stellar populations. On the other hand, Cols 3 and 7 of Table \ref{table:disk_fractions} list the inferred disk fractions associated to each cluster and FOV, whose representativeness is quantified in Cols. 2 and 6 showing the fraction of member stars with JHK photometry. Figure \ref{fig:disk_fractions_comparison} compares the disk fractions derived using FOV$_{2pc}$ versus those derived using FOV$_{20pc}$, illustrating their similarity. The difference between both is below errorbars in most clusters, typically smaller than $\sim$ $^{+10}_{-5}$ $\%$, and with a mean value = 0.05$\%$. Considering errorbars, the difference between disk fractions can reach $^{+14}_{-13}$ $\%$ (except for NGC 1333). Thus, based on this analysis the use of a large FOV does not lead to disk fractions significantly larger than those inferred from compact regions around the clusters' centers, in contrast with the proposal in \citet{Pfalzner14}. The previous results are critically discussed next.

\begin{table*}
\centering
\renewcommand\tabcolsep{3.9pt}
\caption{Class II inner disk fractions from 2MASS color-color diagrams}
\label{table:disk_fractions}
\centering
\begin{tabular}{l|llll|llll}
& \multicolumn{3}{c}{FOV$_{20pc}$} & & \multicolumn{3}{c}{FOV$_{2pc}$} \\
\multicolumn{1}{c}{Name} & N$_{JHK}$/N$_{memb}$ & f$_{disk}^{JHK}$ & N$_{JHK}$/N$_{memb}$ (p$_0$)&  f$_{disk}^{JHK}$ (p$_0$) & N$_{JHK}$/N$_{memb}$ & f$_{disk}^{JHK}$ & N$_{JHK}$/N$_{memb}$ (p$_0$) &  f$_{disk}^{JHK}$ (p$_0$)\\
\hline\hline
\object{NGC 1333} & 91/95 & 8.8$^{+4.4}_{-1.1}$ & 85/88 (0.2) & 9.4$^{+3.5}_{-1.2}$ (0.2) & 26/28 & 19.2$^{+15.4}_{-0.0}$ &26/28 (0.2) & 19.2$^{+15.4}_{-0.0}$ (0.2)\\
\object{IC 348} & 205/209 & 2.4$^{+0.5}_{-0.0}$ & 148/151 (0.6) & 0.7$^{+0.7}_{-0.0}$ (0.6)& 95/97 & 2.1$^{+0.0}_{-0.0}$ & 81/83 (0.6) & 2.5$^{+0.0}_{-0.0}$ (0.6)\\
\object{NGC 1893} & 52/59 & 5.8$^{0.0}_{-0.0}$ & 52/59 (0.5)$^{\dagger}$ & 5.8$^{+0.0}_{-0.0}$ (0.5)$^{\dagger}$ & 7/8 & 0.0$^{+0.0}_{-0.0}$ & 7/8 (0.5)$^{\dagger}$ & 0.0$^{+0.0}_{-0.0}$ (0.5)$^{\dagger}$\\
\object{$\lambda$ Ori} & 366/377 & 4.4$^{+1.9}_{-0.8}$ & 31/31 (0.9) & 12.9$^{+0.0}_{-0.0}$ (0.9)& 61/63 & 3.3$^{+3.3}_{-0.0}$  & 16/16 (0.9) & 12.5$^{+0.0}_{-0.0}$  (0.9)\\
\object{NGC 1960} & 265/273 & 0.0$^{+0.0}_{-0.0}$ & 243/250 (0.8) & 0.0$^{+0.0}_{-0.0}$ (0.8)& 109/114 & 0.0$^{+0.0}_{-0.0}$ & 106/110 (0.8) & 0.0$^{+0.0}_{-0.0}$ (0.8)\\
\object{NGC 2169} & 98/102 & 3.1$^{+0.0}_{-3.1}$ & 59/62 (0.9) & 3.4$^{+0.0}_{-3.4}$ (0.9) & 23/26 & 0.0$^{+0.0}_{-0.0}$ & 21/23 (0.9) & 0.0$^{+0.0}_{-0.0}$ (0.9)\\
\object{NGC 2244} & 407/418 & 6.6$^{+2.5}_{-1.5}$ & 140/147 (0.9) & 7.9$^{+3.6}_{-1.4}$ (0.9) & 34/35 & 8.8$^{+0.0}_{-0.0}$ & 28/28 (0.9) & 7.1$^{+0.0}_{-0.0}$ (0.9) \\
\object{NGC 2264} & 441/452 & 8.4$^{+2.0}_{-1.8}$  & 165/166 (0.8) & 13.9$^{+1.2}_{-2.4}$ (0.8)& 140/142 & 11.4$^{+2.1}_{-2.9}$ & 91/91 (0.8) & 14.3$^{+2.2}_{-4.4}$ (0.8)\\
\object{NGC 2353} & 530/545 & 0.2$^{+1.5}_{-0.2}$ & 236/243 (0.8) & 0.4$^{+0.8}_{-0.4}$ (0.8) & 87/92 & 0.0$^{+0.0}_{-0.0}$ & 75/80 (0.8) &  0.0$^{+0.0}_{-0.0}$ (0.8)\\
\object{NGC 2362} & 461/485 & 2.4$^{+0.6}_{-1.1}$ & 162/170 (0.9) & 2.5$^{+0.6}_{-1.2}$ (0.9) & 80/90 & 2.5$^{+1.2}_{-2.5}$ & 56/63 (0.9) & 1.8$^{+1.8}_{-1.8}$ (0.9)\\
\object{$\gamma$ Vel} & 780/810 & 0.6$^{+2.3}_{-0.4}$ & 399/415 (0.5) & 0.8$^{+2.5}_{-0.5}$ (0.5)& 94/99 &  1.1$^{+2.1}_{-0.0}$ & 74/78 (0.5) & 0.0$^{+2.7}_{-0.0}$  (0.5)\\
\object{ASCC 58} & 150/160 &  0.0$^{+1.3}_{-0.0}$ & 112/118 (0.5)$^{\dagger}$ & 0.0$^{+1.8}_{-0.0}$ (0.5)$^{\dagger}$& 14/16 & 0.0$^{+0.0}_{-0.0}$ & 13/15 (0.5)$^{\dagger}$  & 0.0$^{+0.0}_{-0.0}$ (0.5)$^{\dagger}$\\
\object{Trumpler 15} & 738/870 & 4.9$^{+3.9}_{-0.8}$ & 167/207 (0.8) & 3.6$^{+4.2}_{-1.2}$ (0.8) & 70/94 & 2.9$^{+1.4}_{-0.0}$ & 34/49 (0.8) & 0.0$^{+0.0}_{-0.0}$ (0.8) \\ 
\object{NGC 6231} & 1477/1726 & 4.1$^{+4.6}_{-1.3}$ & 616/714 (0.7) & 4.9$^{+4.1}_{-1.3}$ (0.7)& 225/265 & 3.6$^{+3.6}_{-1.8}$ & 183/214 (0.7) & 2.7$^{+3.8}_{-1.1}$ (0.7)\\
\object{NGC 6531} & 469/696 & 7.5$^{+6.8}_{-2.8}$ & 282/404 (0.6) & 8.2$^{+6.0}_{-2.5}$ (0.6) & 65/91 & 6.2$^{+4.6}_{-4.6}$ & 61/83 (0.6) & 6.6$^{+4.9}_{-4.9}$ (0.6)\\ 
\object{NGC 6611} & 652/809 & 6.1$^{+6.9}_{-2.4}$ & 35/45 (0.99) & 5.7$^{+2.9}_{-0.0}$ (0.99) & 61/72 & 6.6$^{+11.5}_{-4.9}$ & 19/23 (0.99) & 0.0$^{+5.3}_{-0.0}$ (0.99)\\ 
\object{Trumpler 37} & 360/381 & 4.7$^{+3.1}_{-1.7}$ & 199/216 (0.4) & 7.0$^{+3.5}_{-1.5}$ (0.4) & 47/52 & 8.5$^{+6.4}_{-2.1}$ & 38/43 (0.4) & 10.5$^{+7.9}_{-2.6}$ (0.4)\\ 
\object{NGC 7129} & 42/43 & 11.9$^{+2.4}_{-4.8}$ & 42/43 (0.5)$^{\dagger}$ & 11.9$^{+2.4}_{-4.8}$ (0.5)$^{\dagger}$ & 13/14 & 7.7$^{+0.0}_{-0.0}$ & 13/14 (0.5)$^{\dagger}$ & 7.7$^{+0.0}_{-0.0}$ (0.5)$^{\dagger}$\\
\object{NGC 7160} & 185/190 & 1.1$^{+1.6}_{-0.5}$ & 171/176 (0.2) & 1.2$^{+1.8}_{-0.6}$ (0.2) & 27/27 & 0.0$^{+0.0}_{-0.0}$ & 27/27 (0.2) & 0.0$^{+0.0}_{-0.0}$ (0.2) \\
\hline
\hline
\end{tabular}
\begin{minipage}{18cm}

\textbf{Notes.} Columns. 2, 3, 6 and 7: ratio between the number of members with JHK counterparts and the total number of members, and the inner disk fractions (in $\%$) for each cluster and FOV. Rest of the columns: same values for the stars whose membership probability is above a certain threshold p$_0$ (indicated between parentheses). $^{\dagger}$p$_0$ values were arbitrarily set to 0.5, given that the KS test already provided a p-value $>$ 0.05 (see Sect. \ref{Sect:contamination}). Listed disk fractions are valid for comparison purposes but should not be taken as absolute values, given the limitations of our study (see text). 
\end{minipage}
\end{table*}

\section{Analysis and discussion}
\label{Sect:analisis}
\subsection{Consistency}
\label{Sect:consistency}
\subsubsection{Potential contamination}
\label{Sect:contamination}
A proper comparison between disk fractions requires that potential contamination by non-members equally affects the samples within FOV$_{2pc}$ and outside that region. One way to assess this requirement is by having equivalent distributions of membership probabilities for the members identified within FOV$_{2pc}$ and for the rest. Two-sample Kolmogorov-Smirnov (KS) tests were applied to both previous sub-samples in each cluster. The p-value inferred from the KS tests is $<$ 0.05 for most clusters, rejecting the hypothesis that the membership probabilities are drawn from the same parent distribution. The reason explaining those cases is that whereas most FOV$_{2pc}$ members have membership probabilities above a certain value,  the rest of the members located further away have membership probabilities both above and below that value. Filtered samples with membership probabilities above a threshold were built, increasing such a threshold until the KS test provided a p-value $>$ 0.05. The p$_0$ values defining the membership probability thresholds from which the KS tests support that the member sub-samples are drawn from the same parent distribution are indicated in Table \ref{table:disk_fractions}. Considering a more stringent membership probability $>$ p$_0$ (instead of $>$ 0) results in a significantly reduced number of members outside 2 pc, whereas the sample within FOV$_{2pc}$ remains comparatively unaltered in most clusters (see Cols. 2, 4 and 6, 8 in that table). 

\begin{figure}
   \centering
   \includegraphics[width=9cm]{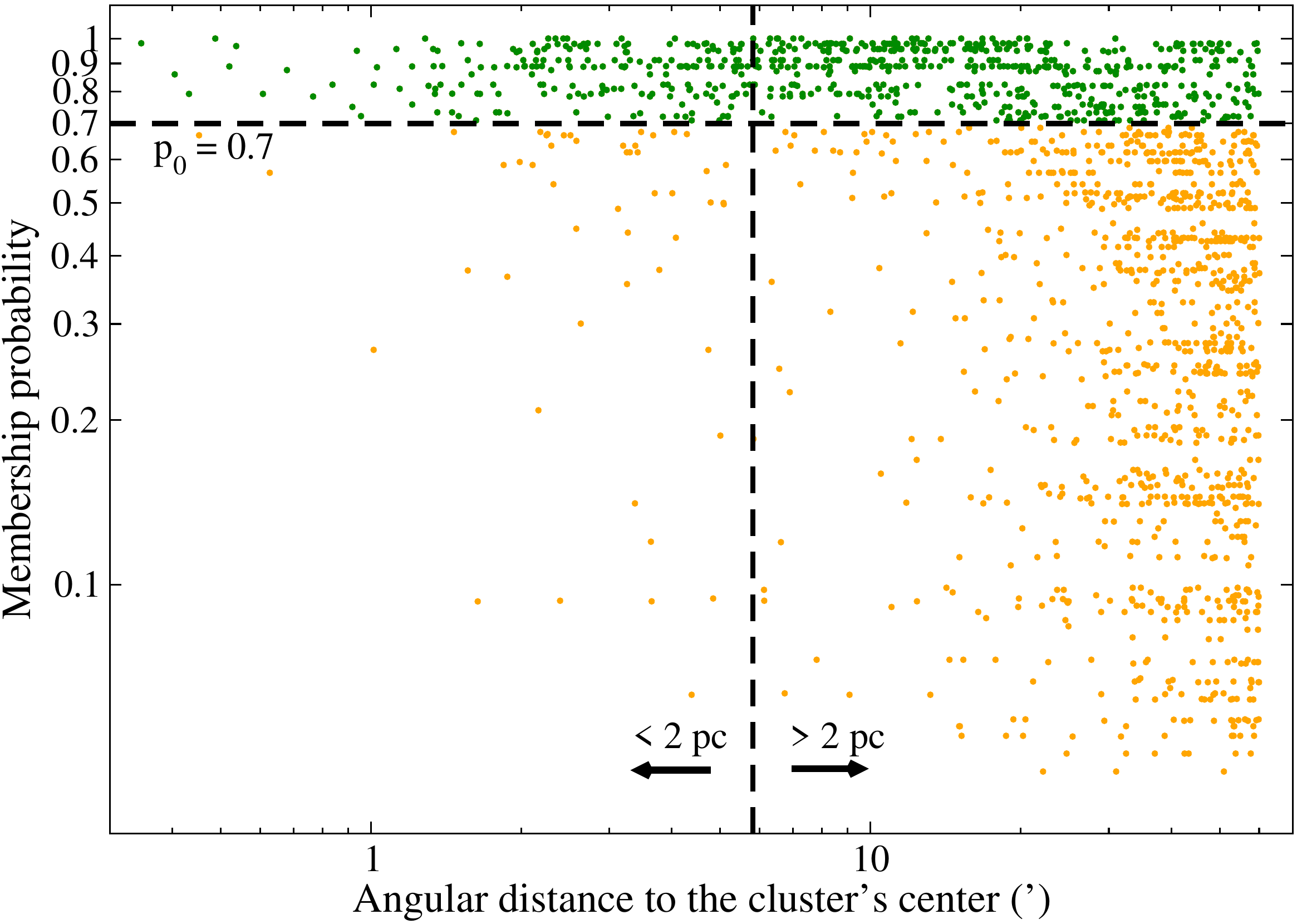}
      \caption{Membership probability against angular distance to the center of NGC 6231 for all members identified in this cluster (log-scale). The vertical dashed line defines two sub-samples closer and further than 2 pc from the cluster's center. The horizontal dashed line indicates the probability threshold p$_0$ below which the KS test rules out that the probability distributions of both previous sub-samples are drawn from the same parent distribution. The resulting high- and low-probability members are indicated with dark green and orange, respectively.} 
         \label{fig:example_KS_test}
   \end{figure}

An example illustrating the previous procedure is shown in Fig. \ref{fig:example_KS_test}. Membership probabilities are plotted against the angular distance to the center of NGC 6231, which is the cluster with the largest number of members identified in this work. Although the membership probabilities assigned by Clusterix only rely on proper motions, the members within 2 pc tend to show comparatively large probabilities, which is not the case for many sources that are further away. The observed difference in the probability distributions is confirmed by the KS test, revealing that such a difference is present until a membership probability threshold p$_0$ = 0.7 is reached. Thus, in order to make a comparison addressing potentially uneven contamination in NGC 6231, disk fractions must be modified by only considering the stars with a membership probability $>$ 0.7. The high-probability members within 2 pc represents 81$\%$ of all members initially identified in that region. In contrast, only 34$\%$ of the members initially identified outside 2 pc are now considered.

Modified disk fractions in Cols. 5 and 9 of Table \ref{table:disk_fractions} are compared in Fig. \ref{fig:disk_fractions_comparison} (green circles). The result obtained in Sect. \ref{Sect:results} based on complete member samples is supported when only the high-probability members are considered. The difference between disk fractions inferred from the compact and extended FOVs is again $\sim$ $^{+10}_{-5}$ $\%$, mostly within $\sim$ 13$\%$ even considering errorbars, and the typical (mean) difference is similar too (-0.8$\%$). The comparison between disk fractions in clusters with low p$_0$ values (e.g. NGC 1333, NGC 7160) does not change significantly if more stringent probability cuts are used, although the statistical significance is reduced because of the smaller sample sizes. 

The previous test supports that if some differential contamination is present this does not significantly affect the result concerning the comparison between disk fractions. Indeed, the mean proper motions inferred from all members within FOV$_{2pc}$ and FOV$_{20pc}$ are equal considering errorbars for all clusters (Table \ref{table:general_stats}), and consistent with previous Gaia-based determinations using narrower FOVs (Appendix \ref{app_comparison}). The direction of motion of the members identified inside and outside 2 pc roughly coincides in all clusters too (Fig. \ref{figure:ppm_sky}).

The fact that the comparison between disk fractions remains essentially unaltered after removing the potential contaminants identified above suggests that these share a similar evolutionary stage, thus supporting that they mostly belong to the clusters too. Indeed, if a large enough number of the low-probability members -mainly located in the periphery- were e.g. MS stars, disk fractions using FOV$_{20pc}$ should be larger once such wrongly introduced members are removed. Therefore, we should not see strong differences between the high-and low- probability members concerning their location on the HR diagram of each cluster. In order to place the members in the Gaia color-magnitude diagram (CMD hereafter), the observed BP-RP colors and G magnitudes should first be corrected from extinction. The Gaia EDR3-based optical extinction values (A$_v$) in \citet{Anders22} are the most recent and complete estimates homogeneously derived, and are best suited for our purposes. Gaia intrinsic colors (BP-RP)$_{0}$ were calculated using such A$_v$ values, the wavelength corrections from \citet{Casagrande21}, and a typical total-to-selective extinction ratio R$_v$ = 3.1. Extinction-corrected absolute MG magnitudes were derived by also using the Gaia EDR3 geometrical distances to each member based on \citet{BailerJones21}. Errorbars were inferred propagating the individual uncertainties in the Gaia magnitudes and parallaxes, and assigning an additional 10$\%$ error associated to extinction.

\begin{figure}
   \centering
   \includegraphics[width=9cm]{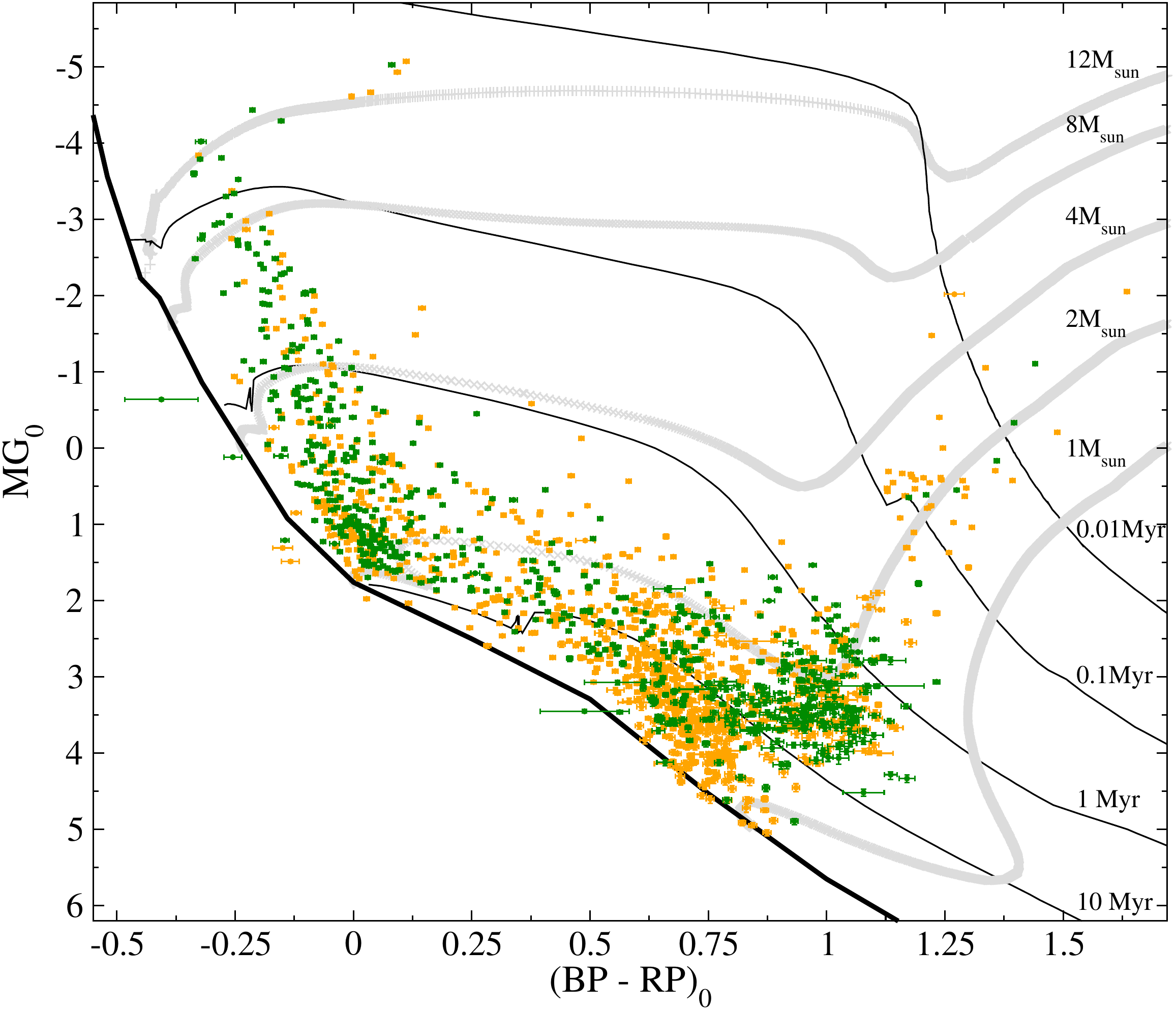}
      \caption{Extinction corrected Gaia CMD of NGC 6231, where the high- and low- probability members are indicated with dark green and orange as identified in Fig. \ref{fig:example_KS_test}. In this case 90$\%$ and 91$\%$ of the high- and low-probability members are shown, given that extinction values are not available for the rest. The ZAMS loci is indicated with the black solid curve. PMS isochrones and evolutionary tracks are overplotted in black and gray for the stellar ages and masses indicated.} 
         \label{fig:CMD_high_low_prob}
   \end{figure}

Fully representative CMDs can be built for NGC 1960, NGC 2353, ASCC 58, and NGC 6231, given that A$_v$ values in \citet{Anders22} are available for $\geq$ 90$\%$ of their members. An example is shown in Fig. \ref{fig:CMD_high_low_prob}, once again focused on NGC 6231 (see Fig. \ref{fig:example_KS_test}). PARSEC isochrones and evolutionary tracks from \citet{Bressan12} are overplotted for reference. High- and low- probability members are similarly distributed, their location in the CMD being above the Zero Age Main Sequence (ZAMS) and consistent with the PMS phase in almost all cases. The majority of the members have ages in between 1 and 10 Myr, in agreement with previous estimates (Table \ref{table:regions_properties}). The $\sim$ 4-12 M$_{\odot}$ and $\sim$ 2 M$_{\odot}$ stars showing ages $<$ 1 Myr, as well as more evolved sources, are both low- and high- probability members. Particularly, the fraction of members with extinction values that are already in the MS is the same for the low- and high-probability sub-samples, $<$ 0.5$\%$. All previous similarities support that the vast majority of the members identified actually belong to the same cluster, explaining why the comparison between disk fractions leads to virtually the same result regardless of the use of the whole sample or the high-probability sub-sample.

The same exercise was applied to the rest of the clusters. It is noted, however, that for most of them extinction values are available only for a fraction of members $<$ 90$\%$, being $<$ 50$\%$ for NGC 1333, IC 348, $\lambda$ Ori, NGC 2264, and NGC 7129. With the previous caveat in mind, we find no evidence indicating that the location of the high- and low- probability members in the Gaia CMDs generally differs. For this reason, and because the overall comparison between disk fractions remains essentially the same (Fig. \ref{fig:disk_fractions_comparison}), all members identified are equally considered in this work\footnote{Membership probabilities are provided for all stars in each cluster in Table \ref{table:members}, for which different probability cuts could be applied by the interested readers.}. Gaia CMDs for all clusters are included in Fig. \ref{figure:CMDs}. The distinction in this figure between the FOV$_{2pc}$ and the FOV$_{20pc}$ members will be used in the remaining discussion. 

\subsubsection{Potential mass and age biases}
\label{Sect:mass_and_age}
   
Disk fraction studies specifically devoted to intermediate-mass young stars are not abundant, but they suggest that disk dissipation is slightly faster than that for lower-mass objects \citep{Hernandez07,Yasui14,Ribas15}. In addition, evolutionary studies based on disk fractions assume that all stars in a given cluster are roughly coeval (see e.g. the references in Sect. \ref{Sect:intro} and Table \ref{table:regions_properties}). Therefore, The interpretation of the result indicating similar disk fractions within each cluster when using FOV$_{2pc}$ and FOV$_{20pc}$ may depend on whether such fractions refer to stars with similar stellar masses and ages or not. 

\begin{figure}
   \centering
   \includegraphics[width=9cm]{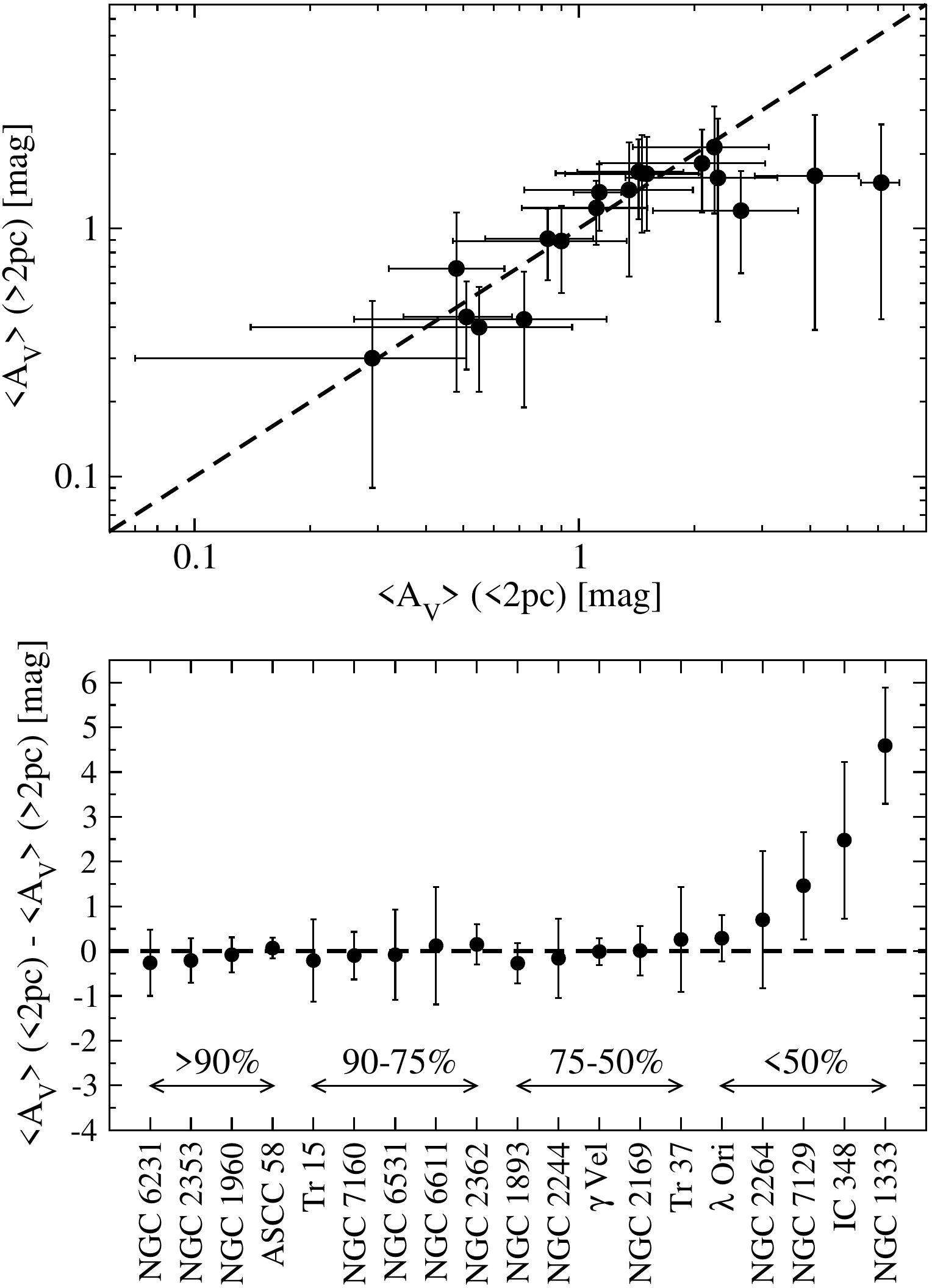}
      \caption{Mean optical extinction from the stars located closer and further than 2 pc from the center of each cluster are plotted against each other in the top panel (log-scale), and their differences in the bottom panel. The dashed lines indicate equal values in both panels. Standard deviations within each cluster are adopted as errorbars in the top panel. Uncertainties in the bottom panel are derived from the propagation of the individual 1$\sigma$ deviations. The fraction of member stars for which extinction values from \citet{Anders22} are available for each cluster is indicated above the arrows in the bottom panel.}
         \label{fig:extinction_comp}
   \end{figure}

Figure \ref{fig:extinction_comp} compares the typical (mean) extinctions of the inner and outer regions of the clusters analysed in this work, as inferred from the individual A$_V$ values of the members from \citet{Anders22}. Errorbars reflect the corresponding standard deviations. For the majority of the clusters there is no difference between the typical extinctions within $\pm$1$\sigma$. Thus, the lower mass limit that could be probed within each of the previous clusters is the same regardless of the FOV. A closer look to the Gaia CMDs (Fig. \ref{figure:CMDs}) of the four clusters with the best extinction characterization (ASCC 58, NGC 1960, NGC 2353, and NGC 6231) confirms that the mass ranges probed by FOV$_{2pc}$ and FOV$_{20pc}$ are indeed the same within each cluster. Therefore, the comparison between the corresponding disk fractions is consistent at least concerning the stellar mass range that they represent. On the other hand, the CMDs reveal that the strongest differences between the smallest stellar mass probed through FOV$_{2pc}$ and FOV$_{20pc}$ are shown by NGC 1333 and IC 348, which results from the strong differential extinction between the inner and outer regions. In such cases (see also e.g. NGC 7129 and Tr 37) the innermost region probes masses $\gtrsim$ 1M$_{\odot}$ whereas the outer region goes down to $\sim$ 0.5M$_{\odot}$. Although small differences in the disk dissipation timescales -and thus on disk fractions- are expected mainly when the mass ranges compared differ by a larger amount \citep[see the references above, but also][]{Hernandez10} these cases should be viewed with caution given the current lack of information concerning extinction.

Regarding the evolutionary stage, although the vast majority of the members identified in each cluster consistently fall within the PMS location in the CMDs, age spread is apparent in several of them. This is not surprising, based on the scatter found in the literature for the clusters in this work analysed from a variety of data and inferred from different sub-populations, as referred in Table \ref{table:regions_properties}. Clusters in that table with narrow age ranges like NGC 1333 are indeed mainly based on single works focused on compact regions and specific sub-samples \citep{Wilking04}, although our data shows that an older population is also present in that cluster (see the corresponding panel in Fig. \ref{figure:CMDs}). More recent Gaia-based works already find considerable age scatter in regions comparable to FOV$_{2pc}$ for many young clusters \citep[e.g.][]{Getman18,Prisinzano19}, confirming previous findings from the pre-Gaia era \citep[e.g.][and references therein]{Getman14a,Getman14b,Jeffries17}.

Nevertheless, the comparison between disk fractions is not compromised as long as the fractions of members with different ages are similar for both FOVs. This is the case for the majority of the clusters, based on their CMDs in Fig. \ref{figure:CMDs}. Following with the same representative example of NGC 6231, $\sim$85$\%$ of the members have ages $>$ 1 Myr (with 0.4$\%$ already in the MS) and the remaining $\sim$ 15$\%$ are younger, such fractions being very similar both for the FOV$_{20pc}$ and a FOV$_{2pc}$ sub-samples. Then the fact that the disk fractions inferred using both FOVs are equal within errorbars can be interpreted in terms of a similar disk dissipation timescale regardless of the distance to the center of the cluster. A relevant exception is NGC 1893. The majority of the members within FOV$_{2pc}$ have ages in between 1 and 10 Myr, but a significant fraction of the members identified in the outer region are younger than 1Myr. This difference would explain why NGC 1893 shows a FOV$_{20pc}$-based disk fraction larger above errorbars than that based on FOV$_{2pc}$ (Fig. \ref{fig:disk_fractions_comparison}). Similarly, the fact that NGC 1333 is the other cluster showing the largest differences between the FOV$_{2pc}$- and FOV$_{20pc}$-based disk fractions is probably related to a significant age spread depending on the radial distance to the center. However, in this case the inner region is probably younger than the outer one. Although such an age difference is motivated by the presence of very young, embedded Class 0/I sources in the inner region of NGC 1333 \citep[e.g.][]{Young15,Fiorellino21} and the observed differential extinction (Fig. \ref{fig:extinction_comp}), that age gradient cannot be assessed solely based on its Gaia CMD. As it was mentioned before, this is barely representative given the lack of enough extinction information.

In other clusters, sub-populations with specific ages present only in FOV$_{20pc}$ are that small that their presence/absence is virtually irrelevant for the disk fraction estimate (e.g. NGC 2353). In particular, the presence of a few MS stars in the Gaia CMDs of some clusters does not affect the disk fraction comparison. Disk fractions were indeed recalculated in all cases showing such evolved sources assuming the percentages of MS stars inferred from the CMDs in Fig. \ref{figure:CMDs}, and that they are contaminants not belonging to the cluster. No significant difference with respect to the values already provided in Table \ref{table:disk_fractions} was found.

\subsection{Limitations}
\label{Sect:limitations}
On top of the already mentioned caveats concerning the methodology described in Sect. \ref{Sect:method}, additional limitations should be considered when the disk fractions associated to the compact and extended regions are compared to each other.

Firstly, because the departure point to identify members are the Gaia magnitudes in the optical, we are probing essentially Class II stars but not most of the embedded Class 0/I sources. Indeed, the ratio between the number of Class II and Class 0/I sources can change depending on the cluster, e.g. a factor $\sim$ 2 in NGC 1333 or $\sim$ 10 in IC 348 \citep[][]{Young15}. Thus, the inferred disk fractions are lower limits for the less evolved clusters hosting Class 0/I sources. However, embedded stars are mainly located in the densest and more extincted parts of the clusters \citep[e.g.][]{Gutermuth08}, which tend to be closer to the center (Fig. \ref{fig:extinction_comp}). Therefore, although the inclusion of these sources would make the absolute disk fractions larger than provided here, such an increase would be mainly related to those obtained using using FOV$_{2pc}$. In this respect, accurate disk fractions based on compact regions should be upper limits to those inferred from extended FOVs, and not the other way around \citep{Pfalzner14}.

Secondly, the use of JHK color-color diagrams allows us to identify "full" disks with IR excess starting at $\sim$ 1.2 -- 2.2 $\mu$m, but not "transitional" disks with excesses only observable at longer wavelengths. The number of transitional disks is larger in older clusters, representing around 10$\%$ to 50$\%$ of disk stars \citep{Currie11,Balog16}. This again indicates that the disk fractions provided here are lower limits, in this case specially affecting the evolved clusters. However, to our knowledge there is no evidence suggesting that the location of transitional disks with respect to the center of the clusters is different than that of full disks. Assuming that transitional and full disks are similarly distributed within the clusters, our result based on the comparison between relative disk fractions inferred from the compact and the extended regions remains unaltered. 

On the other hand, disk dispersal shows roughly similar timescales regardless of the wavelength or the observational tracer used \citep[e.g. the review in][and references therein]{Ercolano17}, but it is not clear if this applies to environmental effects that promote disk destruction. Indeed, the presence of massive stars or stellar interactions in the densest and more massive clusters may lead to a fast disappearance of the outer disk probed at sub-mm and mm wavelengths \citep[e.g.][]{Anderson13,Ansdell17,Vincke18,vanTerwisga19}. The eventual influence of such environmental effects is the subject of ongoing debate \citep[see e.g. the discussion in][and references therein]{Parker21}, but it does not seem to affect the bulk of the disks probed at shorter IR wavelengths \citep[e.g.][]{Richert15}. Such a possible difference with the outer disk dispersal depending on the environment cannot be analysed based on the JHK photometry used in this work, which we remark again that only refers to inner dust disks.

A major limitation concerning the JHK color-color diagram is that although it is very efficient identifying inner disks around intermediate-mass Herbig Ae/Be stars, up to $\sim$ 30$\%$ of classical T Tauri stars (TTs) may fall in the region associated to extincted MS stars, therefore remaining undetected \citep{Lada92,Haisch01,Mamajek09}. This is not problematic for our comparative study of disk fractions, unless the relative number of undetected TTs changes with the FOV. In particular, for certain clusters one may expect that the central regions are more extincted than further away (Fig. \ref{fig:extinction_comp}). Then the relative number of members located between the bands defining extincted MS stars in the JHK diagram should be larger within FOV$_{2pc}$ than within FOV$_{20pc}$. Indeed, Fig. \ref{figure:disk_fractions} shows that this is the case for extincted clusters like NGC 1333, and IC 348. Consequently, the disk fractions inferred using FOV$_{2pc}$ may significantly increase in these clusters once TTs potentially falling within the extinction region for MS stars in the JHK diagram are identified as disk stars. This is once again indicating that disk fractions inferred from FOV$_{20pc}$ cannot generally be significantly larger than those from FOV$_{2pc}$, but the case is probably the opposite. Assuming the worst case scenario in which our disk fractions only account for 70$\%$ of classical TTs (i.e. f$_{disk}^{real}$ = f$_{disk}$/0.7), the maximum differences between disk fractions estimated from both FOVs would increase from $\sim$ 10$\%$ to $\sim$ 14$\%$.

\section{Concluding remarks}
\label{Sect:conclusions}
A representative sample of young clusters has been analysed aiming to compare the inner disk fractions associated to a compact and an extended region with radii $\sim$ 2 and 20 pc from the center of each cluster. This size is larger than that probed in all previous reference surveys studying disk fractions and in more recent Gaia-based studies for the majority of the clusters. Although the density of members in the periphery is smaller than in the compact region, the absolute number is on average factor $\sim$ 5 larger at distances further than 2 pc from the clusters' centers. This motivates future, detailed studies of young clusters including more complete samples by probing extended regions, which may affect e.g. the typical ages or the age dispersion within the clusters. Our analysis also reveals that inner disk fractions based on the extended region are not generally larger than those from the compact region by $\sim$ 50$\%$ or more, contrasting with previous predictions \citep{Pfalzner14}. On the contrary, the found differences are typically $\lesssim$ 10$\%$, and none of the relatively old clusters with negligible disk fractions in the literature (NGC 1960, NGC 2169, NGC 2353, ASCC 58, Trumpler 15, and NGC 7160) shows such a significant increase when the extended region is considered. 

The limitations and caveats affecting our study have been discussed in detail along the text, and must be taken into account when interpreting the previous result. Particularly relevant is that although the presence of massive stars or stellar interactions in the densest and more massive clusters may lead to a fastest disappearance of the outer disks, these cannot be probed based solely on the near-IR photometry used in this work. It is also noted that for this study we limited ourselves to a maximum radius of 20 pc from the clusters' centers, but their true extension could be significantly larger. For instance, a preliminary analysis of NGC 1333 indeed shows new members up to a radial distance of $\sim$ 40 pc, when no additional ones are found based on proper motion analysis with Clusterix.   

Multi-wavelength analysis, potentially combined with data coming from future Gaia releases -e.g. radial velocities-, are necessary to provide accurate disk fractions and better constraint the number and properties of the cluster's members. The database resulting from this work is available online and constitutes a benchmark for such detailed studies.

\begin{acknowledgements}
The authors acknowledge the anonymous referee, whose constructive suggestions have served to improve the manuscript. IM, ES, LBN, NH, and CR acknowledge support by the Spanish MICIN/AEI/10.13039/501100011033. IM, ES, NH, and CR are partially funded by MDM-2017-0737 Unidad de Excelencia {\em Mar\'{\i}a de Maeztu} - Centro de Astrobiolog\'{\i}a (CSIC-INTA). IM is funded by a RyC2019-026992-I grant. ES has been funded through grant PID2020-112949GB-I00. LBN acknowledges support by "ERDF A way of making Europe" by the “European Union” through grant RTI2018-095076-B-C21, and the Institute of Cosmos Sciences University of Barcelona (ICCUB, Unidad de Excelencia ’Mar\'{\i}a de Maeztu’) through grant CEX2019-000918-M. NH has been partially funded by the AEI Project ESP2017-87676-C5-1-R. This research is based on Clusterix 2.0 service at CAB (INTA-CSIC), and has made extensive use of TOPCAT \citep{Taylor05}. This work has made use of data from the European Space Agency (ESA) mission {\it Gaia} (\url{https://www.cosmos.esa.int/gaia}), processed by the {\it Gaia} Data Processing and Analysis Consortium (DPAC, \url{https://www.cosmos.esa.int/web/gaia/dpac/consortium}). Funding for the DPAC has been provided by national institutions, in particular the institutions participating in the {\it Gaia} Multilateral Agreement. This publication makes use of data products from the Two Micron All Sky Survey and the Wide-field Infrared Survey Explorer, which are joint projects of the University of Massachusetts and the Infrared Processing and Analysis Center/California Institute of Technology; and the University of California, Los Angeles, and the Jet Propulsion Laboratory/California Institute of Technology, funded by the National Aeronautics and Space Administration and the National Science Foundation. 
\end{acknowledgements}

%
%
\bibliographystyle{aa}
\bibliography{myrefs.bib}

\begin{thebibliography}{55}
\expandafter\ifx\csname natexlab\endcsname\relax\def\natexlab#1{#1}\fi

\bibitem[{{Anders} {et~al.}(2022){Anders}, {Khalatyan}, {Queiroz}, {Chiappini},
  {Ard{\`e}vol}, {Casamiquela}, {Figueras}, {Jim{\'e}nez-Arranz}, {Jordi},
  {Mongui{\'o}}, {Romero-G{\'o}mez}, {Altamirano}, {Antoja}, {Assaad},
  {Cantat-Gaudin}, {Castro-Ginard}, {Enke}, {Girardi}, {Guiglion}, {Khan},
  {Luri}, {Miglio}, {Minchev}, {Ramos}, {Santiago}, \& {Steinmetz}}]{Anders22}
{Anders}, F., {Khalatyan}, A., {Queiroz}, A.~B.~A., {et~al.} 2022, \aap, 658,
  A91

\bibitem[{{Anderson} {et~al.}(2013){Anderson}, {Adams}, \&
  {Calvet}}]{Anderson13}
{Anderson}, K.~R., {Adams}, F.~C., \& {Calvet}, N. 2013, \apj, 774, 9

\bibitem[{{Ansdell} {et~al.}(2017){Ansdell}, {Williams}, {Manara}, {Miotello},
  {Facchini}, {van der Marel}, {Testi}, \& {van Dishoeck}}]{Ansdell17}
{Ansdell}, M., {Williams}, J.~P., {Manara}, C.~F., {et~al.} 2017, \aj, 153, 240

\bibitem[{{Bailer-Jones} {et~al.}(2021){Bailer-Jones}, {Rybizki}, {Fouesneau},
  {Demleitner}, \& {Andrae}}]{BailerJones21}
{Bailer-Jones}, C.~A.~L., {Rybizki}, J., {Fouesneau}, M., {Demleitner}, M., \&
  {Andrae}, R. 2021, \aj, 161, 147

\bibitem[{{Balaguer-N{\'u}{\~n}ez} {et~al.}(2020){Balaguer-N{\'u}{\~n}ez},
  {L{\'o}pez del Fresno}, {Solano}, {Galad{\'\i}-Enr{\'\i}quez}, {Jordi},
  {Jimenez-Esteban}, {Masana}, {Carbajo-Hijarrubia}, \& {Paunzen}}]{Balaguer20}
{Balaguer-N{\'u}{\~n}ez}, L., {L{\'o}pez del Fresno}, M., {Solano}, E.,
  {et~al.} 2020, \mnras, 492, 5811

\bibitem[{{Balog} {et~al.}(2016){Balog}, {Siegler}, {Rieke}, {Kiss},
  {Muzerolle}, {Gutermuth}, {Bell}, {Vink{\'o}}, {Su}, {Young}, \&
  {G{\'a}sp{\'a}r}}]{Balog16}
{Balog}, Z., {Siegler}, N., {Rieke}, G.~H., {et~al.} 2016, \apj, 832, 87

\bibitem[{{Bressan} {et~al.}(2012){Bressan}, {Marigo}, {Girardi}, {Salasnich},
  {Dal Cero}, {Rubele}, \& {Nanni}}]{Bressan12}
{Bressan}, A., {Marigo}, P., {Girardi}, L., {et~al.} 2012, \mnras, 427, 127

\bibitem[{{Cantat-Gaudin} \& {Anders}(2020)}]{Cantat20}
{Cantat-Gaudin}, T. \& {Anders}, F. 2020, \aap, 633, A99

\bibitem[{{Casagrande} {et~al.}(2021){Casagrande}, {Lin}, {Rains}, {Liu},
  {Buder}, {Horner}, {Asplund}, {Lewis}, {Martell}, {Nordlander}, {Stello},
  {Ting}, {Wittenmyer}, {Bland-Hawthorn}, {Casey}, {De Silva}, {D'Orazi},
  {Freeman}, {Hayden}, {Kos}, {Lind}, {Schlesinger}, {Sharma}, {Simpson},
  {Zucker}, \& {Zwitter}}]{Casagrande21}
{Casagrande}, L., {Lin}, J., {Rains}, A.~D., {et~al.} 2021, \mnras, 507, 2684

\bibitem[{{Currie} \& {Sicilia-Aguilar}(2011)}]{Currie11}
{Currie}, T. \& {Sicilia-Aguilar}, A. 2011, \apj, 732, 24

\bibitem[{{Desch} {et~al.}(2018){Desch}, {Kalyaan}, \& {O'D.
  Alexander}}]{Desch18}
{Desch}, S.~J., {Kalyaan}, A., \& {O'D. Alexander}, C.~M. 2018, \apjs, 238, 11

\bibitem[{{Dias} {et~al.}(2002){Dias}, {Alessi}, {Moitinho}, \&
  {L{\'e}pine}}]{Dias02}
{Dias}, W.~S., {Alessi}, B.~S., {Moitinho}, A., \& {L{\'e}pine}, J.~R.~D. 2002,
  \aap, 389, 871

\bibitem[{{El-Badry} {et~al.}(2021){El-Badry}, {Rix}, \& {Heintz}}]{ElBadry21}
{El-Badry}, K., {Rix}, H.-W., \& {Heintz}, T.~M. 2021, \mnras, 506, 2269

\bibitem[{{Ercolano} \& {Pascucci}(2017)}]{Ercolano17}
{Ercolano}, B. \& {Pascucci}, I. 2017, Royal Society Open Science, 4, 170114

\bibitem[{{Fedele} {et~al.}(2010){Fedele}, {van den Ancker}, {Henning},
  {Jayawardhana}, \& {Oliveira}}]{Fedele10}
{Fedele}, D., {van den Ancker}, M.~E., {Henning}, T., {Jayawardhana}, R., \&
  {Oliveira}, J.~M. 2010, \aap, 510, A72

\bibitem[{{Fiorellino} {et~al.}(2021){Fiorellino}, {Manara}, {Nisini},
  {Ramsay}, {Antoniucci}, {Giannini}, {Biazzo}, {Alcal{\`a}}, \&
  {Fedele}}]{Fiorellino21}
{Fiorellino}, E., {Manara}, C.~F., {Nisini}, B., {et~al.} 2021, \aap, 650, A43

\bibitem[{{Gaia Collaboration} {et~al.}(2021){Gaia Collaboration}, {Brown},
  {Vallenari}, {Prusti}, {de Bruijne}, {Babusiaux}, {Biermann}, {Creevey},
  {Evans}, {Eyer}, {Hutton}, {Jansen}, {Jordi}, {Klioner}, {Lammers},
  {Lindegren}, {Luri}, {Mignard}, {Panem}, {Pourbaix}, {Randich}, {Sartoretti},
  {Soubiran}, {Walton}, {Arenou}, {Bailer-Jones}, {Bastian}, {Cropper},
  {Drimmel}, {Katz}, {Lattanzi}, {van Leeuwen}, {Bakker}, {Cacciari},
  {Casta{\~n}eda}, {De Angeli}, {Ducourant}, {Fabricius}, {Fouesneau},
  {Fr{\'e}mat}, {Guerra}, {Guerrier}, {Guiraud}, {Jean-Antoine Piccolo},
  {Masana}, {Messineo}, {Mowlavi}, {Nicolas}, {Nienartowicz}, {Pailler},
  {Panuzzo}, {Riclet}, {Roux}, {Seabroke}, {Sordo}, {Tanga}, {Th{\'e}venin},
  {Gracia-Abril}, {Portell}, {Teyssier}, {Altmann}, {Andrae}, {Bellas-Velidis},
  {Benson}, {Berthier}, {Blomme}, {Brugaletta}, {Burgess}, {Busso}, {Carry},
  {Cellino}, {Cheek}, {Clementini}, {Damerdji}, {Davidson}, {Delchambre},
  {Dell'Oro}, {Fern{\'a}ndez-Hern{\'a}ndez}, {Galluccio}, {Garc{\'\i}a-Lario},
  {Garcia-Reinaldos}, {Gonz{\'a}lez-N{\'u}{\~n}ez}, {Gosset}, {Haigron},
  {Halbwachs}, {Hambly}, {Harrison}, {Hatzidimitriou}, {Heiter},
  {Hern{\'a}ndez}, {Hestroffer}, {Hodgkin}, {Holl}, {Jan{\ss}en}, {Jevardat de
  Fombelle}, {Jordan}, {Krone-Martins}, {Lanzafame}, {L{\"o}ffler}, {Lorca},
  {Manteiga}, {Marchal}, {Marrese}, {Moitinho}, {Mora}, {Muinonen}, {Osborne},
  {Pancino}, {Pauwels}, {Petit}, {Recio-Blanco}, {Richards}, {Riello},
  {Rimoldini}, {Robin}, {Roegiers}, {Rybizki}, {Sarro}, {Siopis}, {Smith},
  {Sozzetti}, {Ulla}, {Utrilla}, {van Leeuwen}, {van Reeven}, {Abbas}, {Abreu
  Aramburu}, {Accart}, {Aerts}, {Aguado}, {Ajaj}, {Altavilla}, {{\'A}lvarez},
  {{\'A}lvarez Cid-Fuentes}, {Alves}, {Anderson}, {Anglada Varela}, {Antoja},
  {Audard}, {Baines}, {Baker}, {Balaguer-N{\'u}{\~n}ez}, {Balbinot}, {Balog},
  {Barache}, {Barbato}, {Barros}, {Barstow}, {Bartolom{\'e}}, {Bassilana},
  {Bauchet}, {Baudesson-Stella}, {Becciani}, {Bellazzini}, {Bernet}, {Bertone},
  {Bianchi}, {Blanco-Cuaresma}, {Boch}, {Bombrun}, {Bossini}, {Bouquillon},
  {Bragaglia}, {Bramante}, {Breedt}, {Bressan}, {Brouillet}, {Bucciarelli},
  {Burlacu}, {Busonero}, {Butkevich}, {Buzzi}, {Caffau}, {Cancelliere},
  {C{\'a}novas}, {Cantat-Gaudin}, {Carballo}, {Carlucci}, {Carnerero},
  {Carrasco}, {Casamiquela}, {Castellani}, {Castro-Ginard}, {Castro Sampol},
  {Chaoul}, {Charlot}, {Chemin}, {Chiavassa}, {Cioni}, {Comoretto}, {Cooper},
  {Cornez}, {Cowell}, {Crifo}, {Crosta}, {Crowley}, {Dafonte}, {Dapergolas},
  {David}, {David}, {de Laverny}, {De Luise}, {De March}, {De Ridder}, {de
  Souza}, {de Teodoro}, {de Torres}, {del Peloso}, {del Pozo}, {Delbo},
  {Delgado}, {Delgado}, {Delisle}, {Di Matteo}, {Diakite}, {Diener},
  {Distefano}, {Dolding}, {Eappachen}, {Edvardsson}, {Enke}, {Esquej}, {Fabre},
  {Fabrizio}, {Faigler}, {Fedorets}, {Fernique}, {Fienga}, {Figueras},
  {Fouron}, {Fragkoudi}, {Fraile}, {Franke}, {Gai}, {Garabato},
  {Garcia-Gutierrez}, {Garc{\'\i}a-Torres}, {Garofalo}, {Gavras}, {Gerlach},
  {Geyer}, {Giacobbe}, {Gilmore}, {Girona}, {Giuffrida}, {Gomel}, {Gomez},
  {Gonzalez-Santamaria}, {Gonz{\'a}lez-Vidal}, {Granvik},
  {Guti{\'e}rrez-S{\'a}nchez}, {Guy}, {Hauser}, {Haywood}, {Helmi}, {Hidalgo},
  {Hilger}, {H{\l}adczuk}, {Hobbs}, {Holland}, {Huckle}, {Jasniewicz},
  {Jonker}, {Juaristi Campillo}, {Julbe}, {Karbevska}, {Kervella}, {Khanna},
  {Kochoska}, {Kontizas}, {Kordopatis}, {Korn}, {Kostrzewa-Rutkowska},
  {Kruszy{\'n}ska}, {Lambert}, {Lanza}, {Lasne}, {Le Campion}, {Le Fustec},
  {Lebreton}, {Lebzelter}, {Leccia}, {Leclerc}, {Lecoeur-Taibi}, {Liao},
  {Licata}, {Lindstr{\o}m}, {Lister}, {Livanou}, {Lobel}, {Madrero Pardo},
  {Managau}, {Mann}, {Marchant}, {Marconi}, {Marcos Santos}, {Marinoni},
  {Marocco}, {Marshall}, {Martin Polo}, {Mart{\'\i}n-Fleitas}, {Masip},
  {Massari}, {Mastrobuono-Battisti}, {Mazeh}, {McMillan}, {Messina},
  {Michalik}, {Millar}, {Mints}, {Molina}, {Molinaro}, {Moln{\'a}r},
  {Montegriffo}, {Mor}, {Morbidelli}, {Morel}, {Morris}, {Mulone}, {Munoz},
  {Muraveva}, {Murphy}, {Musella}, {Noval}, {Ord{\'e}novic}, {Orr{\`u}},
  {Osinde}, {Pagani}, {Pagano}, {Palaversa}, {Palicio}, {Panahi}, {Pawlak},
  {Pe{\~n}alosa Esteller}, {Penttil{\"a}}, {Piersimoni}, {Pineau}, {Plachy},
  {Plum}, {Poggio}, {Poretti}, {Poujoulet}, {Pr{\v{s}}a}, {Pulone}, {Racero},
  {Ragaini}, {Rainer}, {Raiteri}, {Rambaux}, {Ramos}, {Ramos-Lerate}, {Re
  Fiorentin}, {Regibo}, {Reyl{\'e}}, {Ripepi}, {Riva}, {Rixon}, {Robichon},
  {Robin}, {Roelens}, {Rohrbasser}, {Romero-G{\'o}mez}, {Rowell}, {Royer},
  {Rybicki}, {Sadowski}, {Sagrist{\`a} Sell{\'e}s}, {Sahlmann}, {Salgado},
  {Salguero}, {Samaras}, {Sanchez Gimenez}, {Sanna}, {Santove{\~n}a},
  {Sarasso}, {Schultheis}, {Sciacca}, {Segol}, {Segovia}, {S{\'e}gransan},
  {Semeux}, {Shahaf}, {Siddiqui}, {Siebert}, {Siltala}, {Slezak}, {Smart},
  {Solano}, {Solitro}, {Souami}, {Souchay}, {Spagna}, {Spoto}, {Steele},
  {Steidelm{\"u}ller}, {Stephenson}, {S{\"u}veges}, {Szabados}, {Szegedi-Elek},
  {Taris}, {Tauran}, {Taylor}, {Teixeira}, {Thuillot}, {Tonello}, {Torra},
  {Torra}, {Turon}, {Unger}, {Vaillant}, {van Dillen}, {Vanel}, {Vecchiato},
  {Viala}, {Vicente}, {Voutsinas}, {Weiler}, {Wevers}, {Wyrzykowski}, {Yoldas},
  {Yvard}, {Zhao}, {Zorec}, {Zucker}, {Zurbach}, \& {Zwitter}}]{Brown21}
{Gaia Collaboration}, {Brown}, A.~G.~A., {Vallenari}, A., {et~al.} 2021, \aap,
  649, A1

\bibitem[{{Gaia Collaboration} {et~al.}(2016){Gaia Collaboration}, {Prusti},
  {de Bruijne}, {Brown}, {Vallenari}, {Babusiaux}, {Bailer-Jones}, {Bastian},
  {Biermann}, {Evans}, {Eyer}, {Jansen}, {Jordi}, {Klioner}, {Lammers},
  {Lindegren}, {Luri}, {Mignard}, {Milligan}, {Panem}, {Poinsignon},
  {Pourbaix}, {Randich}, {Sarri}, {Sartoretti}, {Siddiqui}, {Soubiran},
  {Valette}, {van Leeuwen}, {Walton}, {Aerts}, {Arenou}, {Cropper}, {Drimmel},
  {H{\o}g}, {Katz}, {Lattanzi}, {O'Mullane}, {Grebel}, {Holland}, {Huc},
  {Passot}, {Bramante}, {Cacciari}, {Casta{\~n}eda}, {Chaoul}, {Cheek}, {De
  Angeli}, {Fabricius}, {Guerra}, {Hern{\'a}ndez}, {Jean-Antoine-Piccolo},
  {Masana}, {Messineo}, {Mowlavi}, {Nienartowicz}, {Ord{\'o}{\~n}ez-Blanco},
  {Panuzzo}, {Portell}, {Richards}, {Riello}, {Seabroke}, {Tanga},
  {Th{\'e}venin}, {Torra}, {Els}, {Gracia-Abril}, {Comoretto},
  {Garcia-Reinaldos}, {Lock}, {Mercier}, {Altmann}, {Andrae}, {Astraatmadja},
  {Bellas-Velidis}, {Benson}, {Berthier}, {Blomme}, {Busso}, {Carry},
  {Cellino}, {Clementini}, {Cowell}, {Creevey}, {Cuypers}, {Davidson}, {De
  Ridder}, {de Torres}, {Delchambre}, {Dell'Oro}, {Ducourant}, {Fr{\'e}mat},
  {Garc{\'\i}a-Torres}, {Gosset}, {Halbwachs}, {Hambly}, {Harrison}, {Hauser},
  {Hestroffer}, {Hodgkin}, {Huckle}, {Hutton}, {Jasniewicz}, {Jordan},
  {Kontizas}, {Korn}, {Lanzafame}, {Manteiga}, {Moitinho}, {Muinonen},
  {Osinde}, {Pancino}, {Pauwels}, {Petit}, {Recio-Blanco}, {Robin}, {Sarro},
  {Siopis}, {Smith}, {Smith}, {Sozzetti}, {Thuillot}, {van Reeven}, {Viala},
  {Abbas}, {Abreu Aramburu}, {Accart}, {Aguado}, {Allan}, {Allasia},
  {Altavilla}, {{\'A}lvarez}, {Alves}, {Anderson}, {Andrei}, {Anglada Varela},
  {Antiche}, {Antoja}, {Ant{\'o}n}, {Arcay}, {Atzei}, {Ayache}, {Bach},
  {Baker}, {Balaguer-N{\'u}{\~n}ez}, {Barache}, {Barata}, {Barbier}, {Barblan},
  {Baroni}, {Barrado y Navascu{\'e}s}, {Barros}, {Barstow}, {Becciani},
  {Bellazzini}, {Bellei}, {Bello Garc{\'\i}a}, {Belokurov}, {Bendjoya},
  {Berihuete}, {Bianchi}, {Bienaym{\'e}}, {Billebaud}, {Blagorodnova},
  {Blanco-Cuaresma}, {Boch}, {Bombrun}, {Borrachero}, {Bouquillon}, {Bourda},
  {Bouy}, {Bragaglia}, {Breddels}, {Brouillet}, {Br{\"u}semeister},
  {Bucciarelli}, {Budnik}, {Burgess}, {Burgon}, {Burlacu}, {Busonero}, {Buzzi},
  {Caffau}, {Cambras}, {Campbell}, {Cancelliere}, {Cantat-Gaudin}, {Carlucci},
  {Carrasco}, {Castellani}, {Charlot}, {Charnas}, {Charvet}, {Chassat},
  {Chiavassa}, {Clotet}, {Cocozza}, {Collins}, {Collins}, {Costigan}, {Crifo},
  {Cross}, {Crosta}, {Crowley}, {Dafonte}, {Damerdji}, {Dapergolas}, {David},
  {David}, {De Cat}, {de Felice}, {de Laverny}, {De Luise}, {De March}, {de
  Martino}, {de Souza}, {Debosscher}, {del Pozo}, {Delbo}, {Delgado},
  {Delgado}, {di Marco}, {Di Matteo}, {Diakite}, {Distefano}, {Dolding}, {Dos
  Anjos}, {Drazinos}, {Dur{\'a}n}, {Dzigan}, {Ecale}, {Edvardsson}, {Enke},
  {Erdmann}, {Escolar}, {Espina}, {Evans}, {Eynard Bontemps}, {Fabre},
  {Fabrizio}, {Faigler}, {Falc{\~a}o}, {Farr{\`a}s Casas}, {Faye}, {Federici},
  {Fedorets}, {Fern{\'a}ndez-Hern{\'a}ndez}, {Fernique}, {Fienga}, {Figueras},
  {Filippi}, {Findeisen}, {Fonti}, {Fouesneau}, {Fraile}, {Fraser}, {Fuchs},
  {Furnell}, {Gai}, {Galleti}, {Galluccio}, {Garabato}, {Garc{\'\i}a-Sedano},
  {Gar{\'e}}, {Garofalo}, {Garralda}, {Gavras}, {Gerssen}, {Geyer}, {Gilmore},
  {Girona}, {Giuffrida}, {Gomes}, {Gonz{\'a}lez-Marcos},
  {Gonz{\'a}lez-N{\'u}{\~n}ez}, {Gonz{\'a}lez-Vidal}, {Granvik}, {Guerrier},
  {Guillout}, {Guiraud}, {G{\'u}rpide}, {Guti{\'e}rrez-S{\'a}nchez}, {Guy},
  {Haigron}, {Hatzidimitriou}, {Haywood}, {Heiter}, {Helmi}, {Hobbs},
  {Hofmann}, {Holl}, {Holland}, {Hunt}, {Hypki}, {Icardi}, {Irwin}, {Jevardat
  de Fombelle}, {Jofr{\'e}}, {Jonker}, {Jorissen}, {Julbe}, {Karampelas},
  {Kochoska}, {Kohley}, {Kolenberg}, {Kontizas}, {Koposov}, {Kordopatis},
  {Koubsky}, {Kowalczyk}, {Krone-Martins}, {Kudryashova}, {Kull}, {Bachchan},
  {Lacoste-Seris}, {Lanza}, {Lavigne}, {Le Poncin-Lafitte}, {Lebreton},
  {Lebzelter}, {Leccia}, {Leclerc}, {Lecoeur-Taibi}, {Lemaitre}, {Lenhardt},
  {Leroux}, {Liao}, {Licata}, {Lindstr{\o}m}, {Lister}, {Livanou}, {Lobel},
  {L{\"o}ffler}, {L{\'o}pez}, {Lopez-Lozano}, {Lorenz}, {Loureiro},
  {MacDonald}, {Magalh{\~a}es Fernandes}, {Managau}, {Mann}, {Mantelet},
  {Marchal}, {Marchant}, {Marconi}, {Marie}, {Marinoni}, {Marrese},
  {Marschalk{\'o}}, {Marshall}, {Mart{\'\i}n-Fleitas}, {Martino}, {Mary},
  {Matijevi{\v{c}}}, {Mazeh}, {McMillan}, {Messina}, {Mestre}, {Michalik},
  {Millar}, {Miranda}, {Molina}, {Molinaro}, {Molinaro}, {Moln{\'a}r},
  {Moniez}, {Montegriffo}, {Monteiro}, {Mor}, {Mora}, {Morbidelli}, {Morel},
  {Morgenthaler}, {Morley}, {Morris}, {Mulone}, {Muraveva}, {Musella},
  {Narbonne}, {Nelemans}, {Nicastro}, {Noval}, {Ord{\'e}novic},
  {Ordieres-Mer{\'e}}, {Osborne}, {Pagani}, {Pagano}, {Pailler}, {Palacin},
  {Palaversa}, {Parsons}, {Paulsen}, {Pecoraro}, {Pedrosa}, {Pentik{\"a}inen},
  {Pereira}, {Pichon}, {Piersimoni}, {Pineau}, {Plachy}, {Plum}, {Poujoulet},
  {Pr{\v{s}}a}, {Pulone}, {Ragaini}, {Rago}, {Rambaux}, {Ramos-Lerate},
  {Ranalli}, {Rauw}, {Read}, {Regibo}, {Renk}, {Reyl{\'e}}, {Ribeiro},
  {Rimoldini}, {Ripepi}, {Riva}, {Rixon}, {Roelens}, {Romero-G{\'o}mez},
  {Rowell}, {Royer}, {Rudolph}, {Ruiz-Dern}, {Sadowski}, {Sagrist{\`a}
  Sell{\'e}s}, {Sahlmann}, {Salgado}, {Salguero}, {Sarasso}, {Savietto},
  {Schnorhk}, {Schultheis}, {Sciacca}, {Segol}, {Segovia}, {Segransan},
  {Serpell}, {Shih}, {Smareglia}, {Smart}, {Smith}, {Solano}, {Solitro},
  {Sordo}, {Soria Nieto}, {Souchay}, {Spagna}, {Spoto}, {Stampa}, {Steele},
  {Steidelm{\"u}ller}, {Stephenson}, {Stoev}, {Suess}, {S{\"u}veges}, {Surdej},
  {Szabados}, {Szegedi-Elek}, {Tapiador}, {Taris}, {Tauran}, {Taylor},
  {Teixeira}, {Terrett}, {Tingley}, {Trager}, {Turon}, {Ulla}, {Utrilla},
  {Valentini}, {van Elteren}, {Van Hemelryck}, {van Leeuwen}, {Varadi},
  {Vecchiato}, {Veljanoski}, {Via}, {Vicente}, {Vogt}, {Voss}, {Votruba},
  {Voutsinas}, {Walmsley}, {Weiler}, {Weingrill}, {Werner}, {Wevers},
  {Whitehead}, {Wyrzykowski}, {Yoldas}, {{\v{Z}}erjal}, {Zucker}, {Zurbach},
  {Zwitter}, {Alecu}, {Allen}, {Allende Prieto}, {Amorim},
  {Anglada-Escud{\'e}}, {Arsenijevic}, {Azaz}, {Balm}, {Beck}, {Bernstein},
  {Bigot}, {Bijaoui}, {Blasco}, {Bonfigli}, {Bono}, {Boudreault}, {Bressan},
  {Brown}, {Brunet}, {Bunclark}, {Buonanno}, {Butkevich}, {Carret}, {Carrion},
  {Chemin}, {Ch{\'e}reau}, {Corcione}, {Darmigny}, {de Boer}, {de Teodoro}, {de
  Zeeuw}, {Delle Luche}, {Domingues}, {Dubath}, {Fodor}, {Fr{\'e}zouls},
  {Fries}, {Fustes}, {Fyfe}, {Gallardo}, {Gallegos}, {Gardiol}, {Gebran},
  {Gomboc}, {G{\'o}mez}, {Grux}, {Gueguen}, {Heyrovsky}, {Hoar}, {Iannicola},
  {Isasi Parache}, {Janotto}, {Joliet}, {Jonckheere}, {Keil}, {Kim},
  {Klagyivik}, {Klar}, {Knude}, {Kochukhov}, {Kolka}, {Kos}, {Kutka}, {Lainey},
  {LeBouquin}, {Liu}, {Loreggia}, {Makarov}, {Marseille}, {Martayan},
  {Martinez-Rubi}, {Massart}, {Meynadier}, {Mignot}, {Munari}, {Nguyen},
  {Nordlander}, {Ocvirk}, {O'Flaherty}, {Olias Sanz}, {Ortiz}, {Osorio},
  {Oszkiewicz}, {Ouzounis}, {Palmer}, {Park}, {Pasquato}, {Peltzer}, {Peralta},
  {P{\'e}turaud}, {Pieniluoma}, {Pigozzi}, {Poels}, {Prat}, {Prod'homme},
  {Raison}, {Rebordao}, {Risquez}, {Rocca-Volmerange}, {Rosen}, {Ruiz-Fuertes},
  {Russo}, {Sembay}, {Serraller Vizcaino}, {Short}, {Siebert}, {Silva},
  {Sinachopoulos}, {Slezak}, {Soffel}, {Sosnowska}, {Strai{\v{z}}ys}, {ter
  Linden}, {Terrell}, {Theil}, {Tiede}, {Troisi}, {Tsalmantza}, {Tur},
  {Vaccari}, {Vachier}, {Valles}, {Van Hamme}, {Veltz}, {Virtanen}, {Wallut},
  {Wichmann}, {Wilkinson}, {Ziaeepour}, \& {Zschocke}}]{Prusti16}
{Gaia Collaboration}, {Prusti}, T., {de Bruijne}, J.~H.~J., {et~al.} 2016,
  \aap, 595, A1

\bibitem[{{Getman} {et~al.}(2014{\natexlab{a}}){Getman}, {Feigelson}, \&
  {Kuhn}}]{Getman14b}
{Getman}, K.~V., {Feigelson}, E.~D., \& {Kuhn}, M.~A. 2014{\natexlab{a}}, \apj,
  787, 109

\bibitem[{{Getman} {et~al.}(2018){Getman}, {Feigelson}, {Kuhn}, {Bate},
  {Broos}, \& {Garmire}}]{Getman18}
{Getman}, K.~V., {Feigelson}, E.~D., {Kuhn}, M.~A., {et~al.} 2018, \mnras, 476,
  1213

\bibitem[{{Getman} {et~al.}(2014{\natexlab{b}}){Getman}, {Feigelson}, {Kuhn},
  {Broos}, {Townsley}, {Naylor}, {Povich}, {Luhman}, \& {Garmire}}]{Getman14a}
{Getman}, K.~V., {Feigelson}, E.~D., {Kuhn}, M.~A., {et~al.}
  2014{\natexlab{b}}, \apj, 787, 108

\bibitem[{{Gro{\ss}schedl} {et~al.}(2021){Gro{\ss}schedl}, {Alves}, {Meingast},
  \& {Herbst-Kiss}}]{Grosschedl21}
{Gro{\ss}schedl}, J.~E., {Alves}, J., {Meingast}, S., \& {Herbst-Kiss}, G.
  2021, \aap, 647, A91

\bibitem[{{Guarcello} {et~al.}(2021){Guarcello}, {Biazzo}, {Drake}, {Micela},
  {Prisinzano}, {Sciortino}, {Damiani}, {Flaccomio}, {Neiner}, \&
  {Wright}}]{Guarcello21}
{Guarcello}, M.~G., {Biazzo}, K., {Drake}, J.~J., {et~al.} 2021, \aap, 650,
  A157

\bibitem[{{Gutermuth} {et~al.}(2008){Gutermuth}, {Myers}, {Megeath}, {Allen},
  {Pipher}, {Muzerolle}, {Porras}, {Winston}, \& {Fazio}}]{Gutermuth08}
{Gutermuth}, R.~A., {Myers}, P.~C., {Megeath}, S.~T., {et~al.} 2008, \apj, 674,
  336

\bibitem[{{Haisch} {et~al.}(2001){Haisch}, {Lada}, \& {Lada}}]{Haisch01}
{Haisch}, Jr., K.~E., {Lada}, E.~A., \& {Lada}, C.~J. 2001, \apjl, 553, L153

\bibitem[{{Hern{\'a}ndez} {et~al.}(2008){Hern{\'a}ndez}, {Hartmann}, {Calvet},
  {Jeffries}, {Gutermuth}, {Muzerolle}, \& {Stauffer}}]{Hernandez08}
{Hern{\'a}ndez}, J., {Hartmann}, L., {Calvet}, N., {et~al.} 2008, \apj, 686,
  1195

\bibitem[{{Hern{\'a}ndez} {et~al.}(2007){Hern{\'a}ndez}, {Hartmann}, {Megeath},
  {Gutermuth}, {Muzerolle}, {Calvet}, {Vivas}, {Brice{\~n}o}, {Allen},
  {Stauffer}, {Young}, \& {Fazio}}]{Hernandez07}
{Hern{\'a}ndez}, J., {Hartmann}, L., {Megeath}, T., {et~al.} 2007, \apj, 662,
  1067

\bibitem[{{Hern{\'a}ndez} {et~al.}(2010){Hern{\'a}ndez}, {Morales-Calderon},
  {Calvet}, {Hartmann}, {Muzerolle}, {Gutermuth}, {Luhman}, \&
  {Stauffer}}]{Hernandez10}
{Hern{\'a}ndez}, J., {Morales-Calderon}, M., {Calvet}, N., {et~al.} 2010, \apj,
  722, 1226

\bibitem[{{Jeffries}(2017)}]{Jeffries17}
{Jeffries}, R.~D. 2017, \memsai, 88, 637

\bibitem[{{Kharchenko} {et~al.}(2013){Kharchenko}, {Piskunov}, {Schilbach},
  {R{\"o}ser}, \& {Scholz}}]{Kharchenko13}
{Kharchenko}, N.~V., {Piskunov}, A.~E., {Schilbach}, E., {R{\"o}ser}, S., \&
  {Scholz}, R.~D. 2013, \aap, 558, A53

\bibitem[{{Koenig} \& {Leisawitz}(2014)}]{Koenig14}
{Koenig}, X.~P. \& {Leisawitz}, D.~T. 2014, \apj, 791, 131

\bibitem[{{Kuhn} {et~al.}(2019){Kuhn}, {Hillenbrand}, {Sills}, {Feigelson}, \&
  {Getman}}]{Kuhn19}
{Kuhn}, M.~A., {Hillenbrand}, L.~A., {Sills}, A., {Feigelson}, E.~D., \&
  {Getman}, K.~V. 2019, \apj, 870, 32

\bibitem[{{Lada} \& {Adams}(1992)}]{Lada92}
{Lada}, C.~J. \& {Adams}, F.~C. 1992, \apj, 393, 278

\bibitem[{{Lindegren} {et~al.}(2021){Lindegren}, {Klioner}, {Hern{\'a}ndez},
  {Bombrun}, {Ramos-Lerate}, {Steidelm{\"u}ller}, {Bastian}, {Biermann}, {de
  Torres}, {Gerlach}, {Geyer}, {Hilger}, {Hobbs}, {Lammers}, {McMillan},
  {Stephenson}, {Casta{\~n}eda}, {Davidson}, {Fabricius}, {Gracia-Abril},
  {Portell}, {Rowell}, {Teyssier}, {Torra}, {Bartolom{\'e}}, {Clotet},
  {Garralda}, {Gonz{\'a}lez-Vidal}, {Torra}, {Abbas}, {Altmann}, {Anglada
  Varela}, {Balaguer-N{\'u}{\~n}ez}, {Balog}, {Barache}, {Becciani}, {Bernet},
  {Bertone}, {Bianchi}, {Bouquillon}, {Brown}, {Bucciarelli}, {Busonero},
  {Butkevich}, {Buzzi}, {Cancelliere}, {Carlucci}, {Charlot}, {Cioni},
  {Crosta}, {Crowley}, {del Peloso}, {del Pozo}, {Drimmel}, {Esquej}, {Fienga},
  {Fraile}, {Gai}, {Garcia-Reinaldos}, {Guerra}, {Hambly}, {Hauser},
  {Jan{\ss}en}, {Jordan}, {Kostrzewa-Rutkowska}, {Lattanzi}, {Liao}, {Licata},
  {Lister}, {L{\"o}ffler}, {Marchant}, {Masip}, {Mignard}, {Mints}, {Molina},
  {Mora}, {Morbidelli}, {Murphy}, {Pagani}, {Panuzzo}, {Pe{\~n}alosa Esteller},
  {Poggio}, {Re Fiorentin}, {Riva}, {Sagrist{\`a} Sell{\'e}s}, {Sanchez
  Gimenez}, {Sarasso}, {Sciacca}, {Siddiqui}, {Smart}, {Souami}, {Spagna},
  {Steele}, {Taris}, {Utrilla}, {van Reeven}, \& {Vecchiato}}]{Lindegren21}
{Lindegren}, L., {Klioner}, S.~A., {Hern{\'a}ndez}, J., {et~al.} 2021, \aap,
  649, A2

\bibitem[{{Mamajek}(2009)}]{Mamajek09}
{Mamajek}, E.~E. 2009, in American Institute of Physics Conference Series, Vol.
  1158, Exoplanets and Disks: Their Formation and Diversity, ed. T.~{Usuda},
  M.~{Tamura}, \& M.~{Ishii}, 3--10

\bibitem[{{Meingast} {et~al.}(2021){Meingast}, {Alves}, \&
  {Rottensteiner}}]{Meingast2021}
{Meingast}, S., {Alves}, J., \& {Rottensteiner}, A. 2021, \aap, 645, A84

\bibitem[{{Muzerolle} {et~al.}(2010){Muzerolle}, {Allen}, {Megeath},
  {Hern{\'a}ndez}, \& {Gutermuth}}]{Muzerolle10}
{Muzerolle}, J., {Allen}, L.~E., {Megeath}, S.~T., {Hern{\'a}ndez}, J., \&
  {Gutermuth}, R.~A. 2010, \apj, 708, 1107

\bibitem[{{Panja} {et~al.}(2021){Panja}, {Chen}, {Dutta}, {Sun}, {Gao}, \&
  {Mondal}}]{Panja21}
{Panja}, A., {Chen}, W.~P., {Dutta}, S., {et~al.} 2021, \apj, 910, 80

\bibitem[{{Parker} {et~al.}(2021){Parker}, {Alcock}, {Nicholson}, {Pani{\'c}},
  \& {Goodwin}}]{Parker21}
{Parker}, R.~J., {Alcock}, H.~L., {Nicholson}, R.~B., {Pani{\'c}}, O., \&
  {Goodwin}, S.~P. 2021, \apj, 913, 95

\bibitem[{{Pfalzner} {et~al.}(2014){Pfalzner}, {Steinhausen}, \&
  {Menten}}]{Pfalzner14}
{Pfalzner}, S., {Steinhausen}, M., \& {Menten}, K. 2014, \apjl, 793, L34

\bibitem[{{Prisinzano} {et~al.}(2019){Prisinzano}, {Damiani}, {Kalari},
  {Jeffries}, {Bonito}, {Micela}, {Wright}, {Jackson}, {Tognelli}, {Guarcello},
  {Vink}, {Klutsch}, {Jim{\'e}nez-Esteban}, {Roccatagliata},
  {Tautvai{\v{s}}ien{\.{e}}}, {Gilmore}, {Randich}, {Alfaro}, {Flaccomio},
  {Koposov}, {Lanzafame}, {Pancino}, {Bergemann}, {Carraro}, {Franciosini},
  {Frasca}, {Gonneau}, {Hourihane}, {Jofr{\'e}}, {Lewis}, {Magrini}, {Monaco},
  {Morbidelli}, {Sacco}, {Worley}, \& {Zaggia}}]{Prisinzano19}
{Prisinzano}, L., {Damiani}, F., {Kalari}, V., {et~al.} 2019, \aap, 623, A159

\bibitem[{{Ribas} {et~al.}(2015){Ribas}, {Bouy}, \& {Mer{\'\i}n}}]{Ribas15}
{Ribas}, {\'A}., {Bouy}, H., \& {Mer{\'\i}n}, B. 2015, \aap, 576, A52

\bibitem[{{Richert} {et~al.}(2015){Richert}, {Feigelson}, {Getman}, \&
  {Kuhn}}]{Richert15}
{Richert}, A. J.~W., {Feigelson}, E.~D., {Getman}, K.~V., \& {Kuhn}, M.~A.
  2015, \apj, 811, 10

\bibitem[{{Riello} {et~al.}(2021){Riello}, {De Angeli}, {Evans}, {Montegriffo},
  {Carrasco}, {Busso}, {Palaversa}, {Burgess}, {Diener}, {Davidson}, {Rowell},
  {Fabricius}, {Jordi}, {Bellazzini}, {Pancino}, {Harrison}, {Cacciari}, {van
  Leeuwen}, {Hambly}, {Hodgkin}, {Osborne}, {Altavilla}, {Barstow}, {Brown},
  {Castellani}, {Cowell}, {De Luise}, {Gilmore}, {Giuffrida}, {Hidalgo},
  {Holland}, {Marinoni}, {Pagani}, {Piersimoni}, {Pulone}, {Ragaini}, {Rainer},
  {Richards}, {Sanna}, {Walton}, {Weiler}, \& {Yoldas}}]{Riello21}
{Riello}, M., {De Angeli}, F., {Evans}, D.~W., {et~al.} 2021, \aap, 649, A3

\bibitem[{{Saurin} {et~al.}(2012){Saurin}, {Bica}, \& {Bonatto}}]{Saurin12}
{Saurin}, T.~A., {Bica}, E., \& {Bonatto}, C. 2012, \mnras, 421, 3206

\bibitem[{{Schiller} {et~al.}(2018){Schiller}, {Bizzarro}, \&
  {Fernandes}}]{Schiller18}
{Schiller}, M., {Bizzarro}, M., \& {Fernandes}, V.~A. 2018, \nat, 555, 507

\bibitem[{{Skrutskie} {et~al.}(2006){Skrutskie}, {Cutri}, {Stiening},
  {Weinberg}, {Schneider}, {Carpenter}, {Beichman}, {Capps}, {Chester},
  {Elias}, {Huchra}, {Liebert}, {Lonsdale}, {Monet}, {Price}, {Seitzer},
  {Jarrett}, {Kirkpatrick}, {Gizis}, {Howard}, {Evans}, {Fowler}, {Fullmer},
  {Hurt}, {Light}, {Kopan}, {Marsh}, {McCallon}, {Tam}, {Van Dyk}, \&
  {Wheelock}}]{Skrutskie06}
{Skrutskie}, M.~F., {Cutri}, R.~M., {Stiening}, R., {et~al.} 2006, \aj, 131,
  1163

\bibitem[{{Taylor}(2005)}]{Taylor05}
{Taylor}, M.~B. 2005, in Astronomical Society of the Pacific Conference Series,
  Vol. 347, Astronomical Data Analysis Software and Systems XIV, ed.
  P.~{Shopbell}, M.~{Britton}, \& R.~{Ebert}, 29

\bibitem[{{van Terwisga} {et~al.}(2019){van Terwisga}, {Hacar}, \& {van
  Dishoeck}}]{vanTerwisga19}
{van Terwisga}, S.~E., {Hacar}, A., \& {van Dishoeck}, E.~F. 2019, \aap, 628,
  A85

\bibitem[{{Vincke} \& {Pfalzner}(2018)}]{Vincke18}
{Vincke}, K. \& {Pfalzner}, S. 2018, \apj, 868, 1

\bibitem[{{Wang} {et~al.}(2011){Wang}, {Feigelson}, {Townsley}, {Broos},
  {Getman}, {Wolk}, {Preibisch}, {Stassun}, {Moffat}, {Garmire}, {King},
  {McCaughrean}, \& {Zinnecker}}]{Wang11}
{Wang}, J., {Feigelson}, E.~D., {Townsley}, L.~K., {et~al.} 2011, \apjs, 194,
  11

\bibitem[{{Wilking} {et~al.}(2004){Wilking}, {Meyer}, {Greene}, {Mikhail}, \&
  {Carlson}}]{Wilking04}
{Wilking}, B.~A., {Meyer}, M.~R., {Greene}, T.~P., {Mikhail}, A., \& {Carlson},
  G. 2004, \aj, 127, 1131

\bibitem[{{Wright}(2020)}]{Wright20}
{Wright}, N.~J. 2020, \nar, 90, 101549

\bibitem[{{Yasui} {et~al.}(2014){Yasui}, {Kobayashi}, {Tokunaga}, \&
  {Saito}}]{Yasui14}
{Yasui}, C., {Kobayashi}, N., {Tokunaga}, A.~T., \& {Saito}, M. 2014, \mnras,
  442, 2543

\bibitem[{{Young} {et~al.}(2015){Young}, {Young}, {Lai}, {Dunham}, \&
  {Evans}}]{Young15}
{Young}, K.~E., {Young}, C.~H., {Lai}, S.-P., {Dunham}, M.~M., \& {Evans},
  Neal~J., I. 2015, \aj, 150, 40

\end{thebibliography}

\newpage
\begin{appendix} 
\onecolumn
\section{Results. Figures and tables}
\label{appendix_clusters_general}

\begin{figure} [h]
 \centering
    \includegraphics[width=0.37\textwidth]{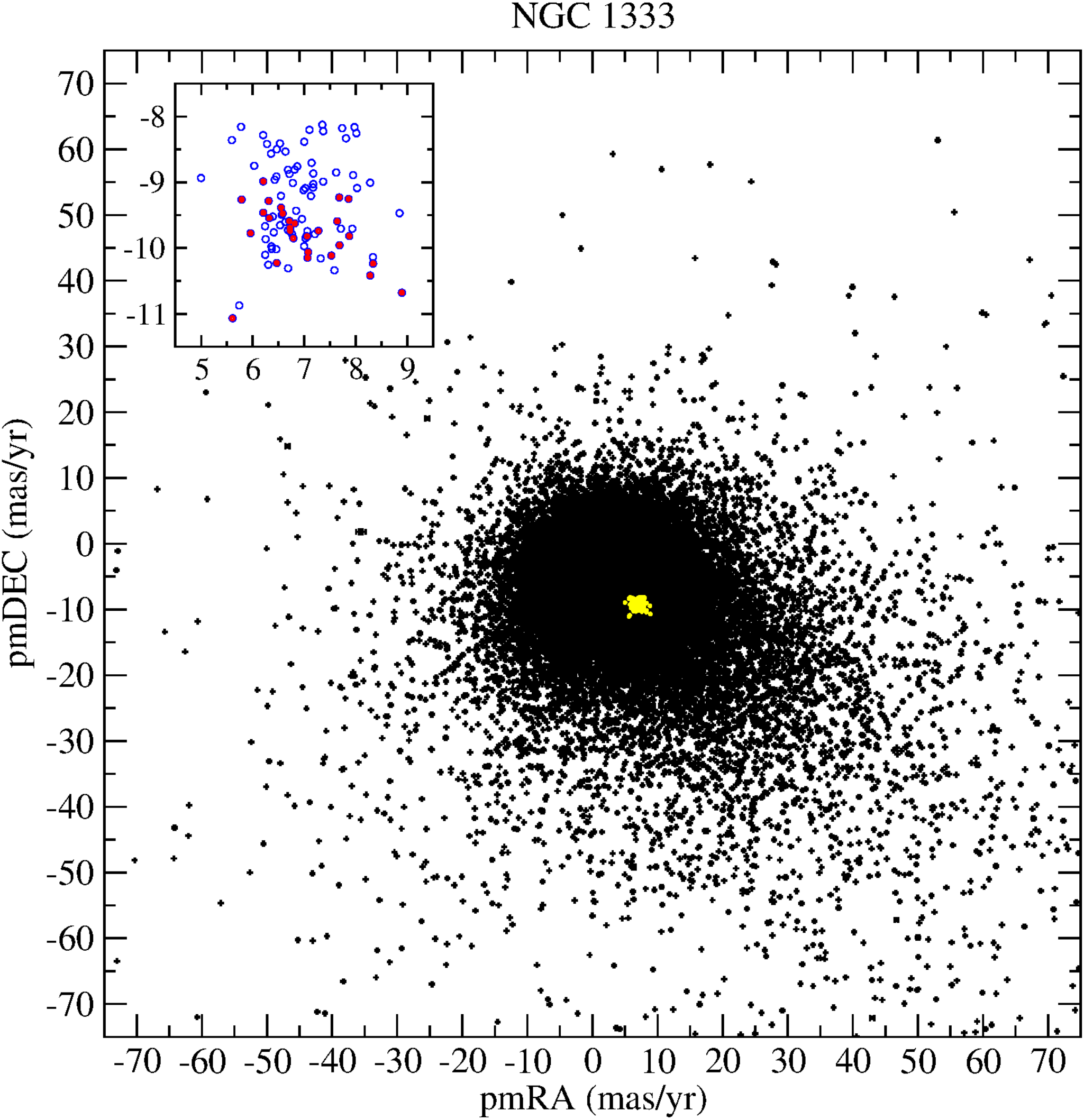}
    \includegraphics[width=0.37\textwidth]{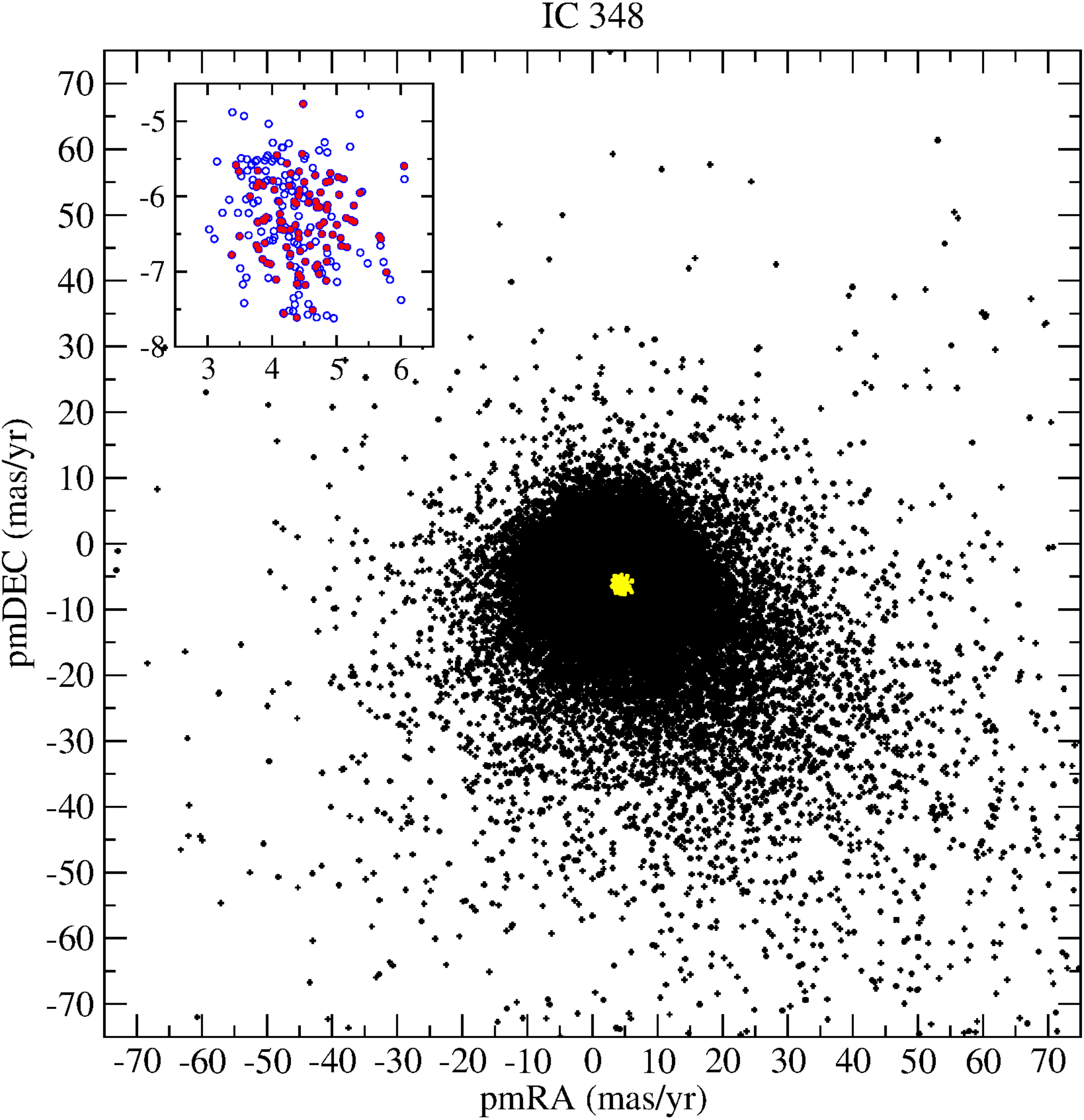}
\end{figure}
\begin{figure} [h]
 \centering
    \includegraphics[width=0.37\textwidth]{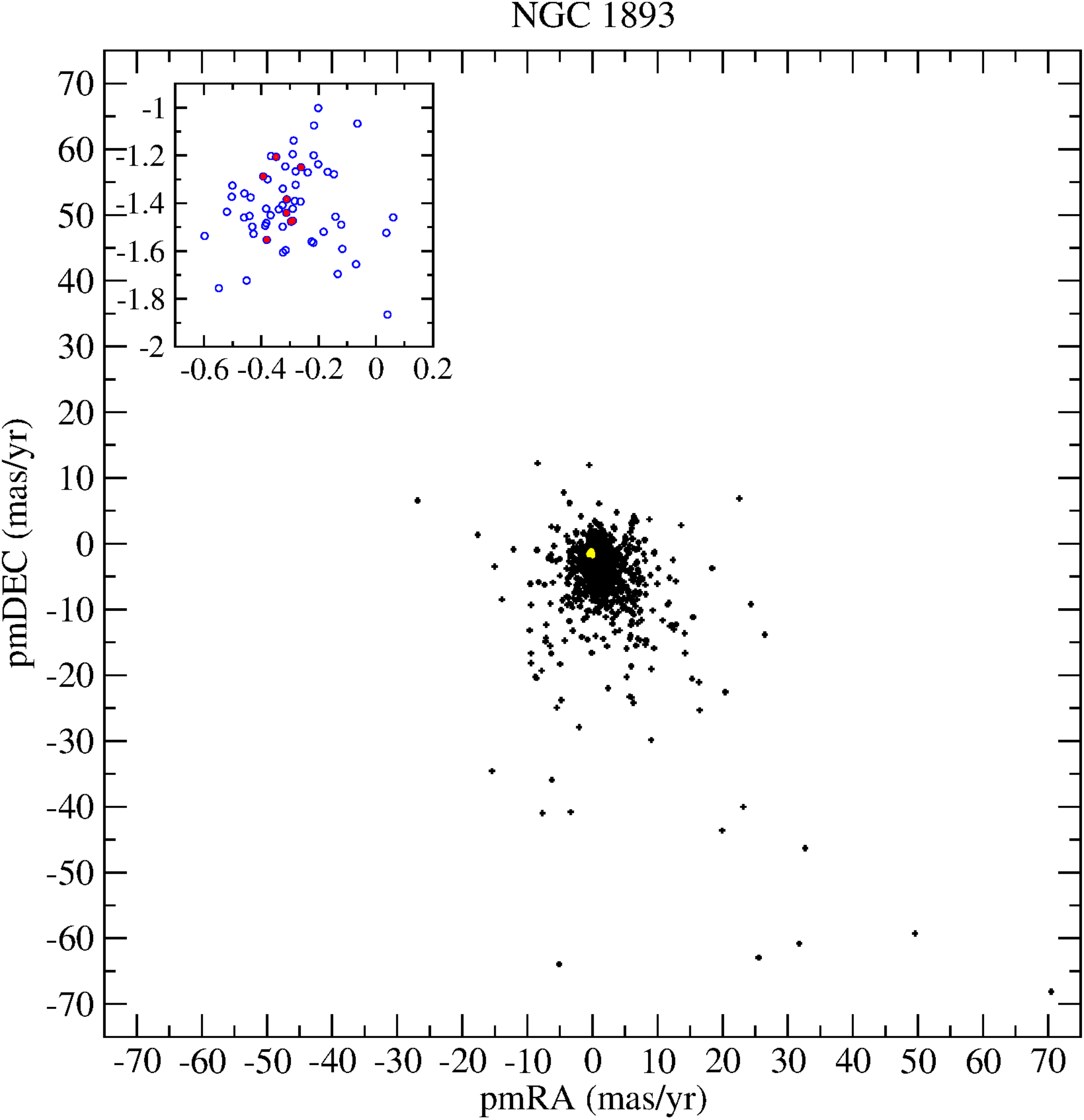}
   \includegraphics[width=0.37\textwidth]{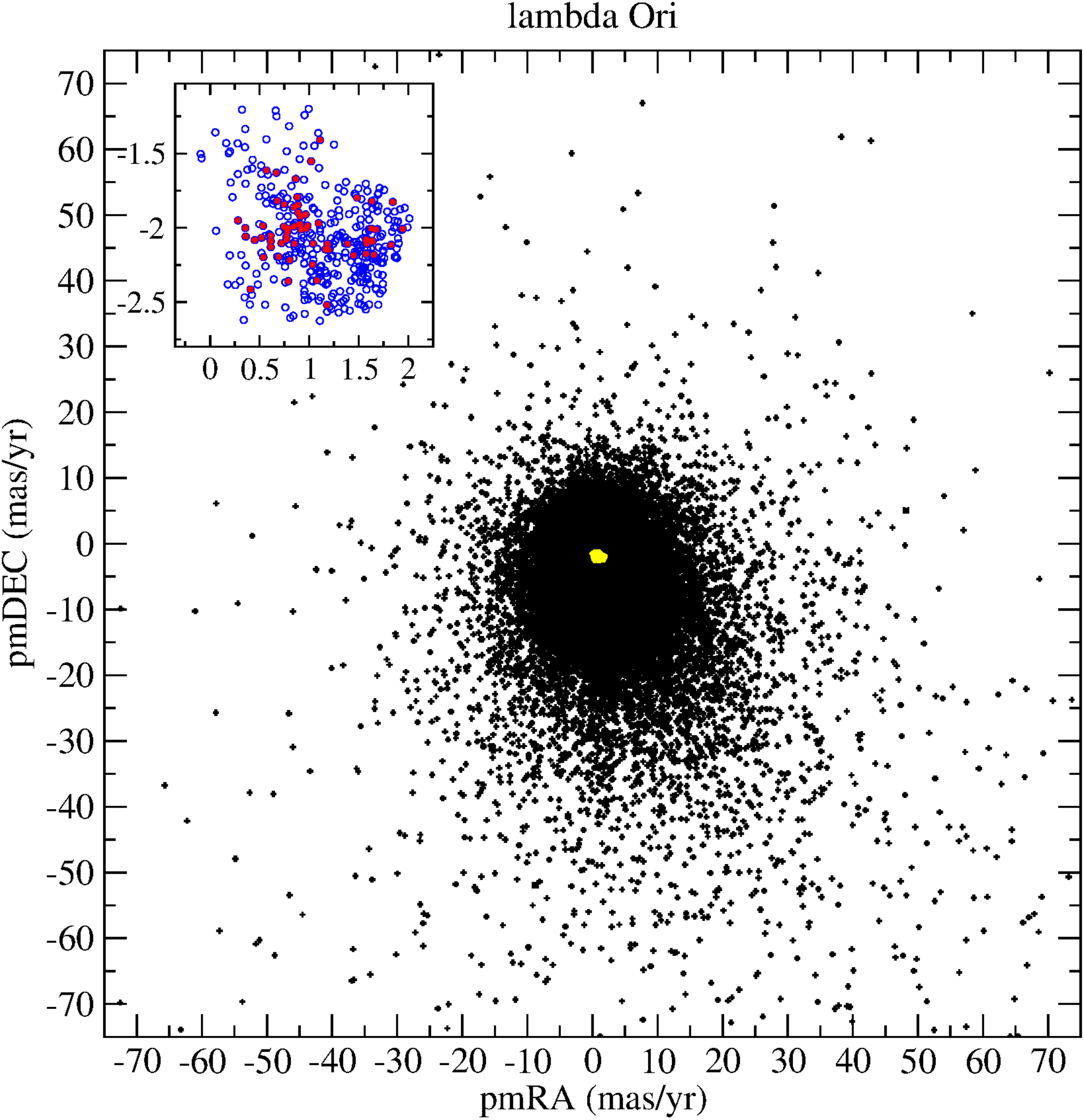} 
\end{figure}
\begin{figure} [h]
 \centering
      \includegraphics[width=0.37\textwidth]{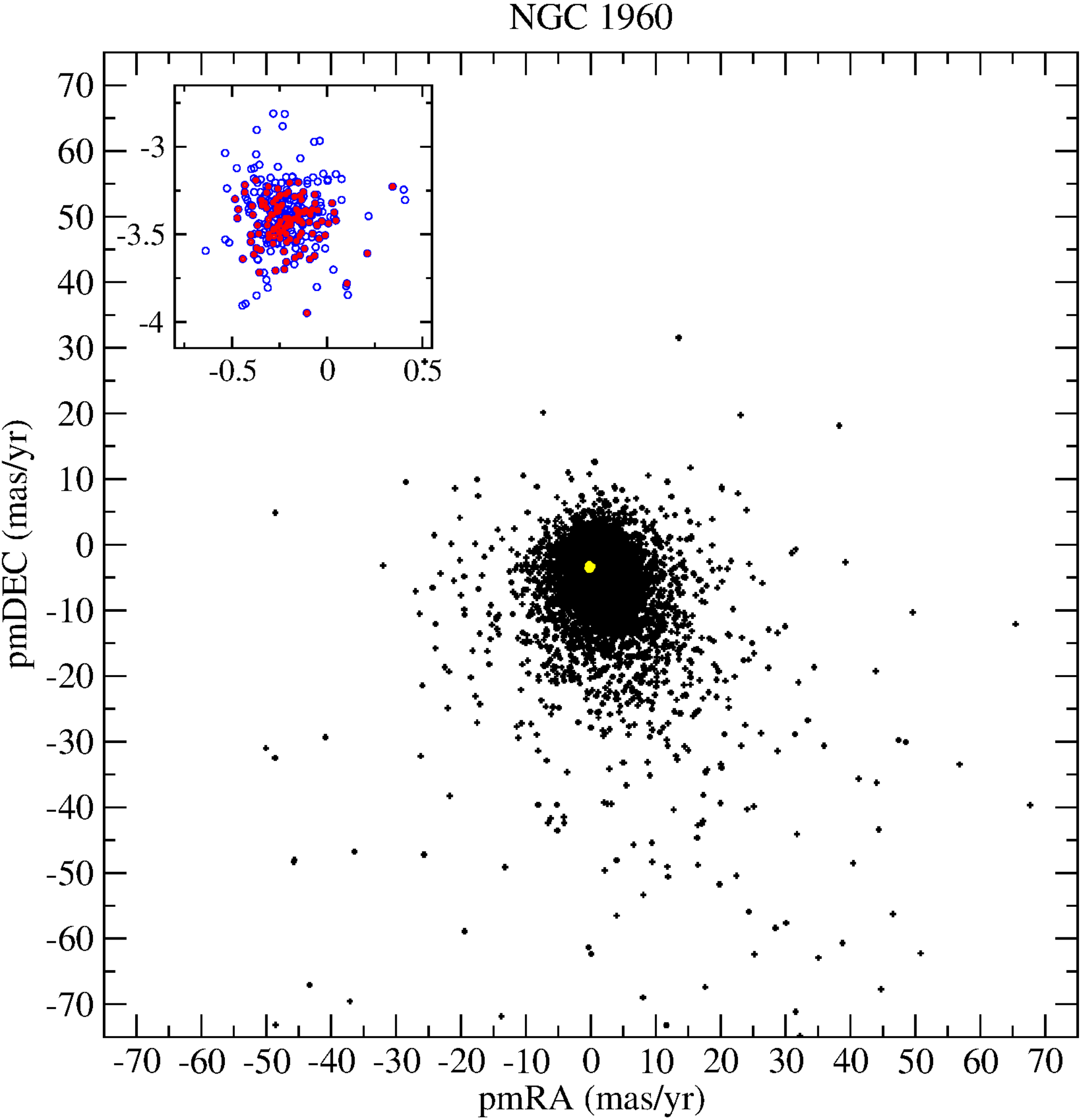}
      \includegraphics[width=0.37\textwidth]{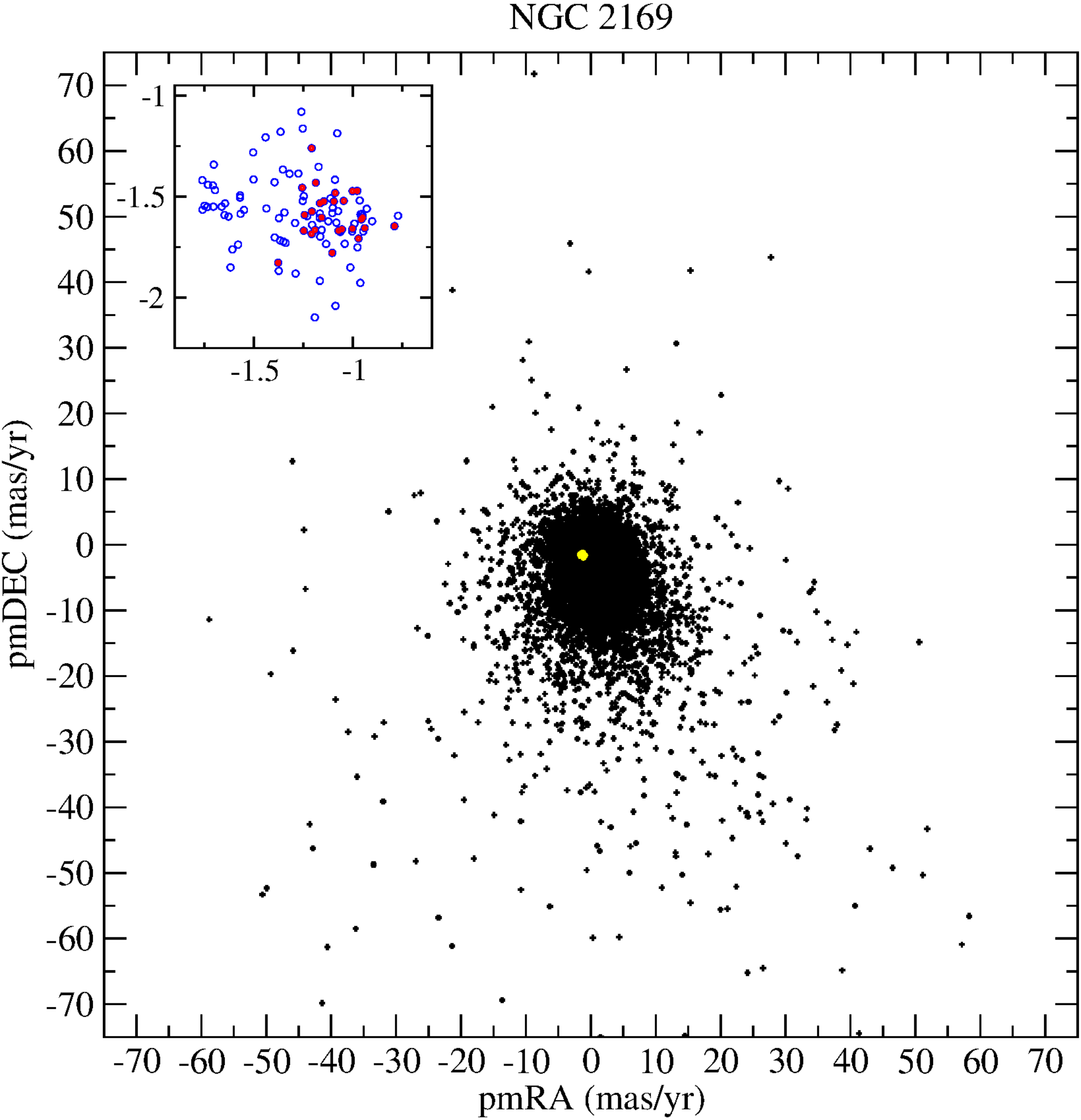}
\end{figure} 
\begin{figure} [h]
 \centering
    \includegraphics[width=0.37\textwidth]{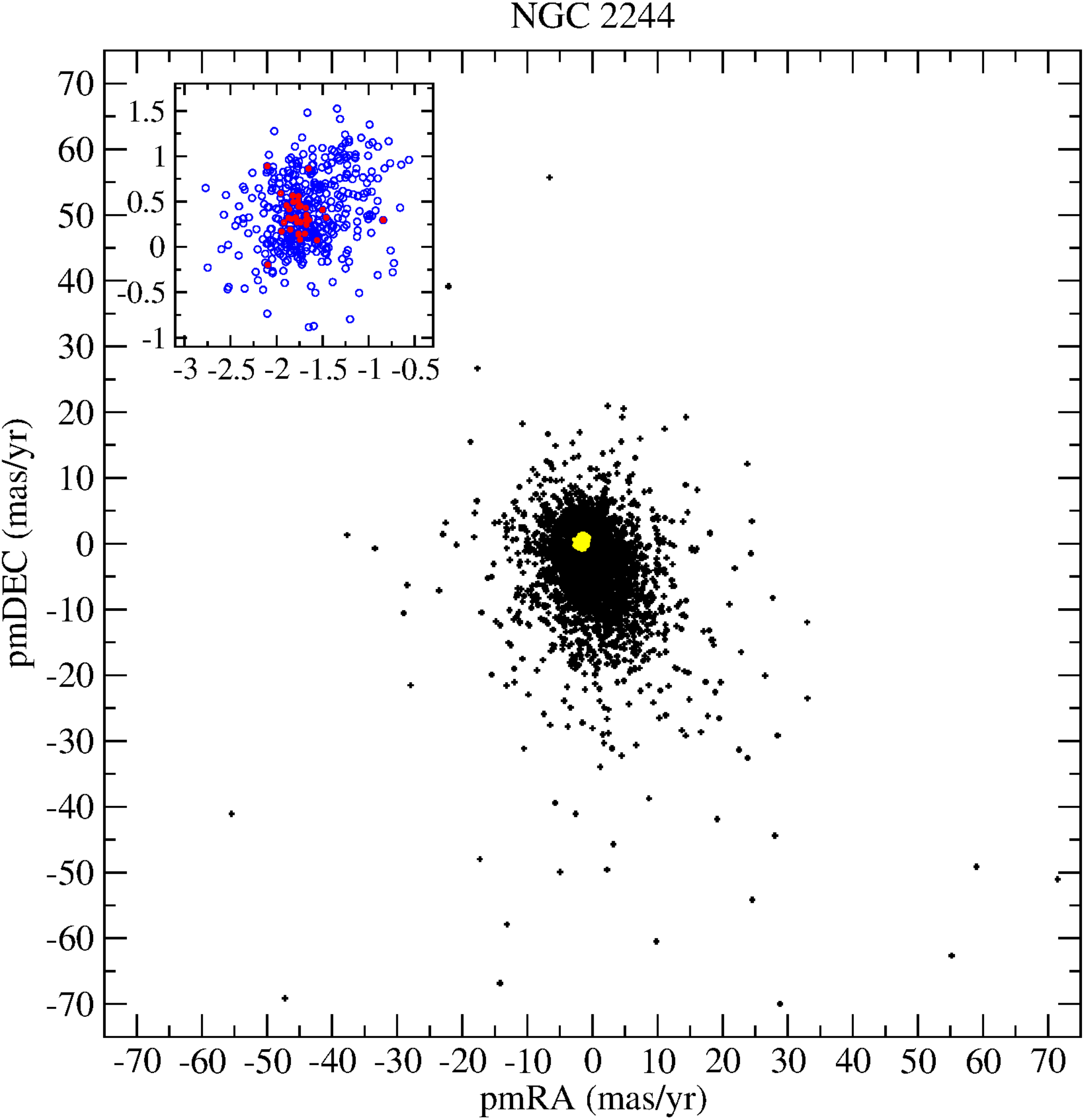}
    \includegraphics[width=0.37\textwidth]{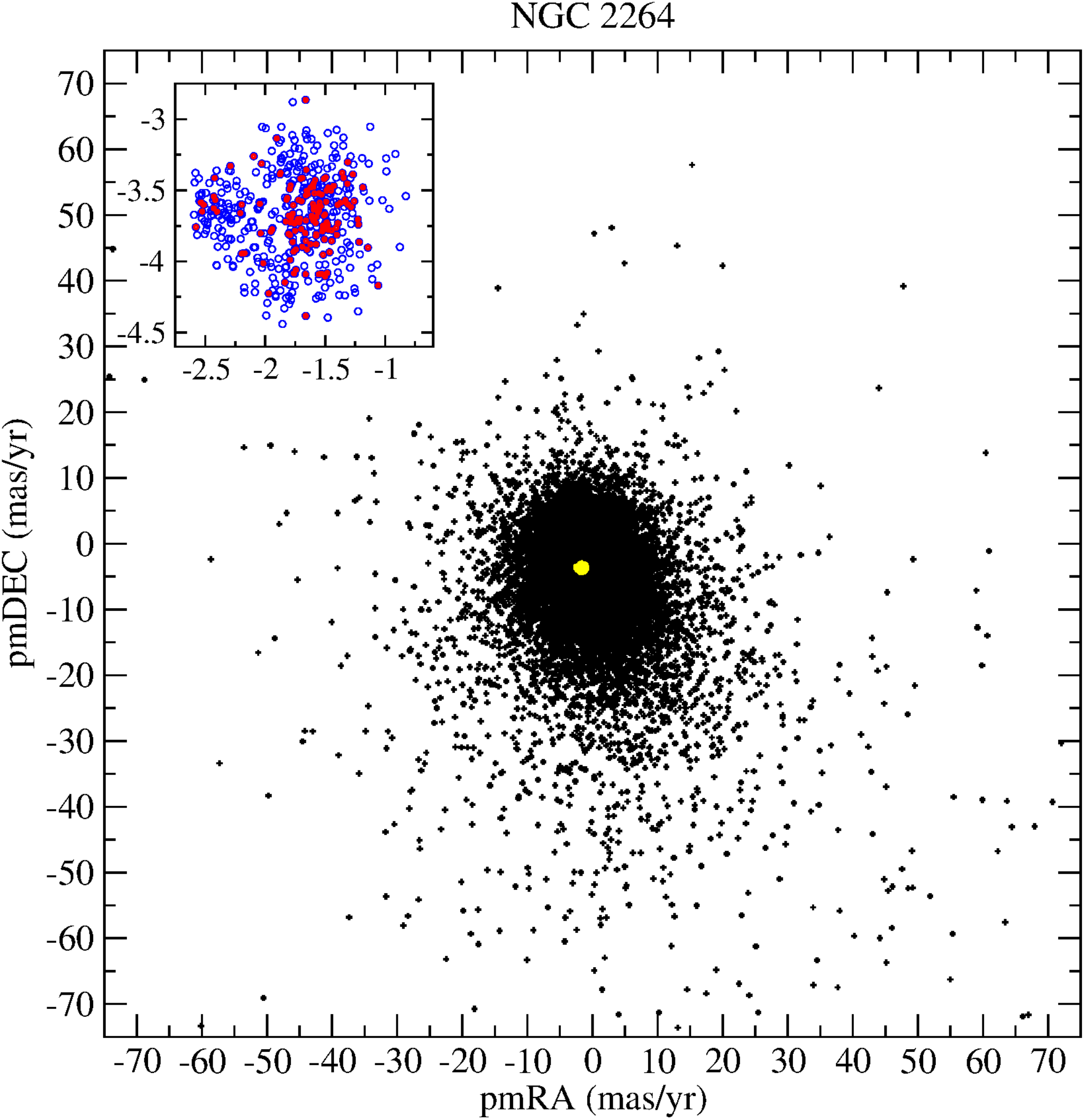}  
\end{figure}
\begin{figure} [h]
 \centering
    \includegraphics[width=0.37\textwidth]{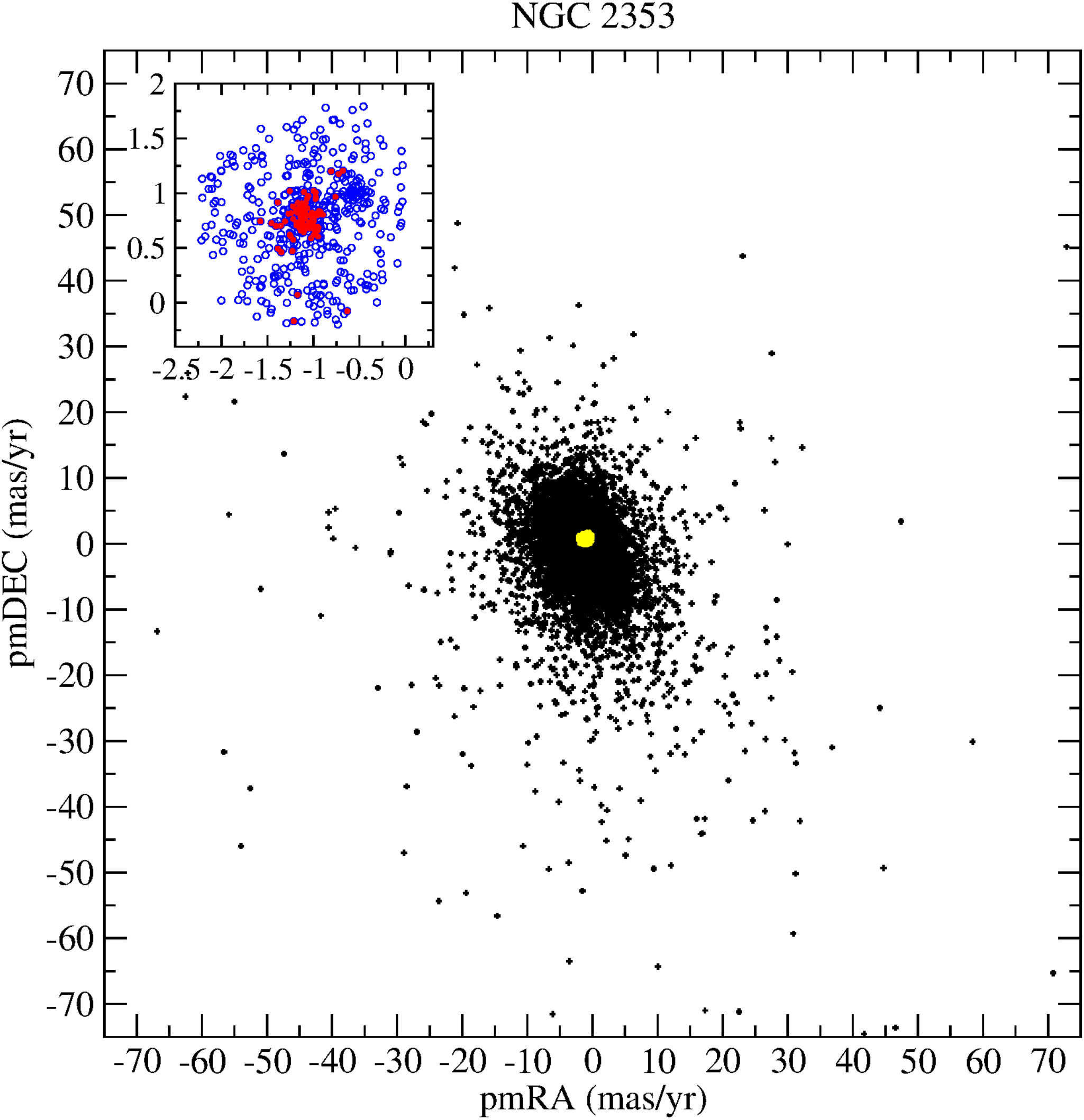}
    \includegraphics[width=0.37\textwidth]{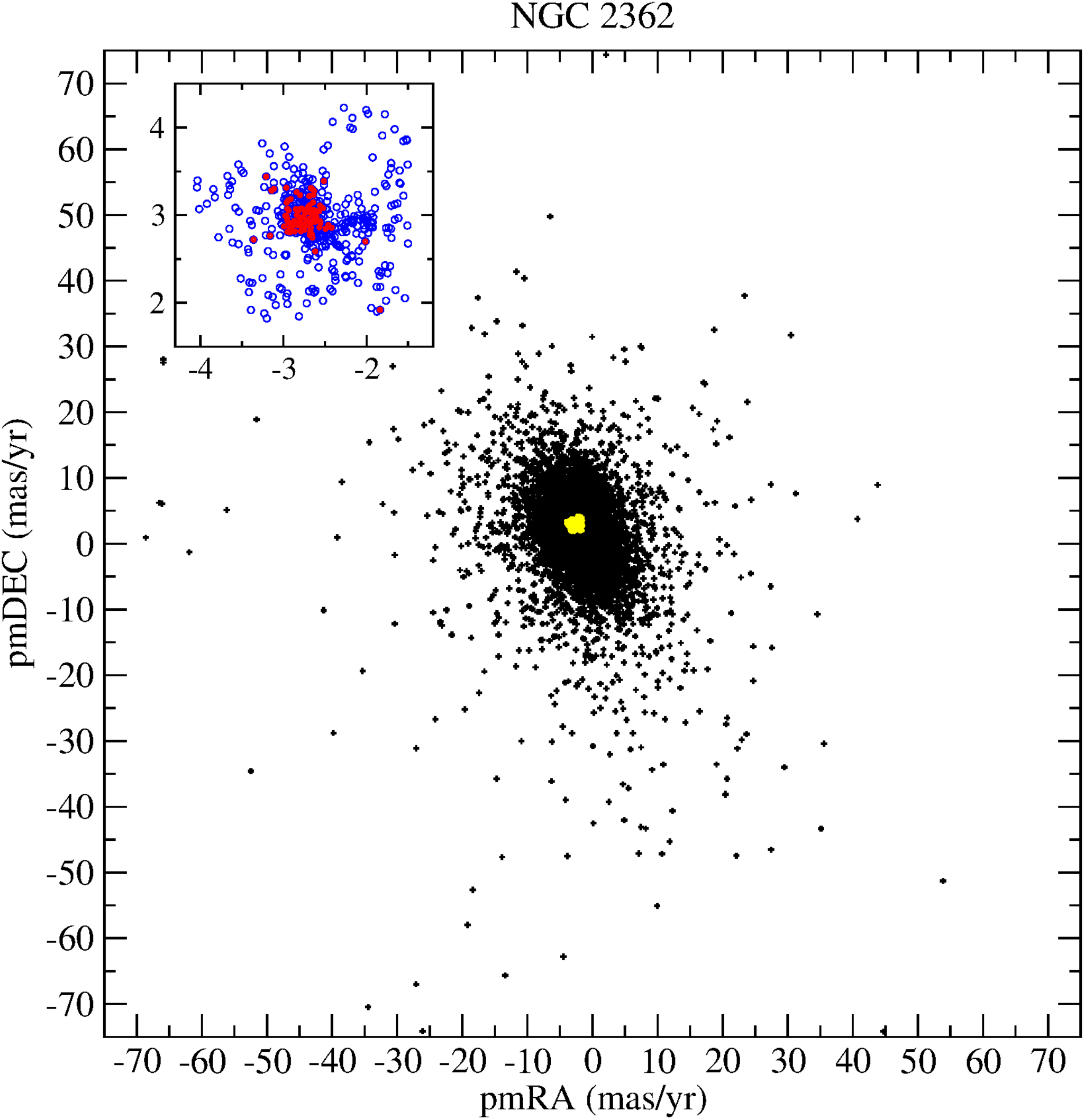}
\end{figure}
\begin{figure} [h]
 \centering
    \includegraphics[width=0.37\textwidth]{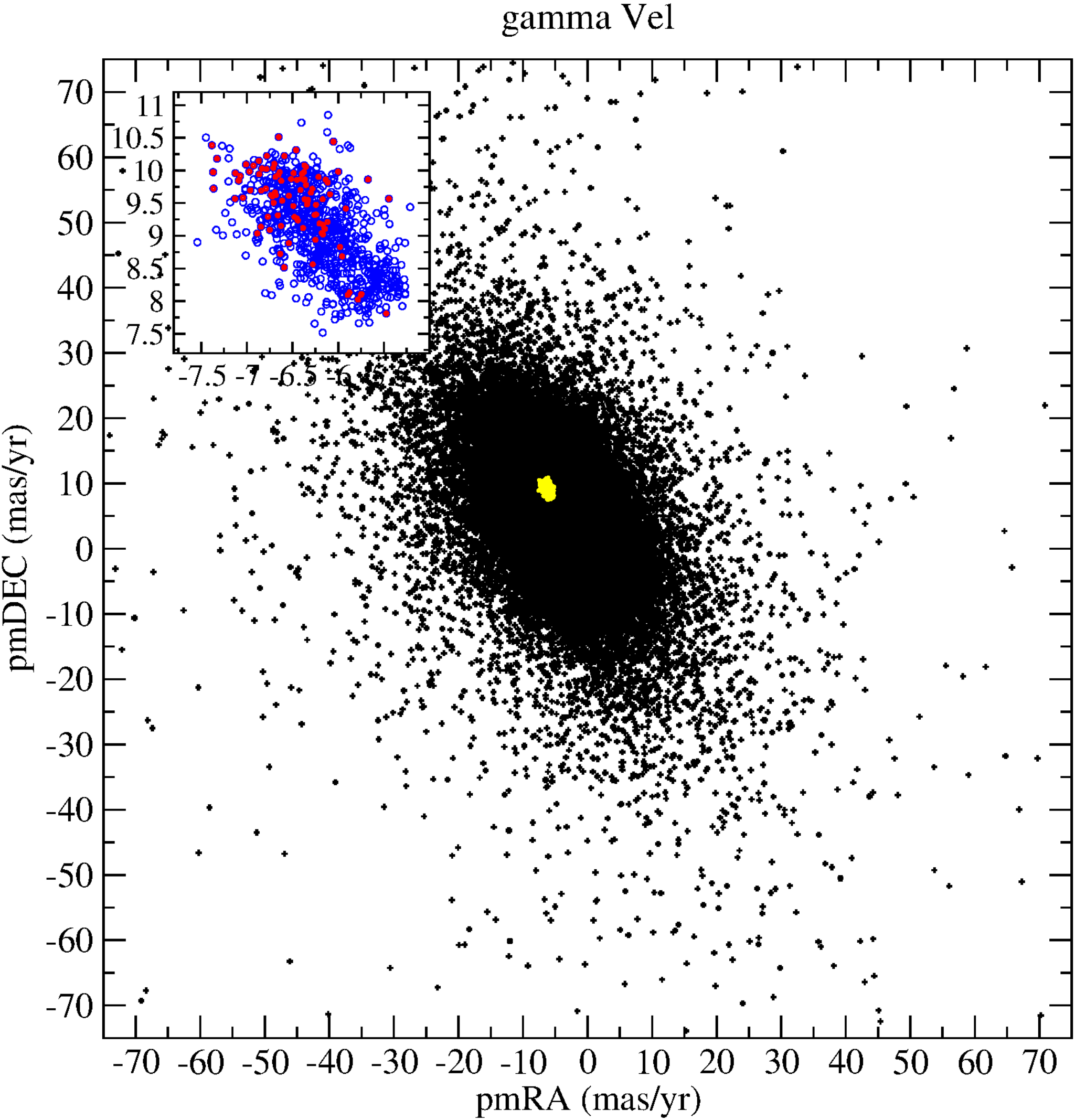}
    \includegraphics[width=0.37\textwidth]{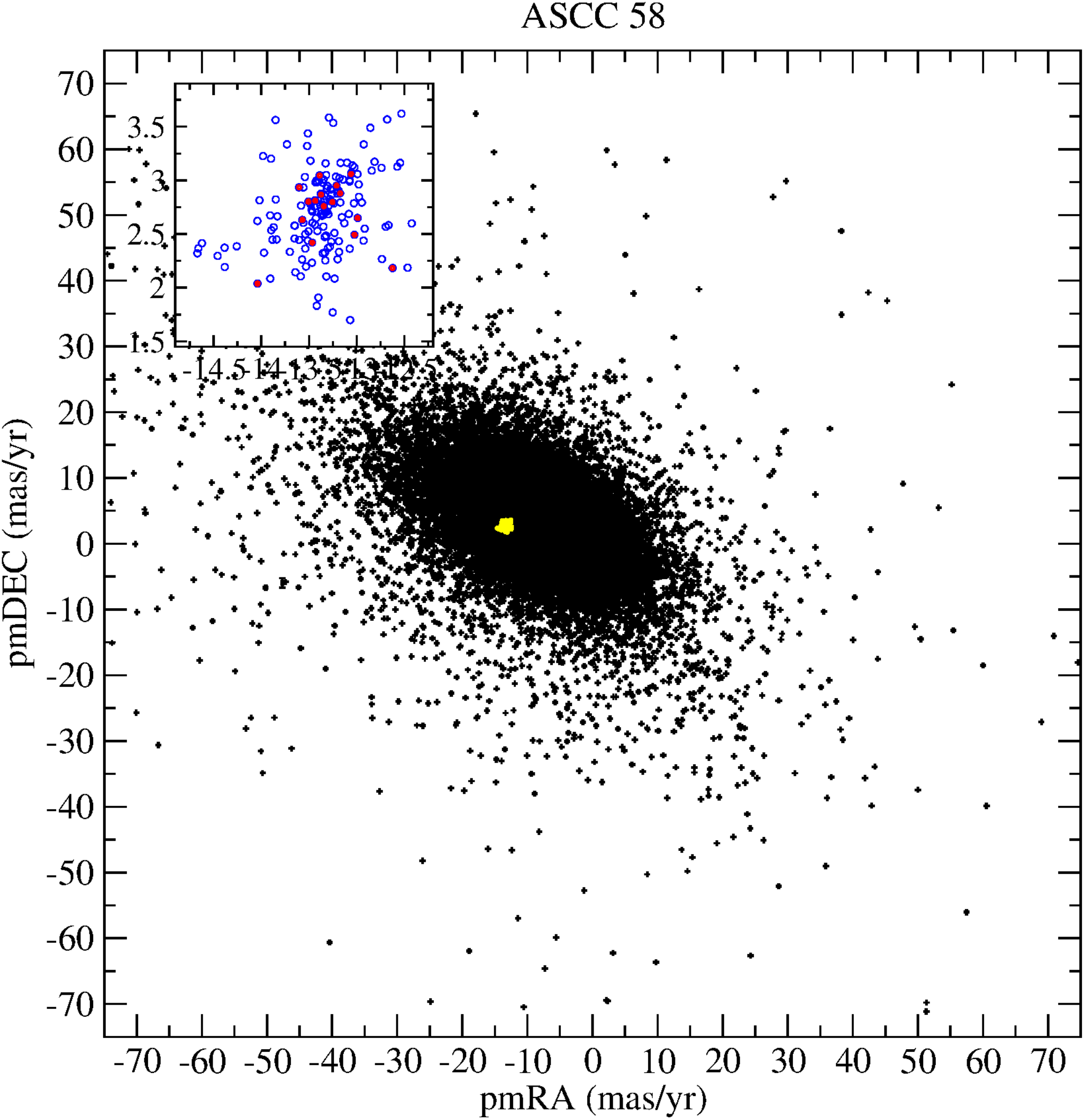}
\end{figure}
\begin{figure} [h]
 \centering
    \includegraphics[width=0.37\textwidth]{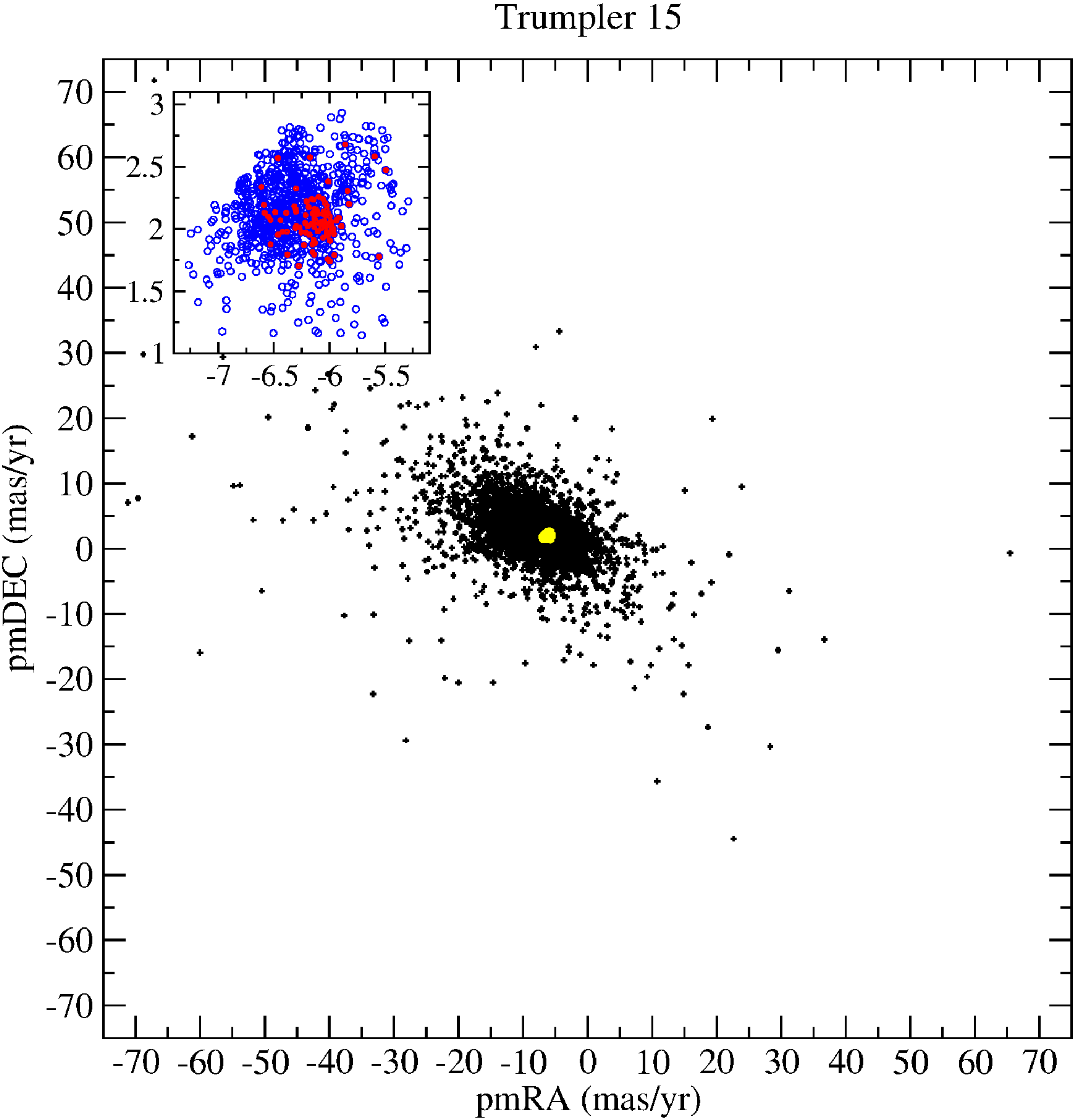}
    \includegraphics[width=0.37\textwidth]{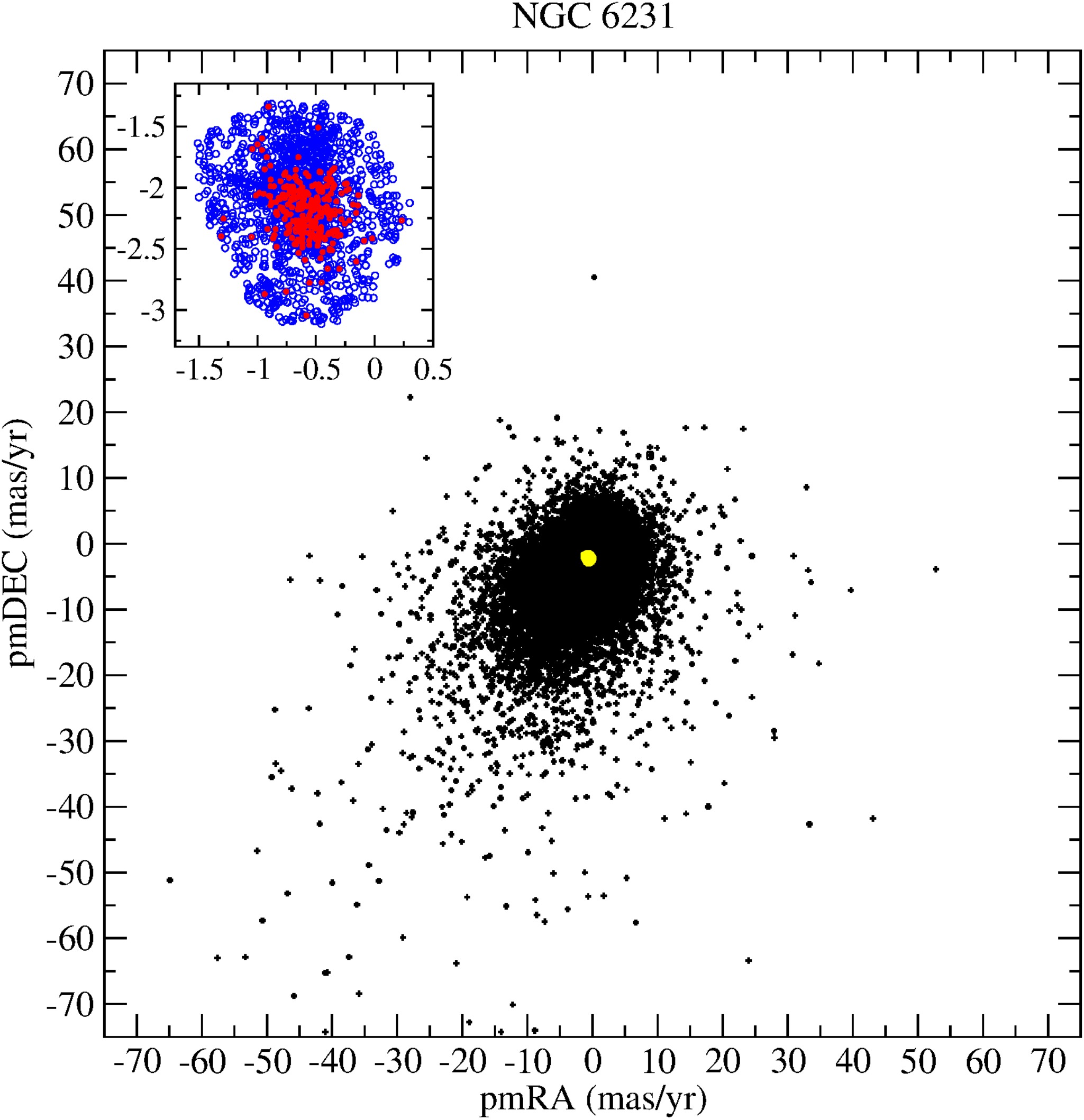}
\end{figure}
\begin{figure} [h]
 \centering
    \includegraphics[width=0.37\textwidth]{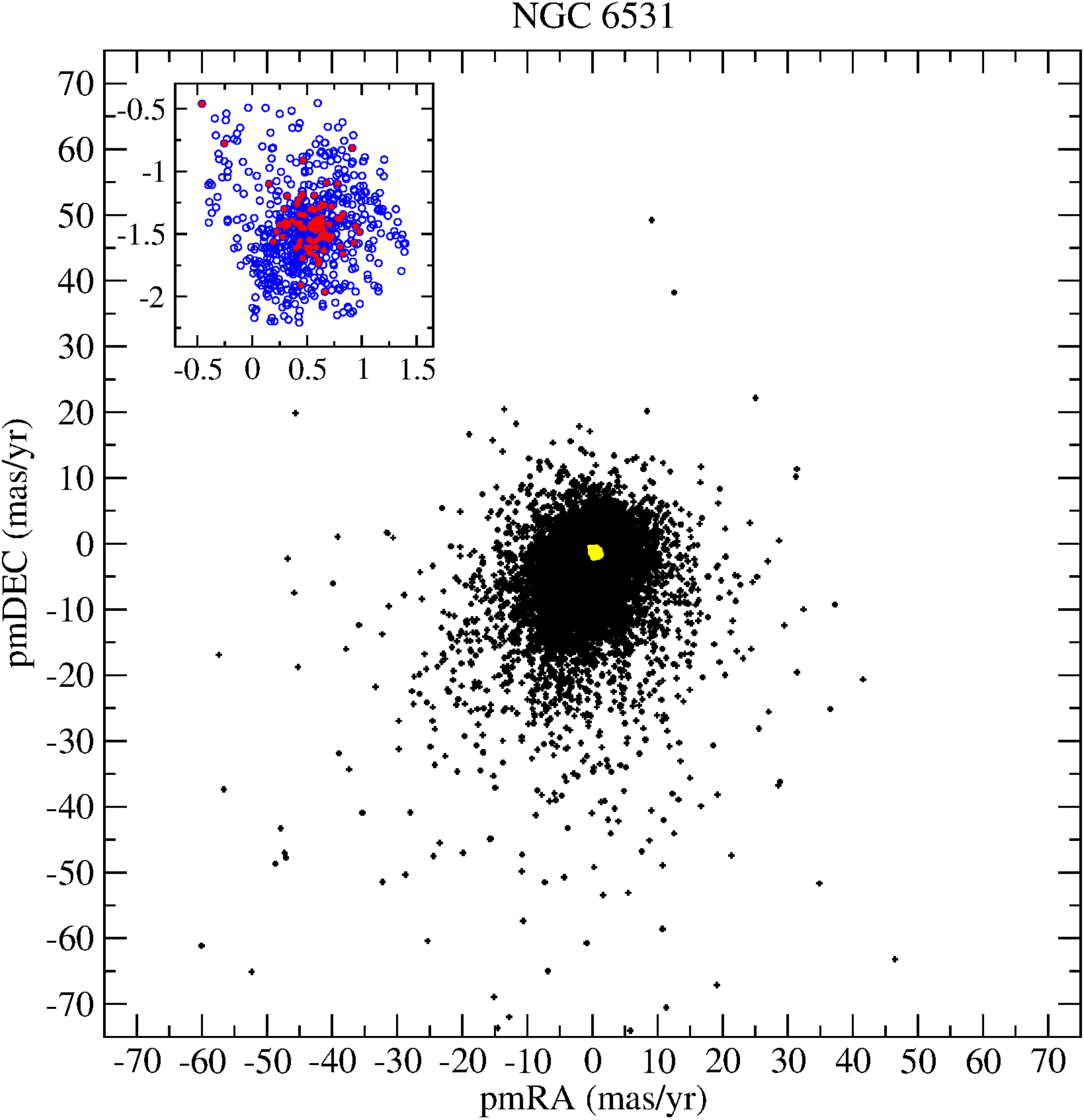}
    \includegraphics[width=0.37\textwidth]{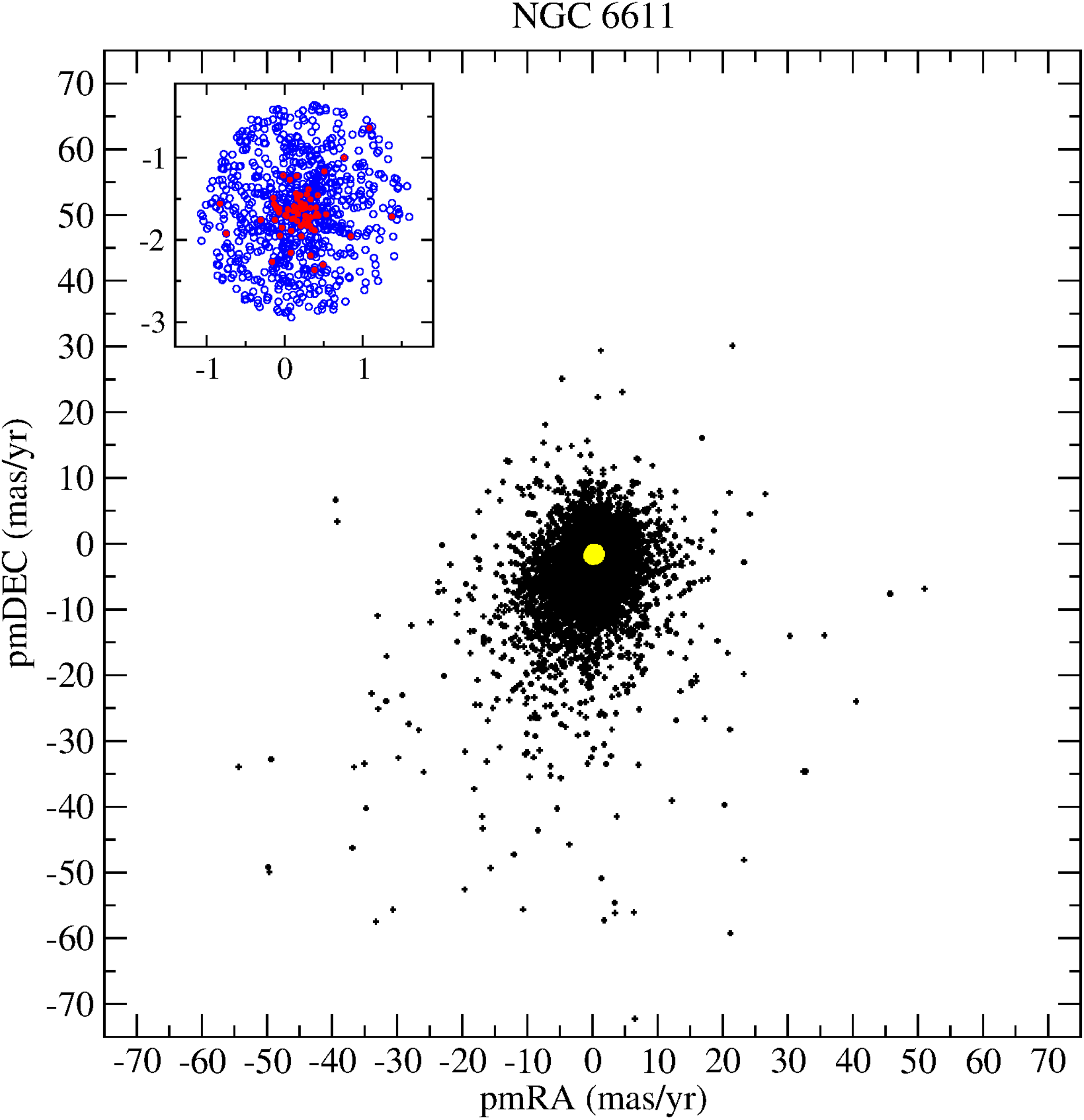}
\end{figure}
\begin{figure} [h]
 \centering
    \includegraphics[width=0.37\textwidth]{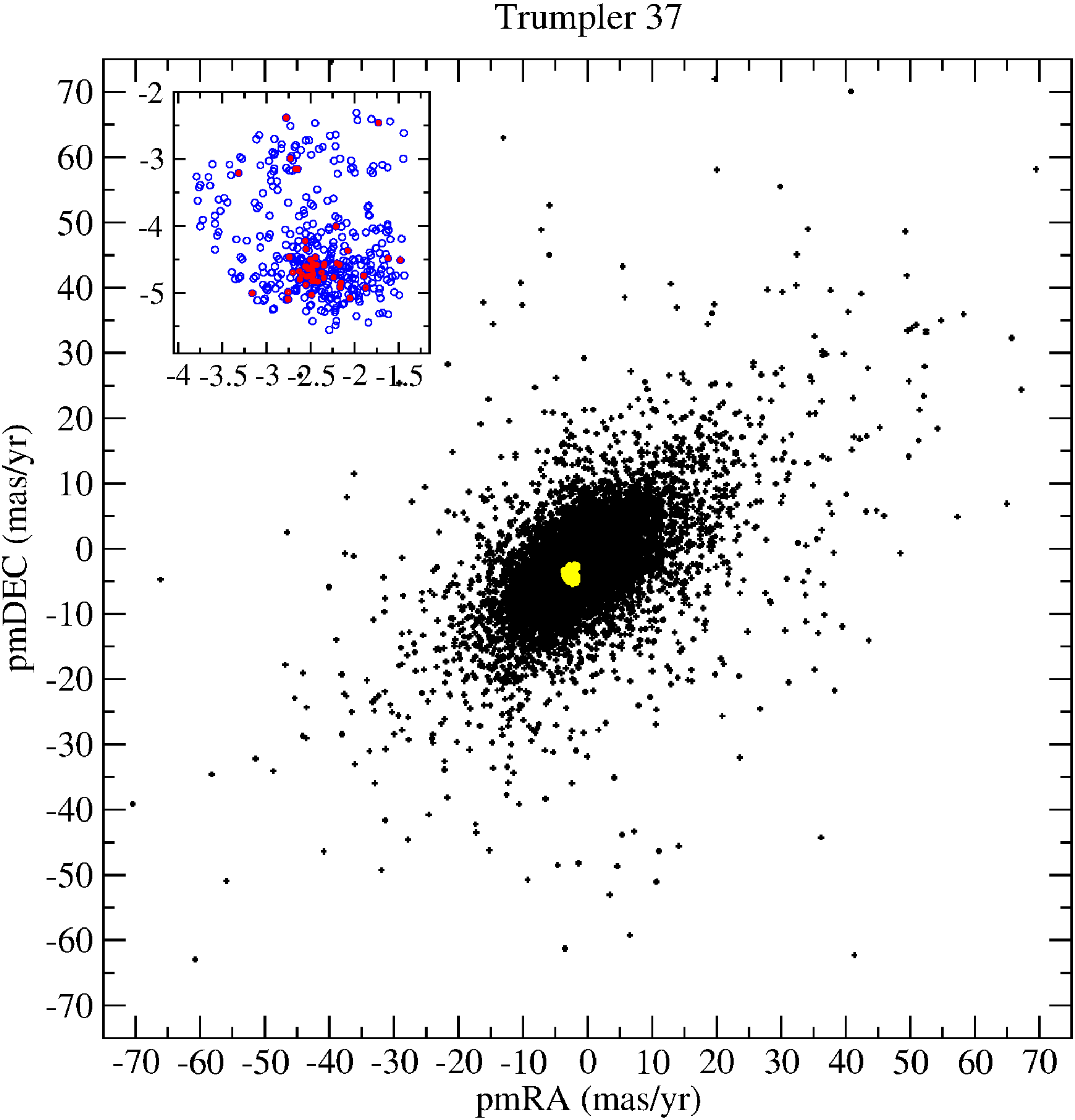}
    \includegraphics[width=0.37\textwidth]{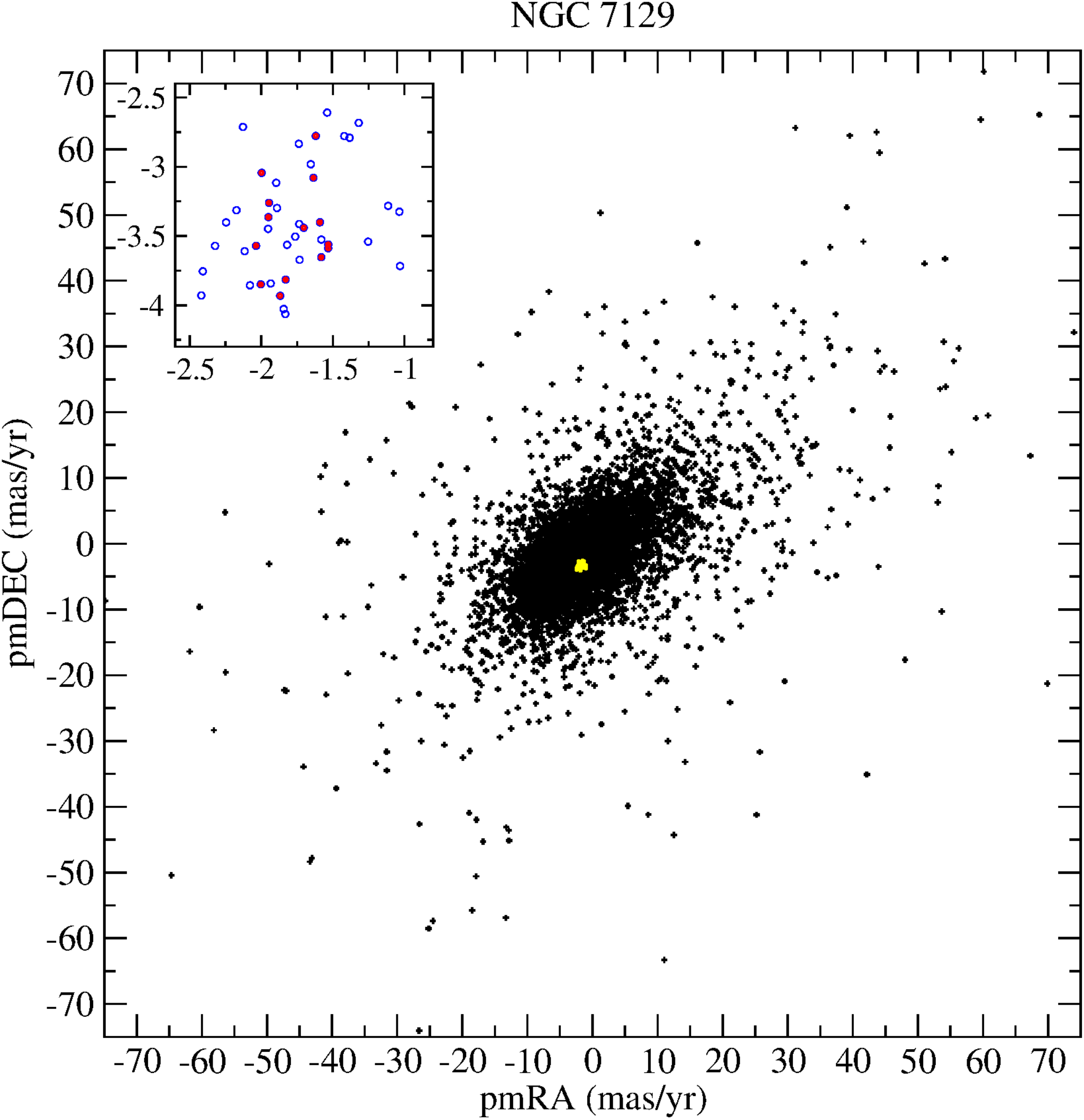}
\end{figure}
\begin{figure} [h]
 \centering
    \includegraphics[width=0.37\textwidth]{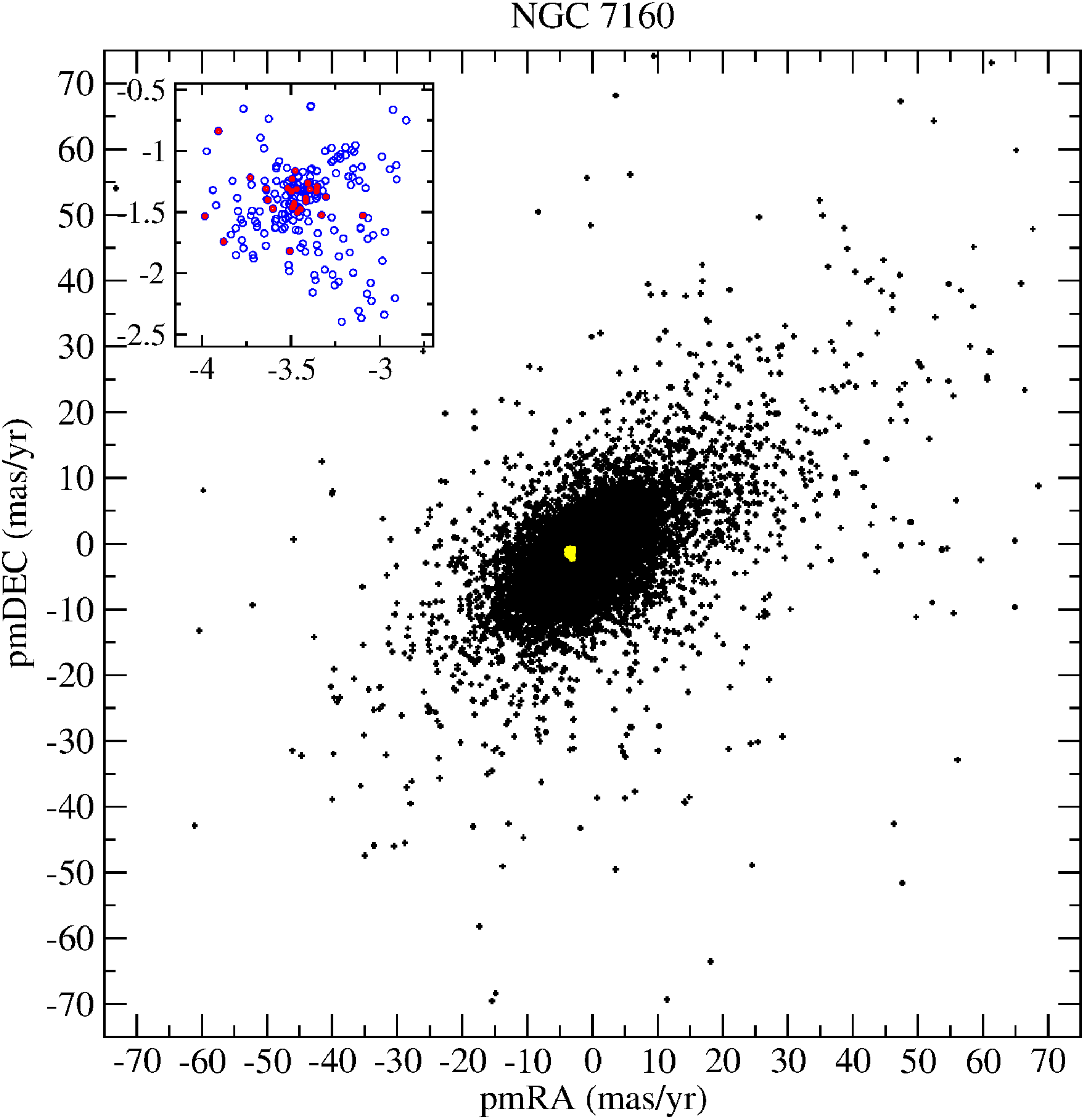} 
            \caption{Proper motion distributions for the different clusters where the field stars within FOV$_{20pc}$ having absolute proper motions $\leq$ 75 mas yr$^{-1}$ and the member stars are indicated in black and yellow, respectively. The top left sub-panels focus on the proper motions of the members, where the blue and red circles correspond to those identified within FOV$_{20pc}$ and FOV$_{2pc}$, respectively. Typical (mean) uncertainties in proper motions are $\sim$ 0.04 mas yr$^{-1}$, roughly comparable to the size of such circles for most cases.}
            \label{figure:ppmra_vs_ppmdec}
\end{figure}

\newpage
\onecolumn
\begin{figure} [h]
 \centering
    \includegraphics[width=0.43\textwidth]{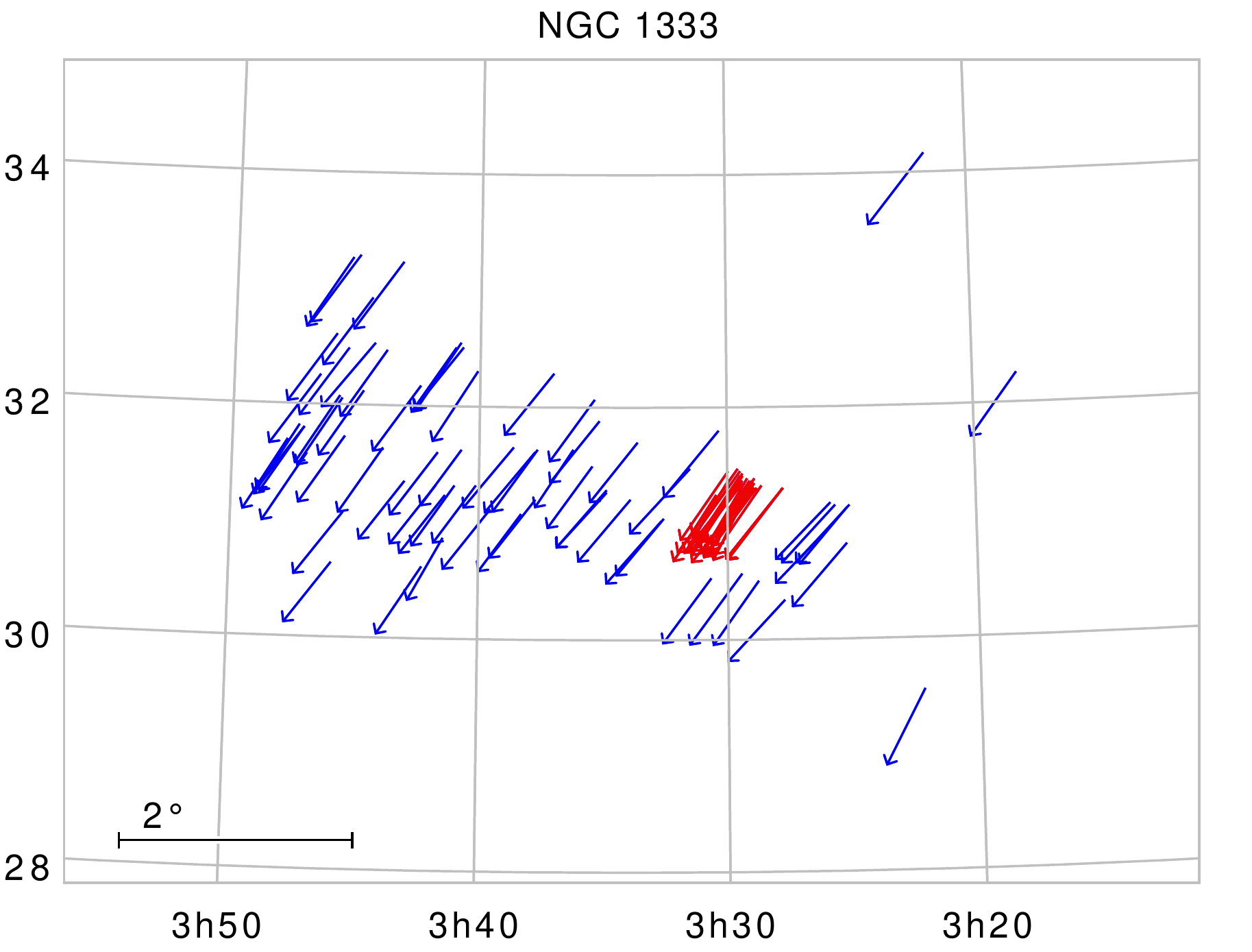}
    \includegraphics[width=0.43\textwidth]{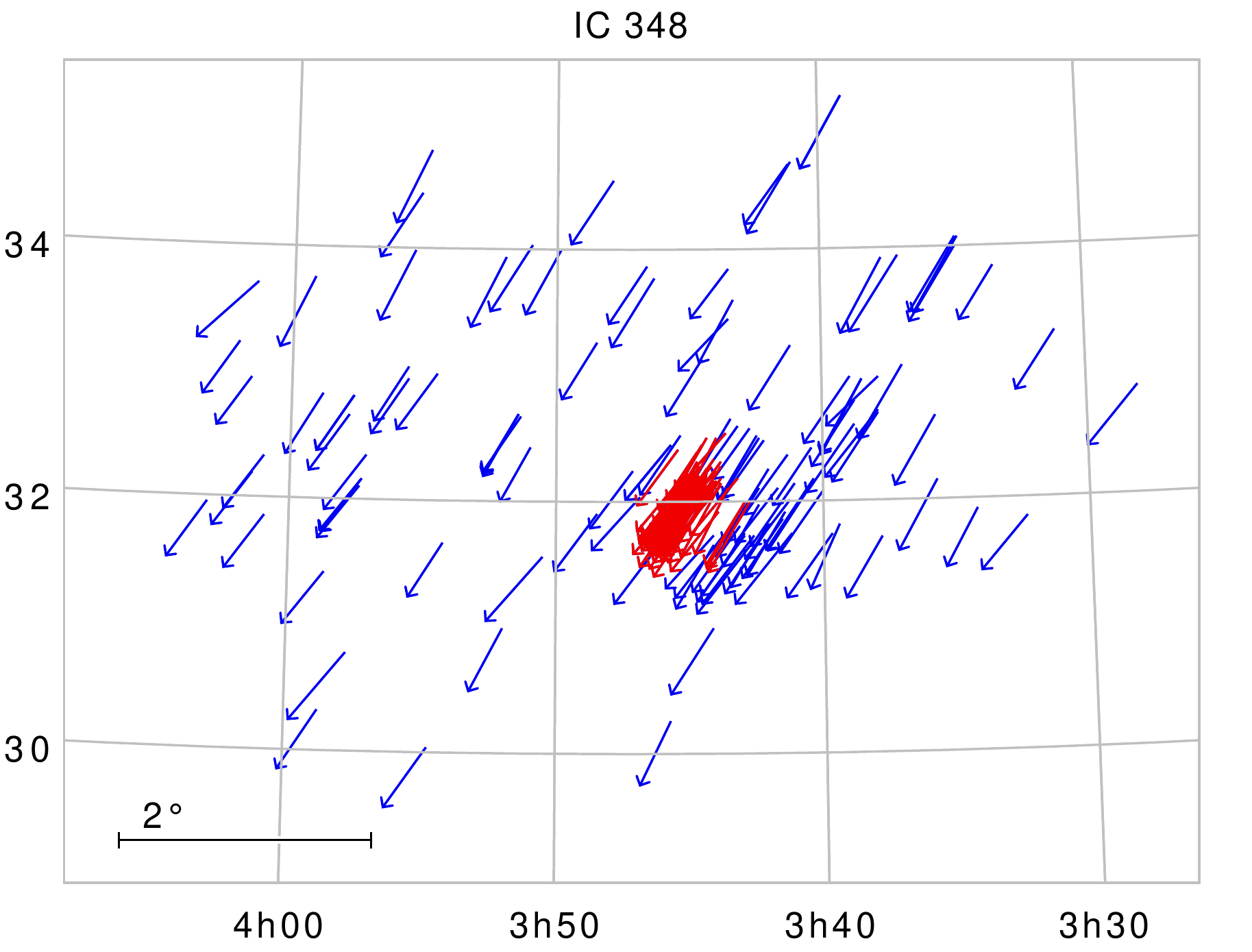}
\end{figure}
\begin{figure} [h]
 \centering
    \includegraphics[width=0.43\textwidth]{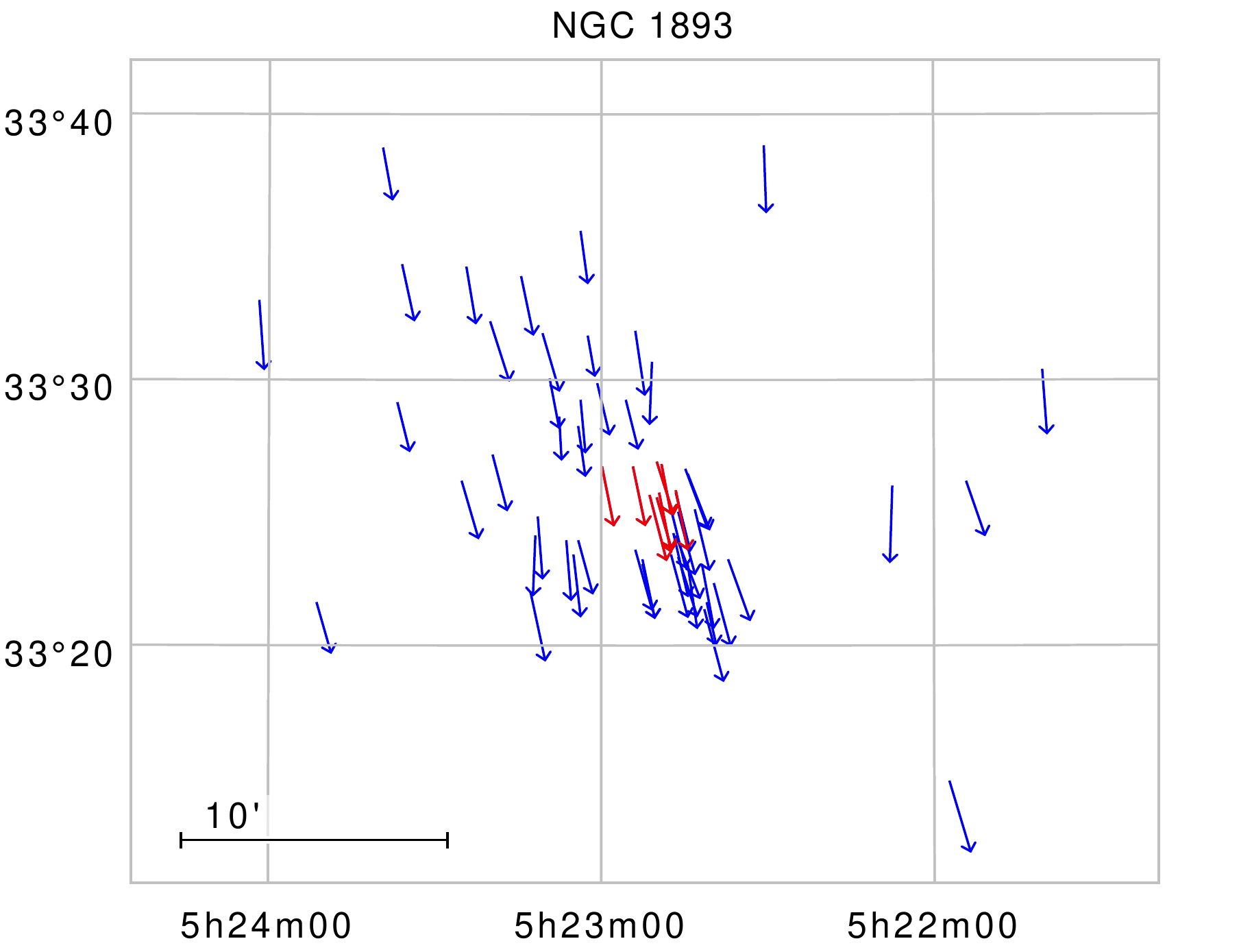}
    \includegraphics[width=0.43\textwidth]{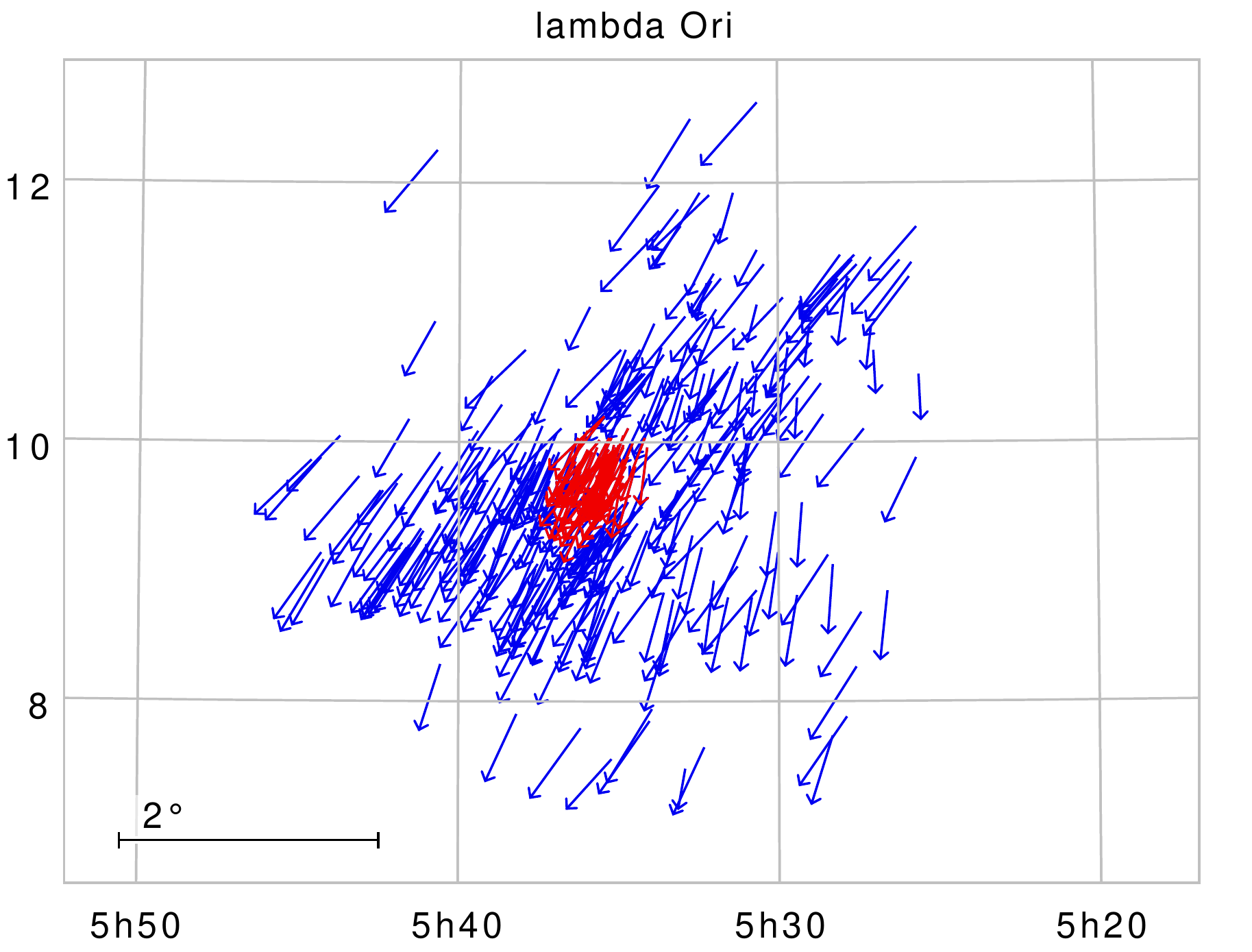} 
\end{figure}
\begin{figure} [h]
 \centering
    \includegraphics[width=0.43\textwidth]{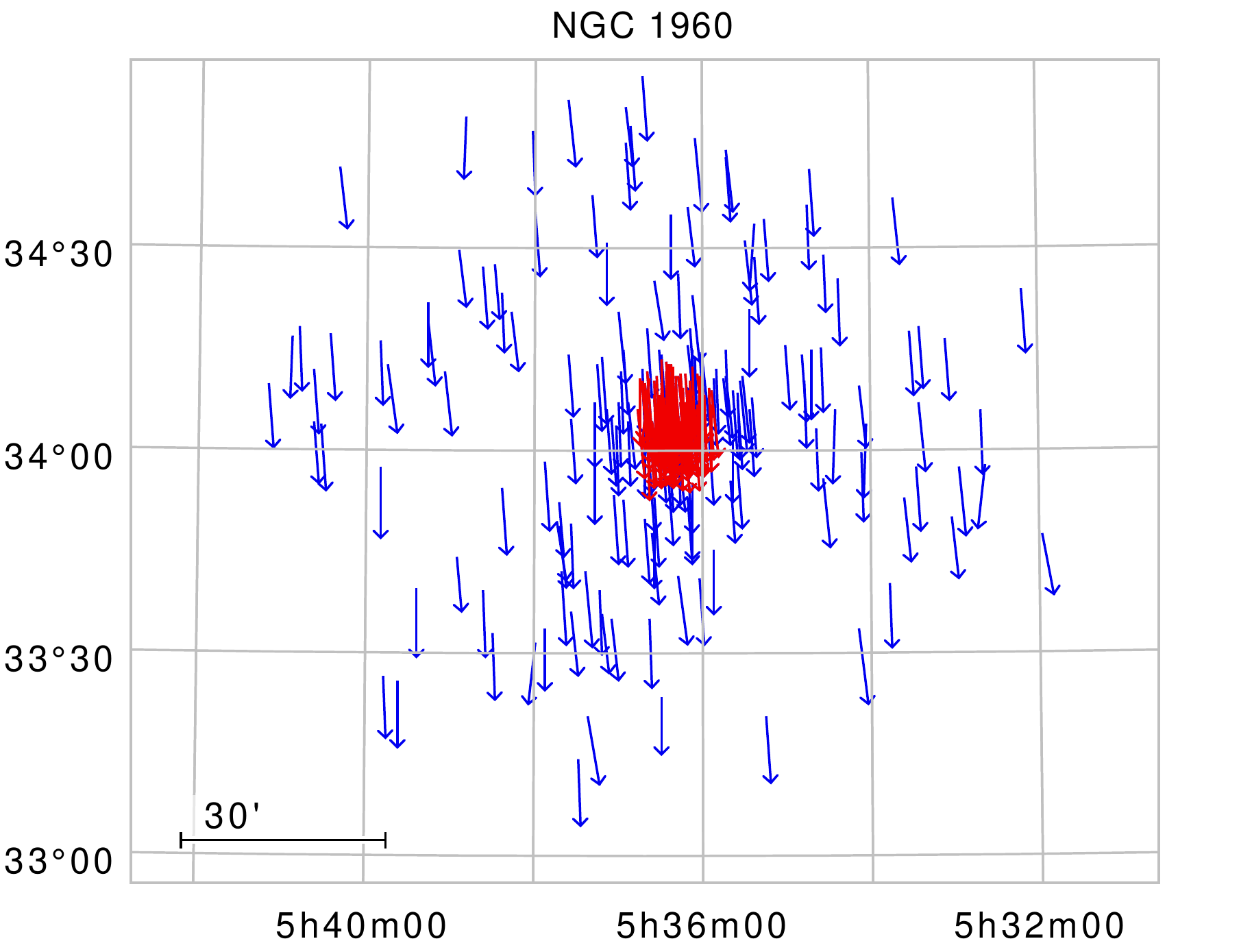}
    \includegraphics[width=0.43\textwidth]{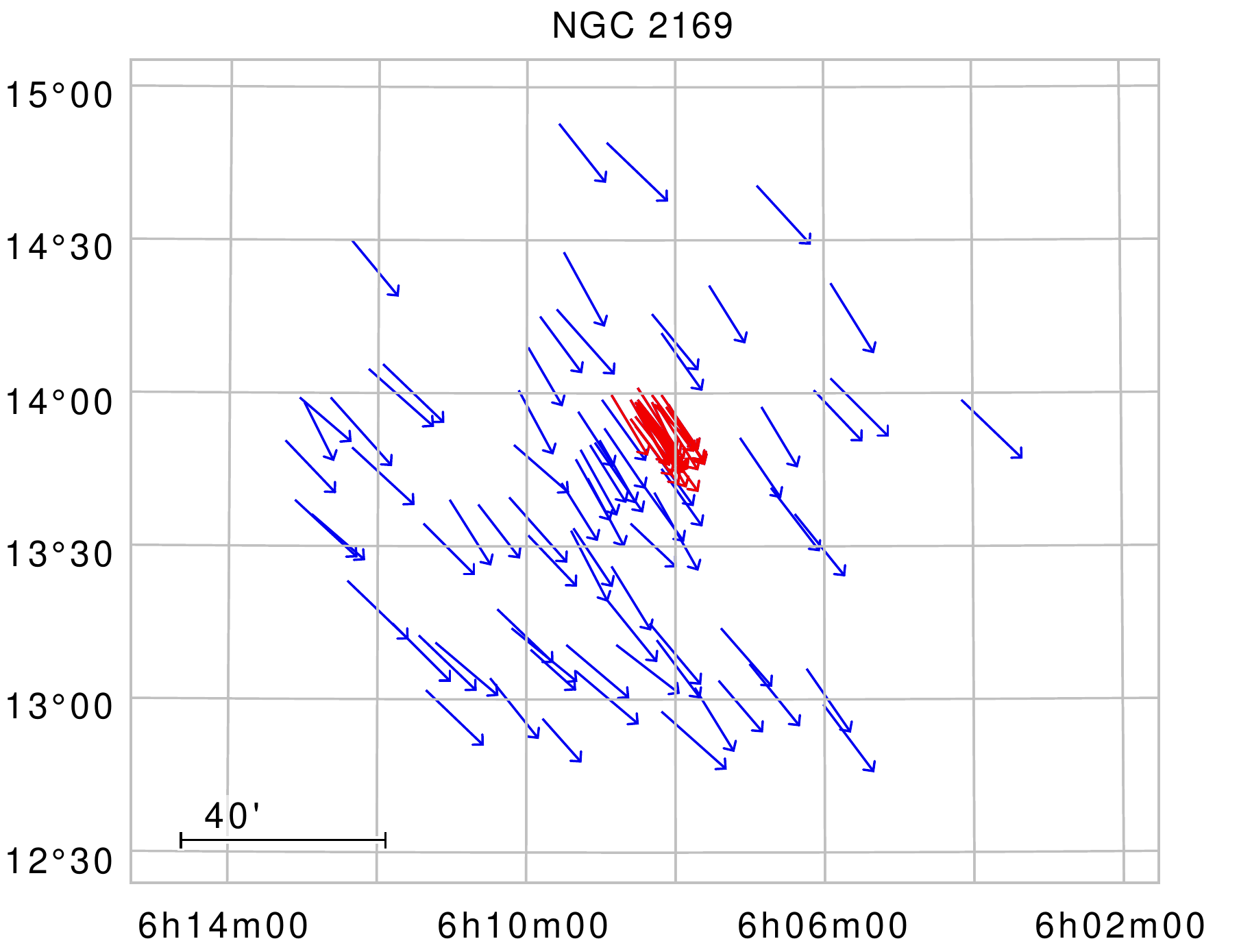}  
\end{figure}   
\begin{figure} [h]
 \centering
    \includegraphics[width=0.43\textwidth]{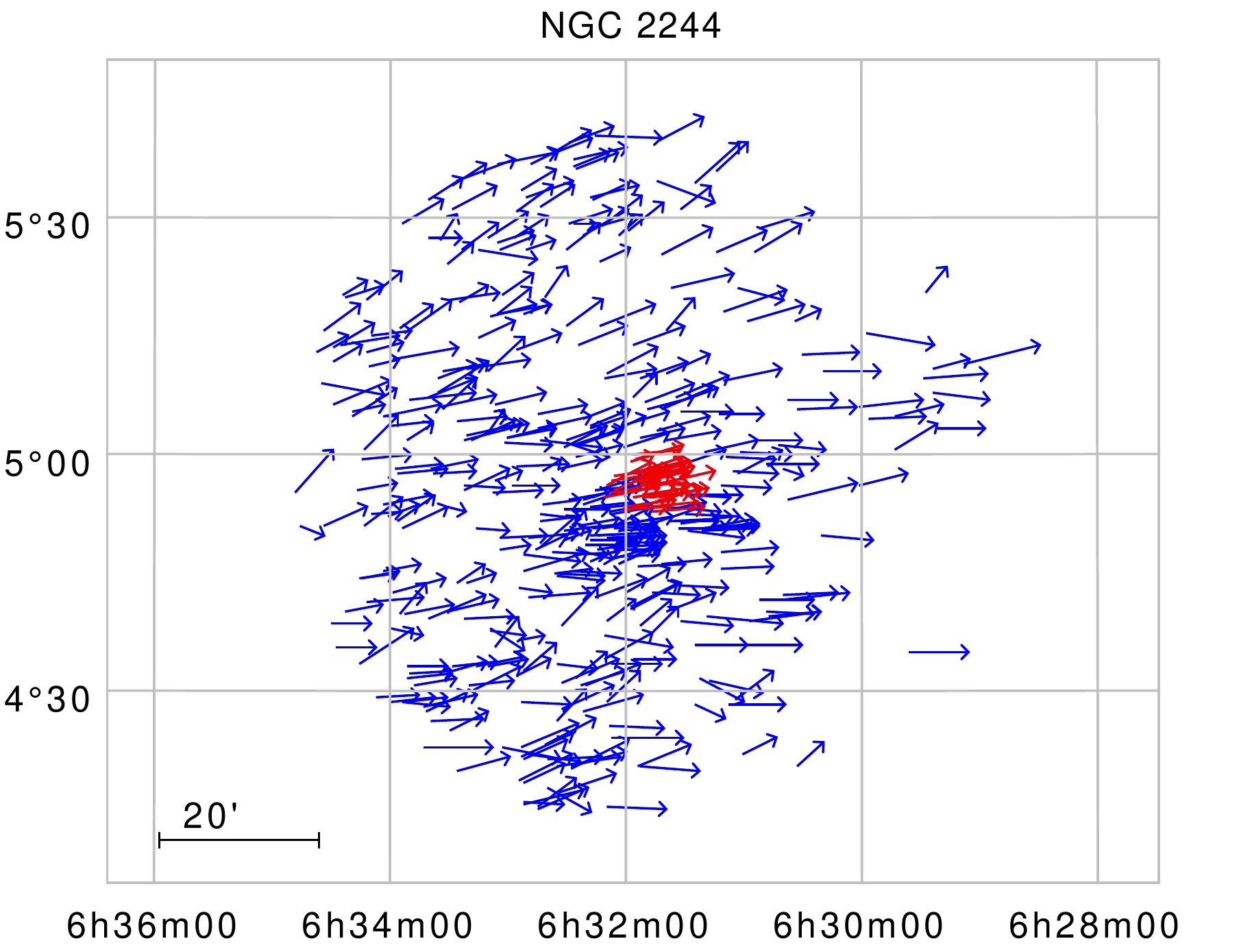}
    \includegraphics[width=0.43\textwidth]{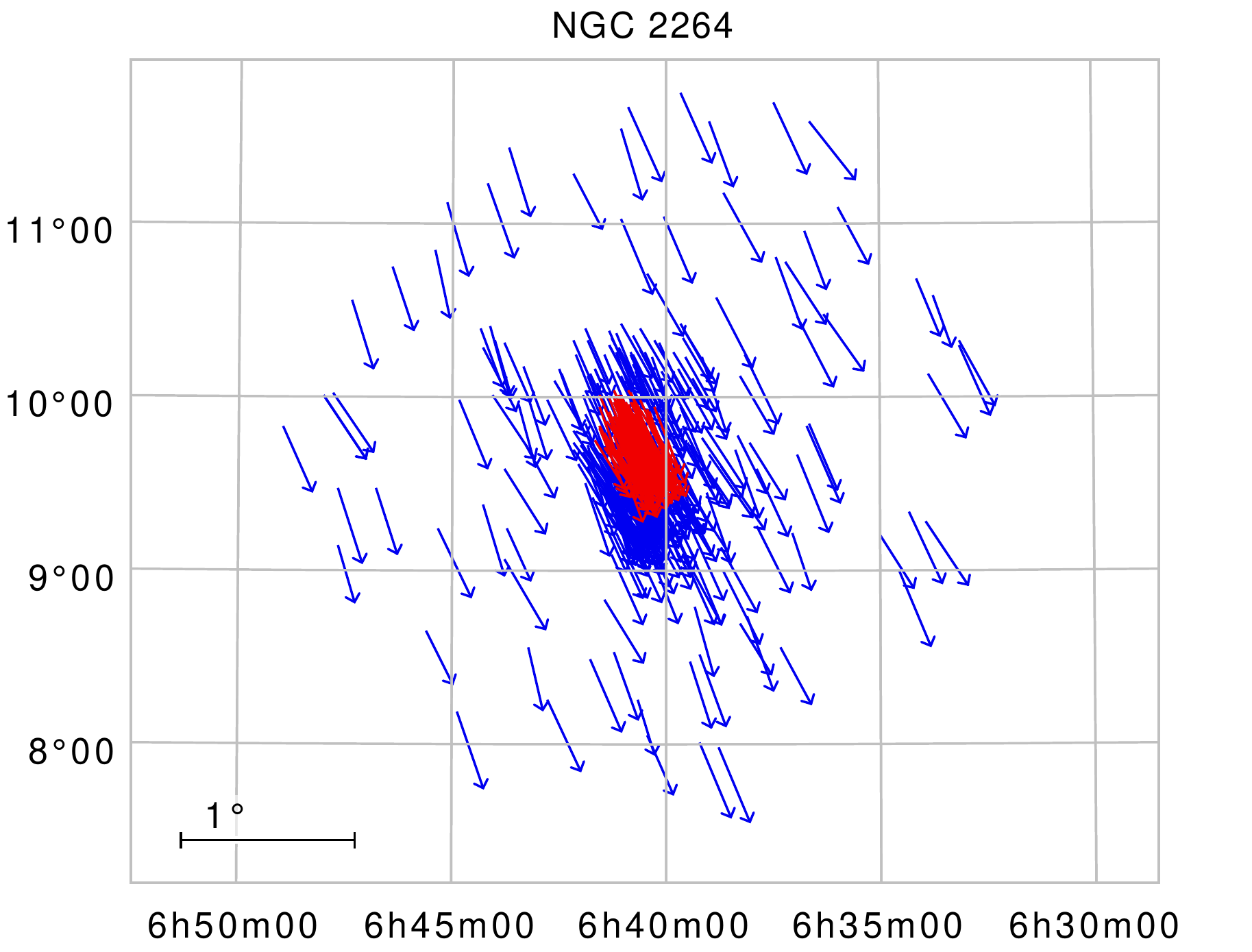}
    \end{figure}
\begin{figure} [h]
 \centering
    \includegraphics[width=0.43\textwidth]{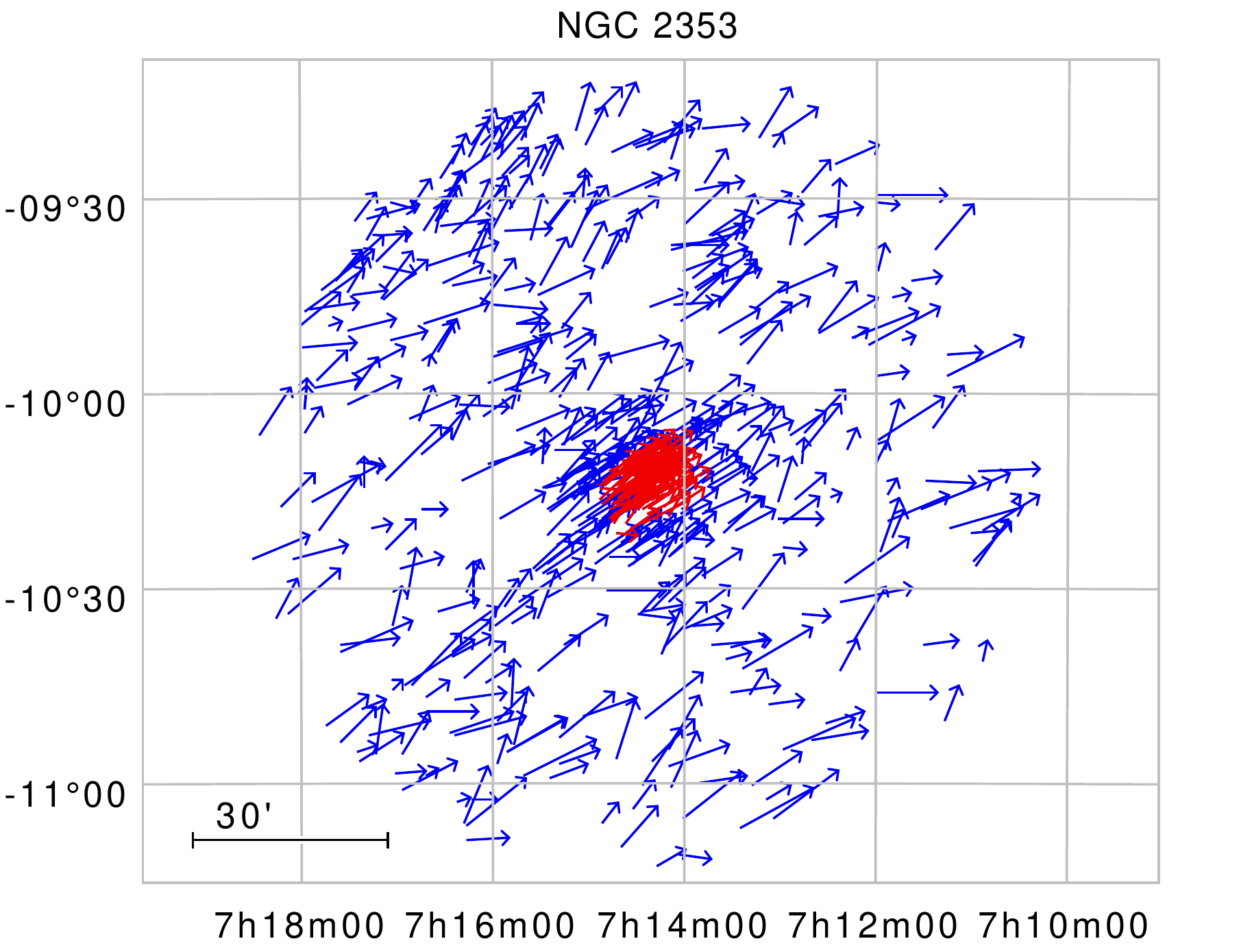}
    \includegraphics[width=0.43\textwidth]{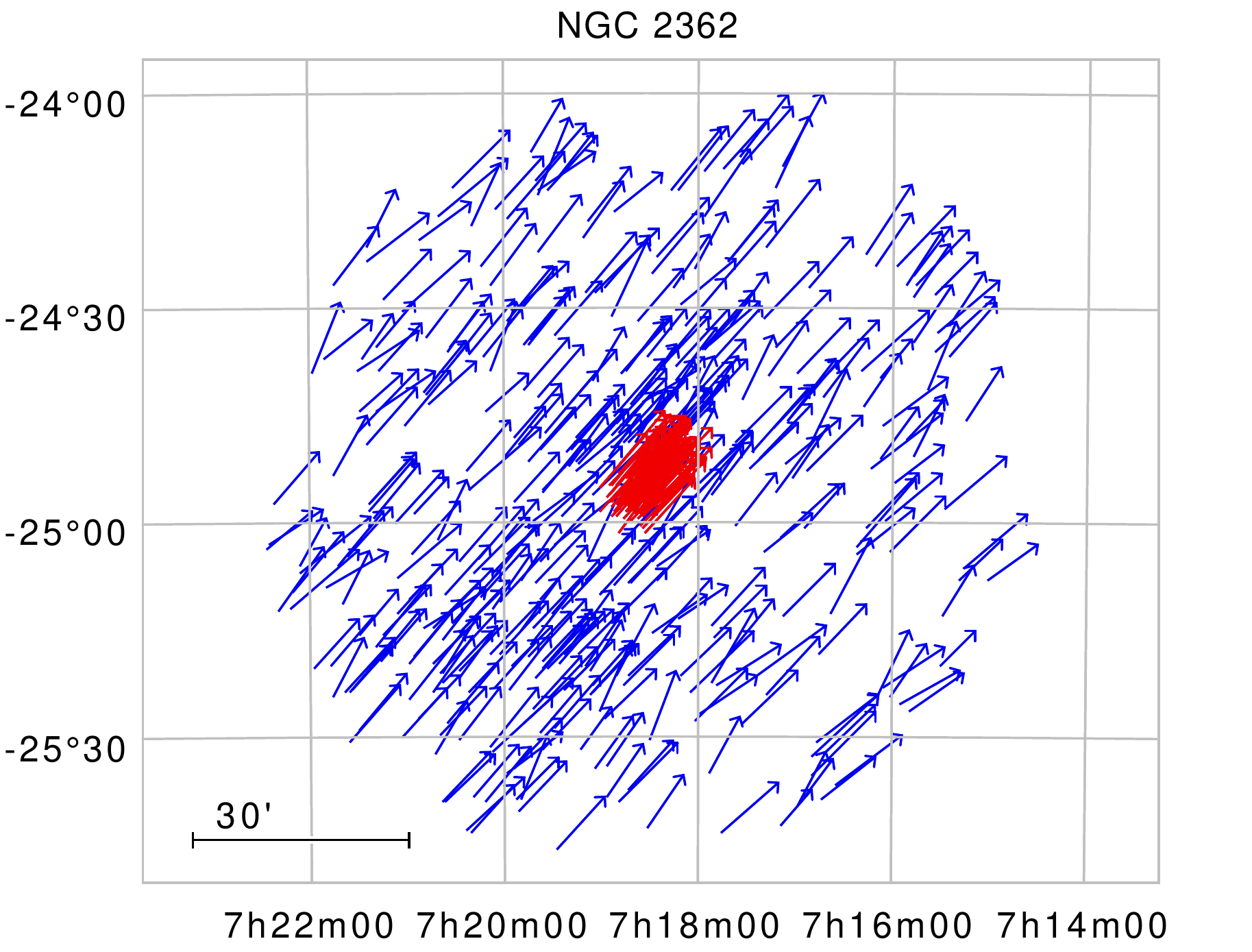}
    \end{figure}    
\begin{figure} [h]
 \centering
    \includegraphics[width=0.43\textwidth]{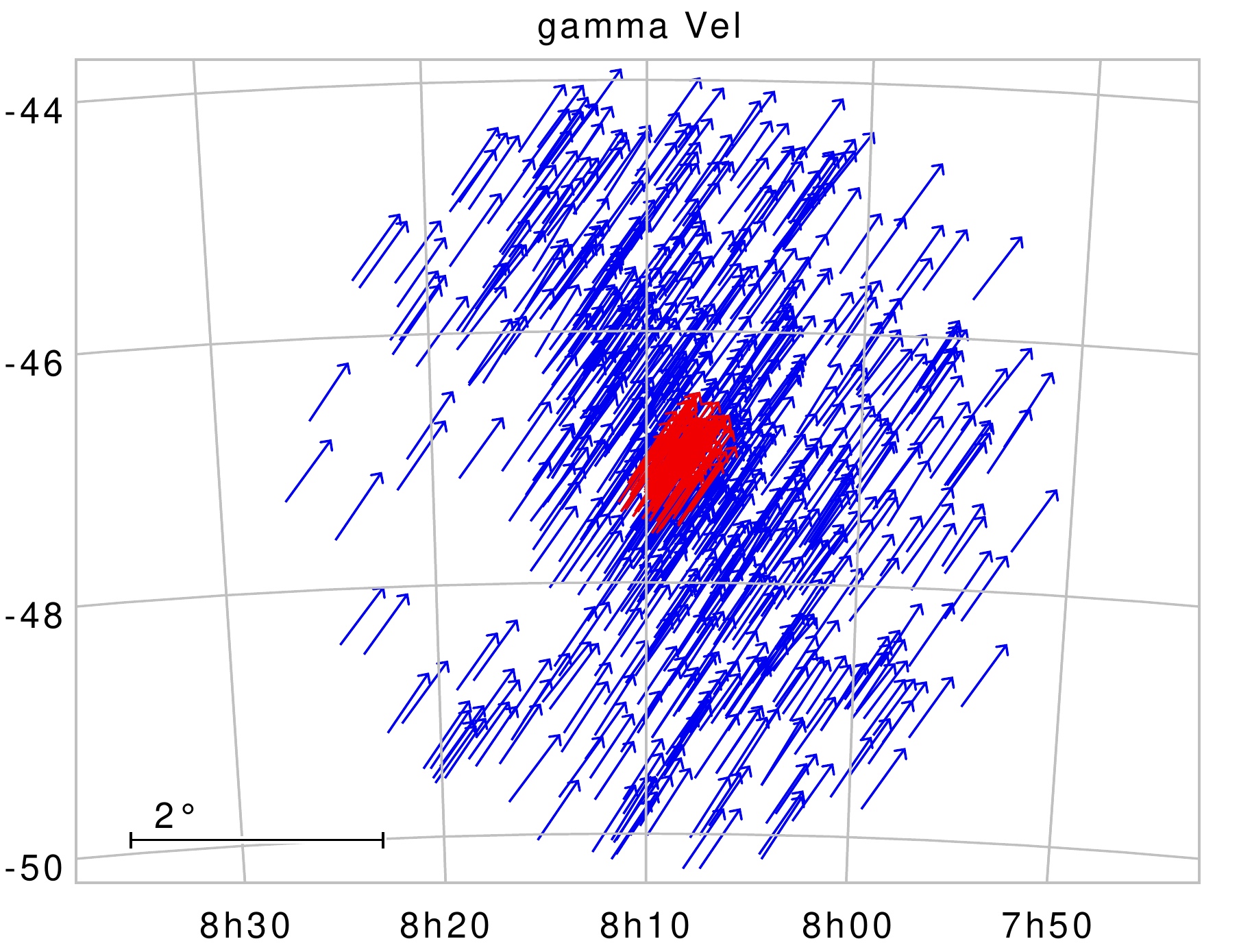}
    \includegraphics[width=0.43\textwidth]{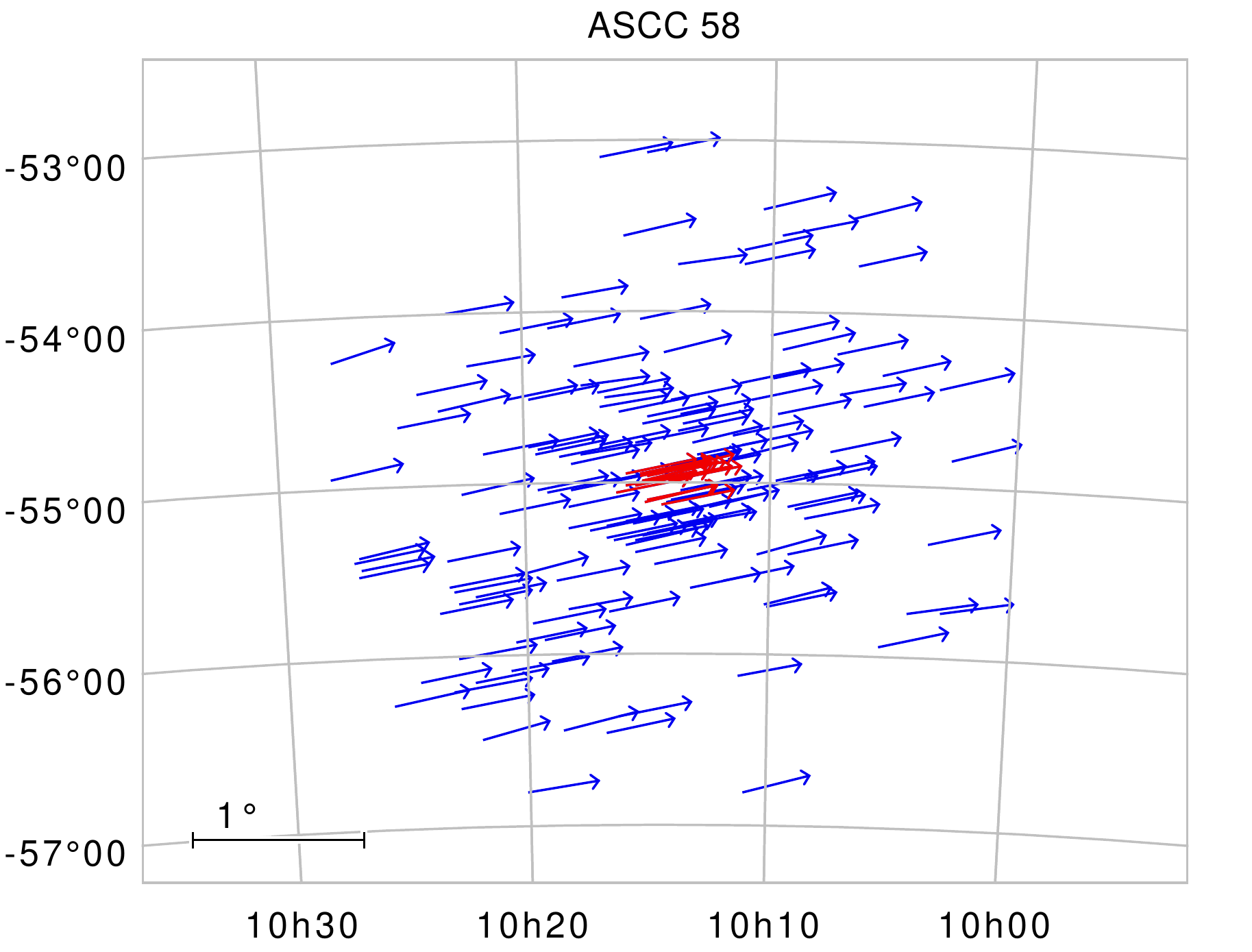}
\end{figure}
\begin{figure} [h]
 \centering
    \includegraphics[width=0.43\textwidth]{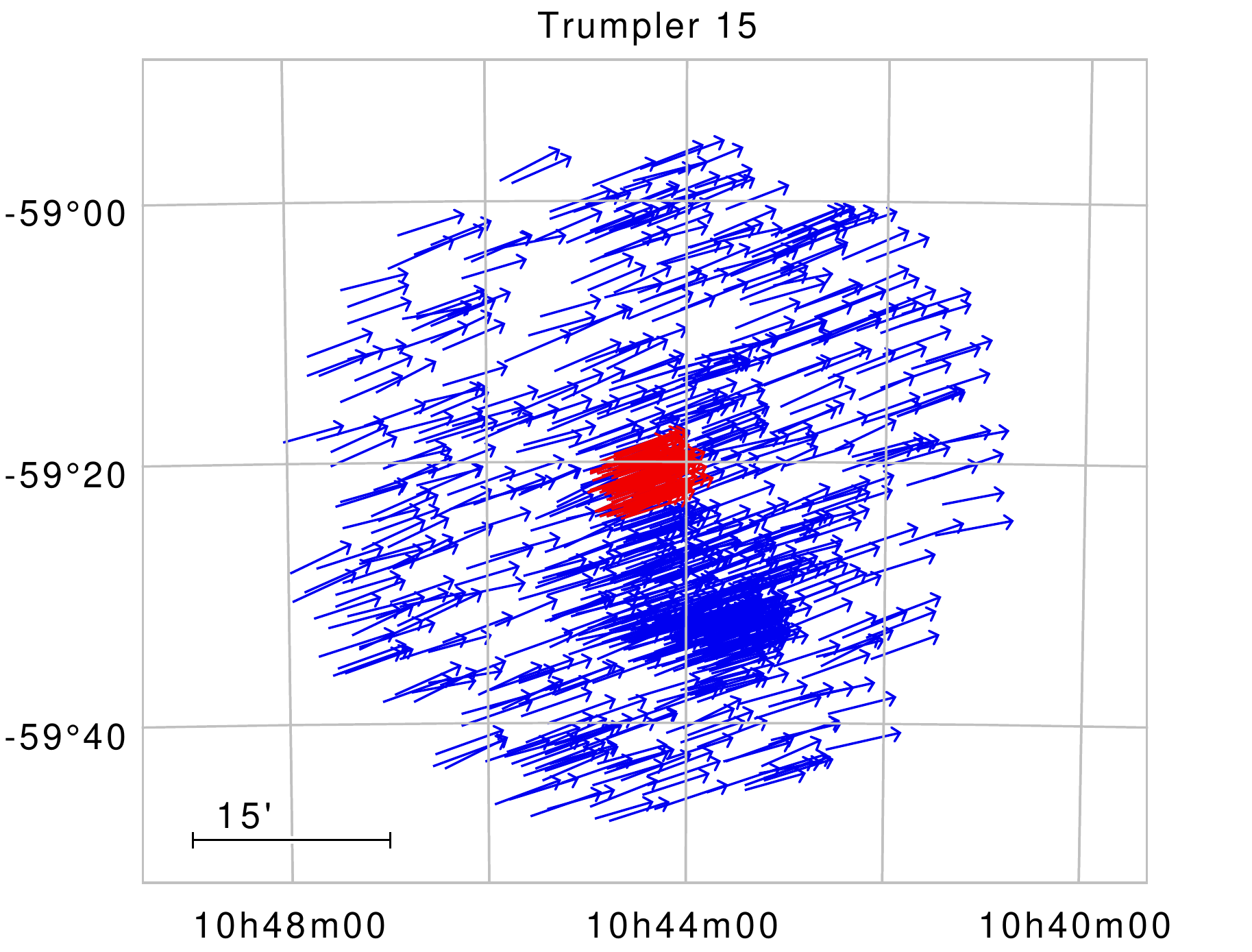}
    \includegraphics[width=0.43\textwidth]{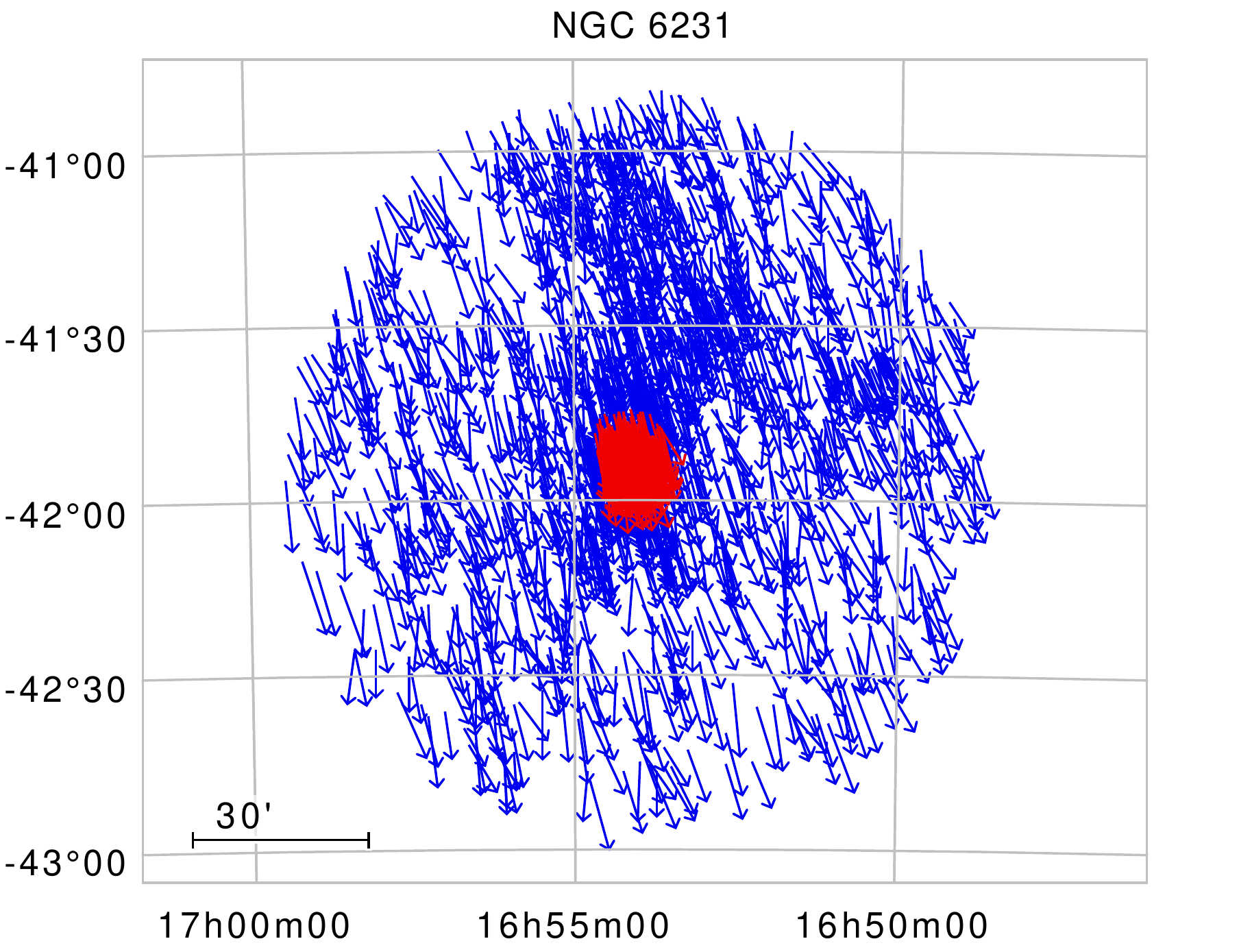}
    \end{figure}
\begin{figure} [h]
 \centering
    \includegraphics[width=0.43\textwidth]{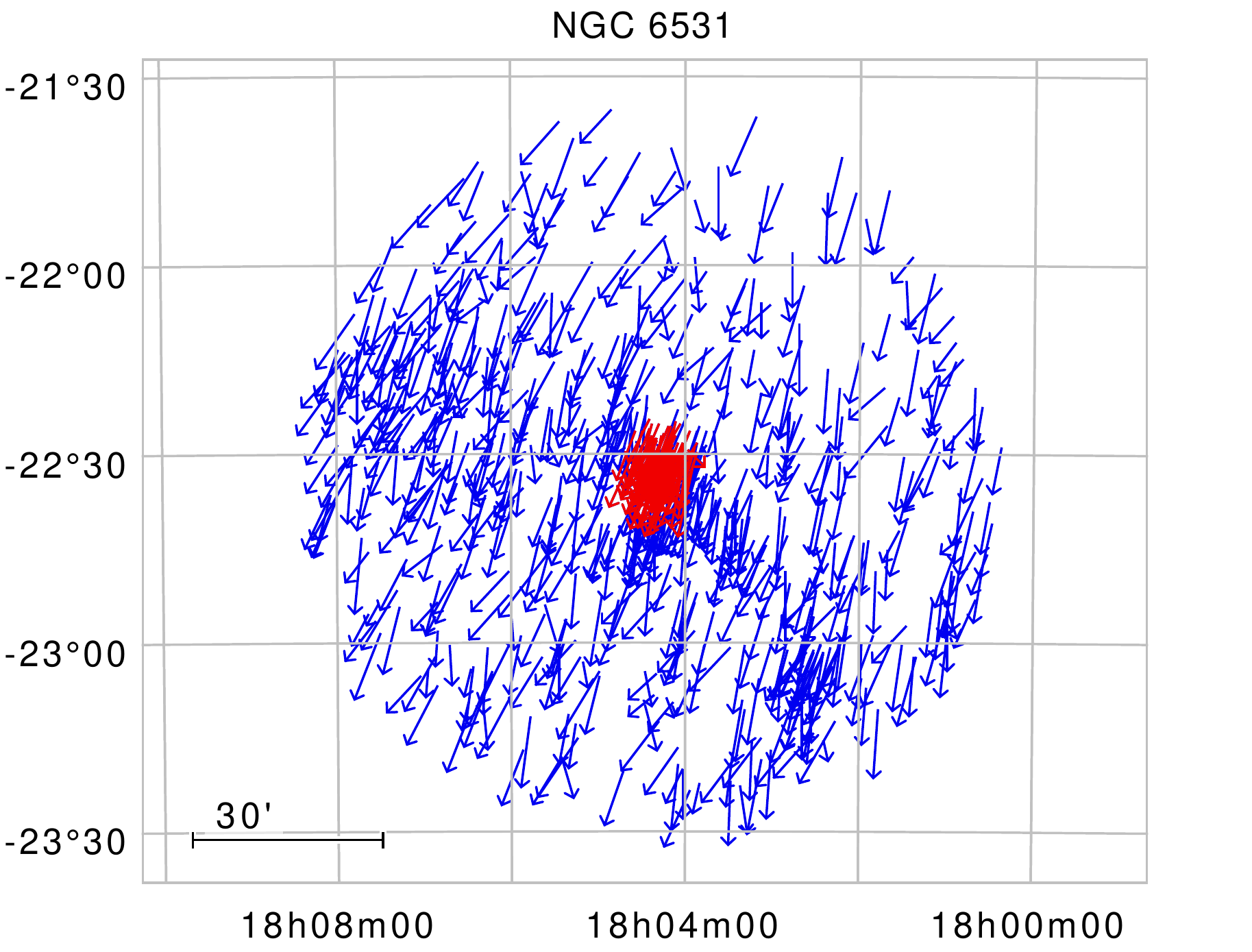}
    \includegraphics[width=0.43\textwidth]{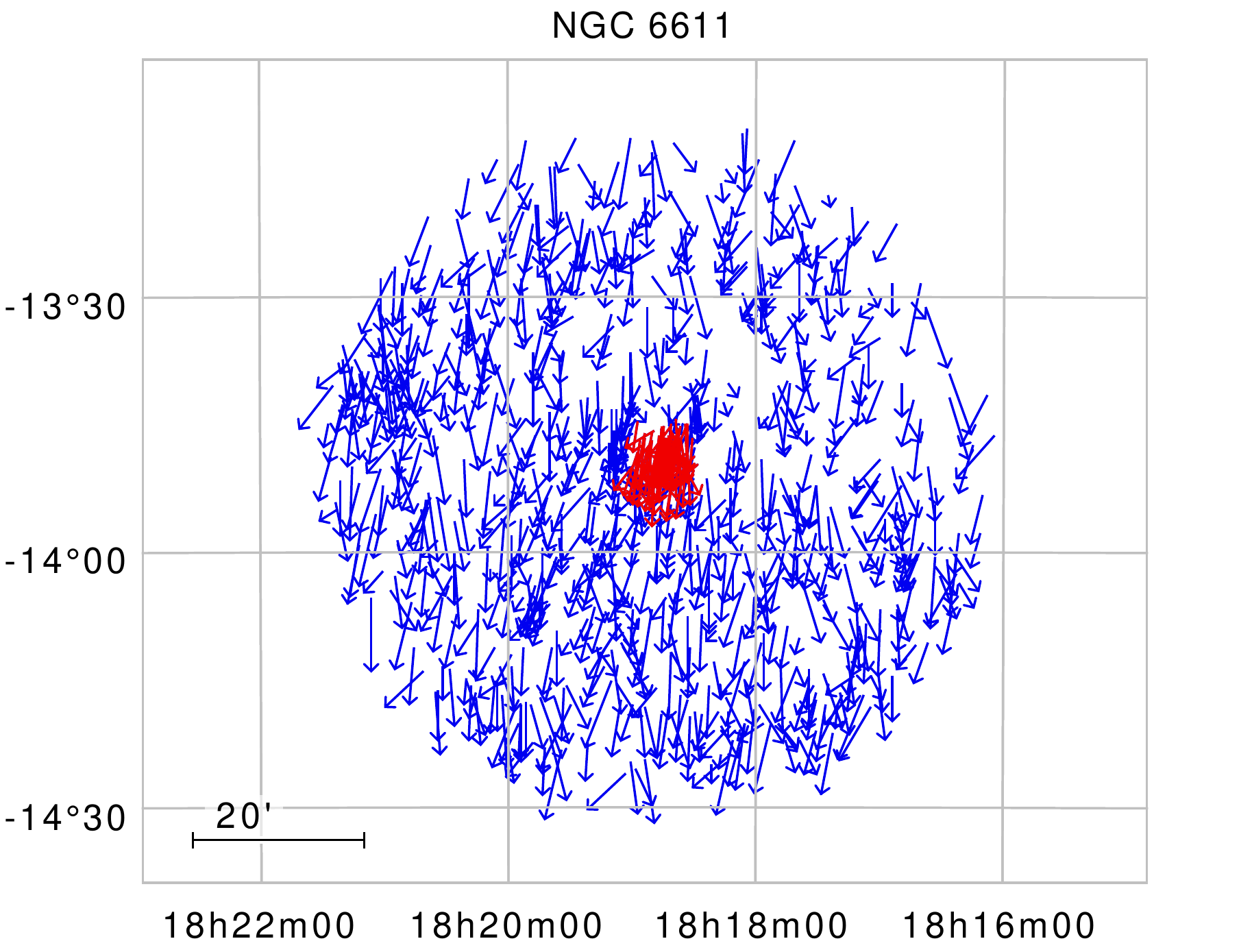}
    \end{figure}
\begin{figure} [h]
 \centering
    \includegraphics[width=0.43\textwidth]{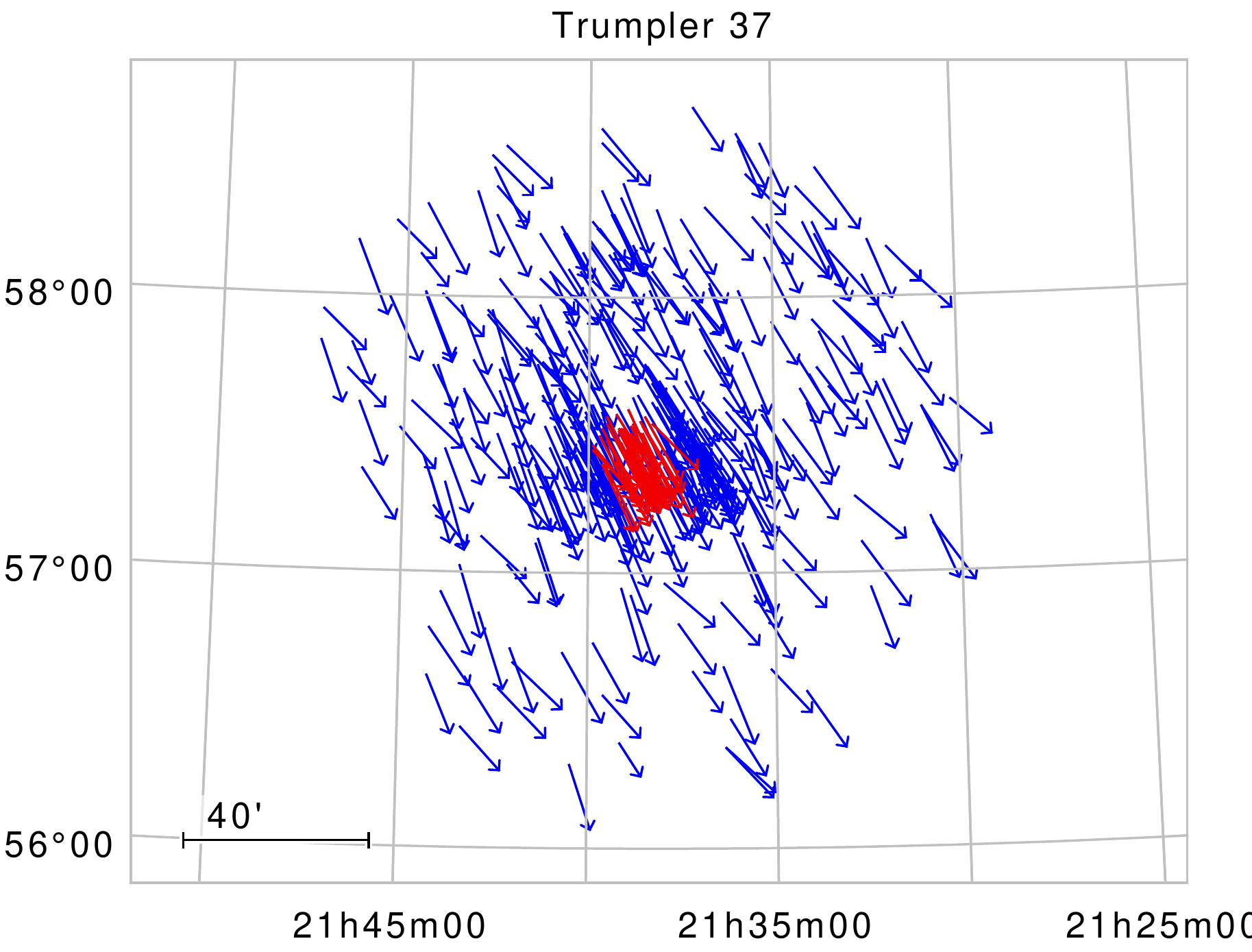}
    \includegraphics[width=0.43\textwidth]{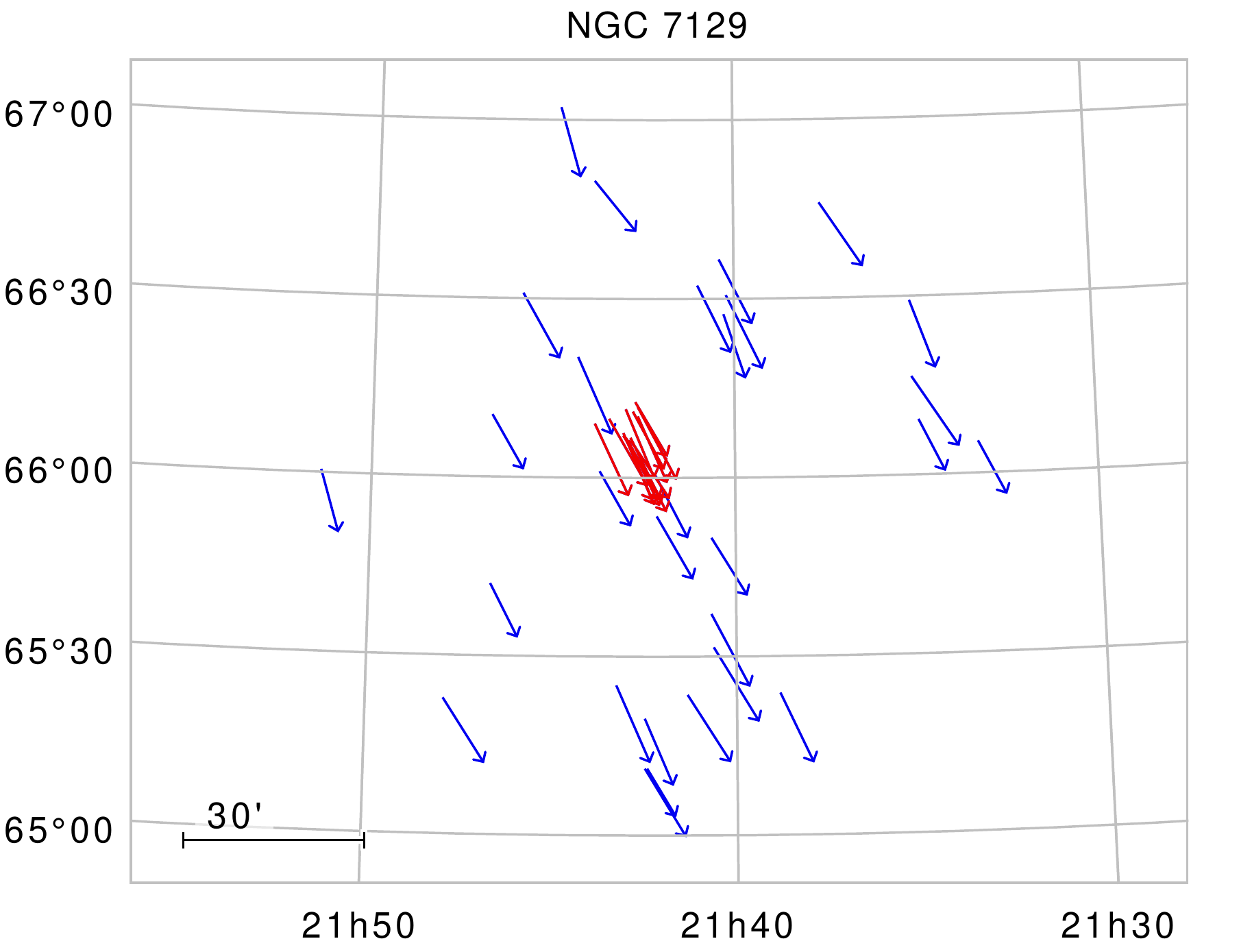}
    \end{figure}
\begin{figure} [h]
 \centering
    \includegraphics[width=0.43\textwidth]{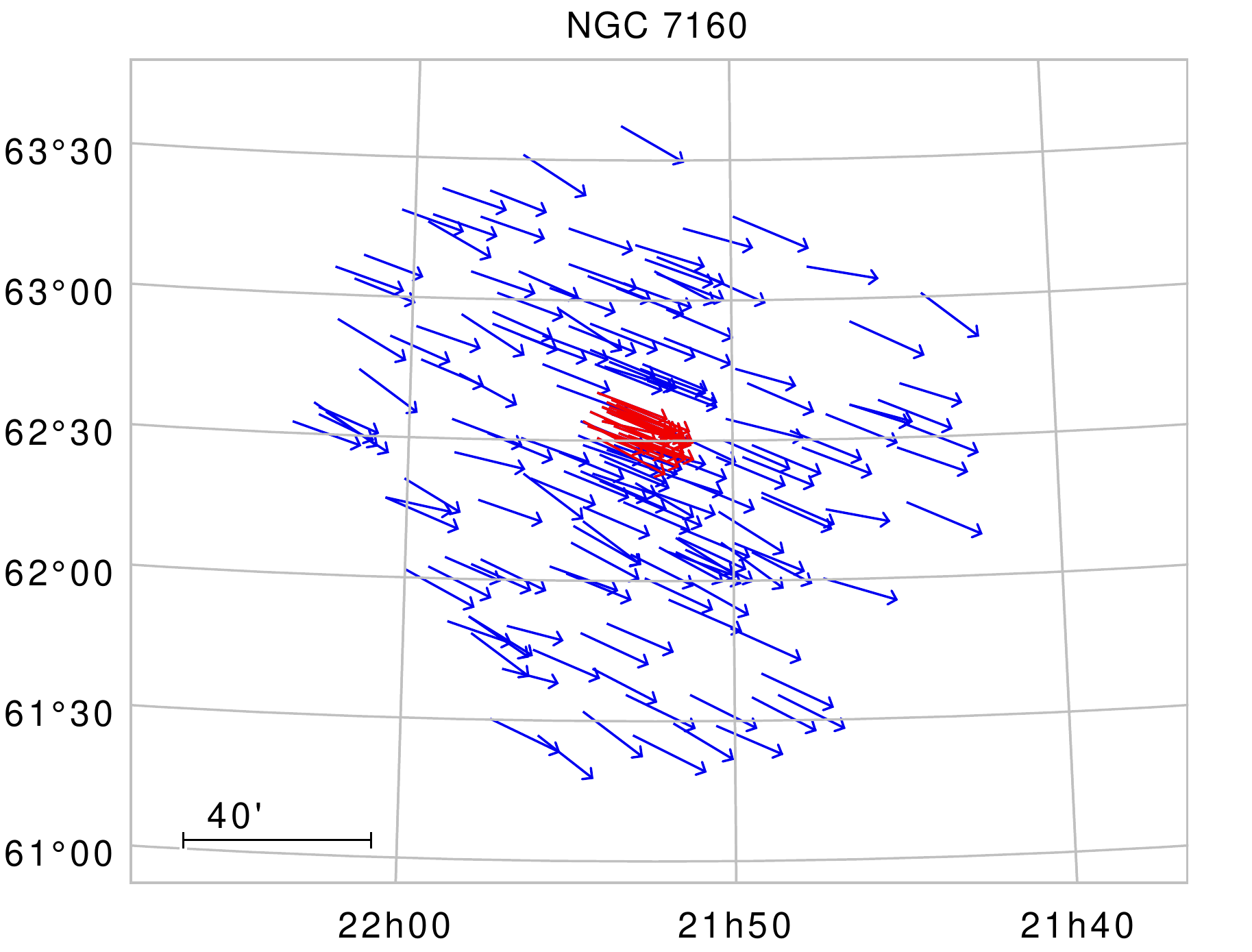}  
            \caption{ Sky projections (North to the top and East to the left) for the different clusters where the blue and red arrows indicate the directions of motion of all members identified within FOV$_{20pc}$ and FOV$_{2pc}$, respectively.}
            \label{figure:ppm_sky}
\end{figure}

\newpage
\onecolumn
\begin{figure} [h]
 \centering
    \includegraphics[width=0.38\textwidth]{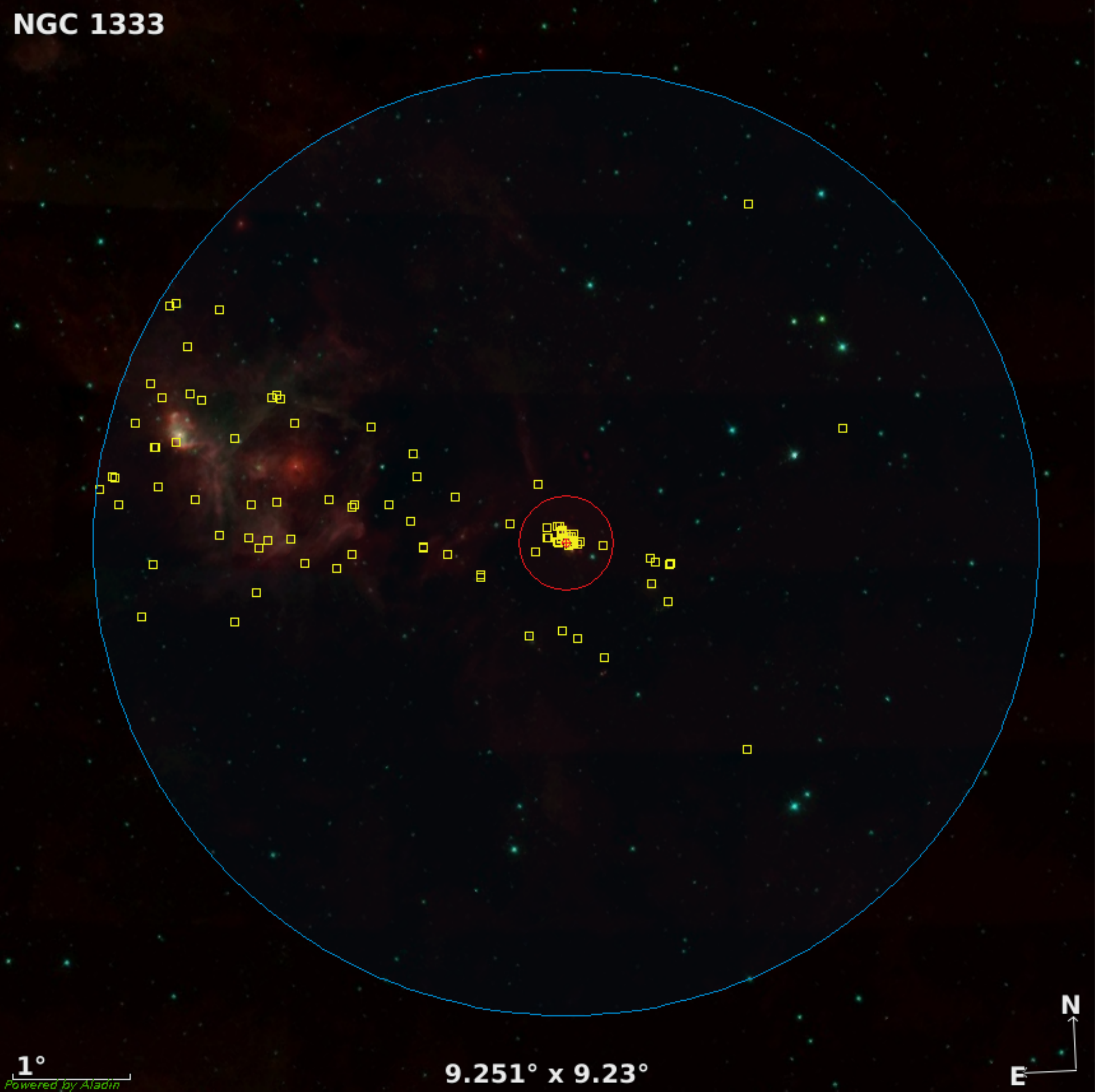}
    \includegraphics[width=0.38\textwidth]{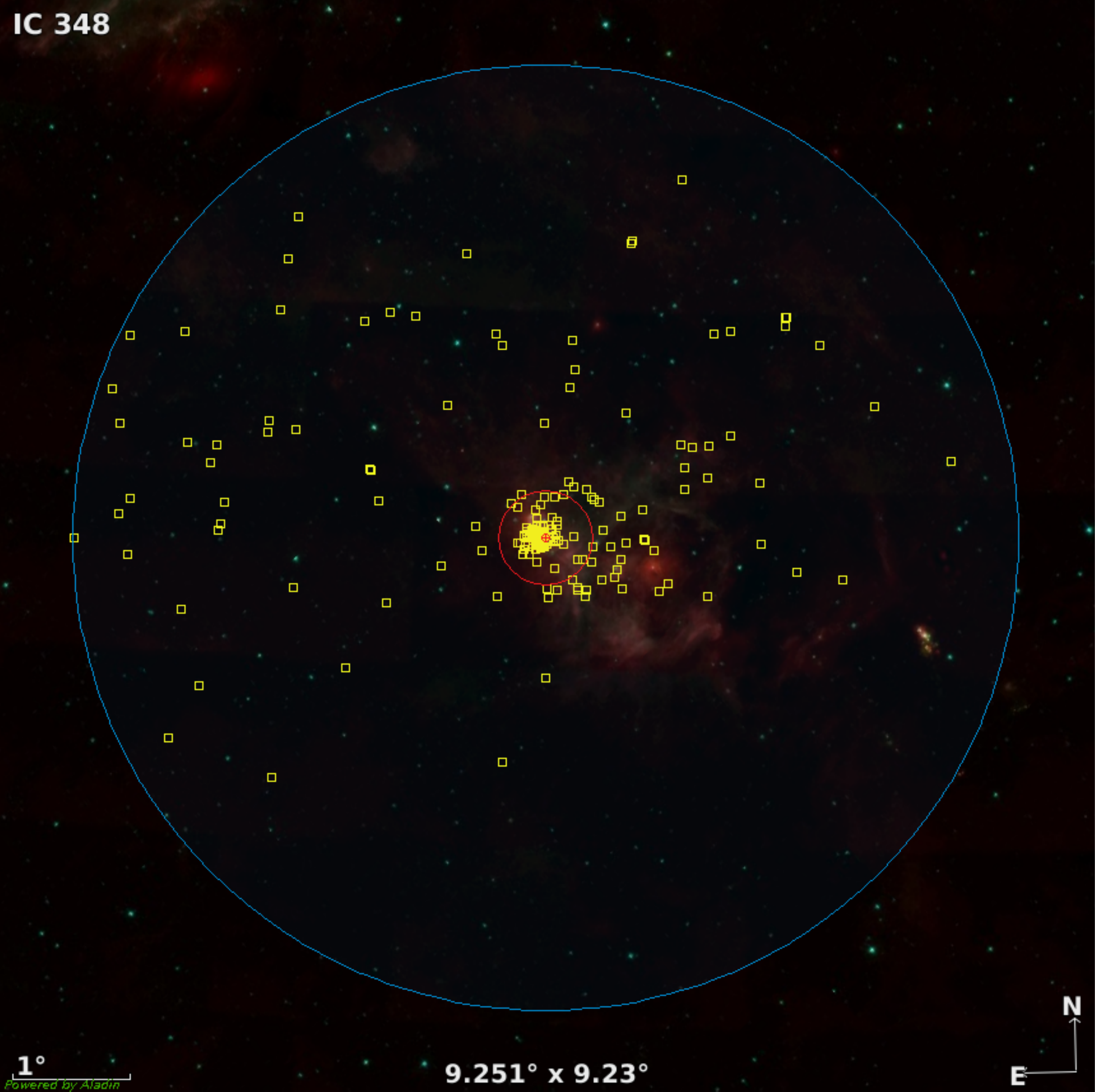}
\end{figure}
\begin{figure} [h]
 \centering
    \includegraphics[width=0.38\textwidth]{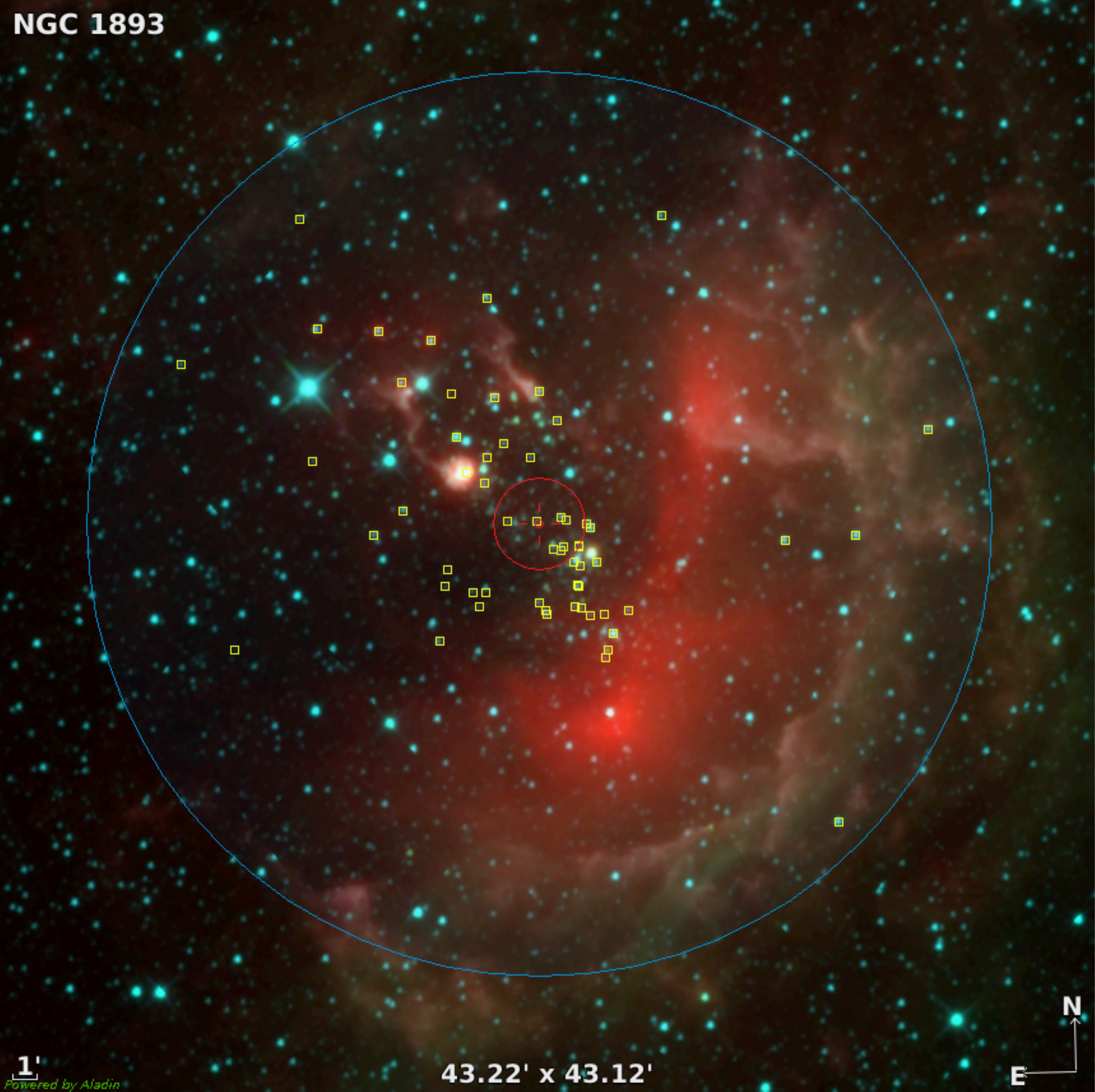}
    \includegraphics[width=0.38\textwidth]{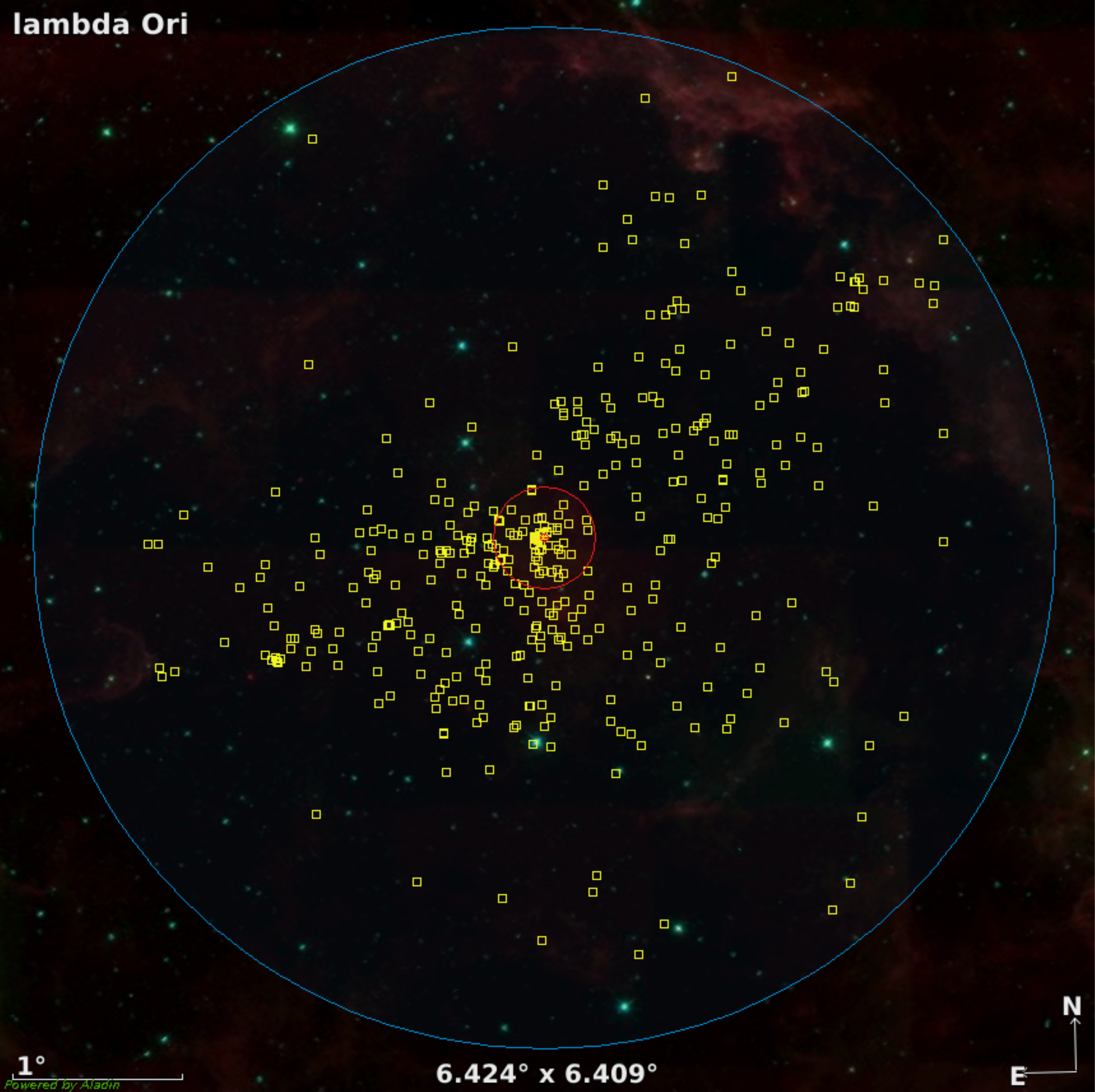}   
\end{figure}
 \begin{figure} [h]
 \centering
 \includegraphics[width=0.38\textwidth]{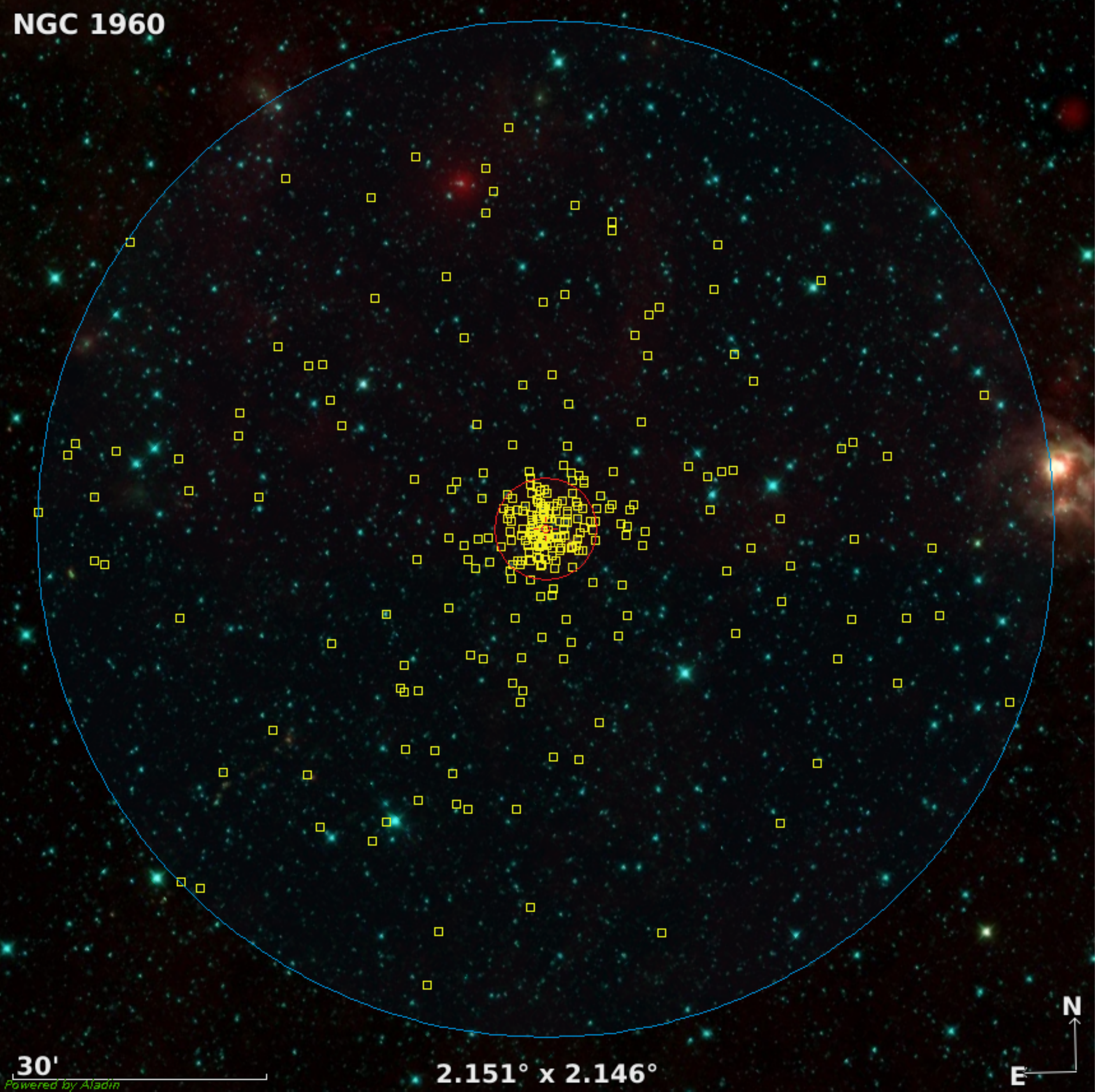}
 \includegraphics[width=0.38\textwidth]{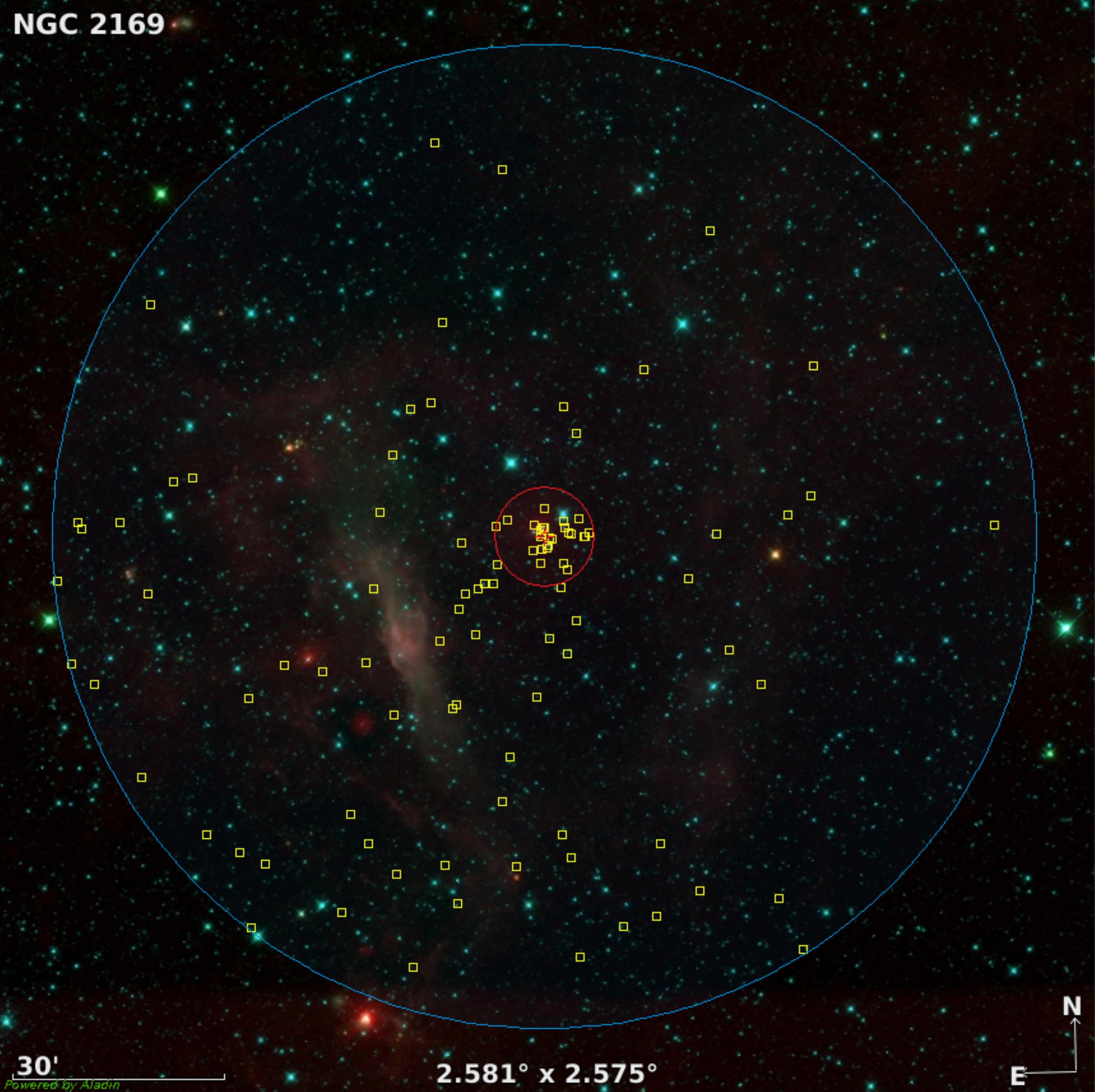}
 \end{figure}
\begin{figure} [h]
 \centering
    \includegraphics[width=0.38\textwidth]{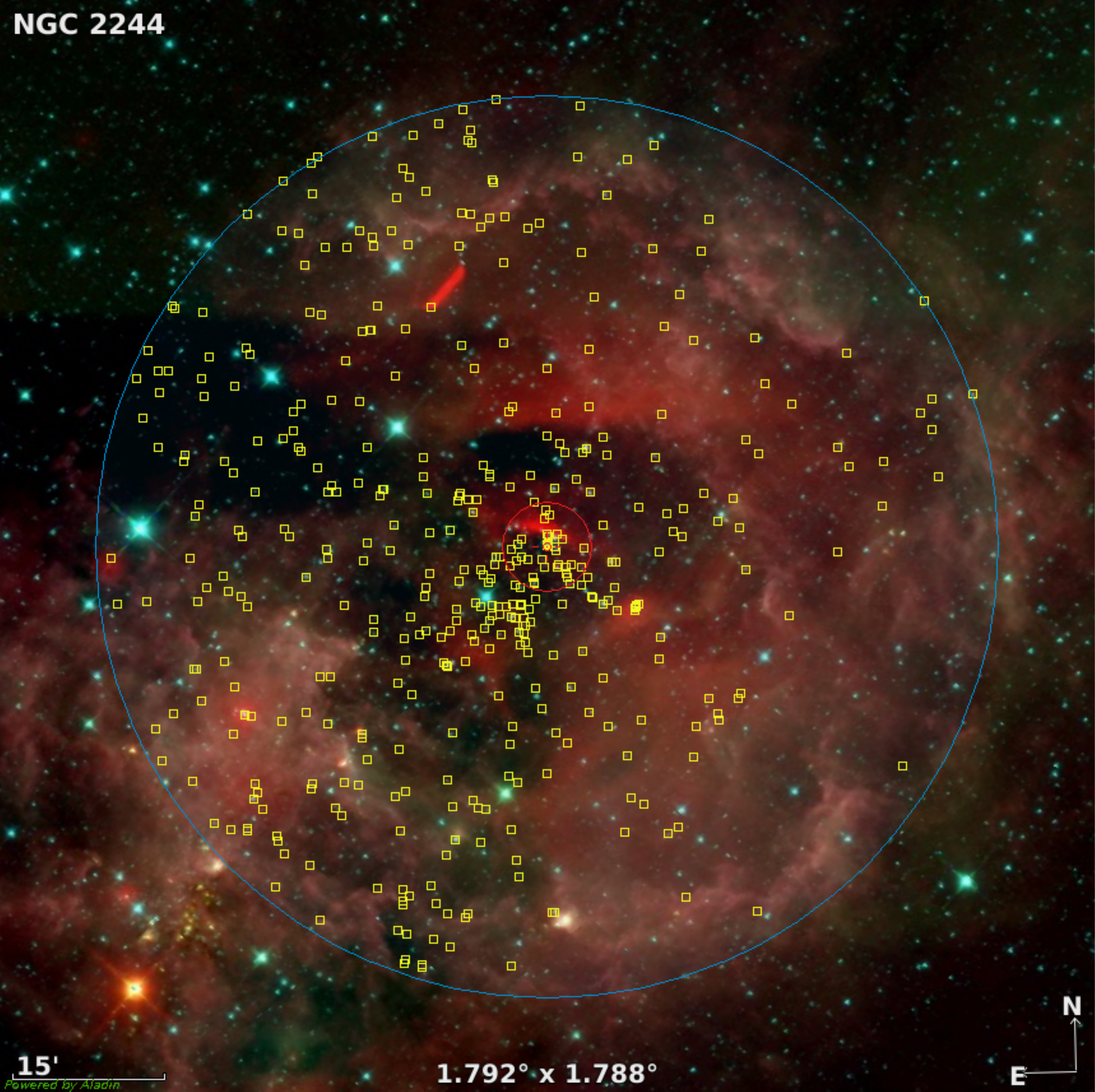}
    \includegraphics[width=0.38\textwidth]{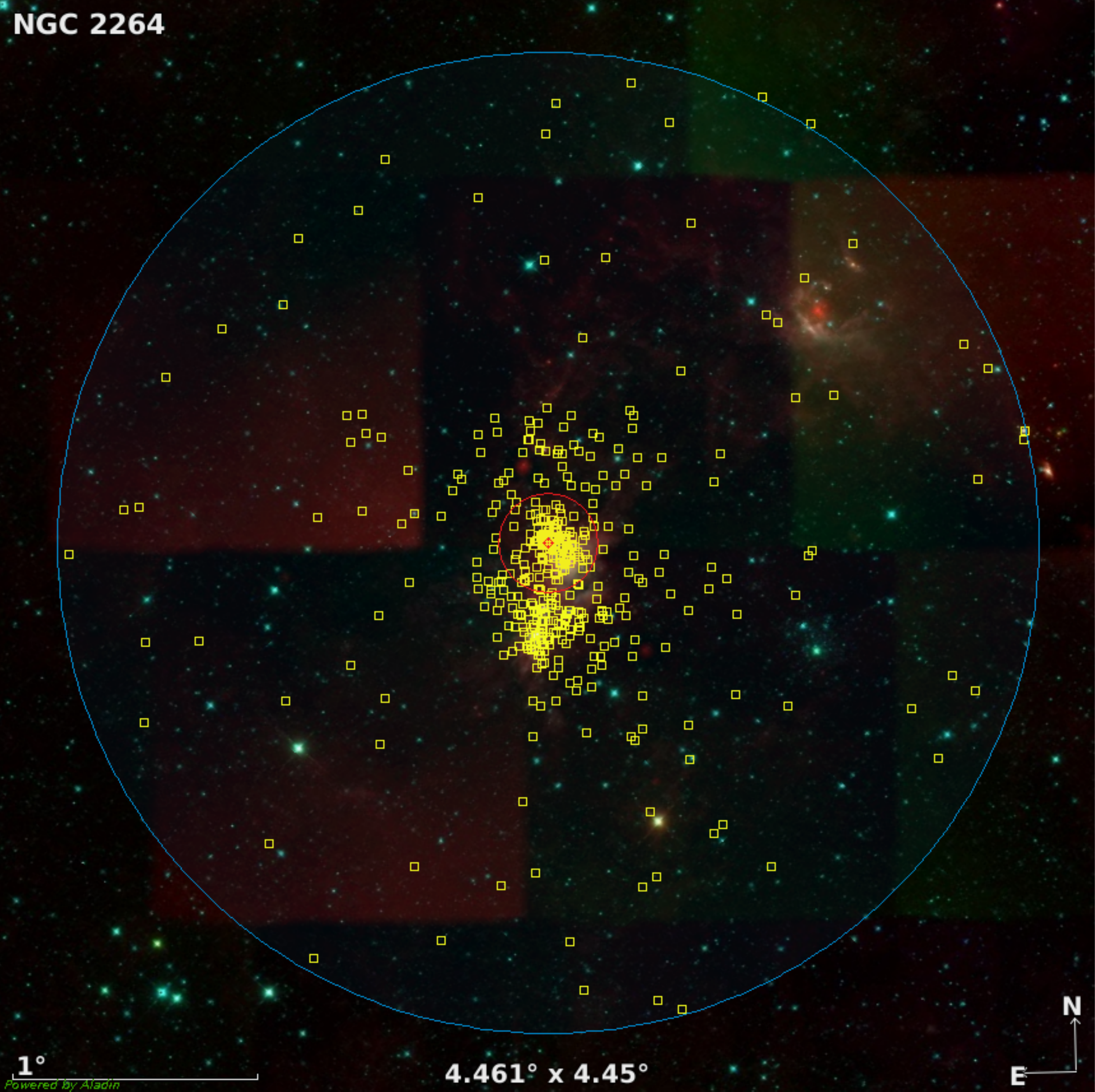} 
\end{figure}
 \begin{figure} [h]
 \centering
    \includegraphics[width=0.38\textwidth]{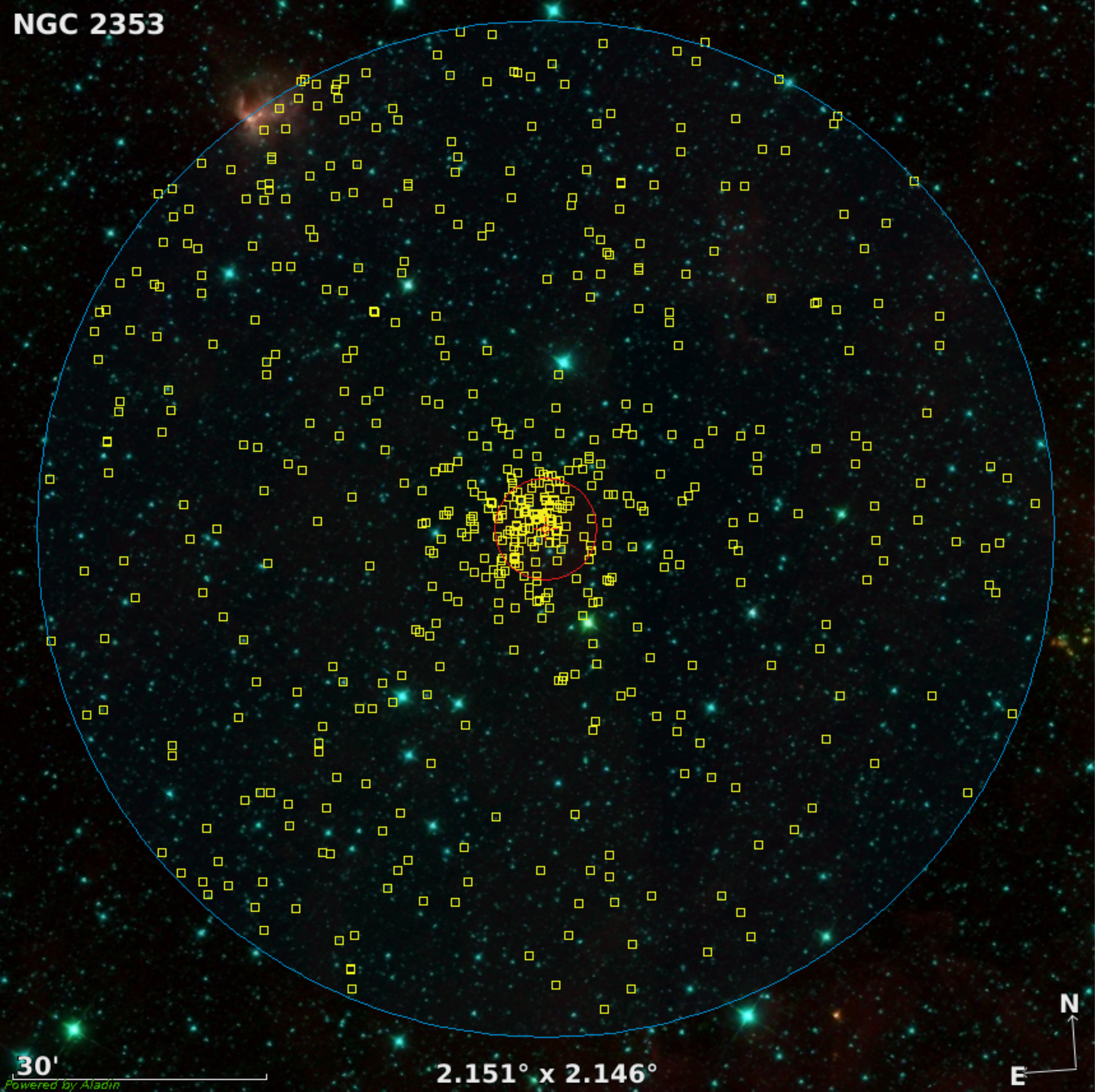}
    \includegraphics[width=0.38\textwidth]{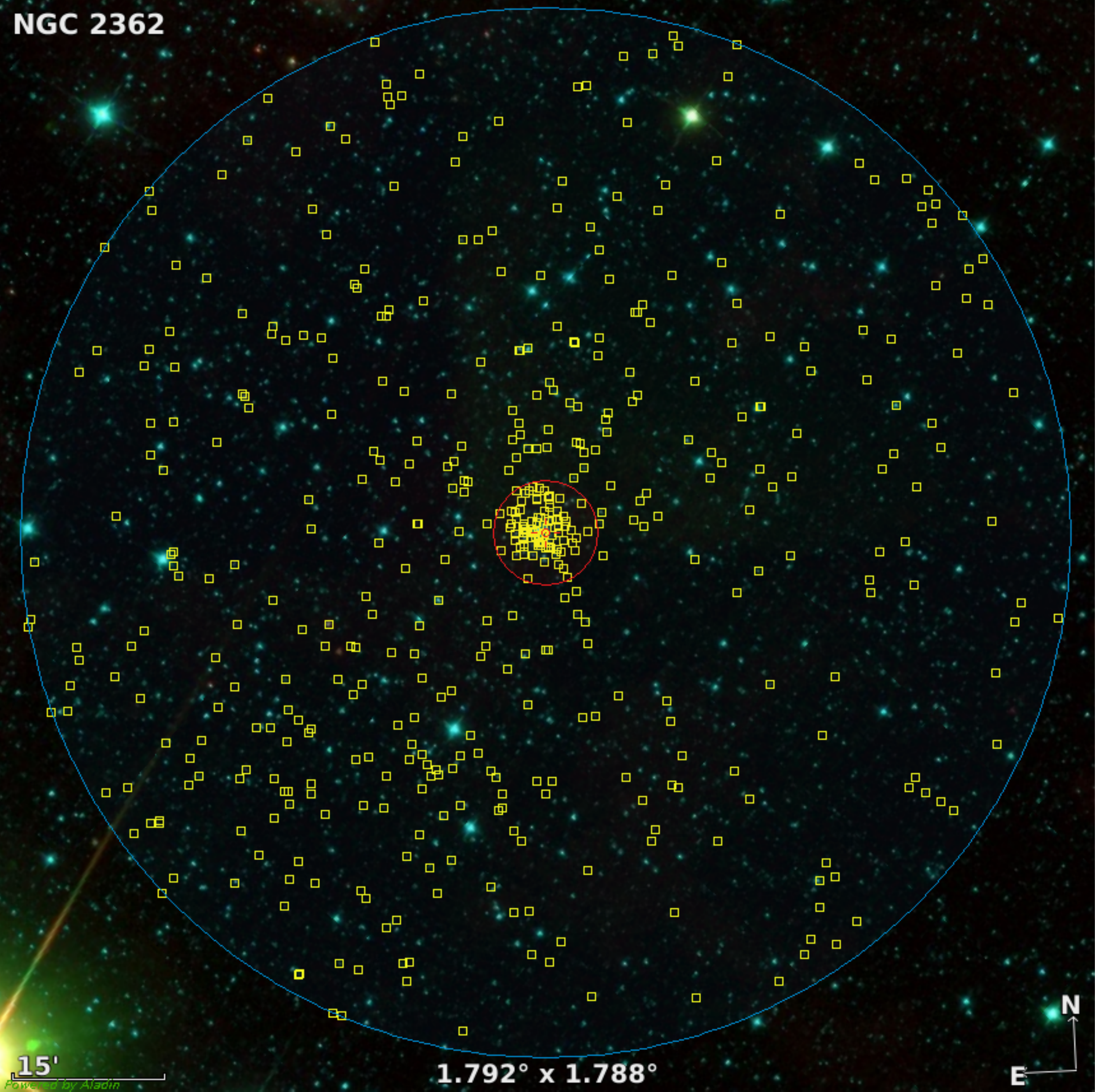} 
\end{figure}
\begin{figure} [h]
 \centering
    \includegraphics[width=0.38\textwidth]{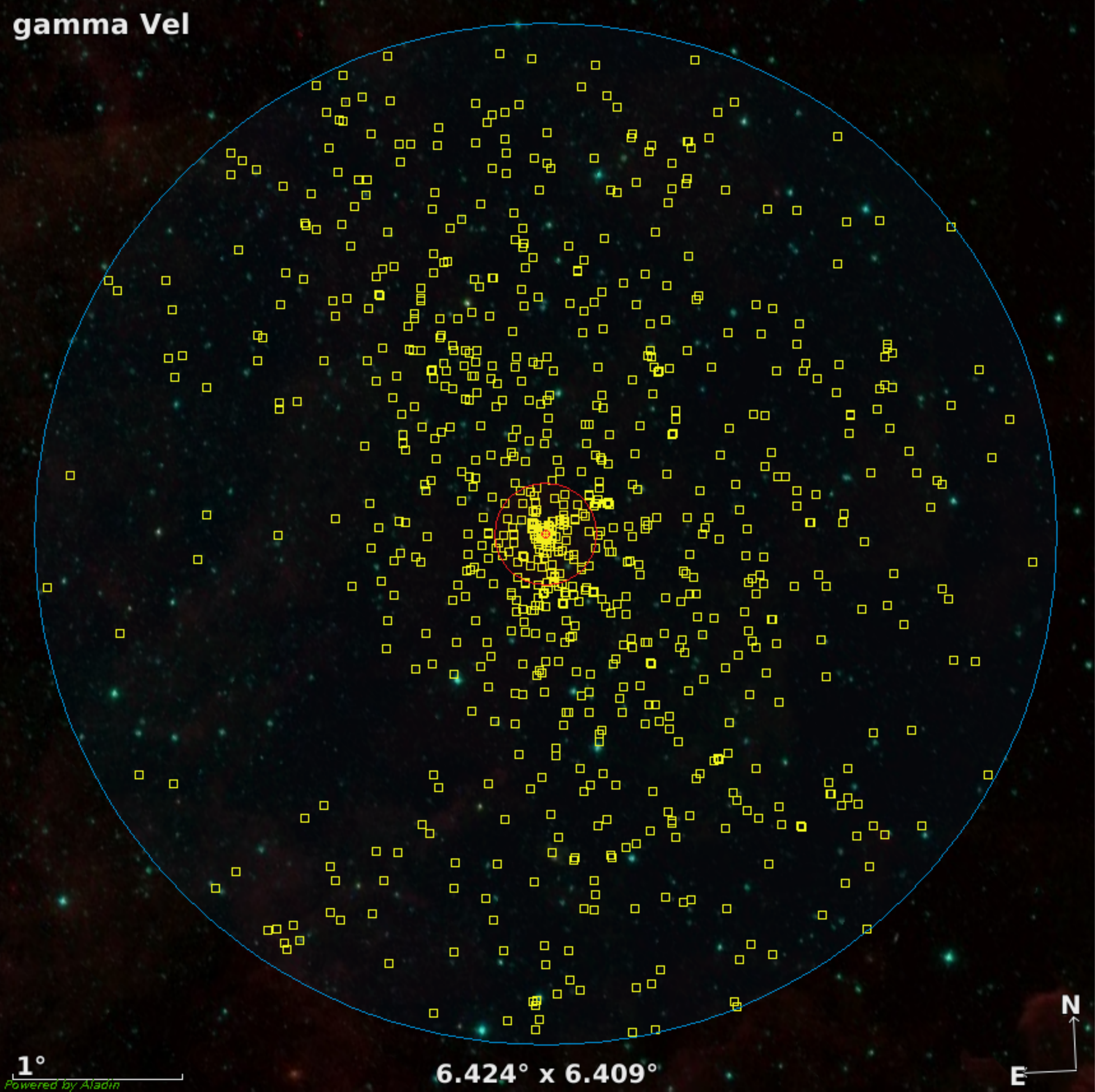}
    \includegraphics[width=0.38\textwidth]{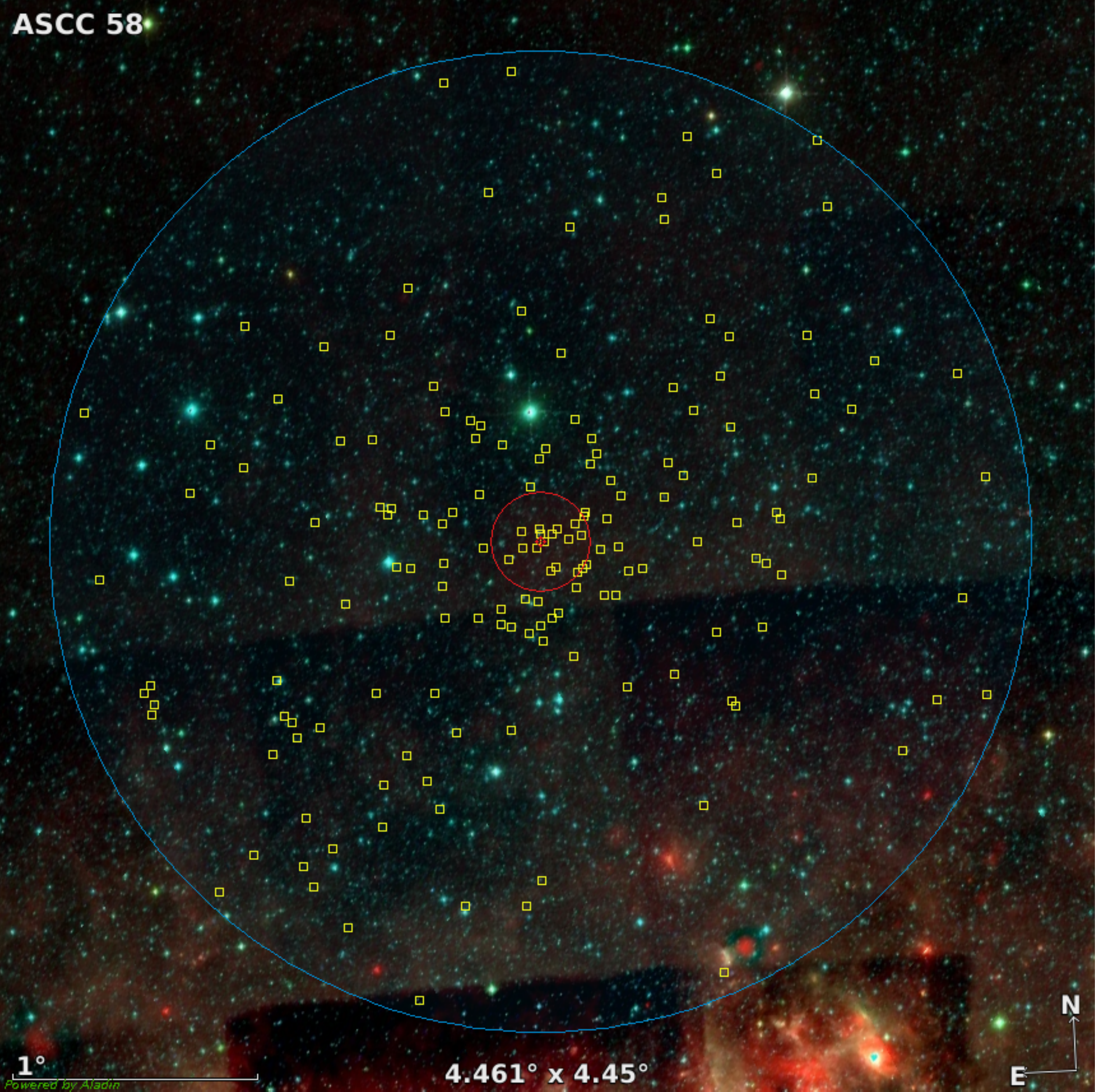}
    \end{figure}
\begin{figure} [h]
 \centering
    \includegraphics[width=0.38\textwidth]{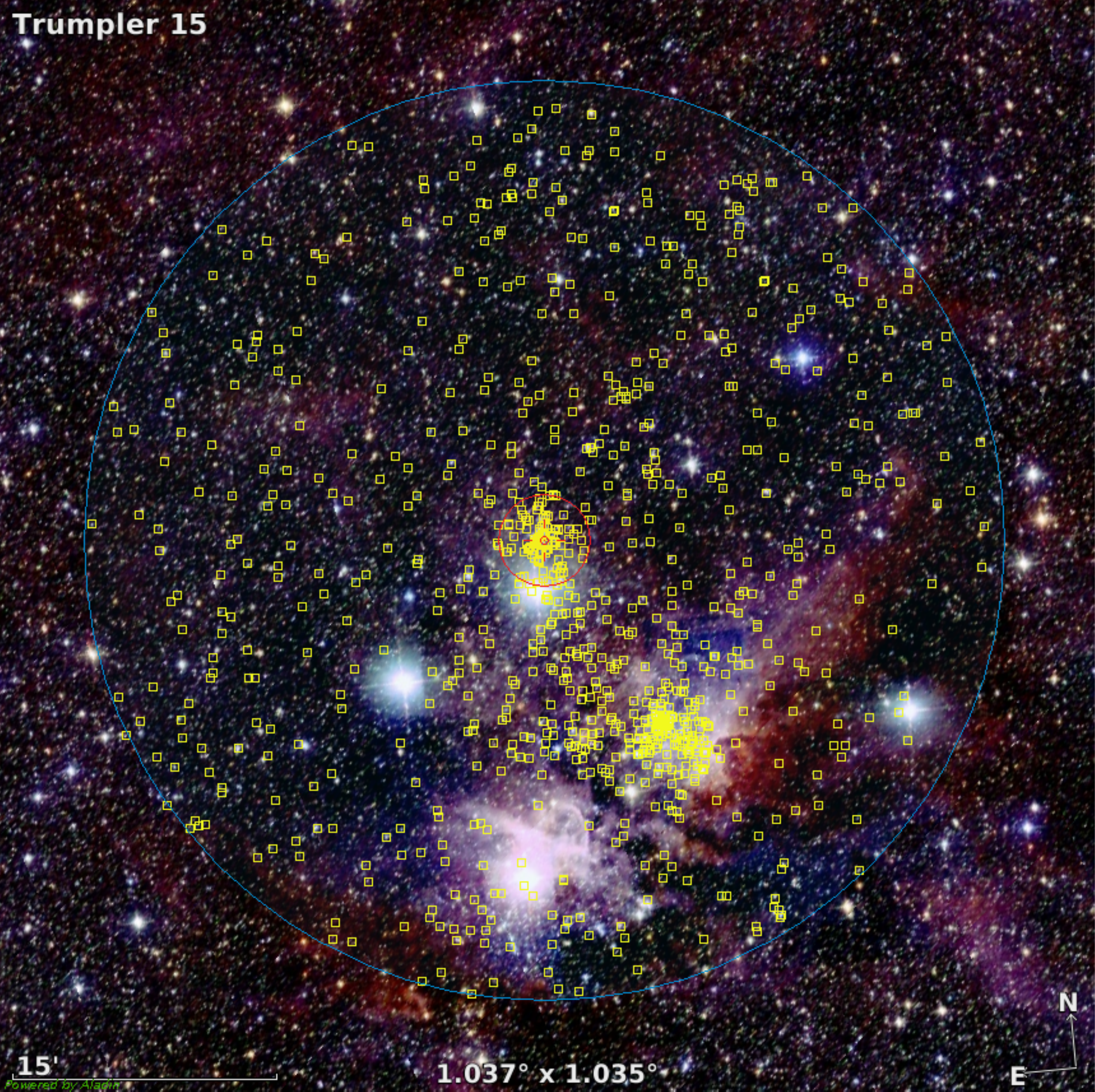}
    \includegraphics[width=0.38\textwidth]{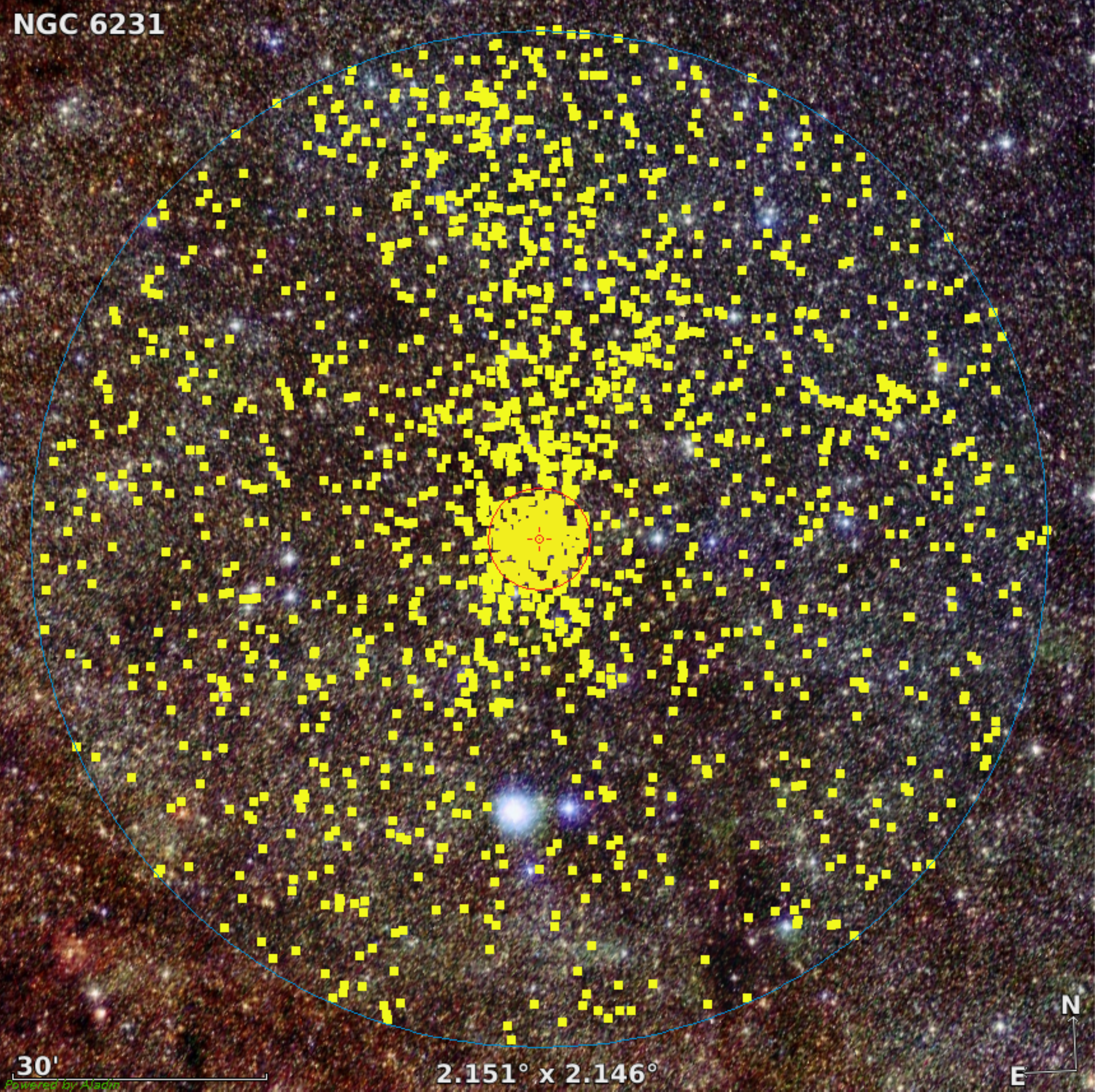}
    \end{figure}
\begin{figure} [h]
 \centering
    \includegraphics[width=0.38\textwidth]{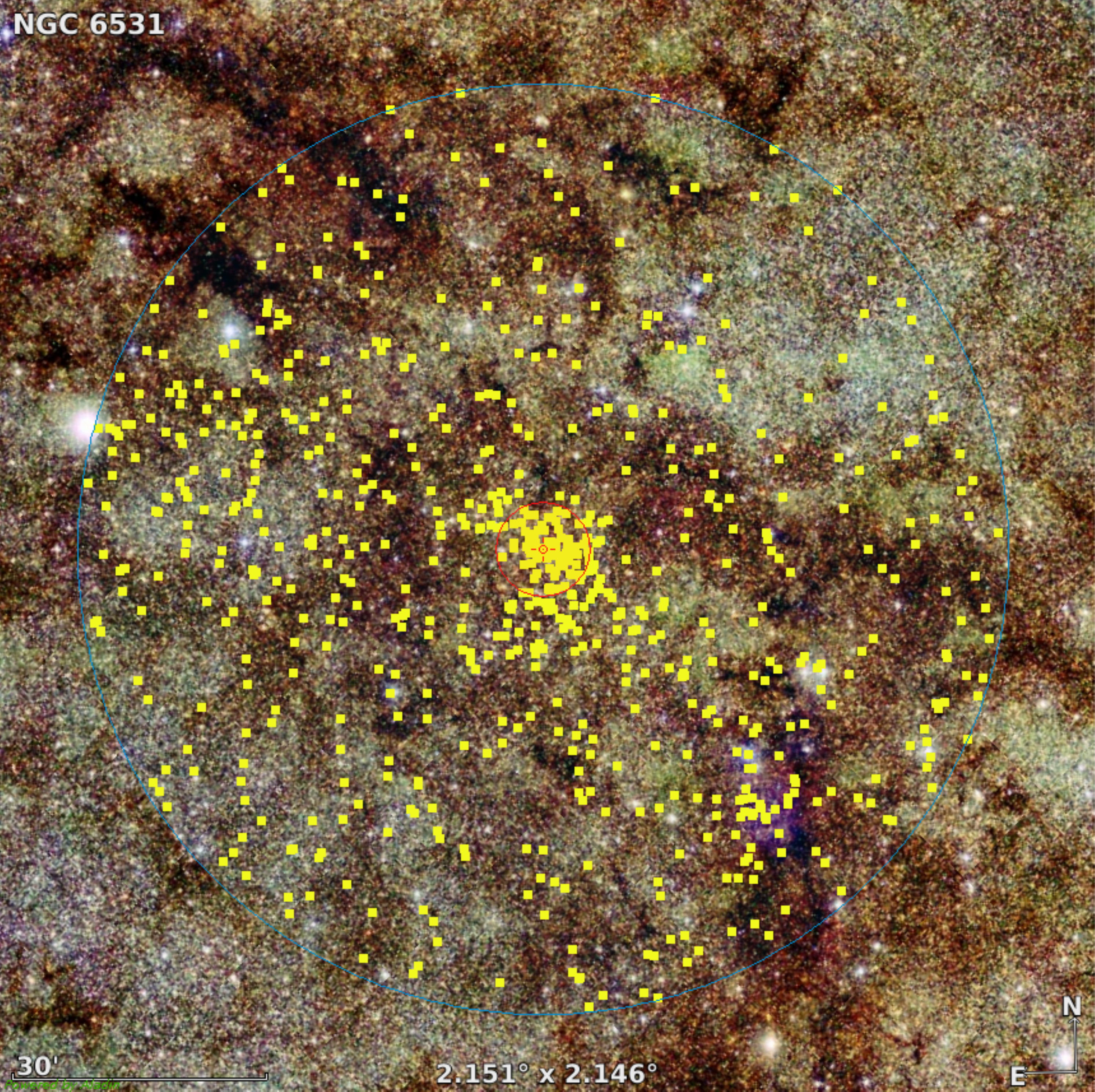}
    \includegraphics[width=0.38\textwidth]{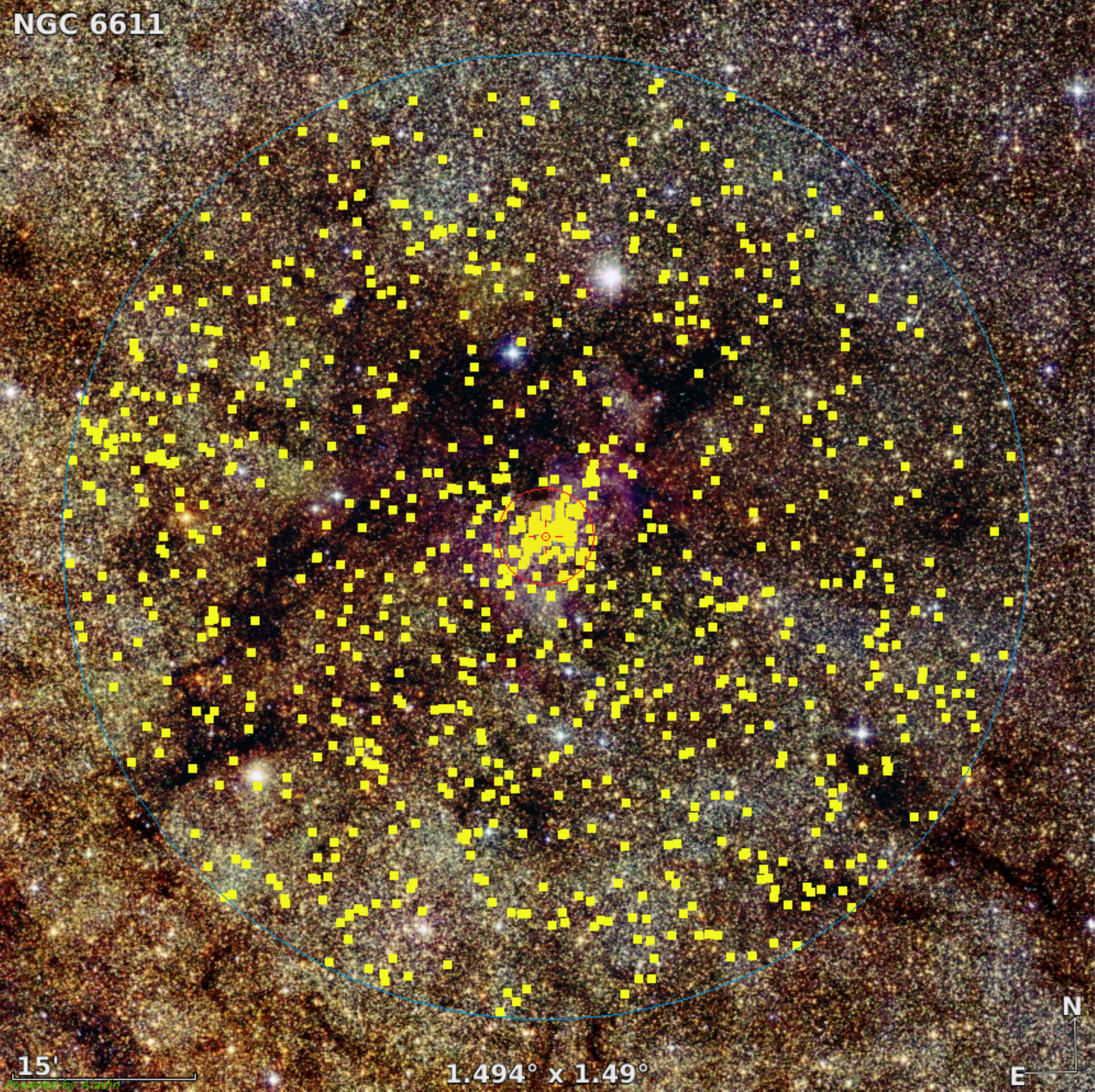}
\end{figure}
\begin{figure} [h]
 \centering
    \includegraphics[width=0.38\textwidth]{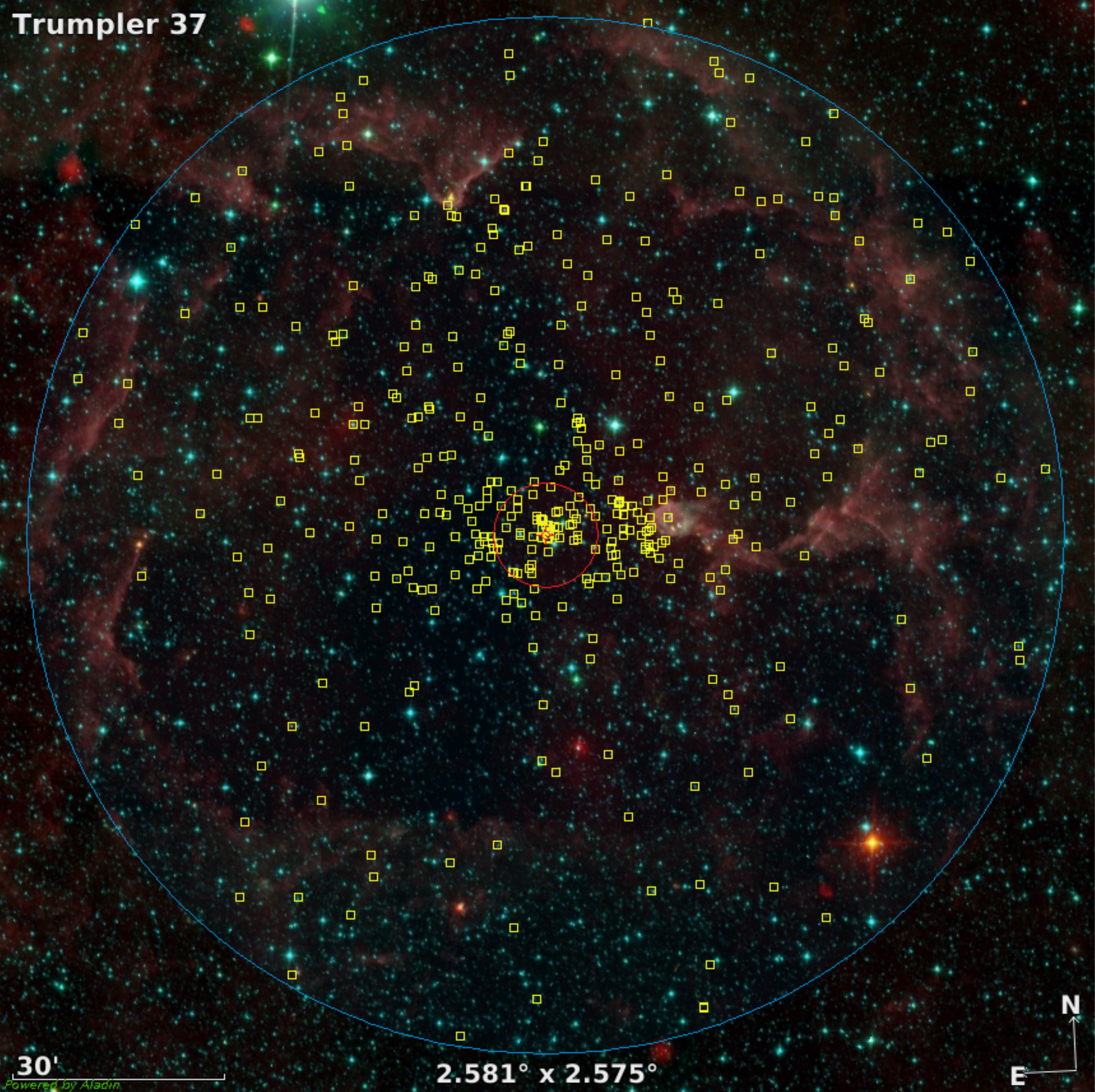}
    \includegraphics[width=0.38\textwidth]{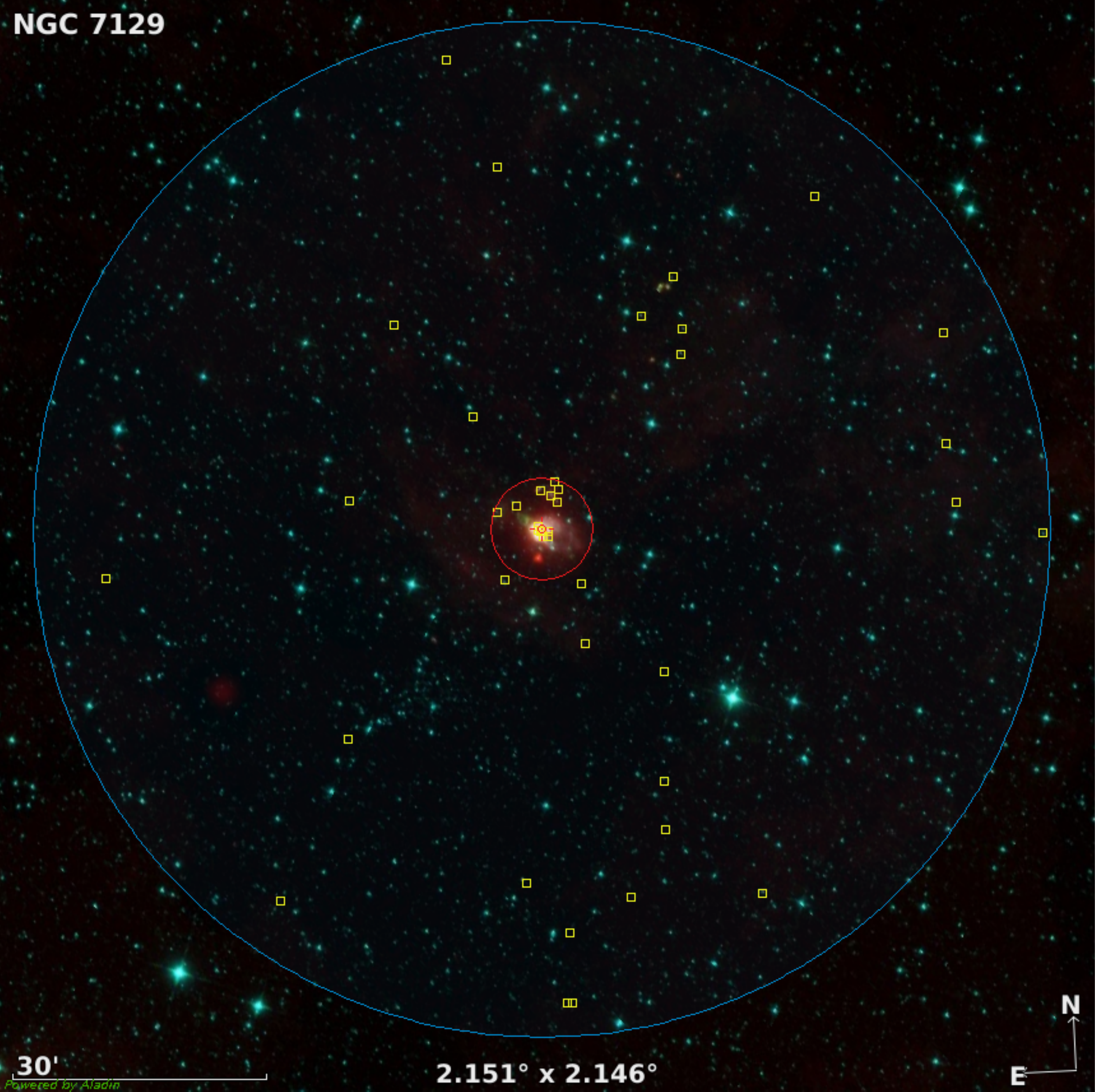}
    \end{figure}
\begin{figure} [h]
 \centering
    \includegraphics[width=0.43\textwidth]{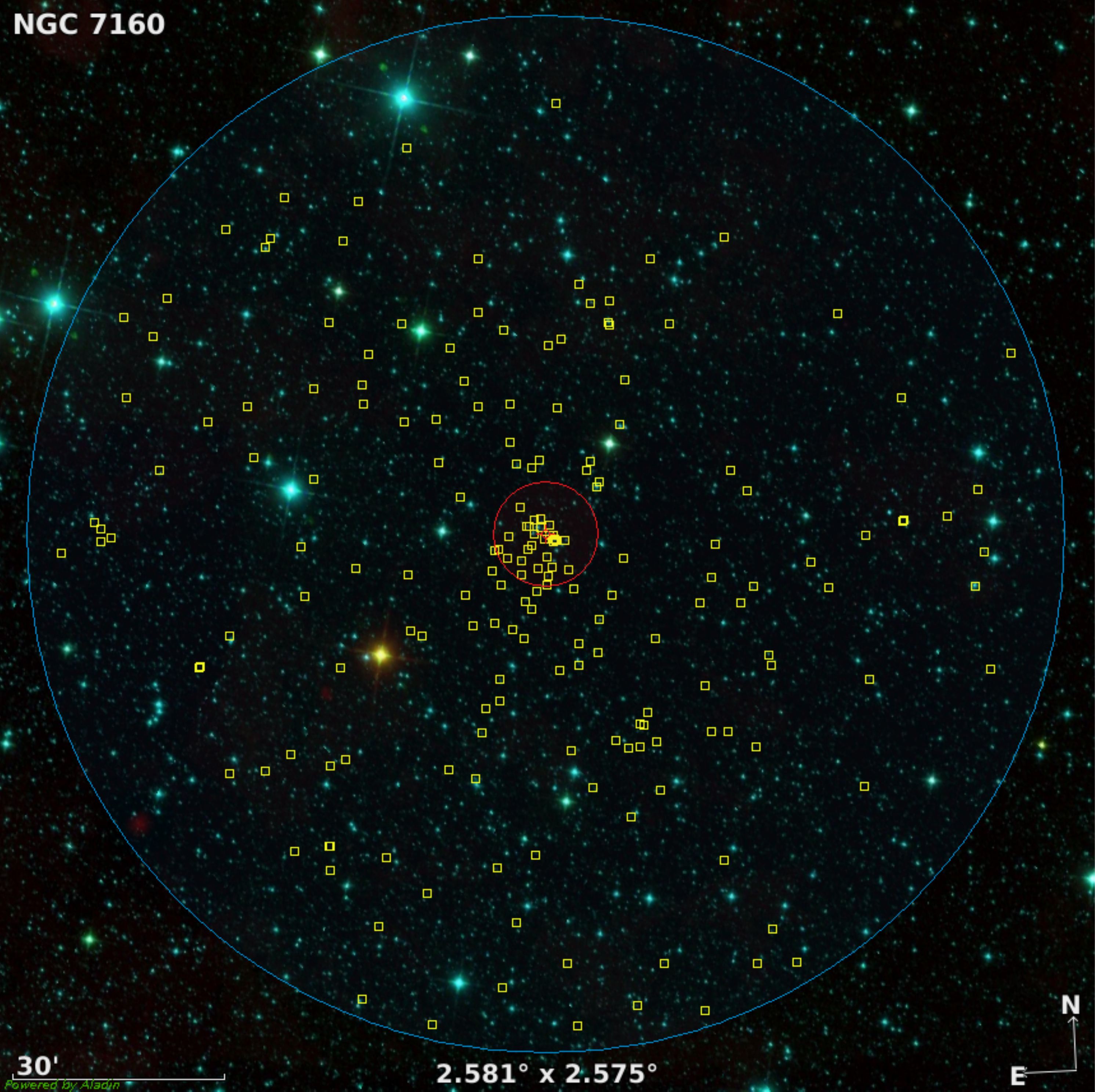}  
            \caption{Sky projections for the different clusters where the yellow squares indicate the positions of all members identified within FOV$_{20pc}$ and FOV$_{2pc}$, whose corresponding sizes are indicated with the blue and red circles, respectively. Background images are colored according to the ALLWISE W1, W2, and W4 bands (except for Trumpler 15, NGC 6231, NGC 6531, and NGC 6611, where 2MASS images are used for a better visualization). }
            \label{figure:FOVS}
\end{figure}

\newpage
\onecolumn
\begin{figure} [h]
 \centering
    \includegraphics[width=0.43\textwidth]{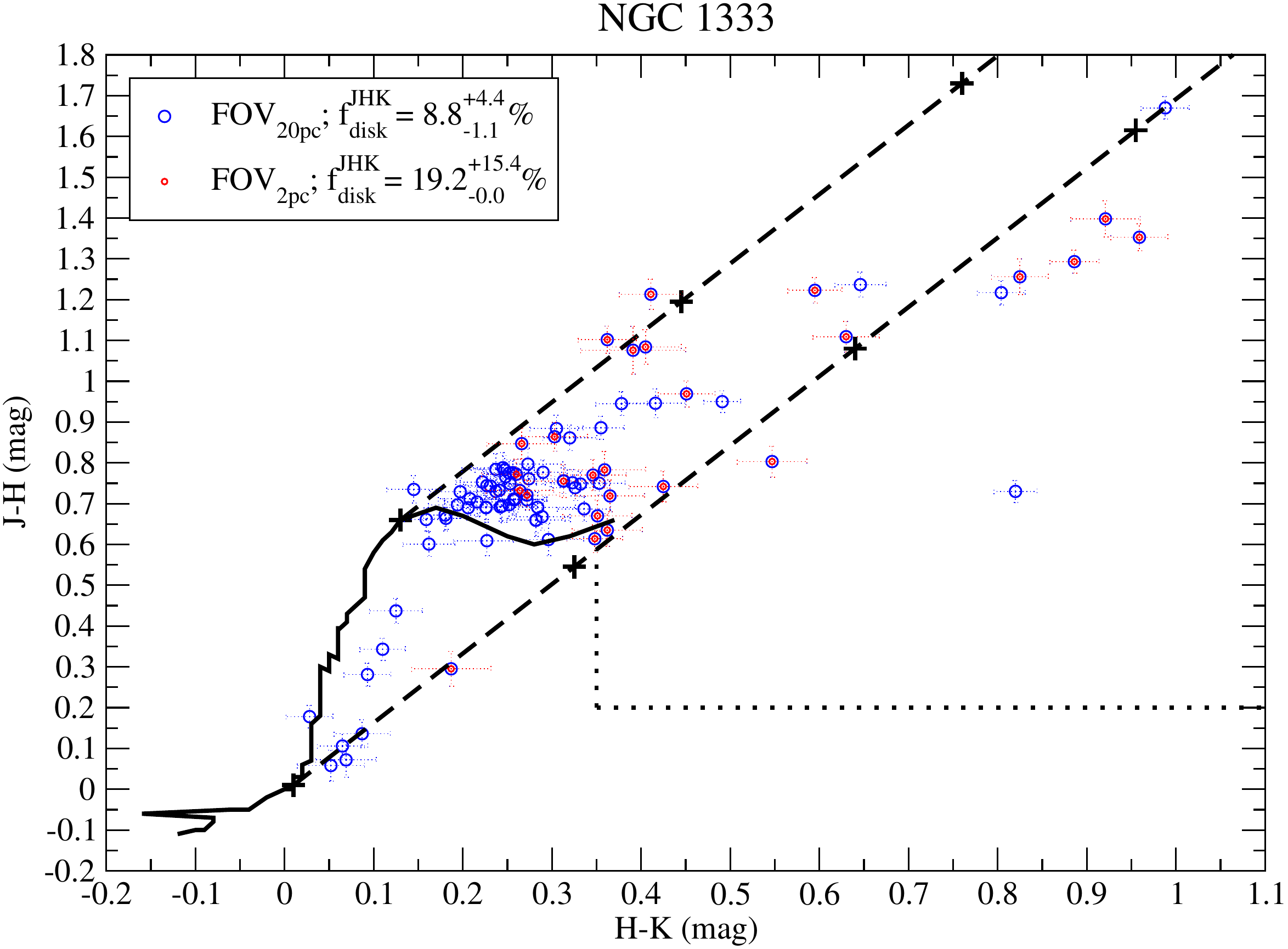}
    \includegraphics[width=0.43\textwidth]{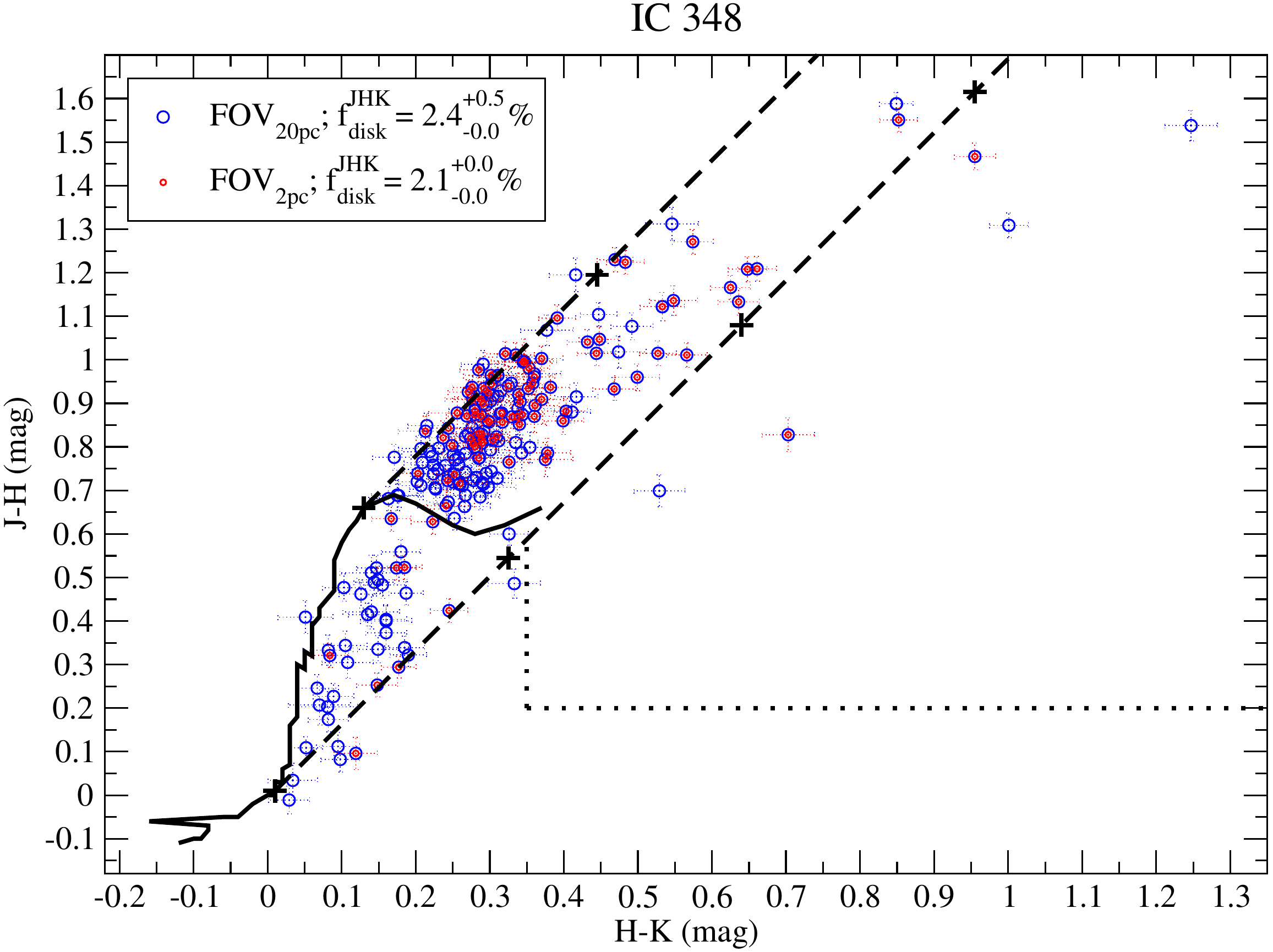}
\end{figure}
\begin{figure} [h]
 \centering
    \includegraphics[width=0.43\textwidth]{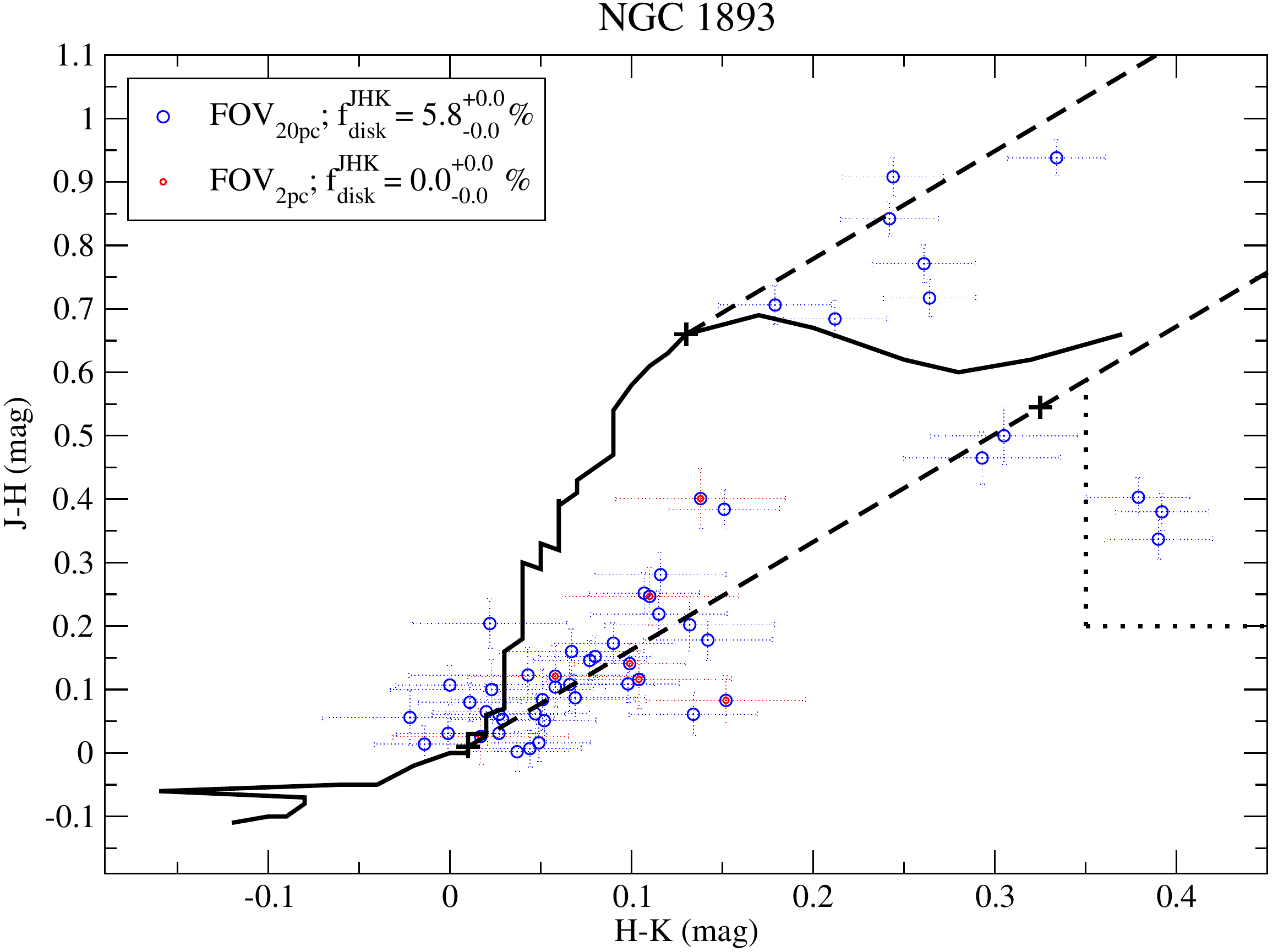}
    \includegraphics[width=0.43\textwidth]{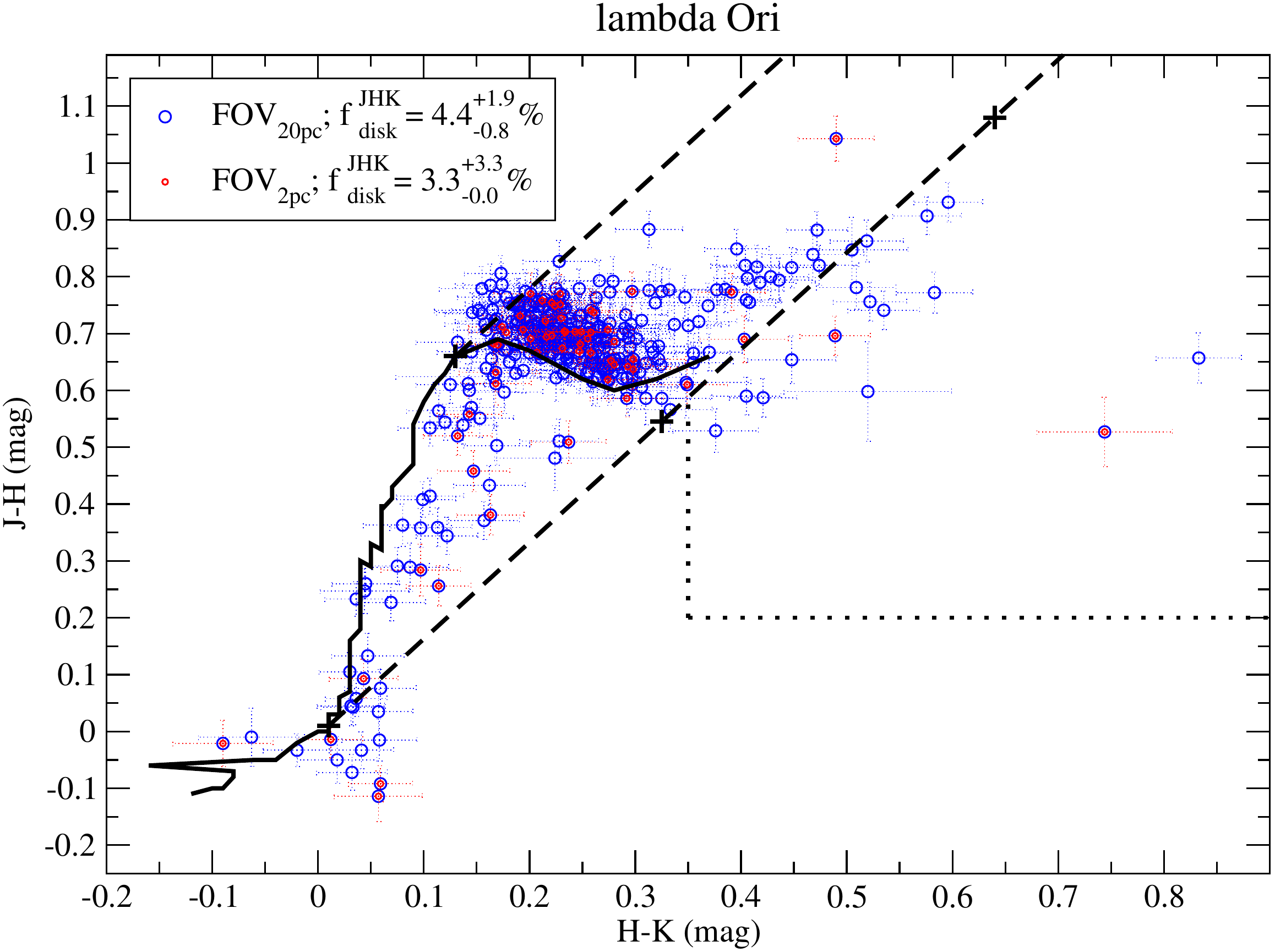} 
\end{figure}
\begin{figure} [h]
 \centering
    \includegraphics[width=0.43\textwidth]{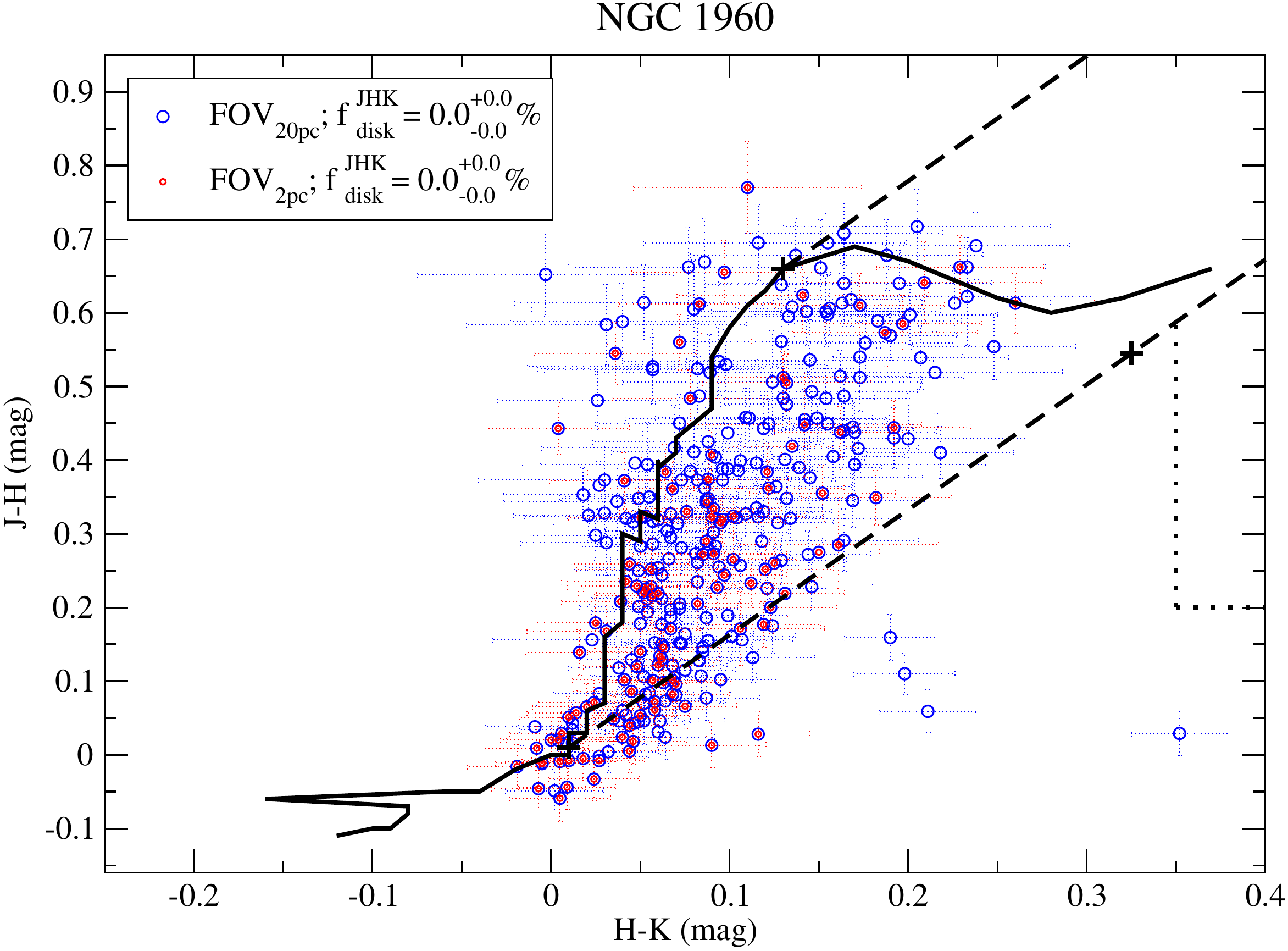}
    \includegraphics[width=0.43\textwidth]{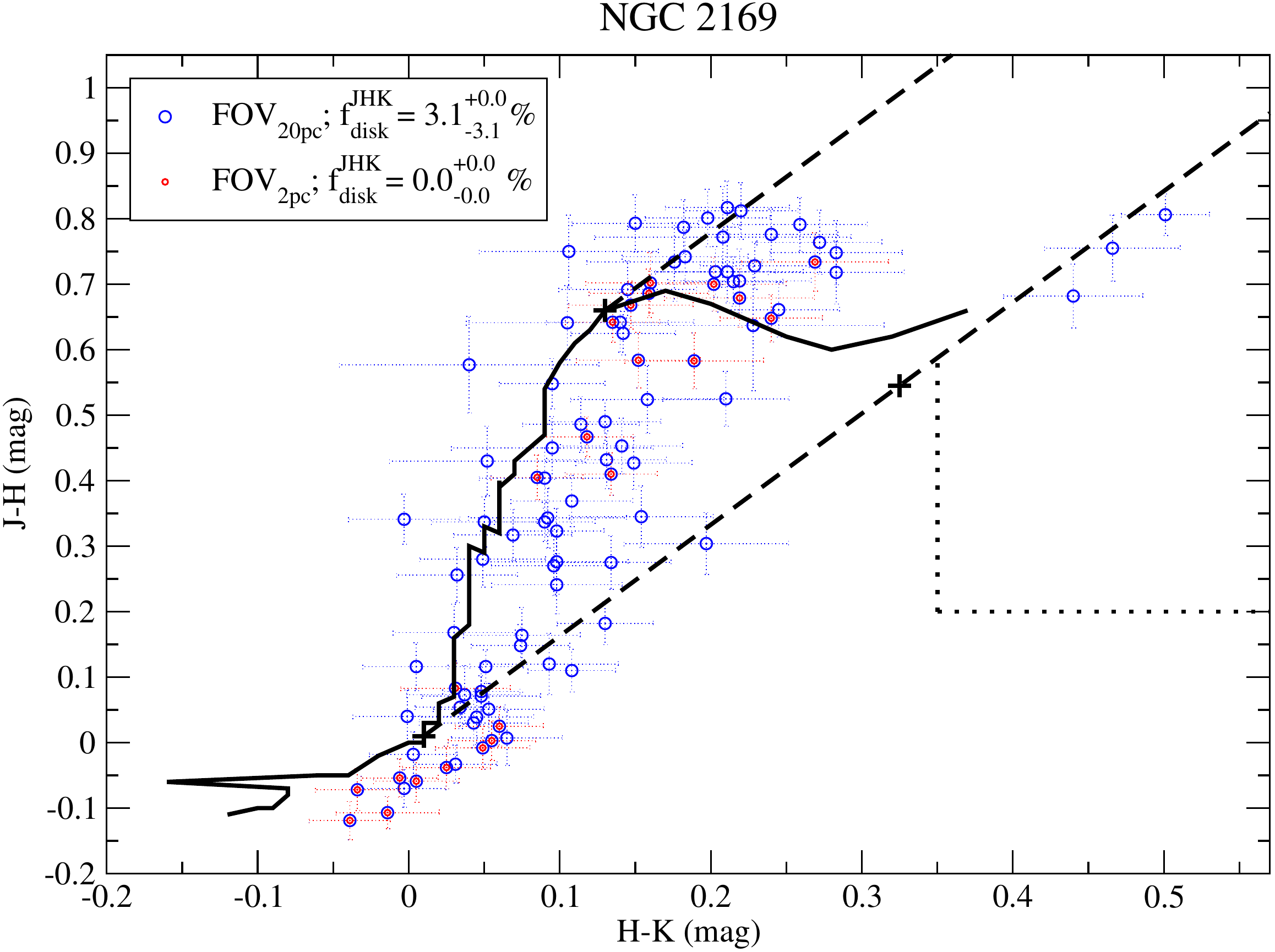}
\end{figure}
\begin{figure} [h]
 \centering
    \includegraphics[width=0.43\textwidth]{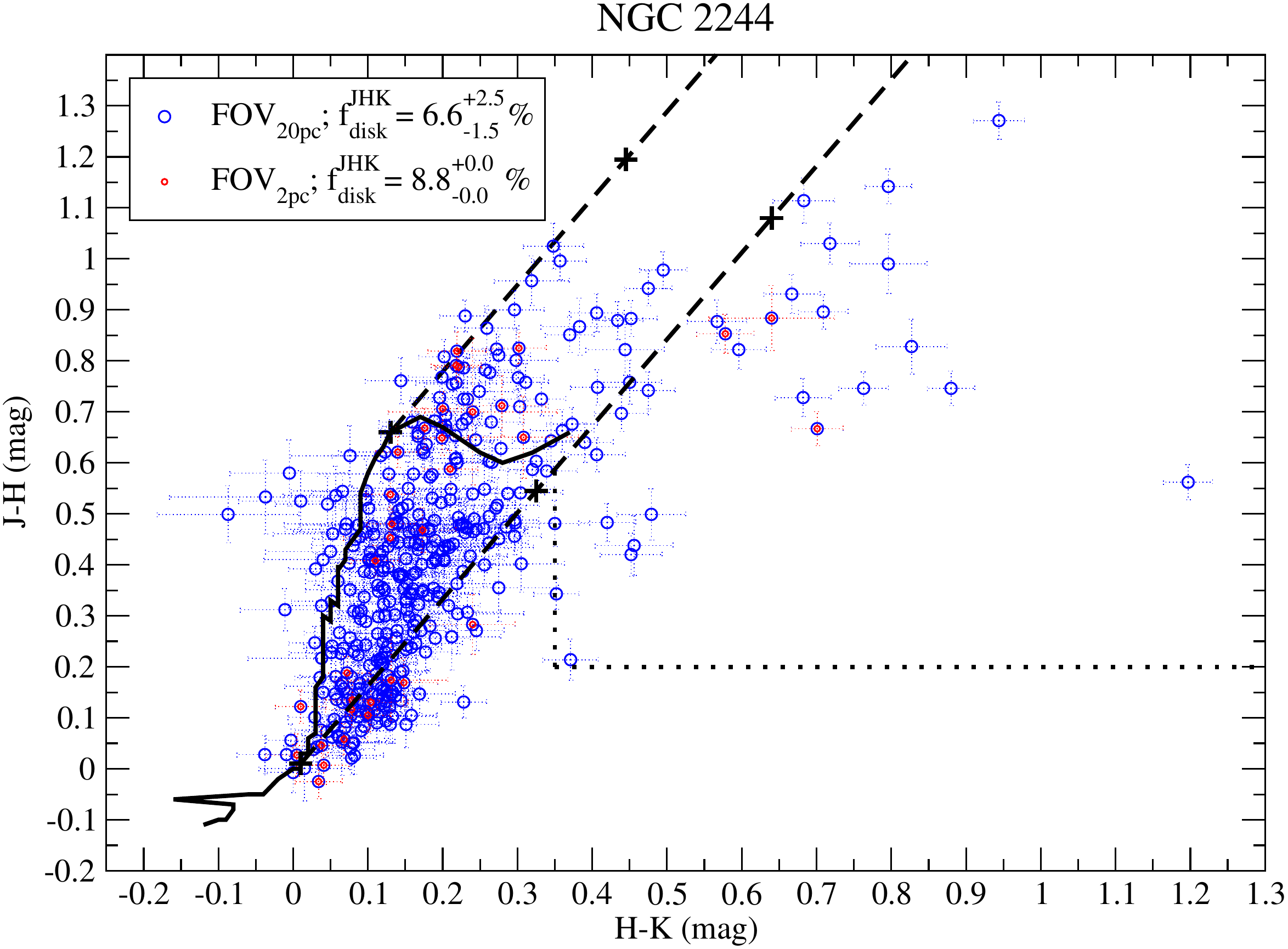}
    \includegraphics[width=0.43\textwidth]{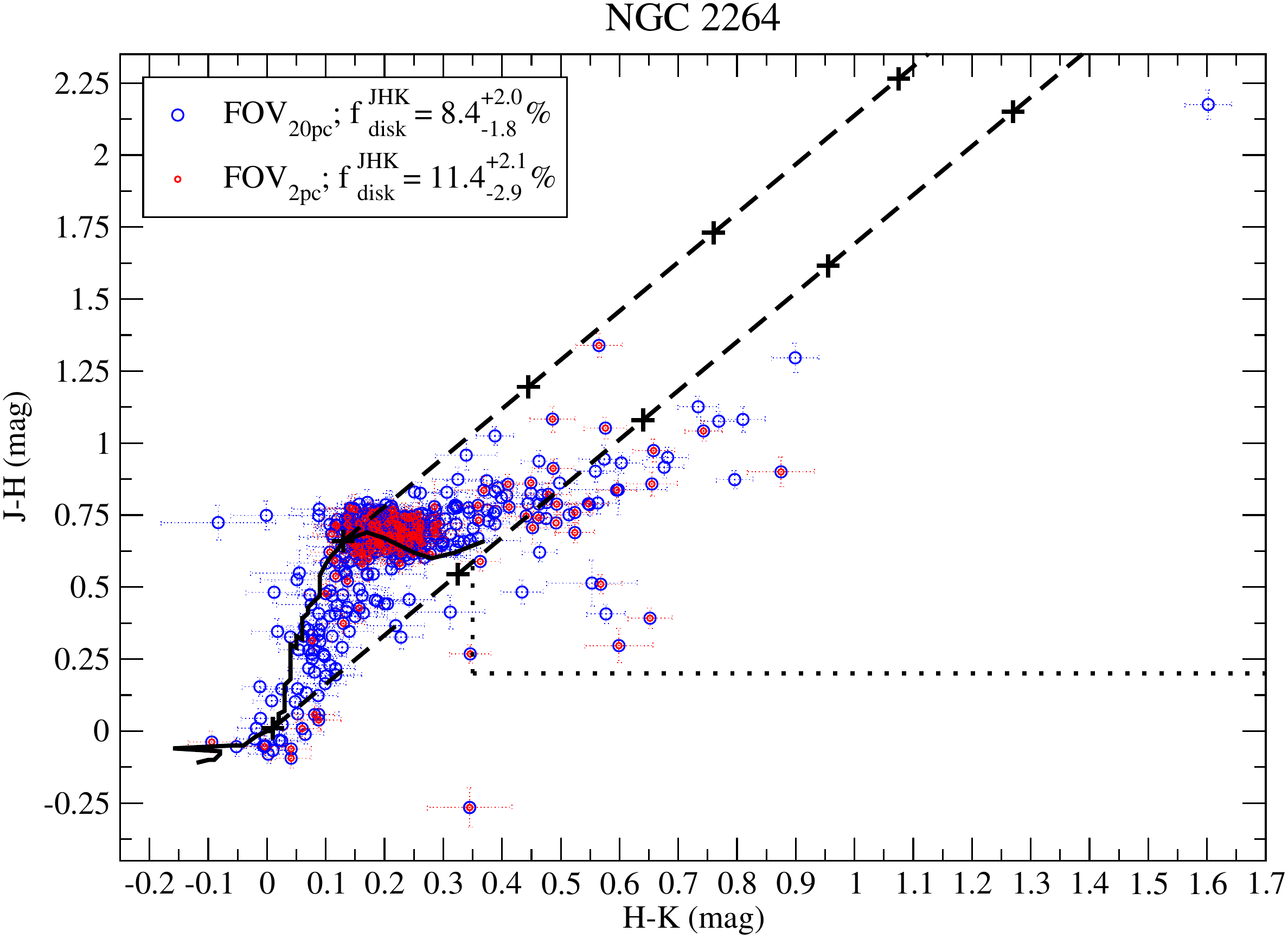}
\end{figure}
\begin{figure} [h]
 \centering
    \includegraphics[width=0.43\textwidth]{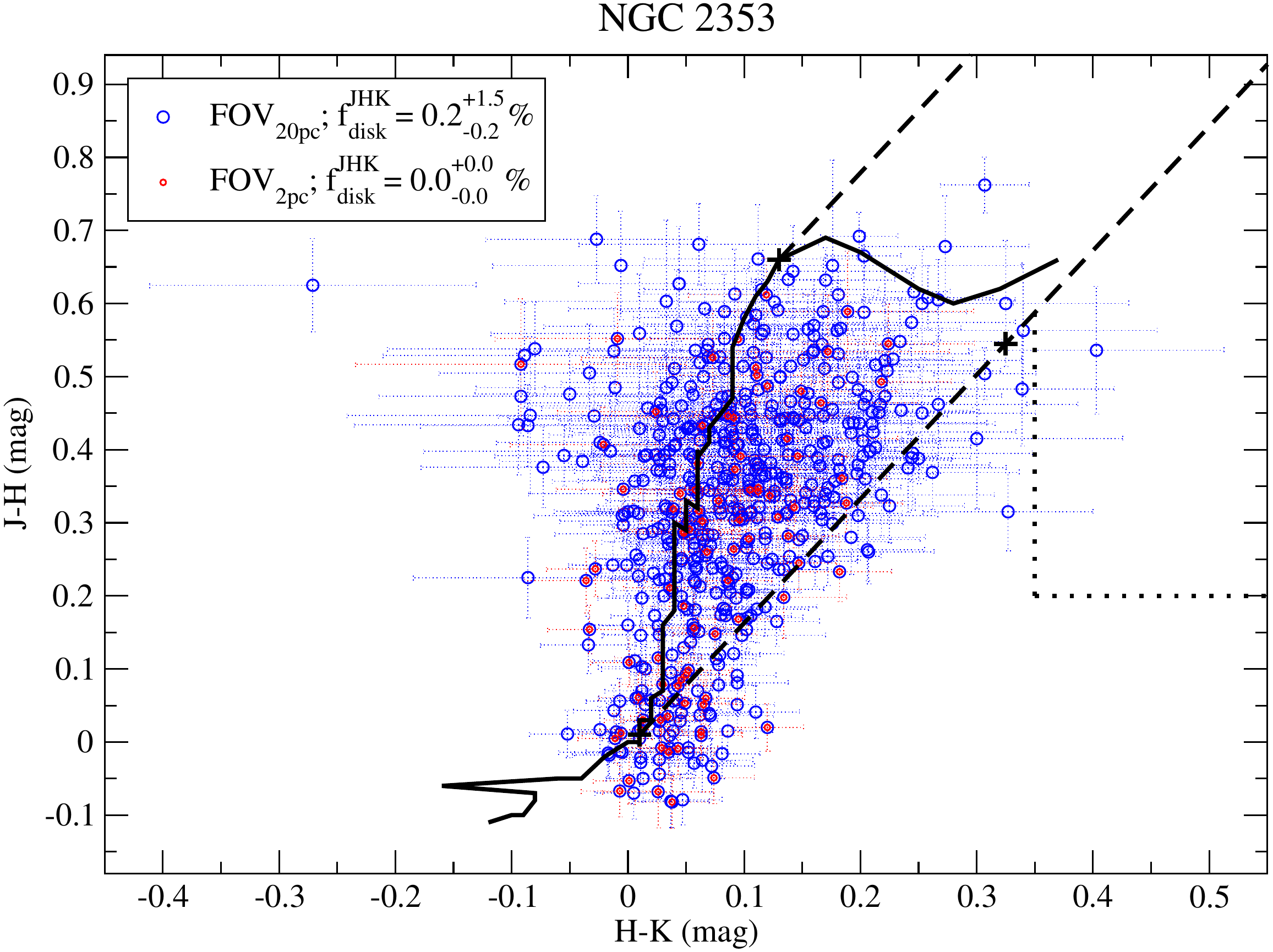}
    \includegraphics[width=0.43\textwidth]{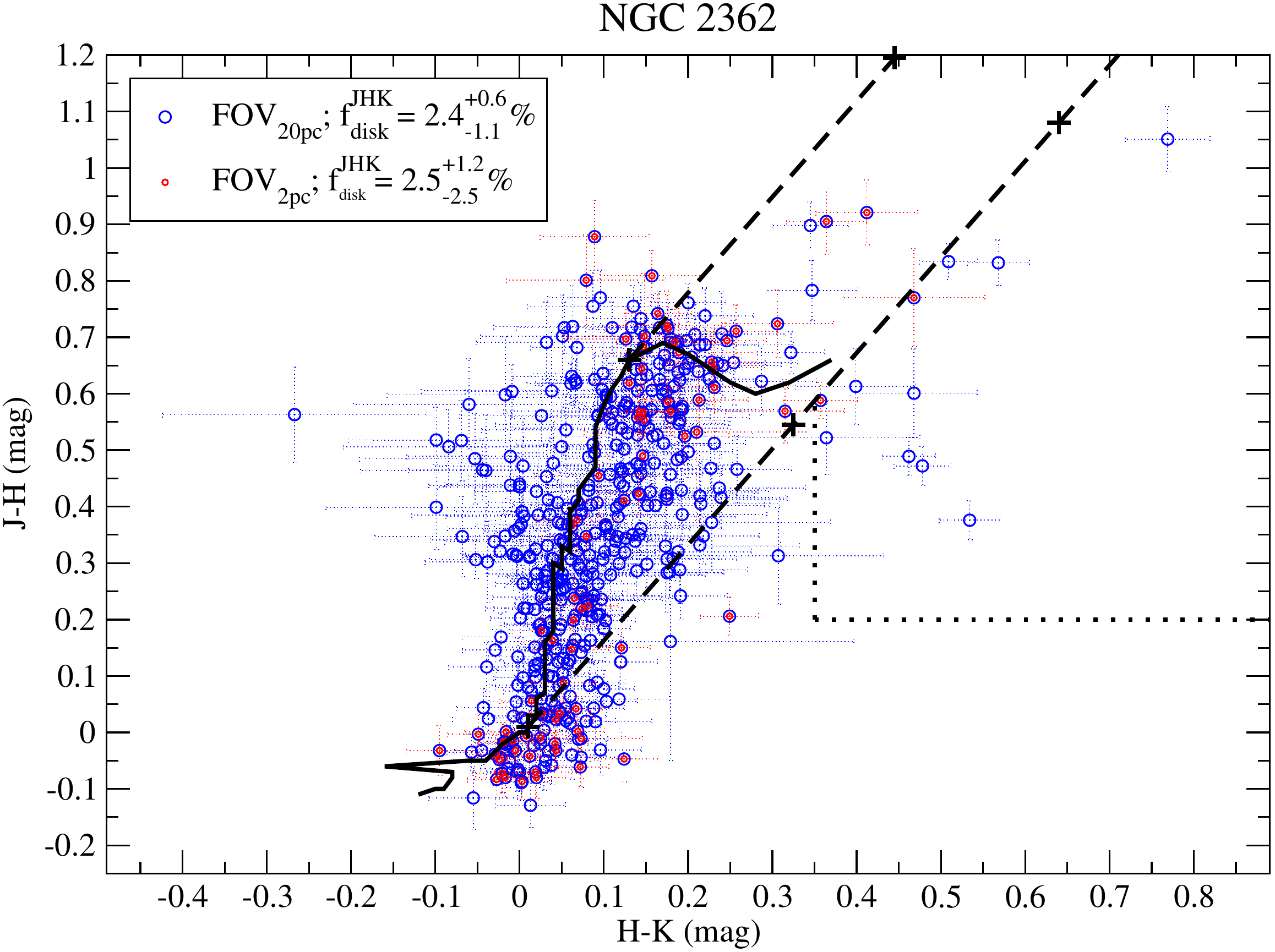}
\end{figure}
\begin{figure} [h]
 \centering
    \includegraphics[width=0.43\textwidth]{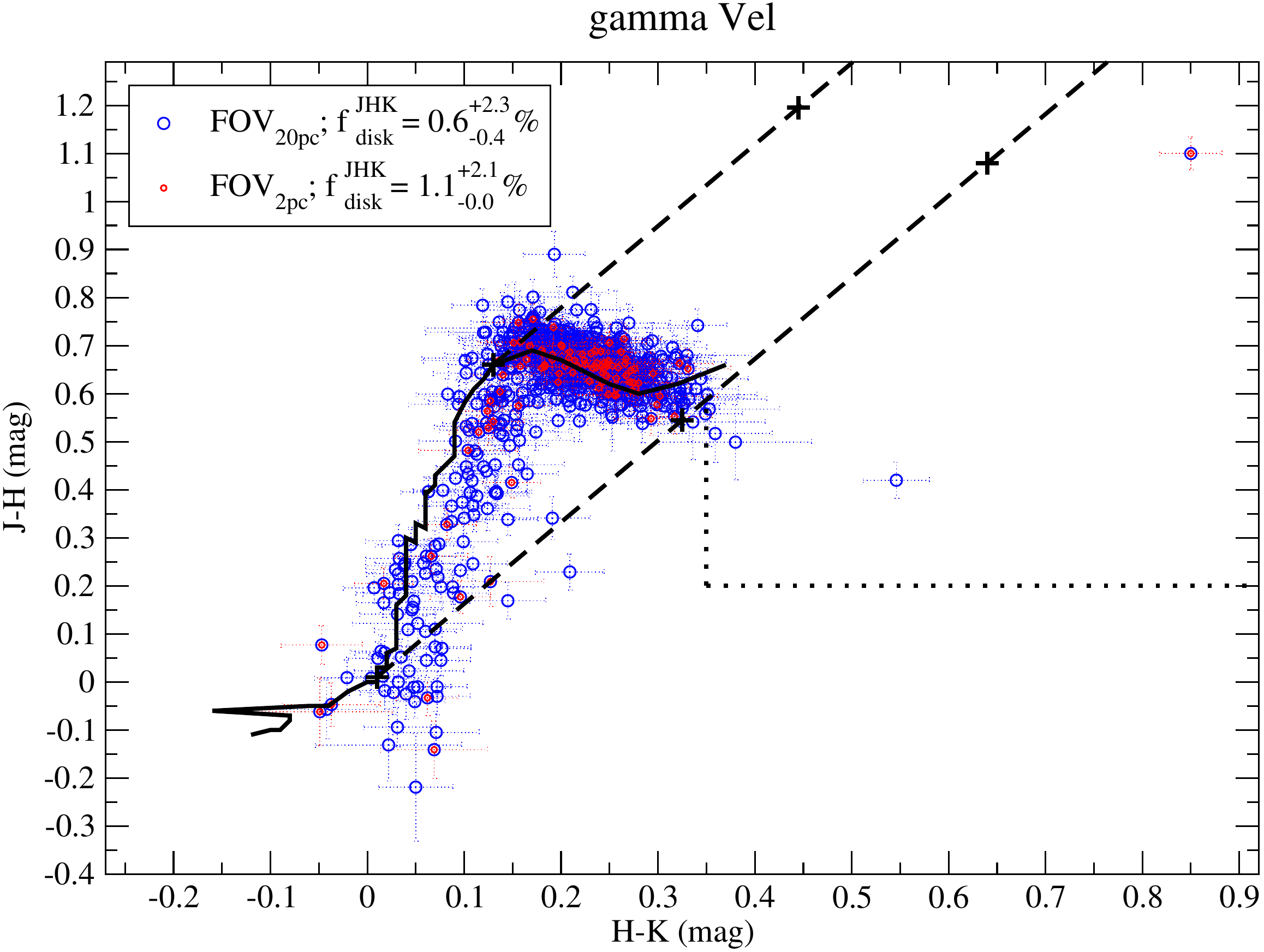}
    \includegraphics[width=0.43\textwidth]{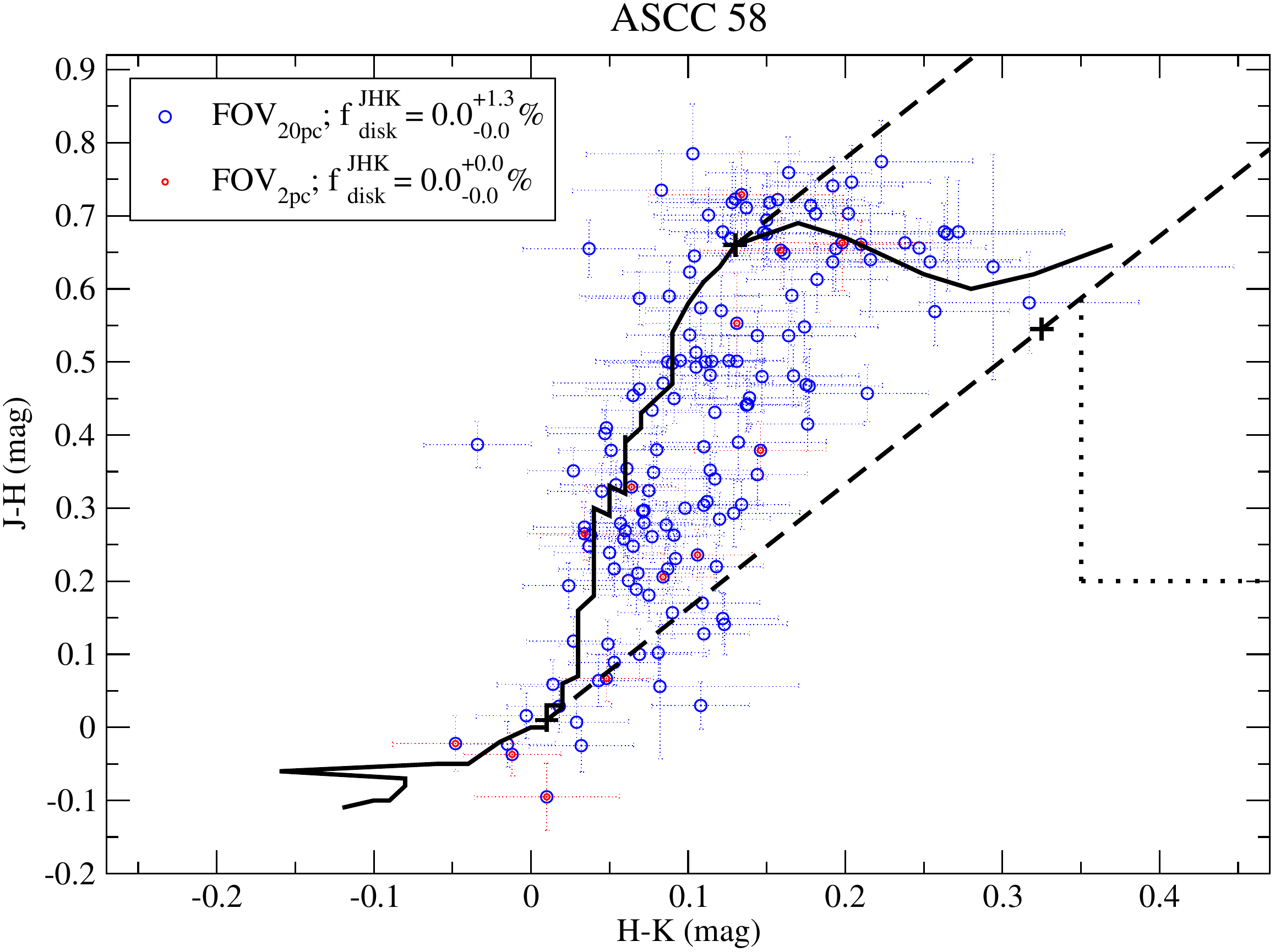}
\end{figure}
\begin{figure} [h]
 \centering
    \includegraphics[width=0.43\textwidth]{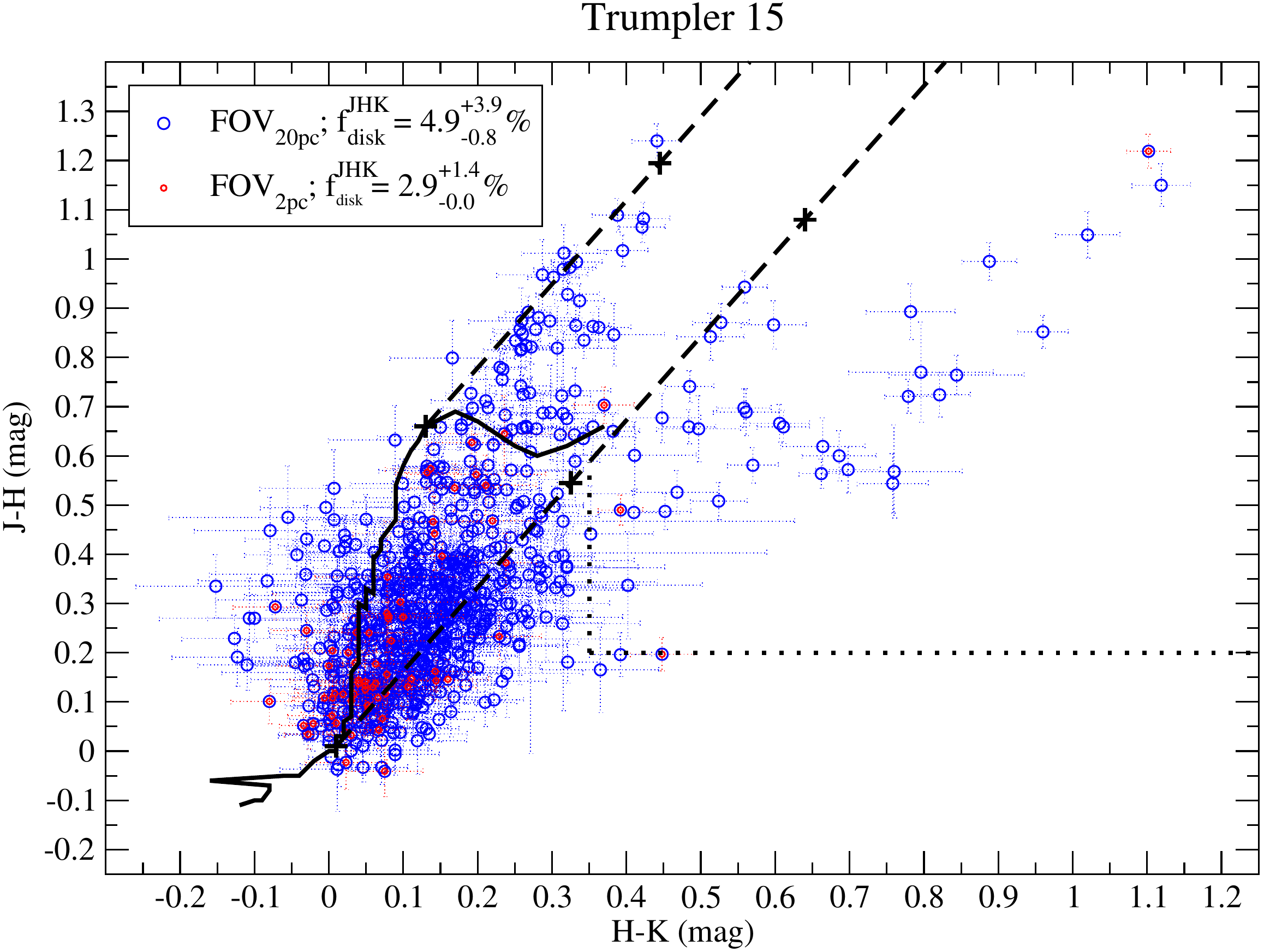}
    \includegraphics[width=0.43\textwidth]{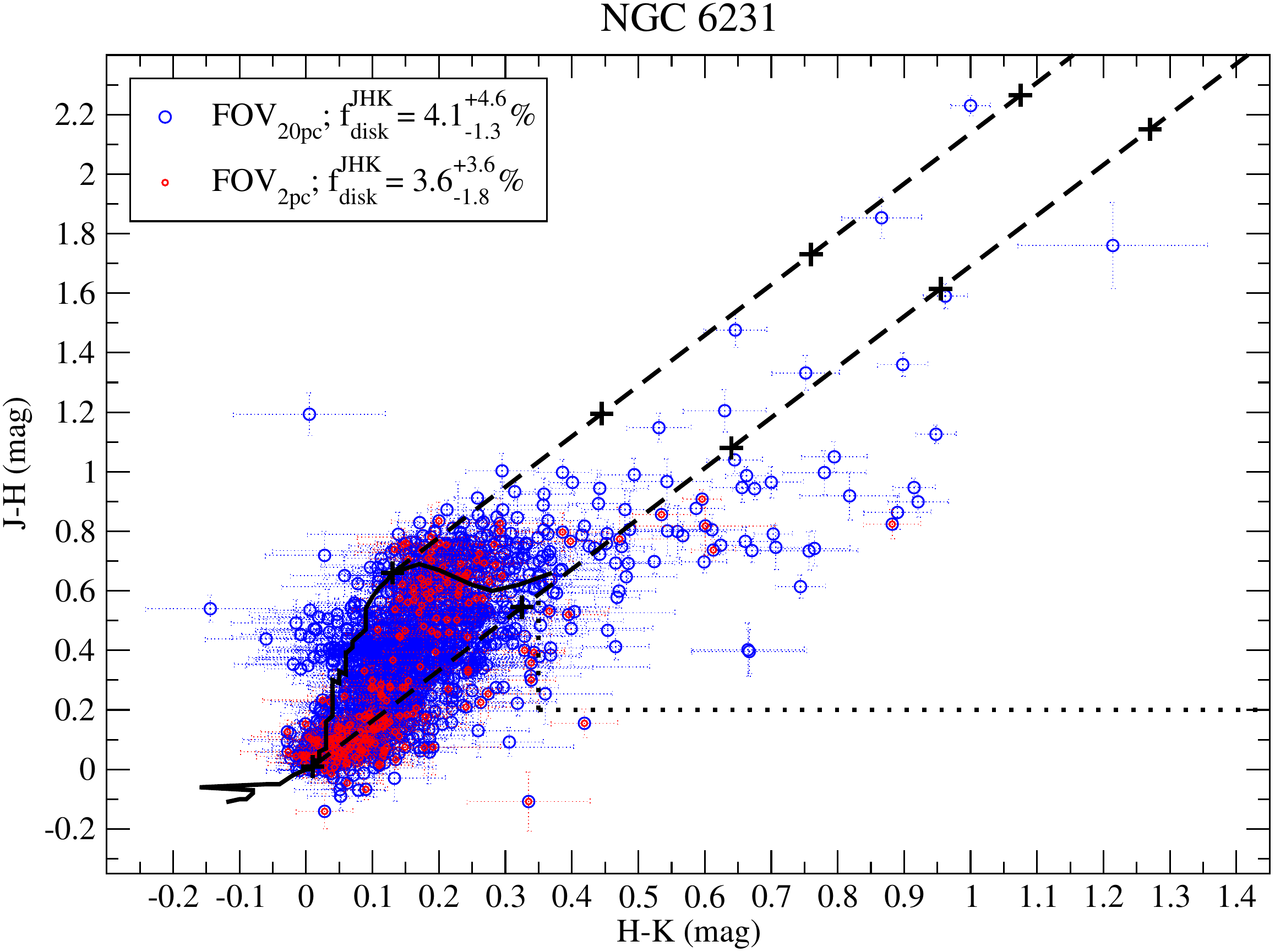}
\end{figure}
\begin{figure} [h]
 \centering
   \includegraphics[width=0.43\textwidth]{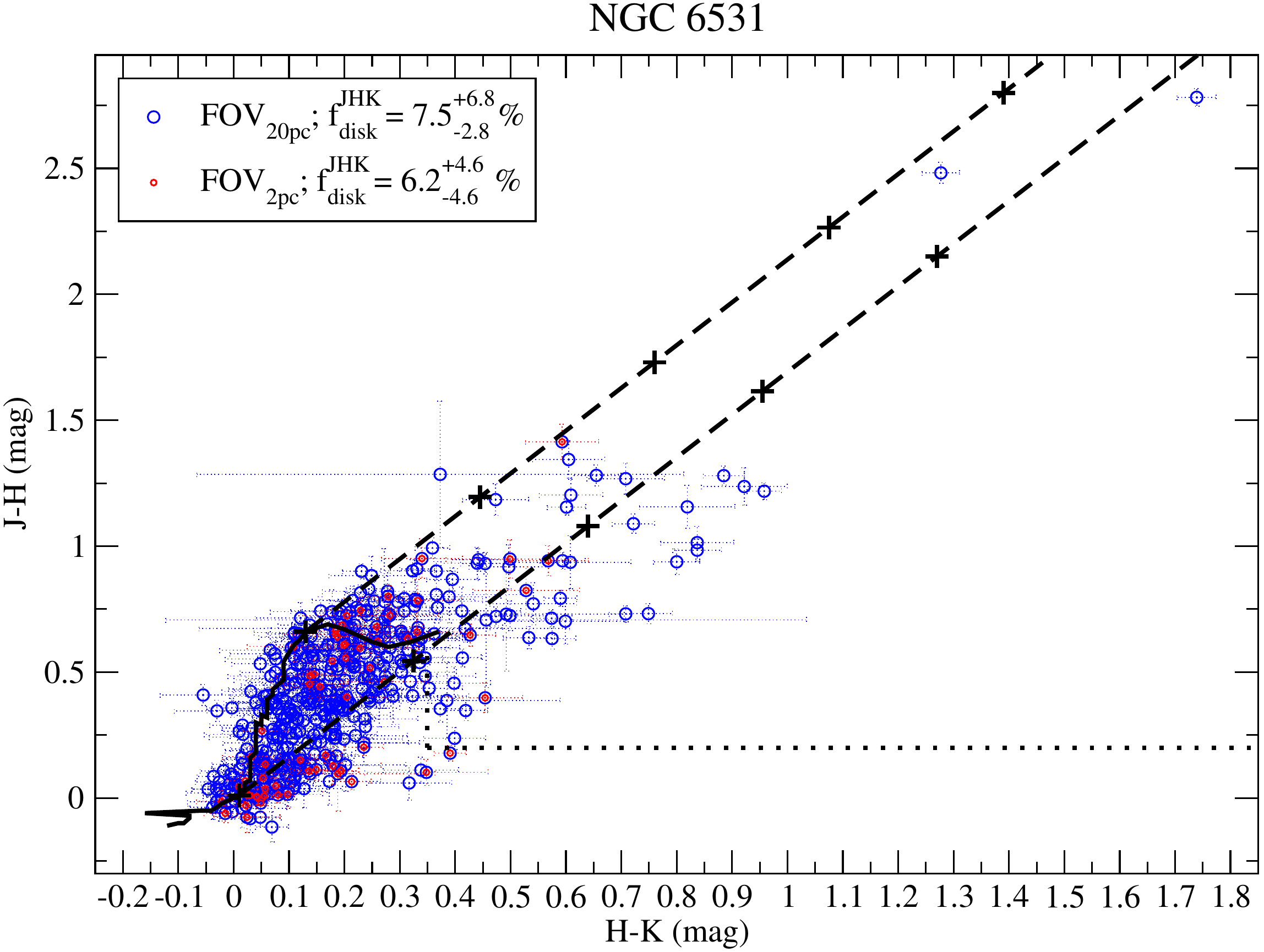}
    \includegraphics[width=0.43\textwidth]{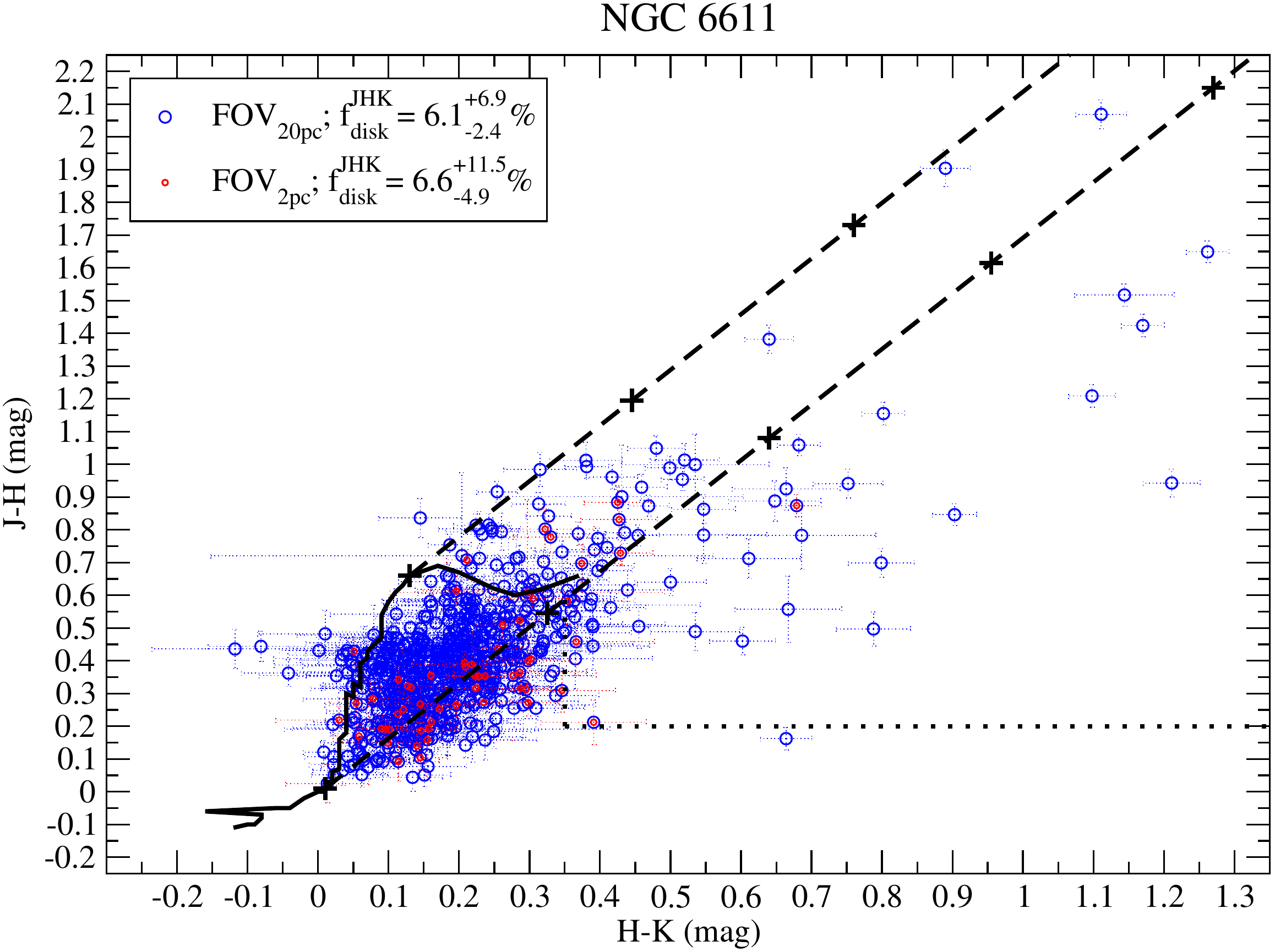}
\end{figure}
\begin{figure} [h]
 \centering
    \includegraphics[width=0.43\textwidth]{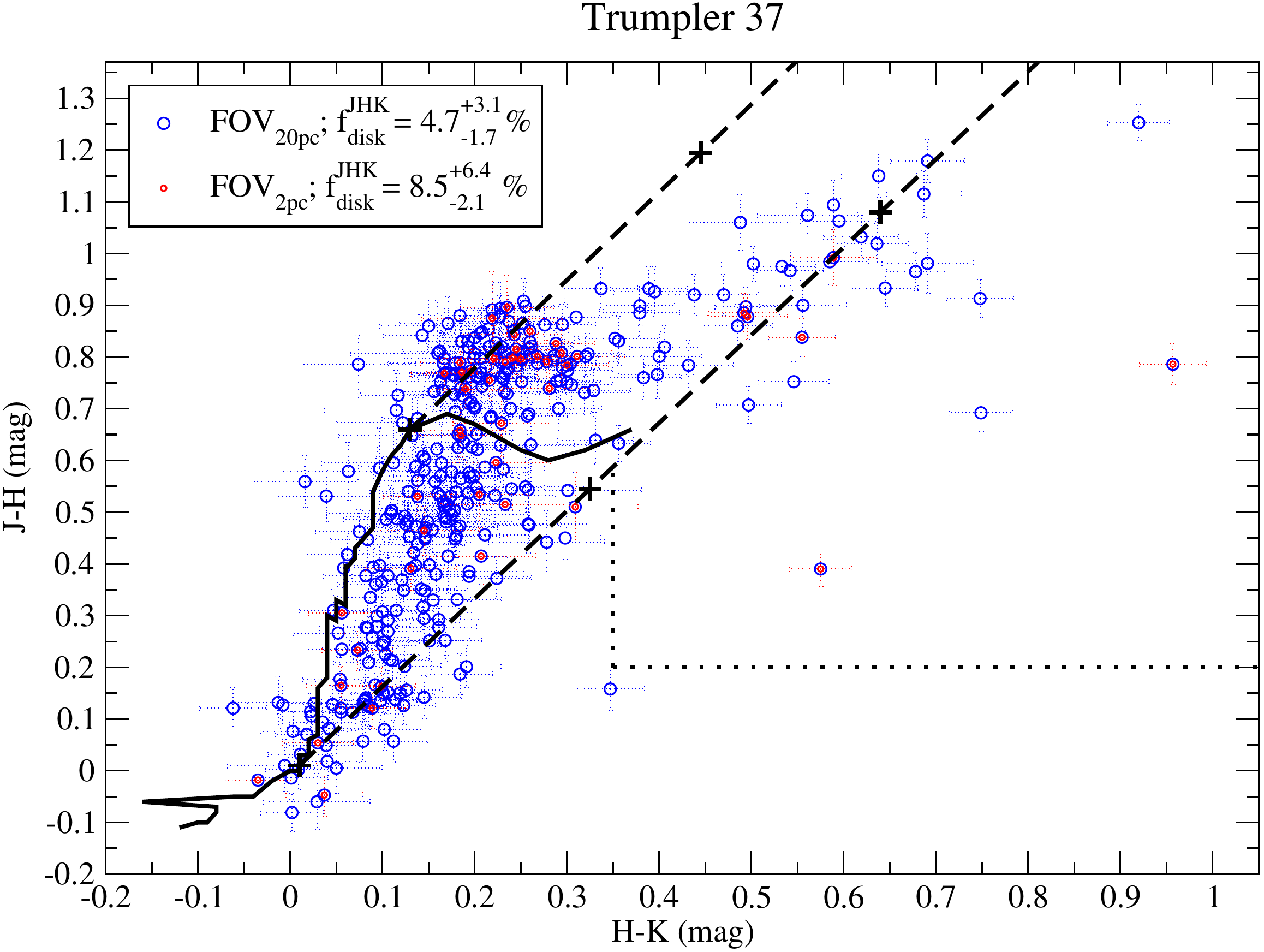}
    \includegraphics[width=0.43\textwidth]{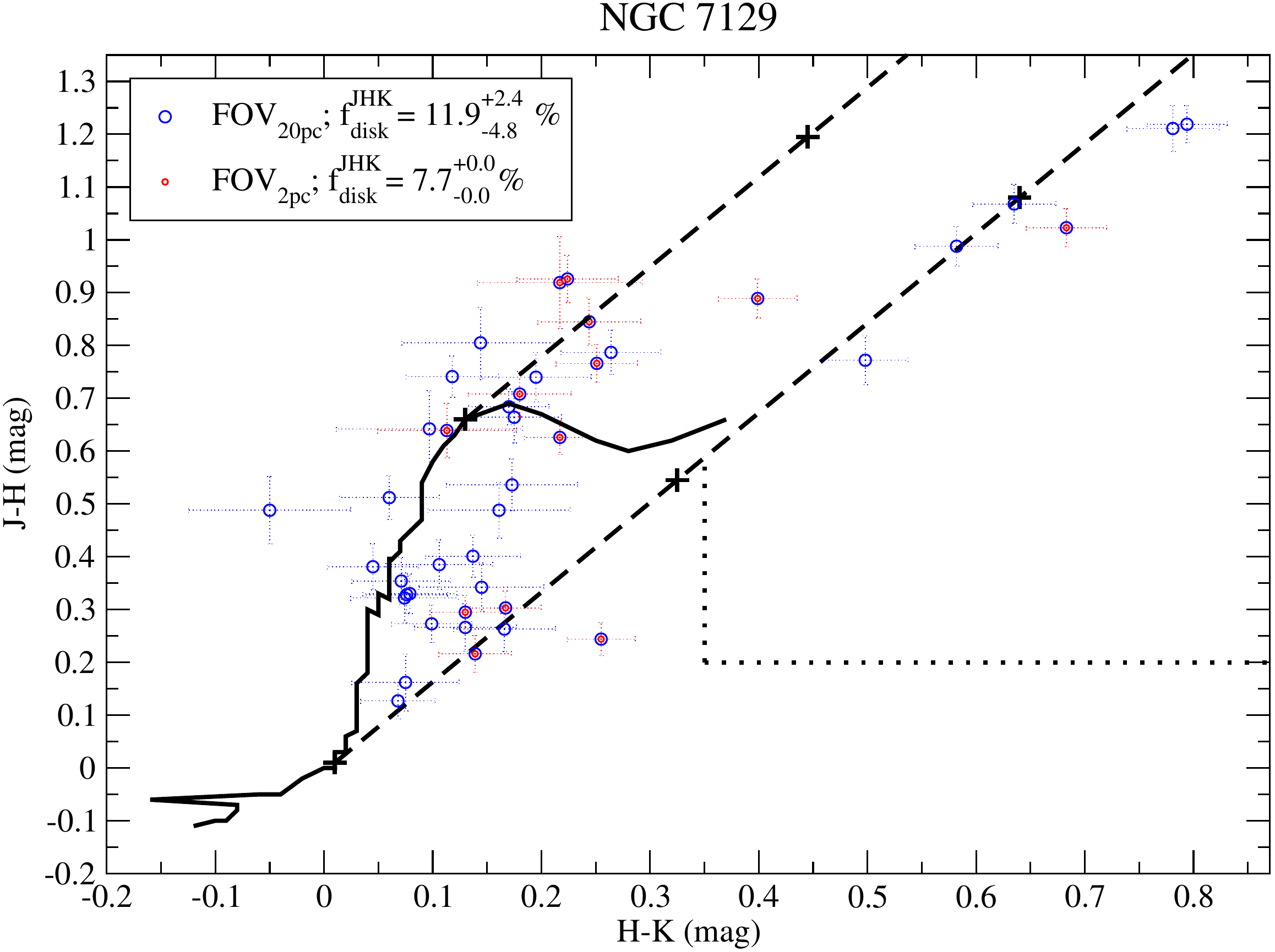}
\end{figure}
\begin{figure} [h]
 \centering
    \includegraphics[width=0.43\textwidth]{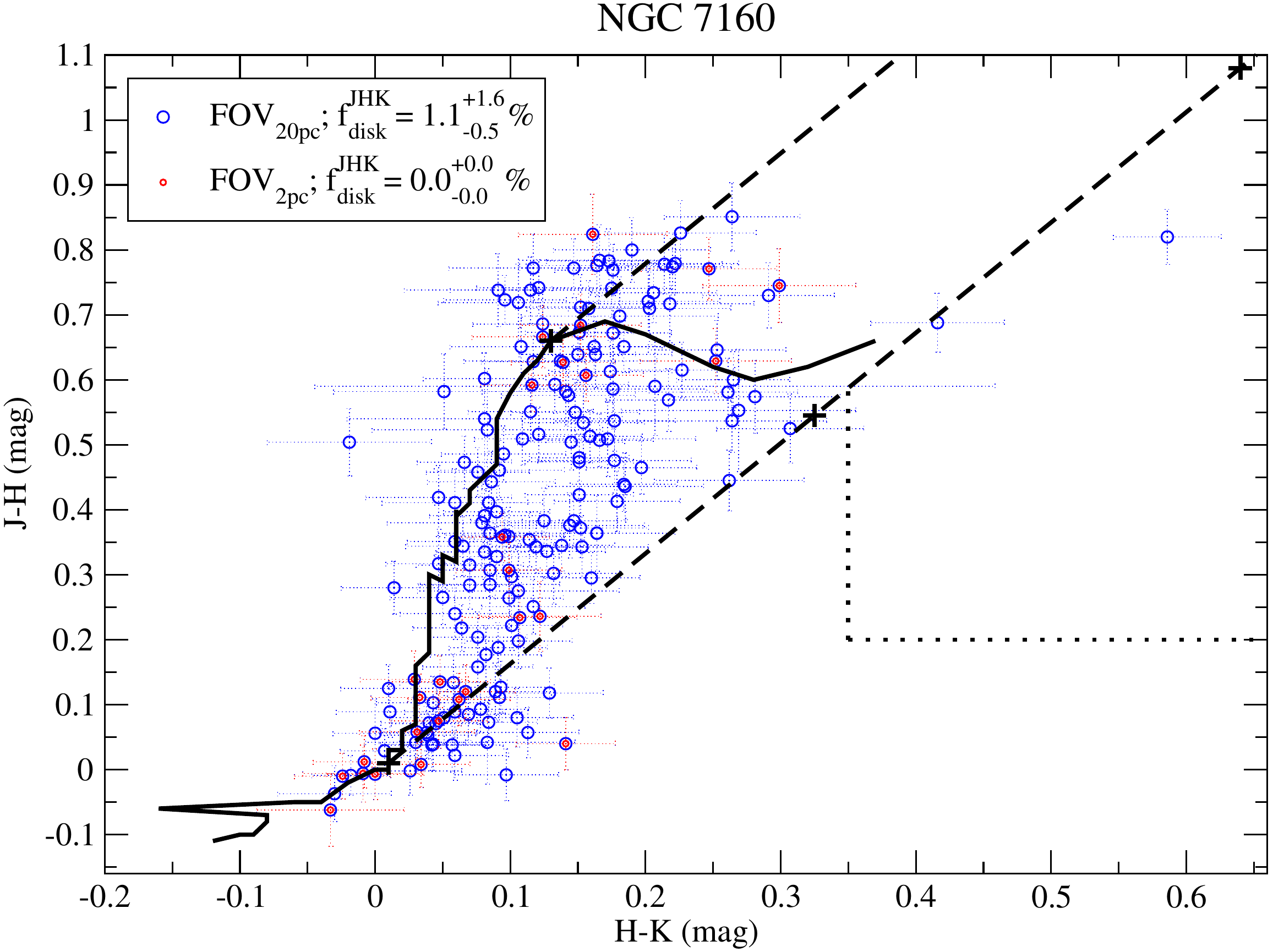}
    \caption{JHK color-color diagrams for the different clusters. In each panel, the solid line represents the expected position of non-extincted MS stars in the diagram, and the dashed lines the direction of the extinction vector and thus the rough boundaries for extincted MS stars (crosses within the dashed lines indicate an optical extinction increase of 5 magnitudes). Sources located to the right of the right hand dashed line and within the dotted lines are considered disk stars. Inner disk fractions are indicated in the legends, as inferred from all identified members within FOV$_{20pc}$ and within FOV$_{2pc}$ (blue and red symbols, respectively)}
            \label{figure:disk_fractions}
\end{figure}

\newpage
\onecolumn
\begin{figure} [h]
 \centering
    \includegraphics[width=0.45\textwidth]{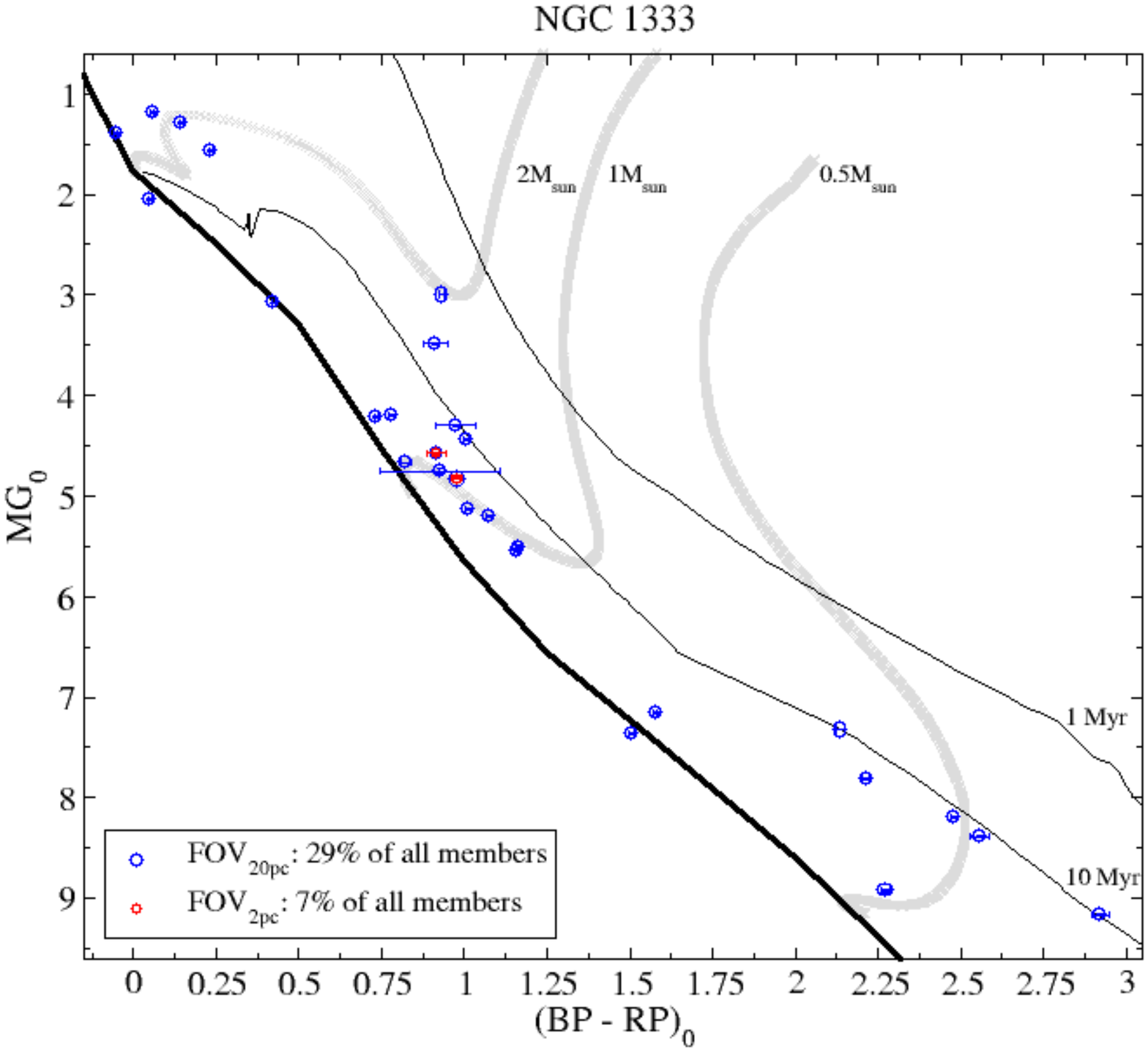}
    \includegraphics[width=0.45\textwidth]{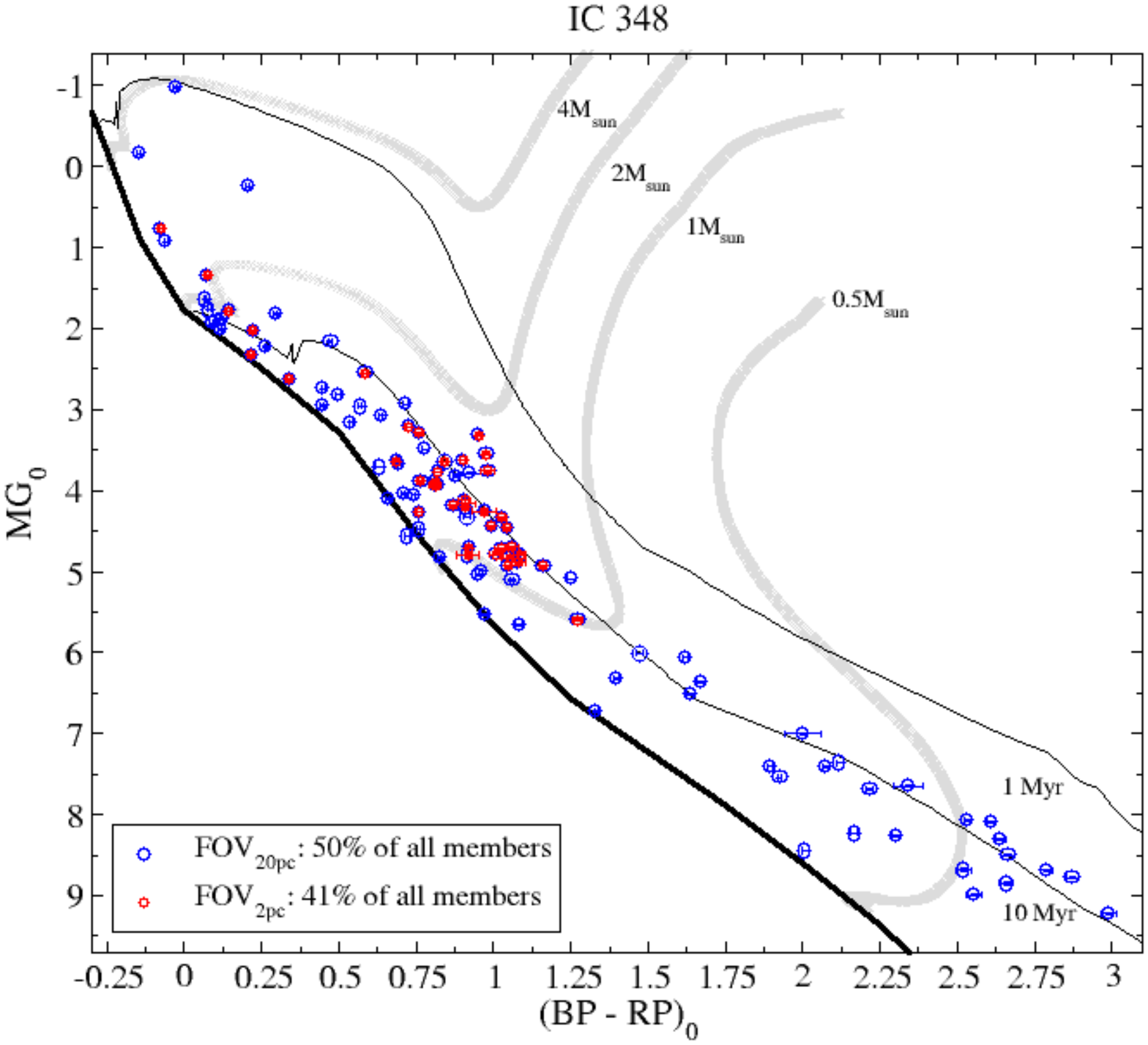}
\end{figure}
\begin{figure} [h]
 \centering
    \includegraphics[width=0.45\textwidth]{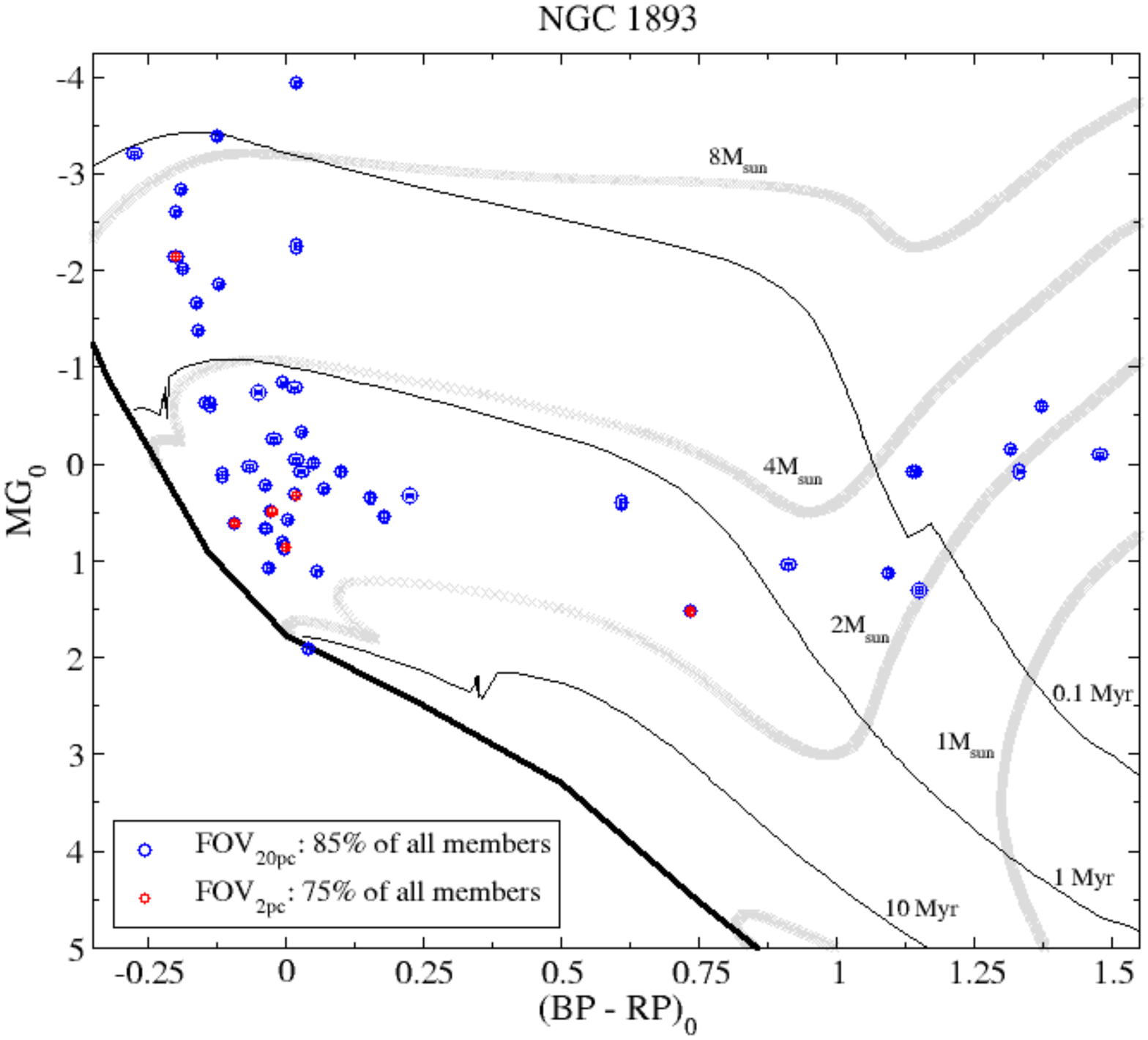}
    \includegraphics[width=0.45\textwidth]{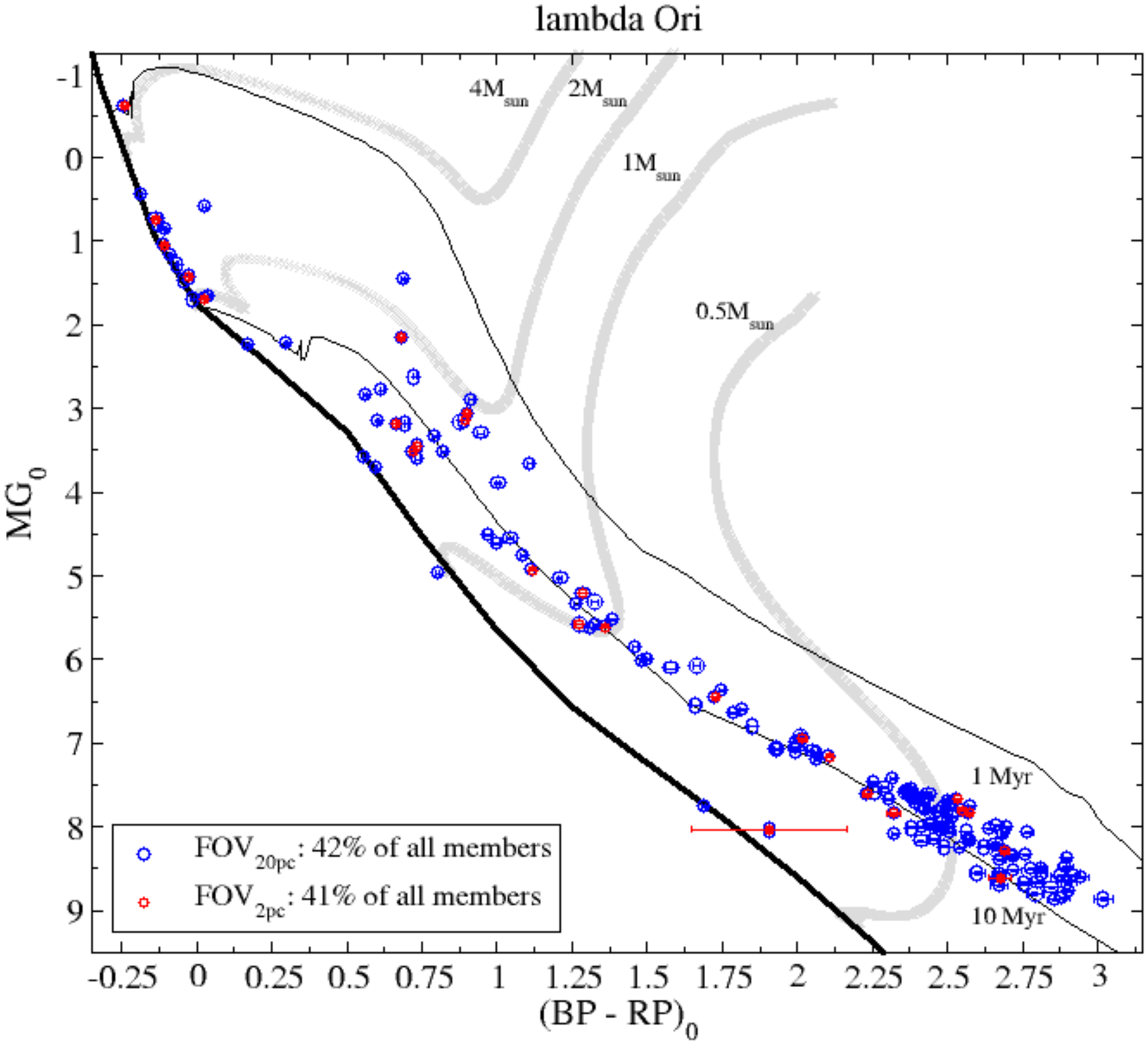} 
\end{figure}
\begin{figure} [h]
 \centering
    \includegraphics[width=0.45\textwidth]{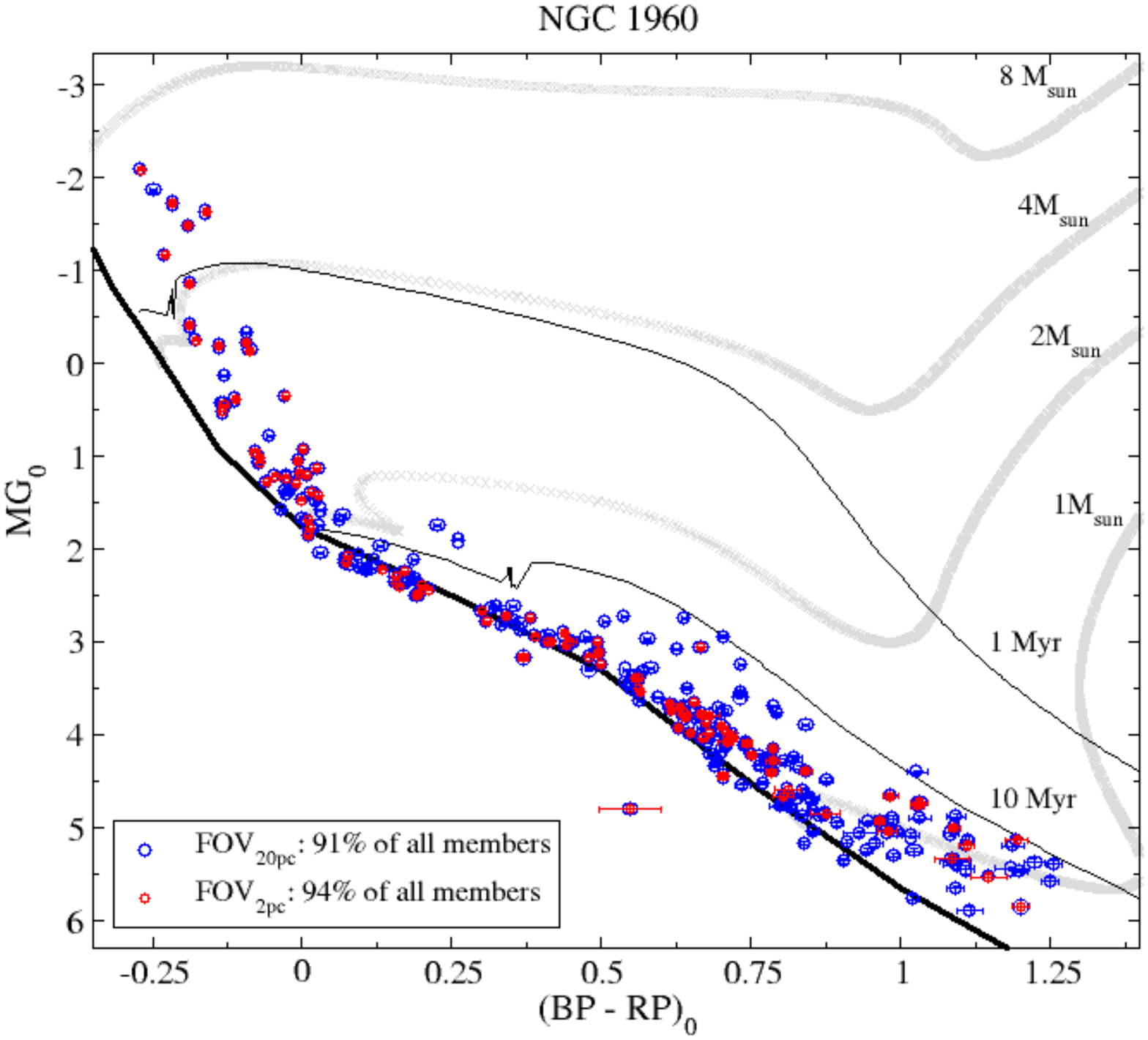}
    \includegraphics[width=0.45\textwidth]{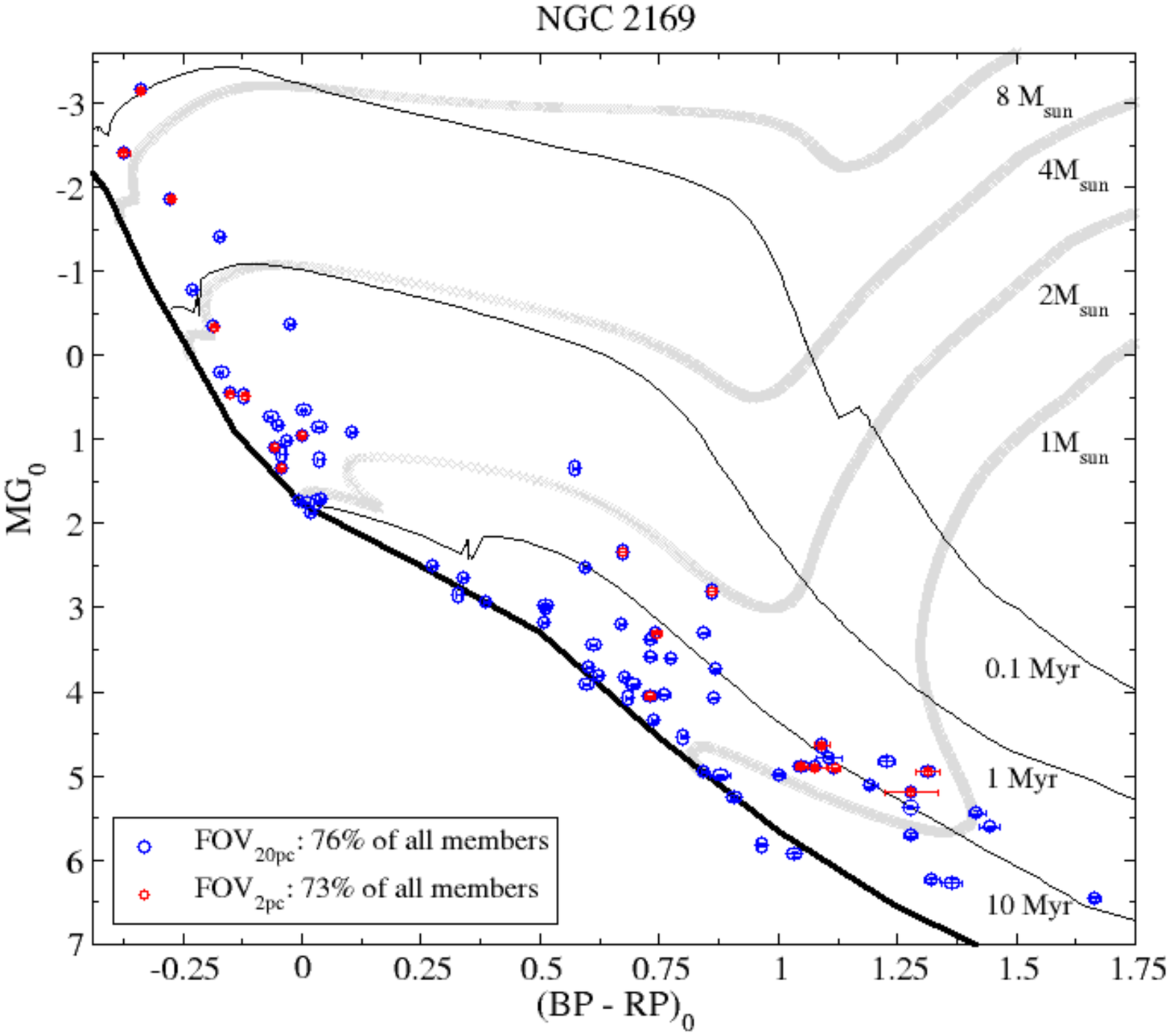}
\end{figure}
\begin{figure} [h]
 \centering
    \includegraphics[width=0.45\textwidth]{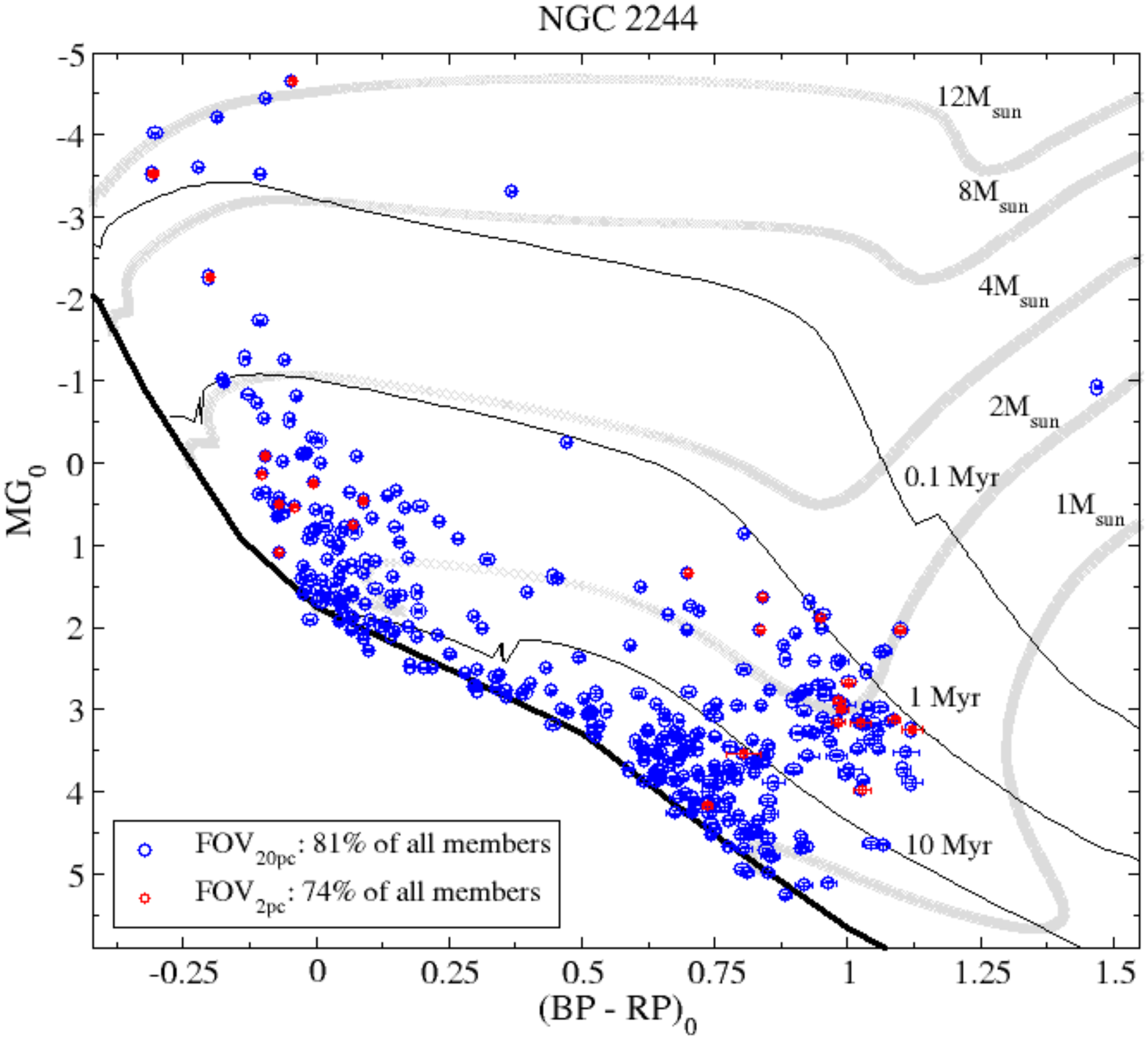}
    \includegraphics[width=0.45\textwidth]{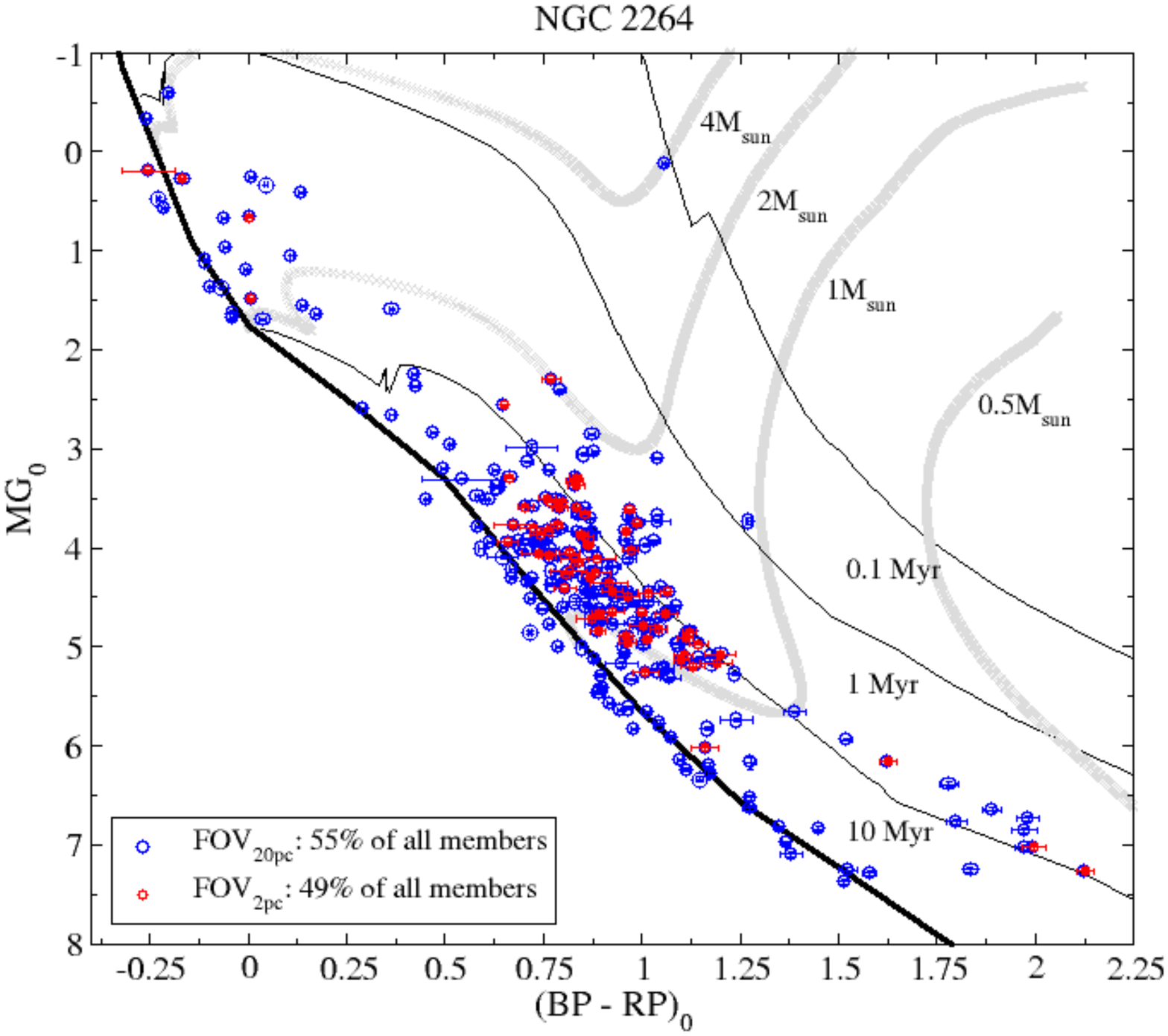}
\end{figure}
\begin{figure} [h]
 \centering
    \includegraphics[width=0.45\textwidth]{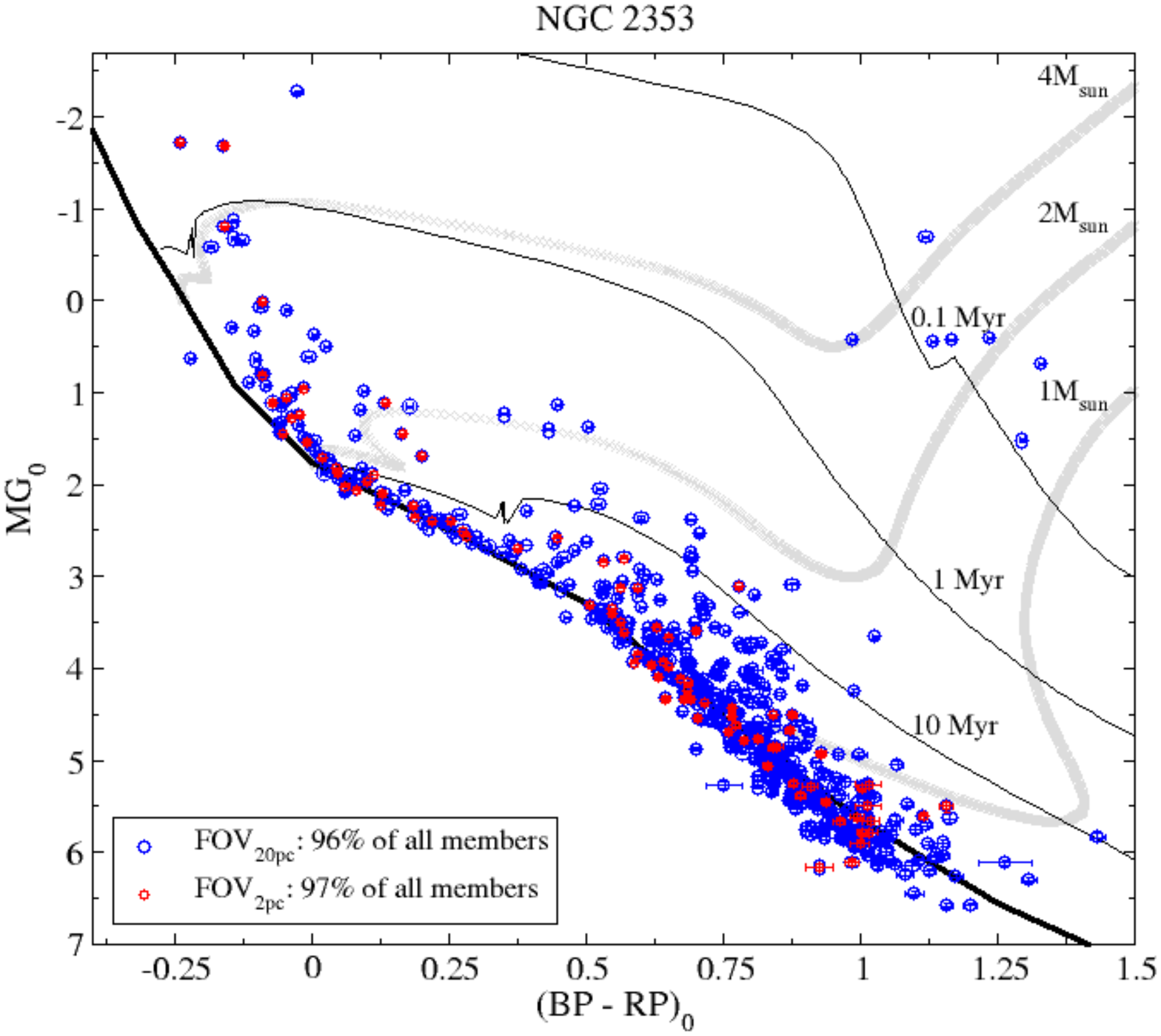}
    \includegraphics[width=0.45\textwidth]{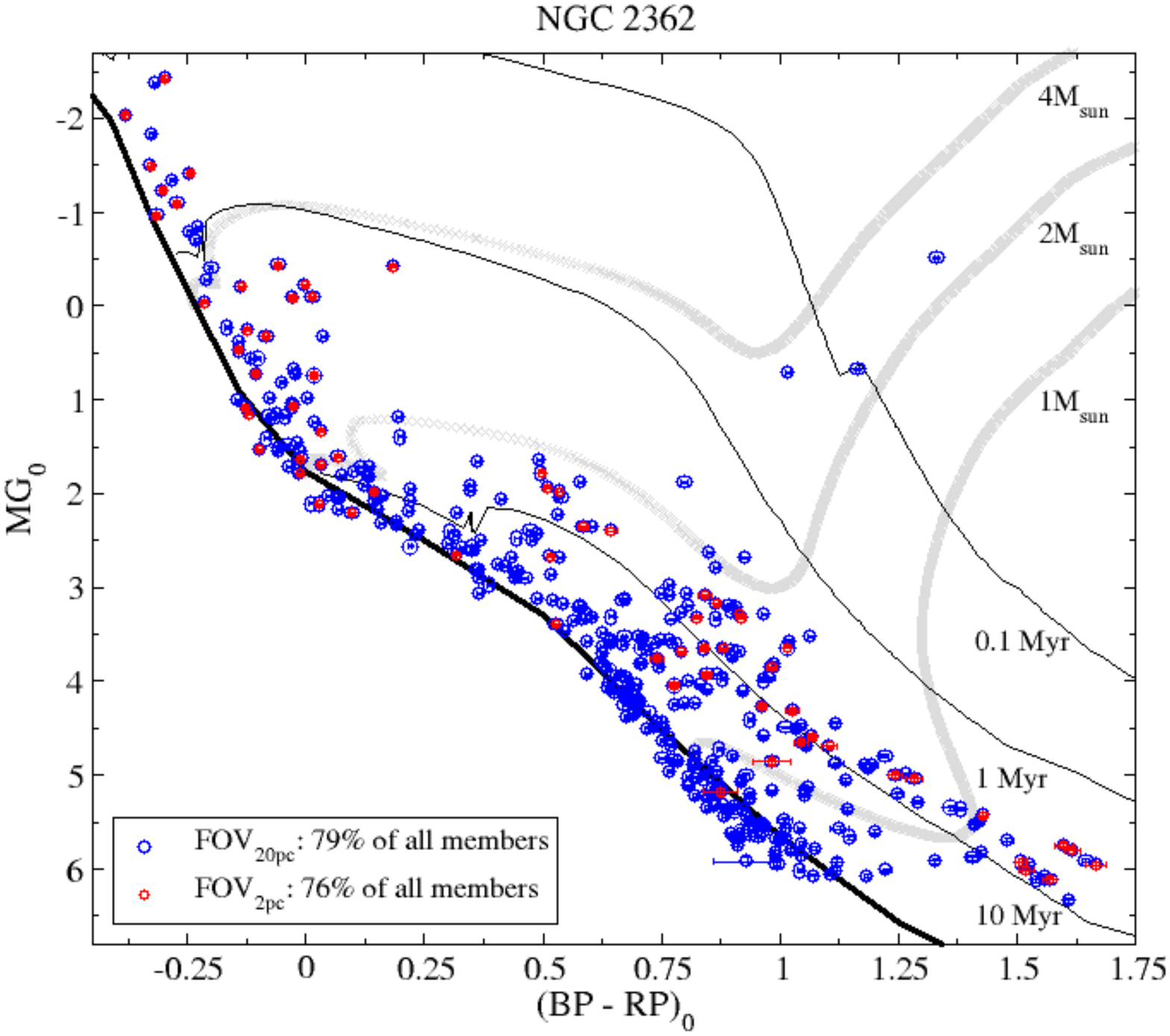}
\end{figure}
\begin{figure} [h]
 \centering
    \includegraphics[width=0.45\textwidth]{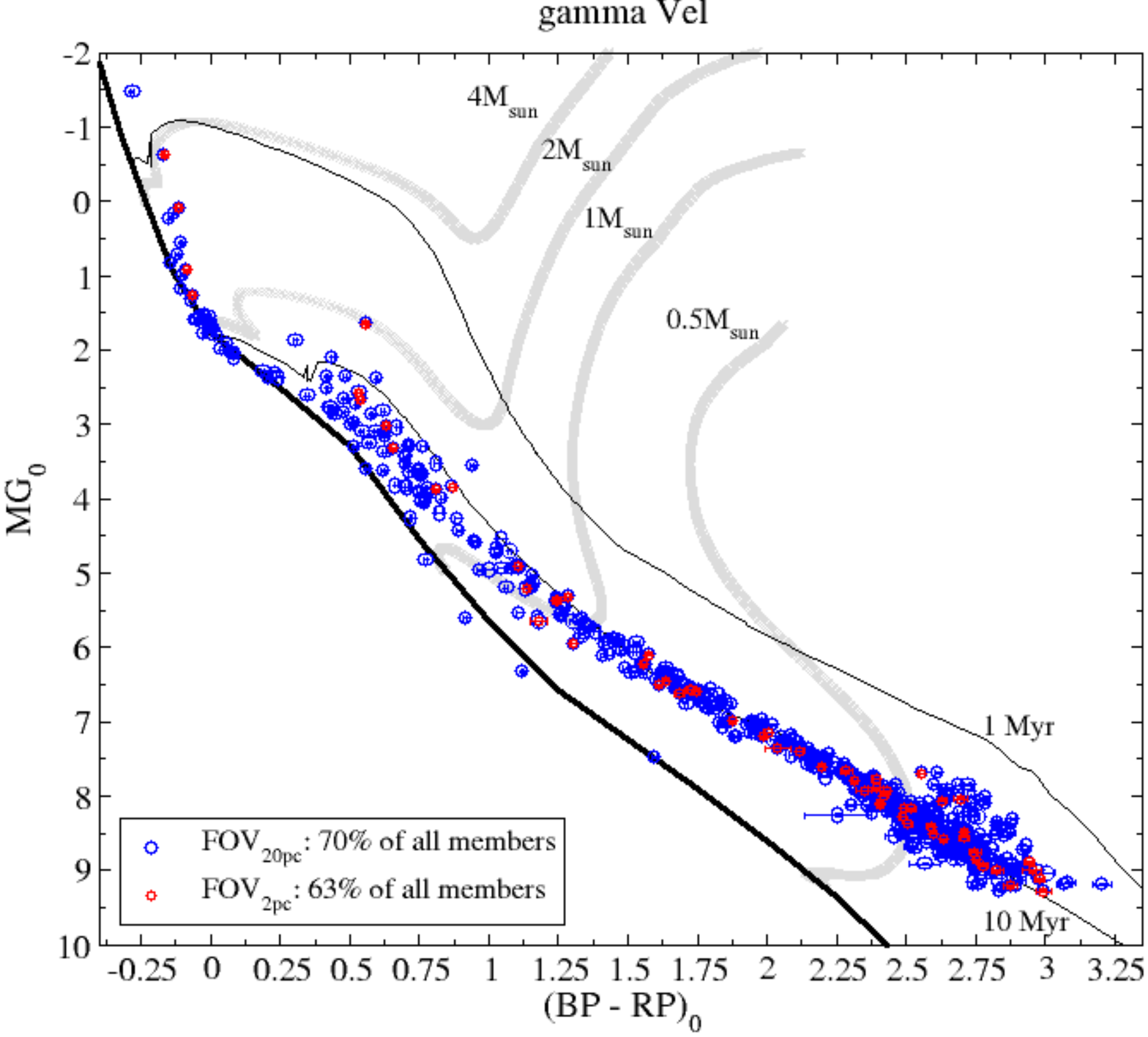}
    \includegraphics[width=0.45\textwidth]{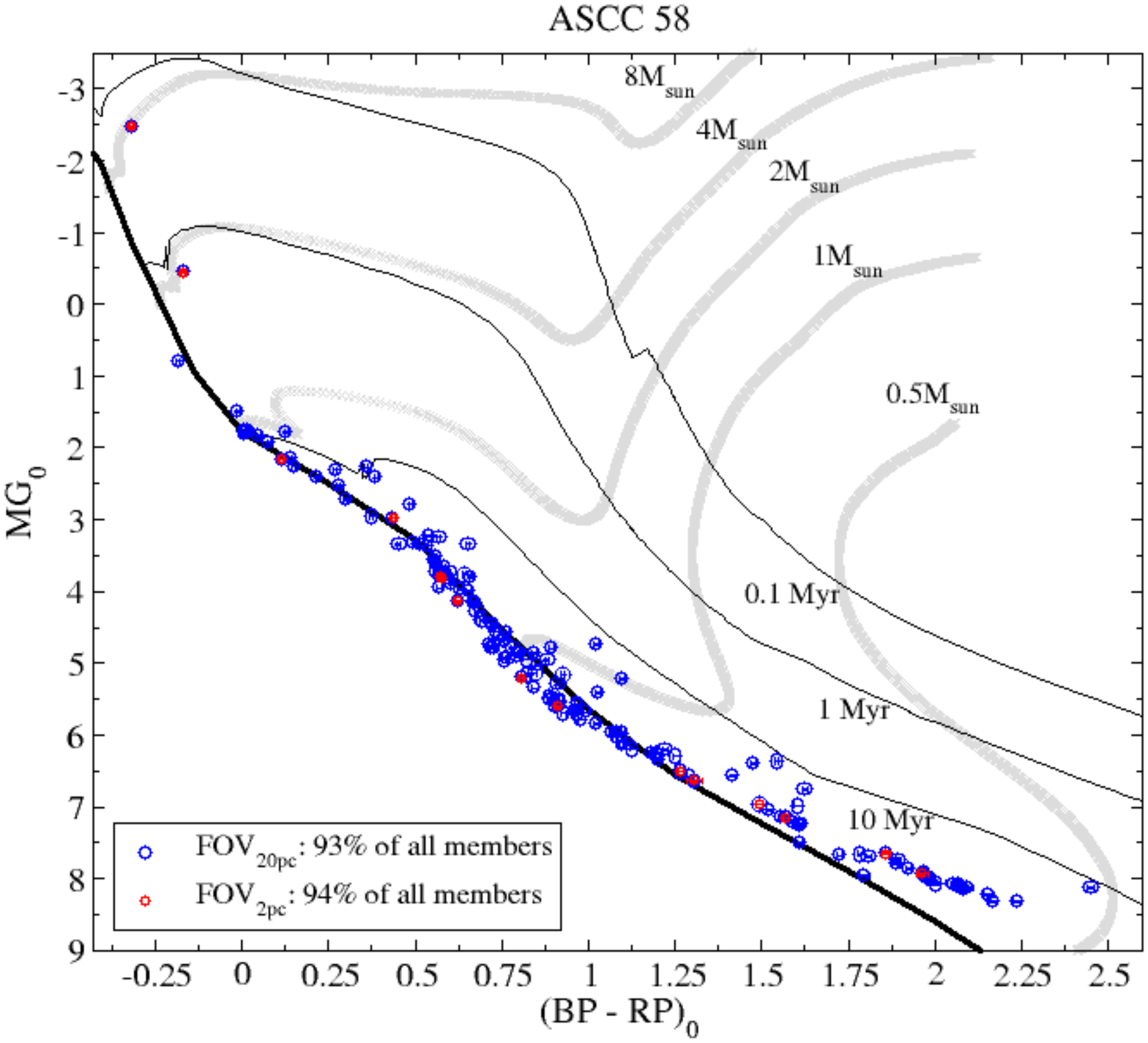}
\end{figure}
\begin{figure} [h]
 \centering
    \includegraphics[width=0.45\textwidth]{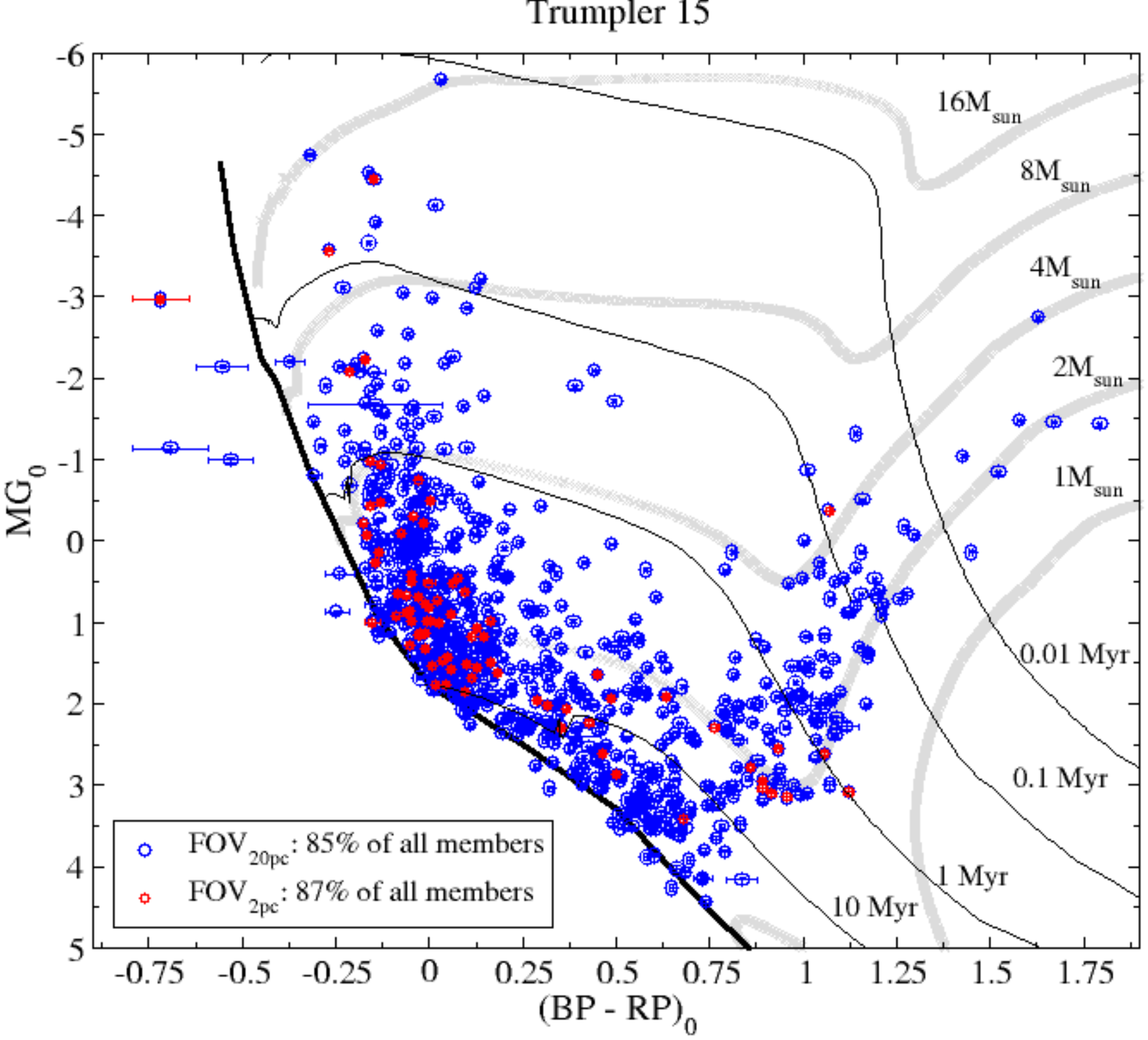}
    \includegraphics[width=0.45\textwidth]{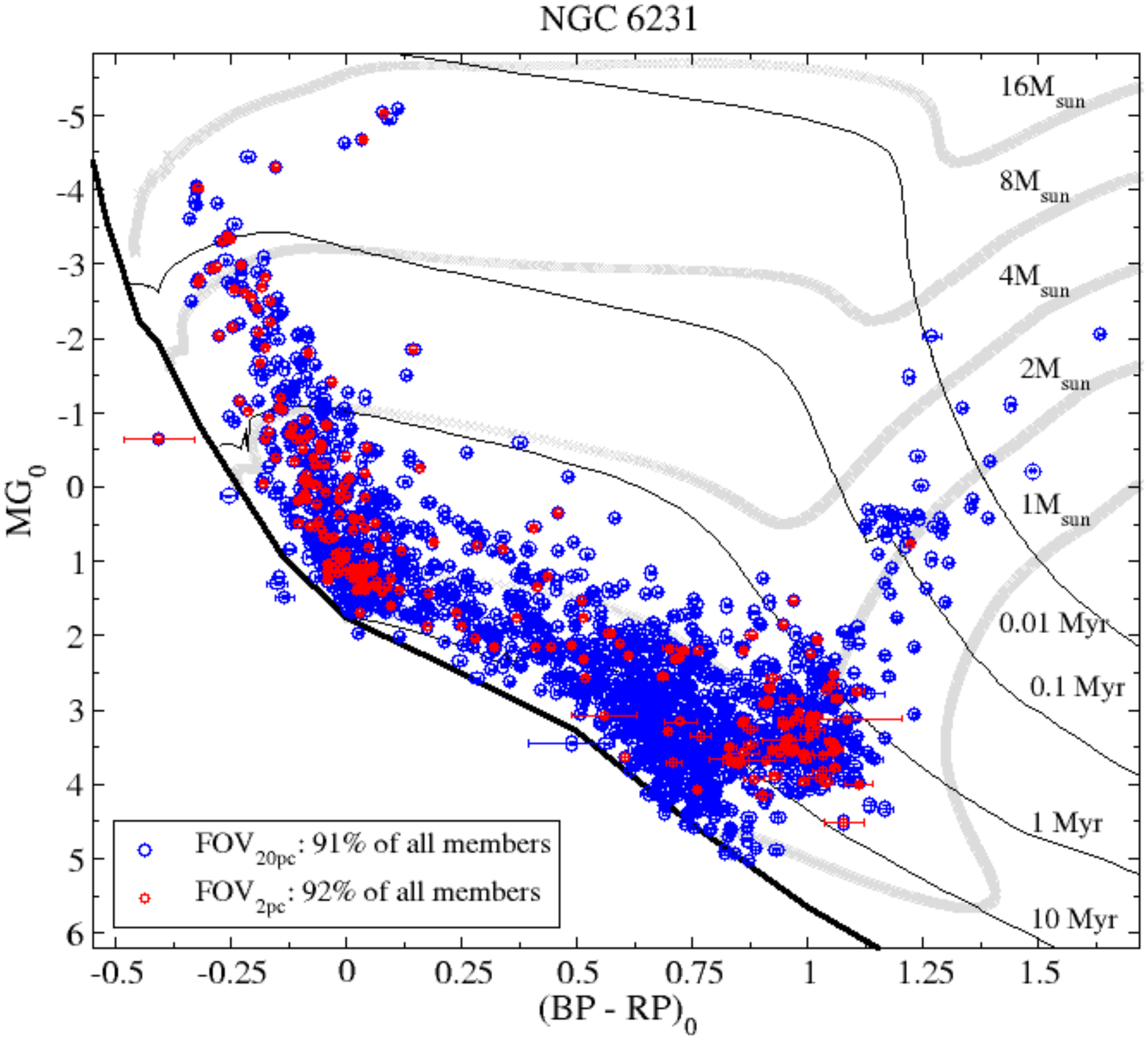}
\end{figure}
\begin{figure} [h]
 \centering
   \includegraphics[width=0.45\textwidth]{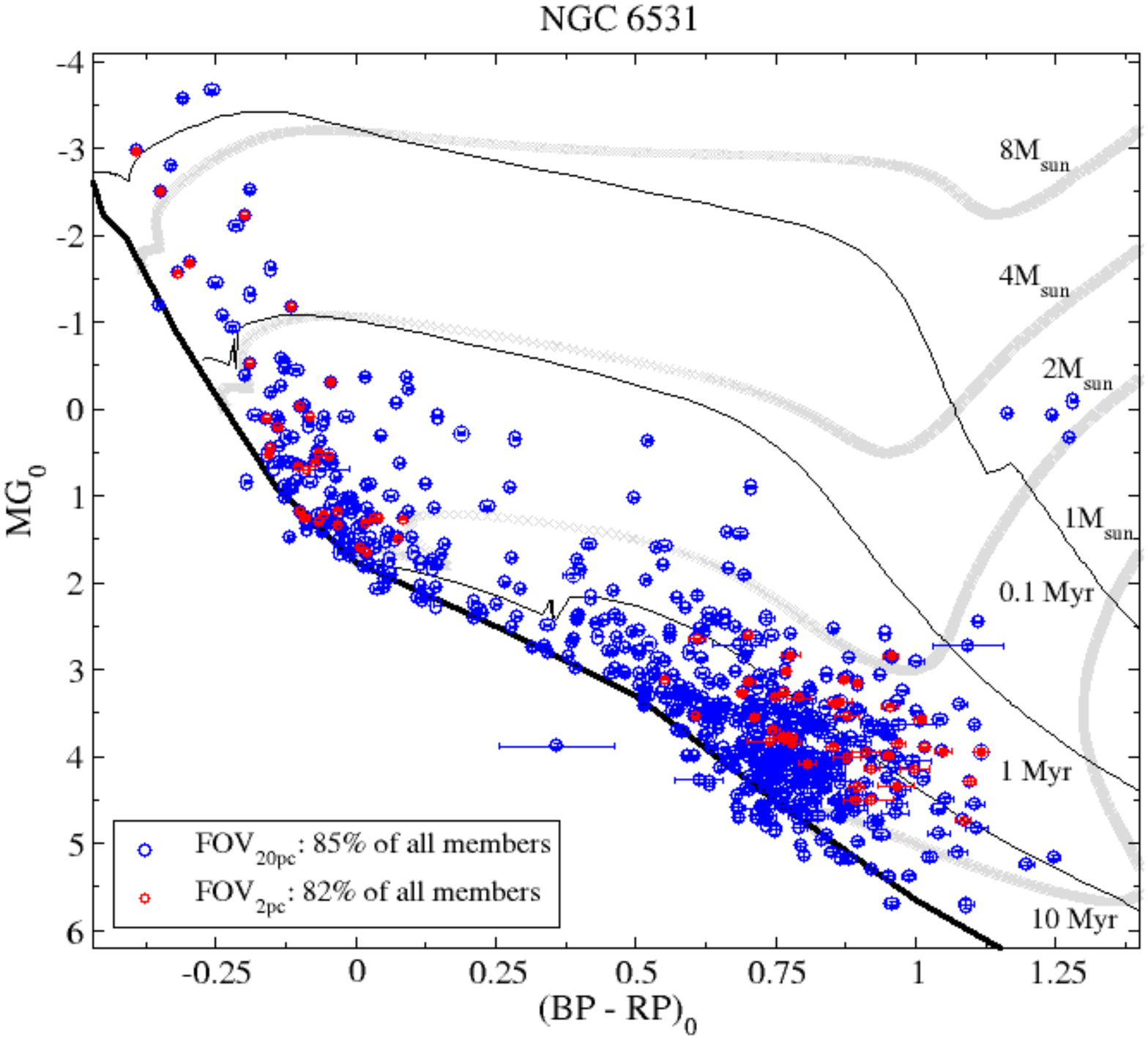}
    \includegraphics[width=0.45\textwidth]{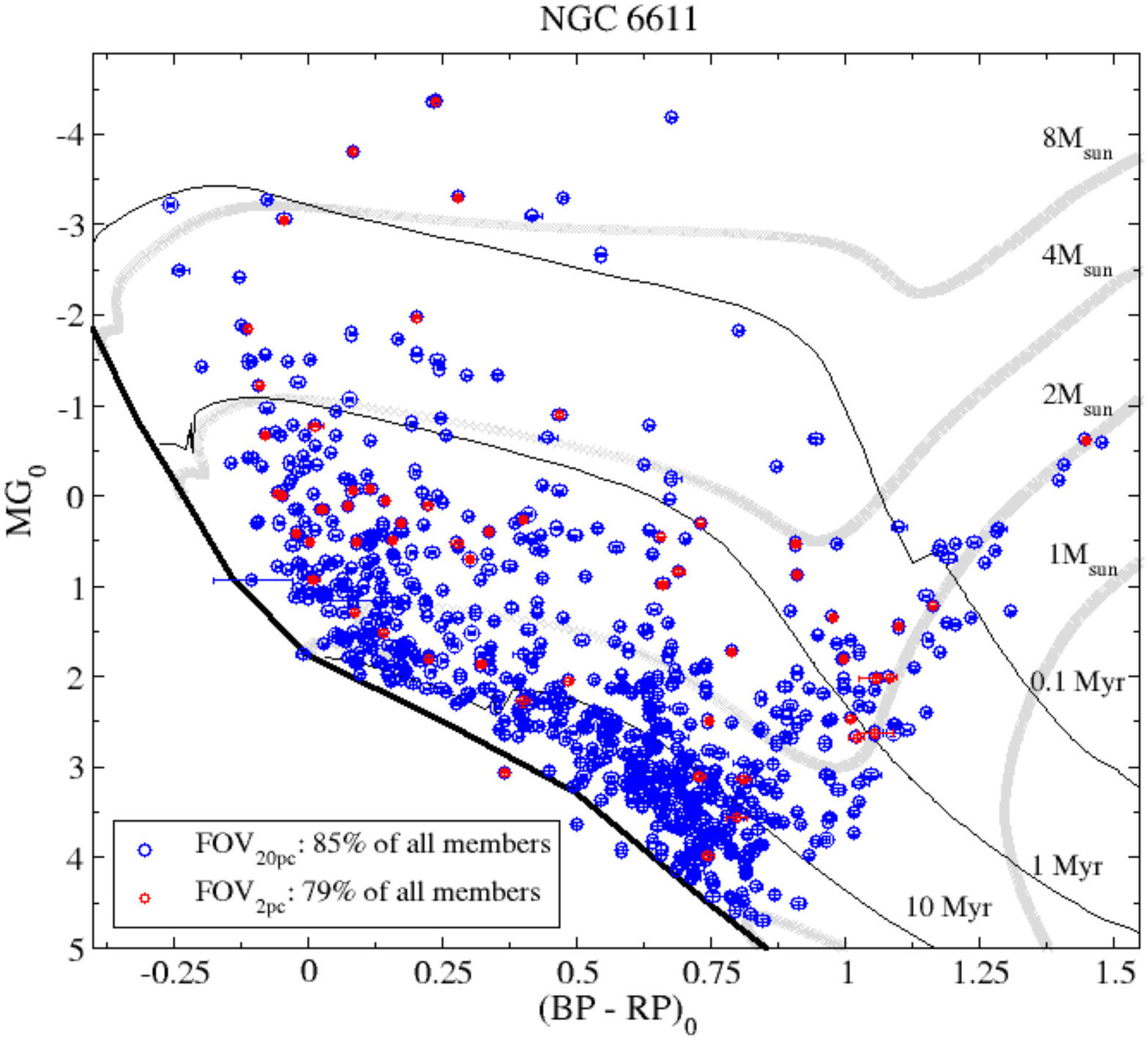}
\end{figure}
\begin{figure} [h]
 \centering
    \includegraphics[width=0.43\textwidth]{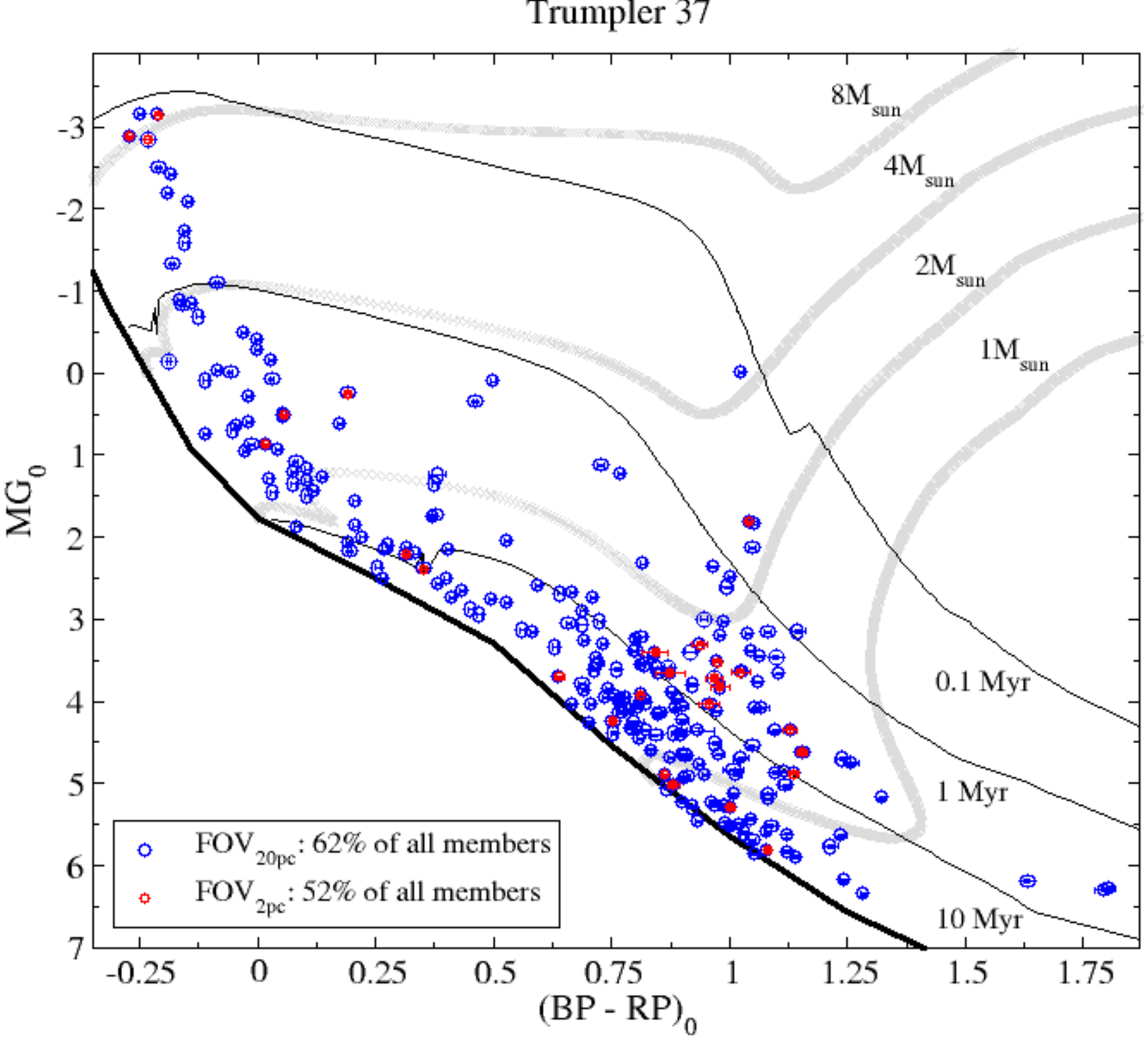}
    \includegraphics[width=0.43\textwidth]{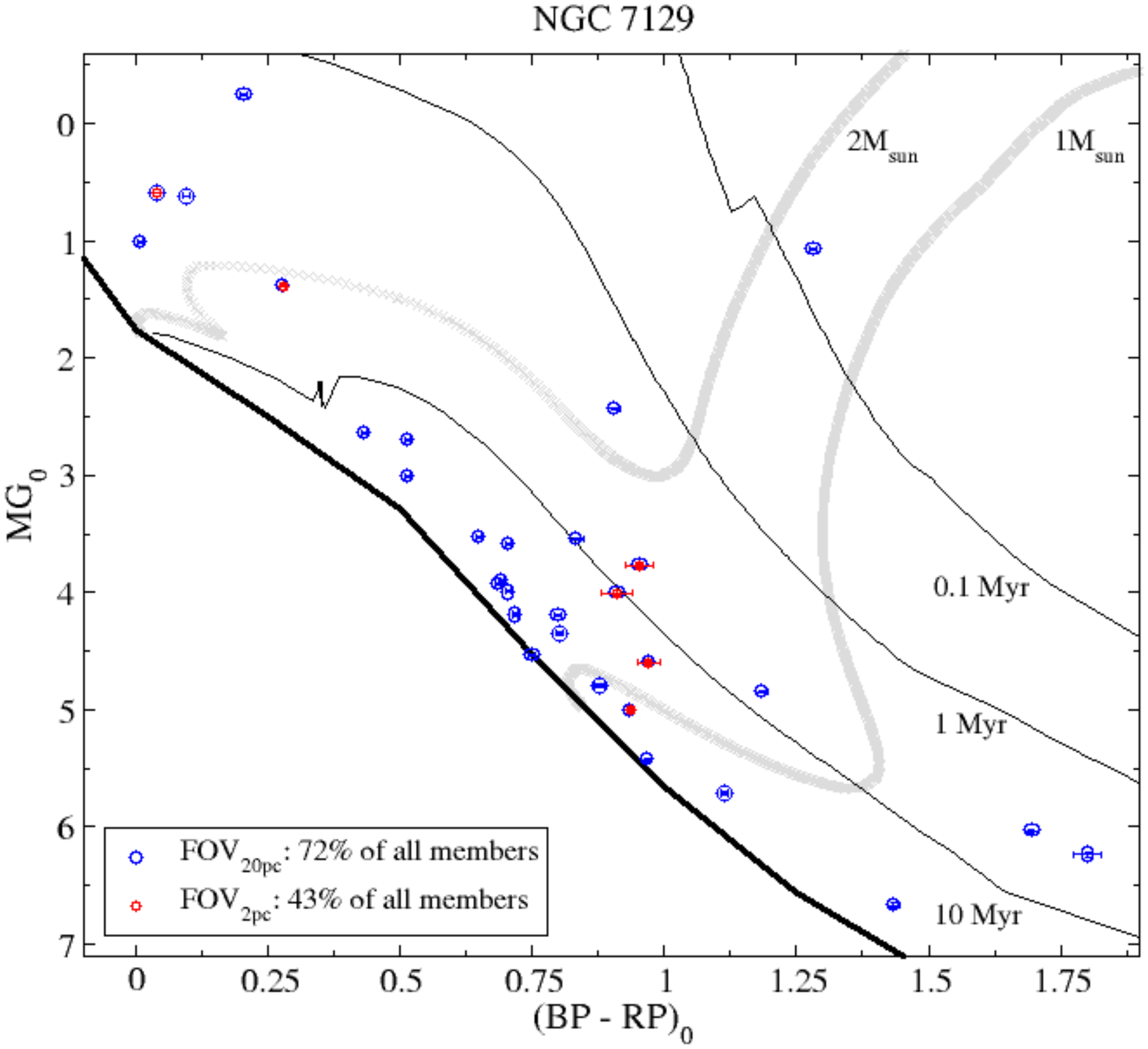}
\end{figure}
\begin{figure} [h]
 \centering
    \includegraphics[width=0.45\textwidth]{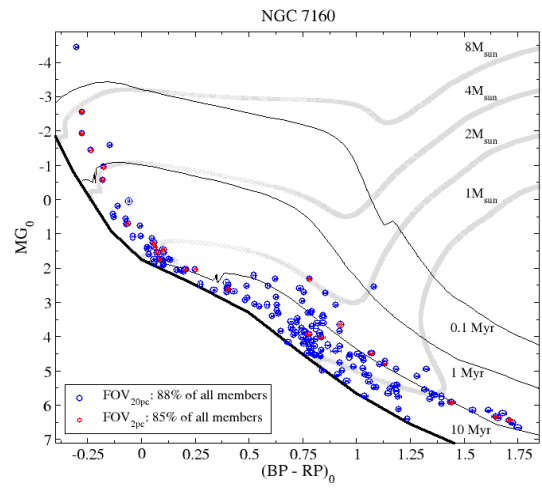}
    \caption{Extinction corrected Gaia CMDs for the different clusters. The ZAMS loci is indicated with the black solid curve. PMS isochrones and evolutionary tracks from \citet{Bressan12} are overplotted in black and gray for the stellar ages and masses indicated. The representativeness of the diagrams is limited to the fraction of all members identified within FOV$_{20pc}$ and FOV$_{2pc}$ having optical extinction estimates in \citet{Anders22}, which is indicated in the legends. }
            \label{figure:CMDs}
\end{figure}

\twocolumn

\begin{landscape}
\begin{table}
\centering
\caption{Young Stellar Clusters. Regions and membership}
\label{table:general_stats}
\renewcommand\tabcolsep{3pt}
\renewcommand{\arraystretch}{1.25}
\begin{tabular}{lll|lllll|lllll}
& & & &  \multicolumn{2}{c}{FOV$_{20pc}$} & & &   \multicolumn{2}{c}{FOV$_{2pc}$} \\
\multicolumn{1}{c}{Name} & RA$_{cent}$ & DEC$_{cent}$ &  R & N$_{memb}$/N$_{tot}$  & $\pi$ & pmRA & pmDEC & R & N$_{memb}$/N$_{tot}$  & $\pi$ & pmRA & pmDEC \\ 
\hline
... & ($^{\circ}$) & ($^{\circ}$) & ($\arcmin$) & ... & (mas) & (mas yr$^{-1}$) & (mas yr$^{-1}$) & ($\arcmin$) & ... & (mas) & (mas yr$^{-1}$) & (mas yr$^{-1}$) \\ 
\hline
\hline
NGC 1333 & 52.2968 & 31.3097 & 240 & 95/43277 & 3.40$\pm$0.08 & 6.94$\pm$0.73 & -9.37$\pm$0.67  & 24 & 28/244 & 3.39$\pm$0.08  & 7.00$\pm$0.80 & -9.79$\pm$0.46 \\
IC 348 & 56.0608 & 32.1560 & 240 & 209/53819 & 3.15$\pm$0.08 & 4.36$\pm$0.59 & -6.30$\pm$0.64  & 24 & 97/312 & 3.14$\pm$0.06 & 4.48$\pm$0.53 & -6.33$\pm$0.52 \\
NGC 1893 & 80.7237 & 33.4438 & 18 & 59/922 & 0.32$\pm$0.04 & -0.30$\pm$0.14 & -1.41$\pm$0.17  & 1.8 & 8/22 & 0.31$\pm$0.03 & -0.32$\pm$0.04 & -1.38$\pm$0.12 \\
$\lambda$ Ori & 83.7802 & 9.9091 & 180 & 377/55430 & 2.50$\pm$0.05 & 1.14$\pm$0.47 & -2.06$\pm$0.30 & 18 & 63/577 & 2.50$\pm$0.05 & 1.00$\pm$0.41 & -2.02$\pm$0.19 \\
NGC 1960 & 84.0840 & 34.1350 & 60 & 273/9263 & 0.85$\pm$0.02 & -0.21$\pm$0.15 & -3.39$\pm$0.17 & 6 & 114/227 & 0.85$\pm$0.02 & -0.21$\pm$0.14 & -3.43$\pm$0.13\\
NGC 2169 & 92.1250 & 13.9510 & 70 & 102/10500 & 1.03$\pm$0.04 & -1.25$\pm$0.25 & -1.58$\pm$0.17 & 7 & 26/153 & 1.03$\pm$0.04 & -1.10$\pm$0.13 & -1.59$\pm$0.12 \\
NGC 2244 & 97.9817 & 4.9390 & 44 & 418/4255 & 0.68$\pm$0.04 & -1.65$\pm$0.38 & 0.39$\pm$0.41 & 4.4 & 35/75 & 0.67$\pm$0.04 & -1.75$\pm$0.21 & 0.35$\pm$0.21 \\
NGC 2264 & 100.2471 & 9.8755 & 120 & 452/32665 & 1.37$\pm$0.06 & -1.78$\pm$0.38 & -3.68$\pm$0.29 & 12 & 142/242 & 1.38$\pm$0.05 & -1.68$\pm$0.30 & -3.68$\pm$0.22 \\
NGC 2353 & 108.6242 & -10.2711 & 60 & 545/11452 & 0.82$\pm$0.08 & -1.05$\pm$0.43 & 0.77$\pm$0.38 & 6 & 92/206 & 0.84$\pm$0.05 & -1.09$\pm$0.16 & 0.76$\pm$0.20 \\
NGC 2362 & 109.6658 & -24.9557 & 52 & 485/12410 & 0.76$\pm$0.05 & -2.61$\pm$0.46 & 2.93$\pm$0.39 & 5.2 & 90/212 & 0.77$\pm$0.03 & -2.75$\pm$0.21 & 2.98$\pm$0.20 \\
$\gamma$ Vel & 122.3831 & -47.3366 & 180 & 810/134095 & 2.76$\pm$0.14 & -6.20$\pm$0.44 & 9.08$\pm$0.64 & 18 & 99/1364 & 2.87$\pm$0.12 & -6.48$\pm$0.40 & 9.53$\pm$0.56 \\
ASCC 58 & 153.8180 & -55.0003 & 120 & 160/99215 & 2.11$\pm$0.07 & -13.42$\pm$0.48 & 2.70$\pm$0.38 & 12 & 16/1068 & 2.08$\pm$0.03 & -13.32$\pm$0.31 & 2.71$\pm$0.28 \\
Trumpler 15 & 161.1743 & -59.3664 & 26 & 870/6216 & 0.40$\pm$0.04 & -6.31$\pm$0.35 & 2.14$\pm$0.31 & 2.6 & 94/139 & 0.39$\pm$0.03 & -6.13$\pm$0.21 & 2.08$\pm$0.18 \\ 
NGC 6231 & 253.5356 & -41.8284 & 60 & 1726/26977 & 0.60$\pm$0.04 & -0.64$\pm$0.32 & -2.12$\pm$0.38 & 6 & 265/546 & 0.60$\pm$0.04 & -0.58$\pm$0.21 & -2.19$\pm$0.22 \\
NGC 6531 & 271.0576 & -22.4935 & 55 & 696/15689 & 0.79$\pm$0.04 & 0.50$\pm$0.34 & -1.46$\pm$0.34 & 5.5 & 91/257 & 0.80$\pm$0.04 & 0.54$\pm$0.21 & -1.42$\pm$0.22 \\
NGC 6611 & 274.6993 & -13.8090 & 40 & 809/7796 & 0.59$\pm$0.05 & 0.16$\pm$0.53 & -1.63$\pm$0.58 & 4 & 72/133 & 0.56$\pm$0.04 & 0.20$\pm$0.31 & -1.66$\pm$0.27 \\
Trumpler 37 & 324.7486 & 57.4782 & 73 & 381/18293 & 1.07$\pm$0.04 & -2.43$\pm$0.52 & -4.28$\pm$0.75 & 7.3 & 52/290 & 1.07$\pm$0.04 & -2.43$\pm$0.33 & -4.47$\pm$0.62 \\
NGC 7129 & 325.7401 & 66.1140 & 60 & 43/8627 & 1.07$\pm$0.06 & -1.77$\pm$0.33 & -3.41$\pm$0.39 & 6 & 14/63 & 1.10$\pm$0.04 & -1.77$\pm$0.18 & -3.45$\pm$0.32 \\
NGC 7160 & 328.4142 & 62.6020 & 72 & 190/20532 & 1.06$\pm$0.05 & -3.42$\pm$0.23 & -1.44$\pm$0.34 & 7.2 & 27/213 & 1.09$\pm$0.04 & -3.51$\pm$0.19 & -1.38$\pm$0.18 \\
\hline
\hline
\end{tabular}
\begin{minipage}{23cm}

\textbf{Notes.} Columns 1 to 3 list the name of the cluster and the coordinates of its central position. Columns 4 to 8 refer to FOV$_{20pc}$, listing the associated angular radius, number of members identified divided by the total number of sources within the FOV having Gaia EDR3 proper motions and reliable parallaxes (as defined in Sect. \ref{Sect:members}), and inferred mean parallax and proper motions (uncertainties refer to the standard deviations). The rest of the columns are equivalent to the previous, but related to FOV$_{2pc}$.
\end{minipage}
\end{table}
\end{landscape}

\newpage
\begin{landscape}
\begin{table}
\centering
\caption{Young Stellar Clusters. Members identified}
\label{table:members}
\renewcommand\tabcolsep{1.8pt}
\begin{tabular}{|l|l|l|l|l|l|l|l|l|l|l|l|l|}
\hline
RA & DEC & pmRA &  pmDEC & $\pi$  & BP & RP & G & {d$_{cent}$} & PROB  & H-K & J-H & disk? \\   
\hline
$^{\circ}$ & $^{\circ}$ & mas yr$^{-1}$ & mas yr$^{-1}$ & mas & mag & mag & mag & $^{\circ}$ & ... & mag & mag & ... \\
\hline
& & & & & & NGC 1333 & & & & & & \\
\hline
55.6003 & 32.1583 & 6.96 $\pm$ 0.08 & -9.56 $\pm$ 0.06 & 3.32 $\pm$ 0.09 & 18.651 $\pm$ 0.023 & 15.577 $\pm$ 0.006 & 16.846 $\pm$ 0.003 & 2.9348 & 1.000 & 0.26 $\pm$ 0.04 & 0.71 $\pm$ 0.04 & \\
52.3399 & 31.3528 & 7.64 $\pm$ 0.06 & -9.59 $\pm$ 0.05 & 3.41 $\pm$ 0.06 & 18.149 $\pm$ 0.015 & 14.594 $\pm$ 0.005 & 15.957 $\pm$ 0.003 & 0.0567 & 0.993 & 0.35 $\pm$ 0.04 & 0.67 $\pm$ 0.04 & \\
52.1815 & 31.2934 & 7.27 $\pm$ 0.13 & -9.74 $\pm$ 0.10 & 3.35 $\pm$ 0.12 & 19.312 $\pm$ 0.088 & 15.196 $\pm$ 0.034 & 16.805 $\pm$ 0.011 & 0.0998 & 0.980 & 0.82 $\pm$ 0.03 & 1.26 $\pm$ 0.04 & \\
\hline
& & & & & & IC 348 & & & & & & \\
\hline
56.1558 & 32.1032 & 4.36 $\pm$ 0.05 & -6.38 $\pm$ 0.03 & 3.10 $\pm$ 0.05 & 16.683 $\pm$ 0.013 & 14.011 $\pm$ 0.005 & 15.205 $\pm$ 0.003 & 0.0962 & 1.000 & 0.28 $\pm$ 0.03 & 0.94 $\pm$ 0.03 & \\
55.3089 & 31.9961 & 4.58 $\pm$ 0.02 & -6.08 $\pm$ 0.02 & 3.28 $\pm$ 0.02 & 12.590 $\pm$ 0.003 & 11.151 $\pm$ 0.004 & 11.945 $\pm$ 0.003 & 0.6569 & 1.000 & 0.18 $\pm$ 0.03 & 0.34 $\pm$ 0.03 & \\
56.0903 & 32.1069 & 4.70 $\pm$ 0.07 & -6.15 $\pm$ 0.05 & 3.12 $\pm$ 0.08 & 17.776 $\pm$ 0.024 & 14.688 $\pm$ 0.009 & 15.984 $\pm$ 0.003 & 0.0551 & 1.000 & 0.29 $\pm$ 0.03 & 0.93 $\pm$ 0.03 & \\
\hline
& & & & & & NGC 1893 & & & & & & \\
\hline
80.6932 & 33.4024 & -0.60 $\pm$ 0.03 & -1.54 $\pm$ 0.02 & 0.30 $\pm$ 0.03 & 14.938 $\pm$ 0.003 & 13.667 $\pm$ 0.004 & 14.395 $\pm$ 0.003 & 0.0486 & 1.000 & 0.30 $\pm$ 0.04 & 0.50 $\pm$ 0.04 & \\
80.6866 & 33.4435 & -0.50 $\pm$ 0.03 & -1.37 $\pm$ 0.02 & 0.34 $\pm$ 0.02 & 11.412 $\pm$ 0.003 & 11.000 $\pm$ 0.004 & 11.274 $\pm$ 0.003 & 0.0310 & 0.998 & 0.03 $\pm$ 0.03 & 0.05 $\pm$ 0.03 & \\
80.6833 & 33.4407 & -0.50 $\pm$ 0.03 & -1.33 $\pm$ 0.02 & 0.37 $\pm$ 0.03 & 10.332 $\pm$ 0.003 & 9.940 $\pm$ 0.004 & 10.204 $\pm$ 0.003 & 0.0338 & 0.996 & 0.05 $\pm$ 0.03 & 0.02 $\pm$ 0.03 & \\
\hline
& & & & & & ... & & & & & & \\
\hline

\end{tabular}
\begin{minipage}{21.5cm}
\textbf{Notes.} Columns 1 to 10 list the Gaia EDR3 coordinates, proper motions, parallax, BP, RP and G magnitudes, angular distance to the center of the cluster, and membership probability (ordered from higher to lower) of each member identified (PROB $>$ 0) within FOV$_{20pc}$. The members within FOV$_{2pc}$ can be retrieved from d$_{cent}$ and the corresponding angular radius (see Col. 9 in Table \ref{table:general_stats}). Columns 11 and 12 list the 2MASS colors, when available. Objects identified as disk stars or potential disk stars based on the JHK colors and errorbars are tagged with "yes" and "yes?" in the last column. The absence of such tags does not necessarily mean that the star is diskless (Sects. \ref{Sect:disk fractions} and \ref{Sect:limitations}). Only the three first rows for the three first clusters are shown. A complete version of this table is available online.
\end{minipage}
\end{table}
\end{landscape}

\section{Virtual Observatory compliant, online catalogue}
\label{app_VO_archive}
In order to help the astronomical community on using the information obtained in this paper, we developed an archive system  that  can  be  accessed  from  a webpage\footnote{http://svocats.cab.inta-csic.es/diskfrac} or  through  a Virtual Observatory ConeSearch\footnote{e.g.  http://svocats.cab.inta-csic.es/diskfrac/cs.php?RA=152.799\&DEC=-54.423\&SR=0.1\&VERB=2}.

The  archive  system  implements  a  very  simple  search interface that allows queries by coordinates and radius as well as by cluster name and other parameters of interest. The user can also select the maximum number of sources (with values from 10 to unlimited). The result of the query is a HTML table with all the sources found in the archive fulfilling the search criteria. The result can also be downloaded as a VOTable or a CSV file. Detailed information on the output fields can be obtained placing the mouse over the question mark located close to the name of the column. The archive also implements the SAMP\footnote{http://www.ivoa.net/documents/SAMP} (Simple  Application  Messaging)  Virtual  Observatory  protocol.  SAMP  allows  Virtual Observatory  applications  to  communicate  with  each  other  in  a  seamless  and transparent manner for the user. This way, the results of a query  can  be  easily  transferred  to  other  VO  applications, such as, for instance, Topcat \citep{Taylor05}.

\section{Comparison with previous Gaia-based results}
\label{app_comparison}

\begin{figure}
   \centering
   \includegraphics[width=9cm]{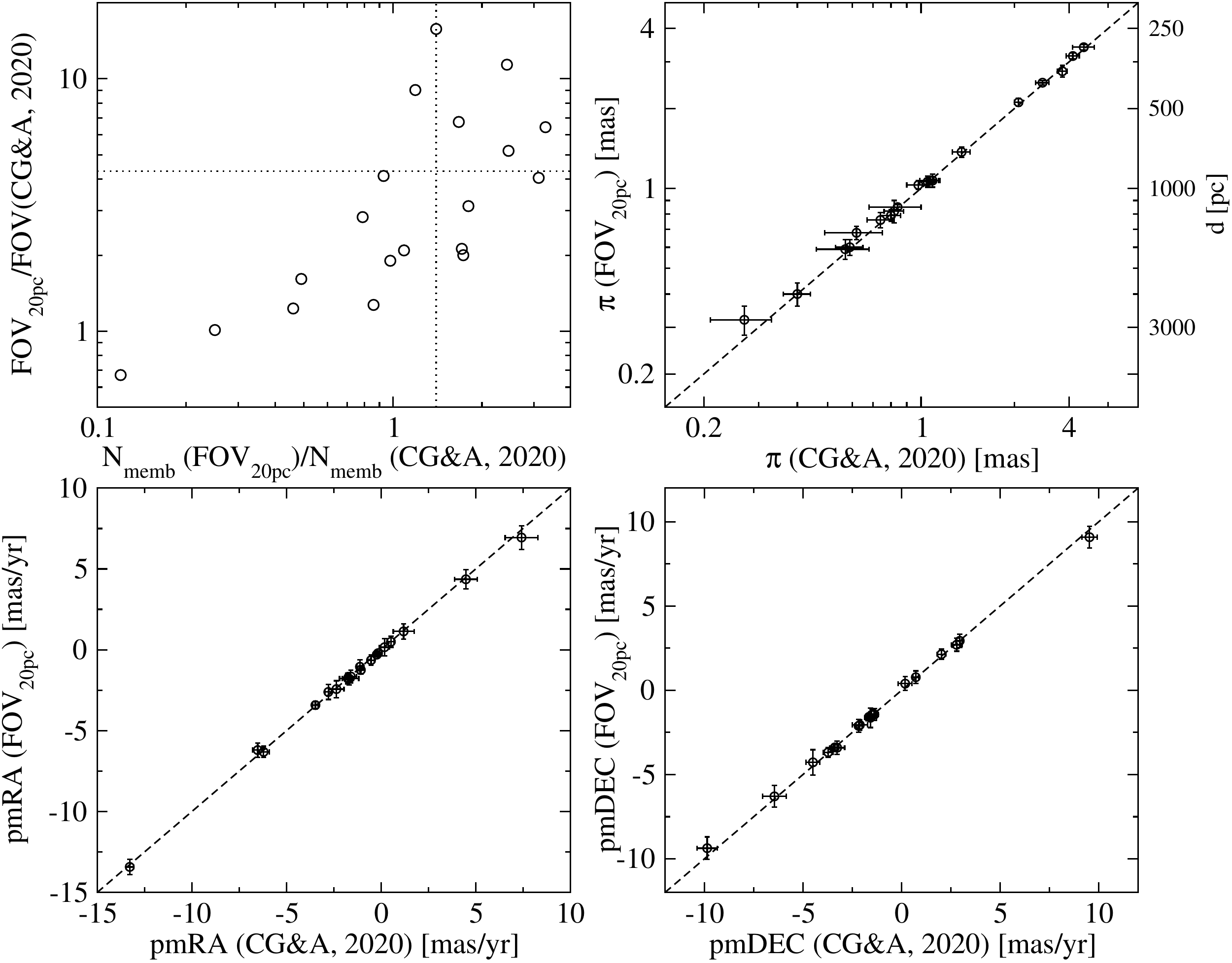}
      \caption{Comparison between data from this work and from \citet{Cantat20} (Top left): ratio between the FOVs used versus ratio between the total number of members identified. The dotted lines indicate the typical (mean) values in each axis. (Top right, bottom left, and bottom right): Comparison between parallaxes and proper motions. The dashed lines indicate equal values. Distances are indicated in the opposite y-axis of the top right panel for reference.} 
         \label{fig:comp_lit}
   \end{figure}

Figure \ref{fig:comp_lit} compares our results with those in \citet{Cantat20}, which provides a complete Gaia DR2-based census of Galactic clusters including all analysed here. \citet{Cantat20} reported the use of FOVs significantly larger than in the previous reference papers by \citet{Dias02} and \citet{Kharchenko13}, although the specific sizes were not systematically listed. Based on the positions of the members located furthest from the clusters' centers, a reasonable estimate of the FOVs used in \citet{Cantat20} can be derived. Such FOVs, as well as the total number of potential members (i.e. with non-null membership probability) from \citet{Cantat20}, are compared to those from this work in the top left panel of Fig. \ref{fig:comp_lit}. Our FOV$_{20pc}$ and number of potential members are $\sim$ 4.3 and $\sim$ 1.4 times larger on average. It is noted that this is a rough comparison that does not take into account the different methodologies for membership identification or the limiting magnitudes considered. In this particular, although \citet{Cantat20} probed more sources by including stars 1 Gaia G-magnitude fainter than here, we still identify a larger number of members for the majority of the clusters because of the use of a larger FOV. The rest of the panels in Fig. \ref{fig:comp_lit} compare the parallaxes and proper motions. Values and errorbars refer to the means and standard deviations derived from FOV$_{20pc}$ (see Table \ref{table:general_stats}) and to the ones tabulated in \citet{Cantat20}. Despite of the above mentioned differences concerning methodologies, FOVs, and the number of members identified, parallaxes and proper motions coincide within $\pm$1$\sigma$ for all clusters. Mean distances to each cluster were derived from the ones of all members within FOV$_{20pc}$, based on the parallax-distance geometrical transformation in \citet{BailerJones21}. Distances are indicated in the top right panel of Fig. \ref{fig:comp_lit}, and were listed in Table \ref{table:regions_properties} along with the corresponding standard deviations.

\end{appendix}

\end{document}